\documentclass[longauth]{aa}

\usepackage{graphicx}
\usepackage{natbib}
\usepackage{scalerel}

\newcommand{\dotdeg}{\rlap{.}^\circ}
\usepackage{multirow}
\usepackage{multicol}
\usepackage{array}
\usepackage{makecell}

\usepackage[table]{xcolor}

\usepackage{txfonts}
\usepackage[pdfencoding=auto,psdextra]{hyperref}
\hypersetup{
    colorlinks=true,
    linkcolor=blue,
    filecolor=magenta,      
    urlcolor=blue,
    citecolor=blue
}
\makeatletter
\renewcommand*\aa@pageof{, page \thepage{} of \pageref*{LastPage}}
\makeatother

\usepackage[utf8]{inputenc}

\usepackage[switch, modulo]{lineno}

\usepackage{euclid}

\newcommand{\corr}[1]{\textcolor{black}{#1}}

\newcommand{\corrtwo}[1]{\textcolor{black}{#1}}

\newcommand{\corrthree}[1]{\textcolor{black}{#1}}

\begin{document}

%
%

\title{\textit{Euclid} preparation}
\subtitle{XLVIII. The pre-launch Science Ground Segment simulation framework}

\newcommand{\orcid}[1]{} 
\author{Euclid Collaboration: S.~Serrano\orcid{0000-0002-0211-2861}\thanks{\email{serrano@ieec.cat}}\inst{\ref{aff1},\ref{aff2},\ref{aff3}}
\and P.~Hudelot\inst{\ref{aff4}}
\and G.~Seidel\orcid{0000-0003-2907-353X}\inst{\ref{aff5}}
\and J.~E.~Pollack\inst{\ref{aff6},\ref{aff7}}
\and E.~Jullo\orcid{0000-0002-9253-053X}\inst{\ref{aff8}}
\and F.~Torradeflot\orcid{0000-0003-1160-1517}\inst{\ref{aff9},\ref{aff10}}
\and D.~Benielli\inst{\ref{aff11}}
\and R.~Fahed\orcid{0009-0006-3518-2854}\inst{\ref{aff12}}
\and T.~Auphan\orcid{0009-0008-9988-3646}\inst{\ref{aff11}}
\and J.~Carretero\orcid{0000-0002-3130-0204}\inst{\ref{aff13},\ref{aff9}}
\and H.~Aussel\orcid{0000-0002-1371-5705}\inst{\ref{aff14}}
\and P.~Casenove\inst{\ref{aff15}}
\and F.~J.~Castander\orcid{0000-0001-7316-4573}\inst{\ref{aff2},\ref{aff1}}
\and J.~E.~Davies\orcid{0000-0002-5079-9098}\inst{\ref{aff5}}
\and N.~Fourmanoit\inst{\ref{aff11}}
\and S.~Huot\inst{\ref{aff12}}
\and A.~Kara\inst{\ref{aff11}}
\and E.~Keih\"anen\orcid{0000-0003-1804-7715}\inst{\ref{aff16}}
\and S.~Kermiche\orcid{0000-0002-0302-5735}\inst{\ref{aff11}}
\and K.~Okumura\inst{\ref{aff6}}
\and J.~Zoubian\inst{\ref{aff11}}
\and A.~Ealet\inst{\ref{aff17}}
\and A.~Boucaud\orcid{0000-0001-7387-2633}\inst{\ref{aff7}}
\and H.~Bretonni\`ere\inst{\ref{aff18}}
\and R.~Casas\orcid{0000-0002-8165-5601}\inst{\ref{aff1},\ref{aff2}}
\and B.~Cl\'ement\orcid{0000-0002-7966-3661}\inst{\ref{aff19}}
\and C.~A.~J.~Duncan\inst{\ref{aff20},\ref{aff21}}
\and K.~George\orcid{0000-0002-1734-8455}\inst{\ref{aff22}}
\and K.~Kiiveri\inst{\ref{aff16}}
\and H.~Kurki-Suonio\orcid{0000-0002-4618-3063}\inst{\ref{aff23},\ref{aff24}}
\and M.~K\"ummel\orcid{0000-0003-2791-2117}\inst{\ref{aff22}}
\and D.~Laugier\orcid{0000-0002-2517-0204}\inst{\ref{aff11}}
\and G.~Mainetti\inst{\ref{aff25}}
\and J.~J.~Mohr\orcid{0000-0002-6875-2087}\inst{\ref{aff22},\ref{aff26}}
\and A.~Montoro\orcid{0000-0003-4730-8590}\inst{\ref{aff2},\ref{aff1}}
\and C.~Neissner\inst{\ref{aff13},\ref{aff9}}
\and C.~Rosset\orcid{0000-0003-0286-2192}\inst{\ref{aff7}}
\and M.~Schirmer\orcid{0000-0003-2568-9994}\inst{\ref{aff5}}
\and P.~Tallada-Cresp\'{i}\orcid{0000-0002-1336-8328}\inst{\ref{aff10},\ref{aff9}}
\and N.~Tonello\orcid{0000-0003-0550-1667}\inst{\ref{aff13},\ref{aff9}}
\and A.~Venhola\orcid{0000-0001-6071-4564}\inst{\ref{aff27}}
\and A.~Verderi\orcid{0000-0002-2704-3620}\inst{\ref{aff2},\ref{aff1}}
\and A.~Zacchei\orcid{0000-0003-0396-1192}\inst{\ref{aff28},\ref{aff29}}
\and N.~Aghanim\inst{\ref{aff30}}
\and B.~Altieri\orcid{0000-0003-3936-0284}\inst{\ref{aff31}}
\and A.~Amara\inst{\ref{aff32}}
\and S.~Andreon\orcid{0000-0002-2041-8784}\inst{\ref{aff33}}
\and N.~Auricchio\orcid{0000-0003-4444-8651}\inst{\ref{aff34}}
\and R.~Azzollini\orcid{0000-0002-0438-0886}\inst{\ref{aff35}}
\and C.~Baccigalupi\orcid{0000-0002-8211-1630}\inst{\ref{aff36},\ref{aff29},\ref{aff28},\ref{aff37}}
\and M.~Baldi\orcid{0000-0003-4145-1943}\inst{\ref{aff38},\ref{aff34},\ref{aff39}}
\and S.~Bardelli\orcid{0000-0002-8900-0298}\inst{\ref{aff34}}
\and A.~Basset\inst{\ref{aff15}}
\and P.~Battaglia\orcid{0000-0002-7337-5909}\inst{\ref{aff34}}
\and F.~Bernardeau\inst{\ref{aff40},\ref{aff4}}
\and C.~Bodendorf\inst{\ref{aff26}}
\and D.~Bonino\inst{\ref{aff41}}
\and E.~Branchini\orcid{0000-0002-0808-6908}\inst{\ref{aff42},\ref{aff43}}
\and M.~Brescia\orcid{0000-0001-9506-5680}\inst{\ref{aff44},\ref{aff45}}
\and J.~Brinchmann\orcid{0000-0003-4359-8797}\inst{\ref{aff46}}
\and S.~Camera\orcid{0000-0003-3399-3574}\inst{\ref{aff47},\ref{aff48},\ref{aff41}}
\and G.~P.~Candini\orcid{0000-0001-9481-8206}\inst{\ref{aff35}}
\and V.~Capobianco\orcid{0000-0002-3309-7692}\inst{\ref{aff41}}
\and C.~Carbone\orcid{0000-0003-0125-3563}\inst{\ref{aff49}}
\and S.~Casas\orcid{0000-0002-4751-5138}\inst{\ref{aff50}}
\and M.~Castellano\orcid{0000-0001-9875-8263}\inst{\ref{aff51}}
\and G.~Castignani\orcid{0000-0001-6831-0687}\inst{\ref{aff52},\ref{aff34}}
\and S.~Cavuoti\orcid{0000-0002-3787-4196}\inst{\ref{aff45},\ref{aff53}}
\and A.~Cimatti\inst{\ref{aff54}}
\and R.~Cledassou\orcid{0000-0002-8313-2230}\thanks{Deceased}\inst{\ref{aff15},\ref{aff55}}
\and C.~Colodro-Conde\inst{\ref{aff56}}
\and G.~Congedo\orcid{0000-0003-2508-0046}\inst{\ref{aff57}}
\and C.~J.~Conselice\inst{\ref{aff21}}
\and L.~Conversi\orcid{0000-0002-6710-8476}\inst{\ref{aff58},\ref{aff31}}
\and Y.~Copin\orcid{0000-0002-5317-7518}\inst{\ref{aff17}}
\and L.~Corcione\orcid{0000-0002-6497-5881}\inst{\ref{aff41}}
\and F.~Courbin\orcid{0000-0003-0758-6510}\inst{\ref{aff19}}
\and H.~M.~Courtois\orcid{0000-0003-0509-1776}\inst{\ref{aff59}}
\and M.~Crocce\inst{\ref{aff60},\ref{aff2}}
\and M.~Cropper\orcid{0000-0003-4571-9468}\inst{\ref{aff35}}
\and A.~Da~Silva\orcid{0000-0002-6385-1609}\inst{\ref{aff61},\ref{aff62}}
\and H.~Degaudenzi\orcid{0000-0002-5887-6799}\inst{\ref{aff63}}
\and G.~De~Lucia\orcid{0000-0002-6220-9104}\inst{\ref{aff28}}
\and A.~M.~Di~Giorgio\inst{\ref{aff64}}
\and J.~Dinis\inst{\ref{aff62},\ref{aff61}}
\and F.~Dubath\orcid{0000-0002-6533-2810}\inst{\ref{aff63}}
\and X.~Dupac\inst{\ref{aff31}}
\and S.~Dusini\orcid{0000-0002-1128-0664}\inst{\ref{aff65}}
\and M.~Farina\inst{\ref{aff64}}
\and S.~Farrens\orcid{0000-0002-9594-9387}\inst{\ref{aff14}}
\and S.~Ferriol\inst{\ref{aff17}}
\and M.~Frailis\orcid{0000-0002-7400-2135}\inst{\ref{aff28}}
\and E.~Franceschi\orcid{0000-0002-0585-6591}\inst{\ref{aff34}}
\and P.~Franzetti\inst{\ref{aff49}}
\and S.~Galeotta\orcid{0000-0002-3748-5115}\inst{\ref{aff28}}
\and B.~Garilli\orcid{0000-0001-7455-8750}\inst{\ref{aff49}}
\and W.~Gillard\orcid{0000-0003-4744-9748}\inst{\ref{aff11}}
\and B.~Gillis\orcid{0000-0002-4478-1270}\inst{\ref{aff57}}
\and C.~Giocoli\orcid{0000-0002-9590-7961}\inst{\ref{aff34},\ref{aff39}}
\and B.~R.~Granett\orcid{0000-0003-2694-9284}\inst{\ref{aff33}}
\and A.~Grazian\orcid{0000-0002-5688-0663}\inst{\ref{aff66}}
\and F.~Grupp\inst{\ref{aff26},\ref{aff22}}
\and L.~Guzzo\inst{\ref{aff67},\ref{aff33},\ref{aff68}}
\and S.~V.~H.~Haugan\orcid{0000-0001-9648-7260}\inst{\ref{aff69}}
\and J.~Hoar\inst{\ref{aff31}}
\and H.~Hoekstra\orcid{0000-0002-0641-3231}\inst{\ref{aff70}}
\and W.~Holmes\inst{\ref{aff71}}
\and I.~Hook\orcid{0000-0002-2960-978X}\inst{\ref{aff72}}
\and F.~Hormuth\inst{\ref{aff73}}
\and A.~Hornstrup\orcid{0000-0002-3363-0936}\inst{\ref{aff74},\ref{aff75}}
\and K.~Jahnke\orcid{0000-0003-3804-2137}\inst{\ref{aff5}}
\and B.~Joachimi\orcid{0000-0001-7494-1303}\inst{\ref{aff76}}
\and A.~Kiessling\orcid{0000-0002-2590-1273}\inst{\ref{aff71}}
\and T.~Kitching\orcid{0000-0002-4061-4598}\inst{\ref{aff35}}
\and R.~Kohley\inst{\ref{aff31}}
\and M.~Kunz\orcid{0000-0002-3052-7394}\inst{\ref{aff77}}
\and Q.~Le~Boulc'h\inst{\ref{aff25}}
\and P.~Liebing\inst{\ref{aff35}}
\and S.~Ligori\orcid{0000-0003-4172-4606}\inst{\ref{aff41}}
\and P.~B.~Lilje\orcid{0000-0003-4324-7794}\inst{\ref{aff69}}
\and V.~Lindholm\orcid{0000-0003-2317-5471}\inst{\ref{aff23},\ref{aff24}}
\and I.~Lloro\inst{\ref{aff78}}
\and D.~Maino\inst{\ref{aff67},\ref{aff49},\ref{aff68}}
\and E.~Maiorano\orcid{0000-0003-2593-4355}\inst{\ref{aff34}}
\and O.~Mansutti\orcid{0000-0001-5758-4658}\inst{\ref{aff28}}
\and S.~Marcin\inst{\ref{aff79}}
\and O.~Marggraf\orcid{0000-0001-7242-3852}\inst{\ref{aff80}}
\and K.~Markovic\orcid{0000-0001-6764-073X}\inst{\ref{aff71}}
\and M.~Martinelli\orcid{0000-0002-6943-7732}\inst{\ref{aff51},\ref{aff81}}
\and N.~Martinet\orcid{0000-0003-2786-7790}\inst{\ref{aff8}}
\and F.~Marulli\inst{\ref{aff52},\ref{aff34},\ref{aff39}}
\and R.~Massey\orcid{0000-0002-6085-3780}\inst{\ref{aff82}}
\and S.~Maurogordato\inst{\ref{aff83}}
\and E.~Medinaceli\orcid{0000-0002-4040-7783}\inst{\ref{aff34}}
\and S.~Mei\inst{\ref{aff7}}
\and M.~Melchior\inst{\ref{aff79}}
\and Y.~Mellier\inst{\ref{aff12},\ref{aff4},\ref{aff6}}
\and M.~Meneghetti\orcid{0000-0003-1225-7084}\inst{\ref{aff34},\ref{aff39}}
\and E.~Merlin\orcid{0000-0001-6870-8900}\inst{\ref{aff51}}
\and G.~Meylan\inst{\ref{aff19}}
\and M.~Moresco\orcid{0000-0002-7616-7136}\inst{\ref{aff52},\ref{aff34}}
\and P.~Morris\orcid{0000-0002-5186-4381}\inst{\ref{aff84}}
\and L.~Moscardini\orcid{0000-0002-3473-6716}\inst{\ref{aff52},\ref{aff34},\ref{aff39}}
\and E.~Munari\orcid{0000-0002-1751-5946}\inst{\ref{aff28}}
\and R.~Nakajima\inst{\ref{aff80}}
\and S.-M.~Niemi\inst{\ref{aff85}}
\and T.~Nutma\inst{\ref{aff86},\ref{aff70}}
\and C.~Padilla\orcid{0000-0001-7951-0166}\inst{\ref{aff13}}
\and S.~Paltani\inst{\ref{aff63}}
\and F.~Pasian\inst{\ref{aff28}}
\and K.~Pedersen\inst{\ref{aff87}}
\and W.~J.~Percival\orcid{0000-0002-0644-5727}\inst{\ref{aff88},\ref{aff89},\ref{aff90}}
\and V.~Pettorino\inst{\ref{aff14}}
\and S.~Pires\orcid{0000-0002-0249-2104}\inst{\ref{aff14}}
\and G.~Polenta\orcid{0000-0003-4067-9196}\inst{\ref{aff91}}
\and M.~Poncet\inst{\ref{aff15}}
\and L.~A.~Popa\inst{\ref{aff92}}
\and L.~Pozzetti\orcid{0000-0001-7085-0412}\inst{\ref{aff34}}
\and F.~Raison\orcid{0000-0002-7819-6918}\inst{\ref{aff26}}
\and R.~Rebolo\inst{\ref{aff56},\ref{aff93}}
\and A.~Renzi\orcid{0000-0001-9856-1970}\inst{\ref{aff94},\ref{aff65}}
\and J.~Rhodes\inst{\ref{aff71}}
\and G.~Riccio\inst{\ref{aff45}}
\and E.~Romelli\orcid{0000-0003-3069-9222}\inst{\ref{aff28}}
\and M.~Roncarelli\orcid{0000-0001-9587-7822}\inst{\ref{aff34}}
\and E.~Rossetti\inst{\ref{aff38}}
\and B.~Rusholme\orcid{0000-0001-7648-4142}\inst{\ref{aff95}}
\and R.~Saglia\orcid{0000-0003-0378-7032}\inst{\ref{aff22},\ref{aff26}}
\and Z.~Sakr\orcid{0000-0002-4823-3757}\inst{\ref{aff96},\ref{aff97},\ref{aff98}}
\and A.~G.~S\'anchez\orcid{0000-0003-1198-831X}\inst{\ref{aff26}}
\and D.~Sapone\orcid{0000-0001-7089-4503}\inst{\ref{aff99}}
\and B.~Sartoris\inst{\ref{aff22},\ref{aff28}}
\and M.~Sauvage\orcid{0000-0002-0809-2574}\inst{\ref{aff14}}
\and P.~Schneider\orcid{0000-0001-8561-2679}\inst{\ref{aff80}}
\and T.~Schrabback\orcid{0000-0002-6987-7834}\inst{\ref{aff100}}
\and M.~Scodeggio\inst{\ref{aff49}}
\and A.~Secroun\orcid{0000-0003-0505-3710}\inst{\ref{aff11}}
\and C.~Sirignano\orcid{0000-0002-0995-7146}\inst{\ref{aff94},\ref{aff65}}
\and G.~Sirri\orcid{0000-0003-2626-2853}\inst{\ref{aff39}}
\and J.~Skottfelt\orcid{0000-0003-1310-8283}\inst{\ref{aff101}}
\and L.~Stanco\orcid{0000-0002-9706-5104}\inst{\ref{aff65}}
\and J.-L.~Starck\orcid{0000-0003-2177-7794}\inst{\ref{aff14}}
\and J.~Steinwagner\inst{\ref{aff26}}
\and A.~N~Taylor\inst{\ref{aff57}}
\and H.~Teplitz\orcid{0000-0002-7064-5424}\inst{\ref{aff102}}
\and I.~Tereno\inst{\ref{aff61},\ref{aff103}}
\and R.~Toledo-Moreo\orcid{0000-0002-2997-4859}\inst{\ref{aff104}}
\and I.~Tutusaus\orcid{0000-0002-3199-0399}\inst{\ref{aff98}}
\and E.~A.~Valentijn\inst{\ref{aff86}}
\and L.~Valenziano\orcid{0000-0002-1170-0104}\inst{\ref{aff34},\ref{aff105}}
\and T.~Vassallo\orcid{0000-0001-6512-6358}\inst{\ref{aff22},\ref{aff28}}
\and A.~Veropalumbo\orcid{0000-0003-2387-1194}\inst{\ref{aff33}}
\and Y.~Wang\orcid{0000-0002-4749-2984}\inst{\ref{aff102}}
\and J.~Weller\orcid{0000-0002-8282-2010}\inst{\ref{aff22},\ref{aff26}}
\and G.~Zamorani\orcid{0000-0002-2318-301X}\inst{\ref{aff34}}
\and E.~Zucca\orcid{0000-0002-5845-8132}\inst{\ref{aff34}}
\and A.~Biviano\orcid{0000-0002-0857-0732}\inst{\ref{aff28},\ref{aff29}}
\and E.~Bozzo\orcid{0000-0002-8201-1525}\inst{\ref{aff63}}
\and D.~Di~Ferdinando\inst{\ref{aff39}}
\and R.~Farinelli\inst{\ref{aff34}}
\and J.~Graci\'{a}-Carpio\inst{\ref{aff26}}
\and N.~Mauri\orcid{0000-0001-8196-1548}\inst{\ref{aff54},\ref{aff39}}
\and V.~Scottez\inst{\ref{aff12},\ref{aff106}}
\and M.~Tenti\orcid{0000-0002-4254-5901}\inst{\ref{aff105}}
\and Y.~Akrami\orcid{0000-0002-2407-7956}\inst{\ref{aff107},\ref{aff108},\ref{aff109},\ref{aff110},\ref{aff111}}
\and V.~Allevato\orcid{0000-0001-7232-5152}\inst{\ref{aff45},\ref{aff112}}
\and M.~Ballardini\orcid{0000-0003-4481-3559}\inst{\ref{aff113},\ref{aff114},\ref{aff34}}
\and A.~Blanchard\orcid{0000-0001-8555-9003}\inst{\ref{aff98}}
\and S.~Borgani\orcid{0000-0001-6151-6439}\inst{\ref{aff28},\ref{aff115},\ref{aff37},\ref{aff29}}
\and A.~S.~Borlaff\inst{\ref{aff116},\ref{aff117}}
\and S.~Bruton\orcid{0000-0002-6503-5218}\inst{\ref{aff118}}
\and C.~Burigana\orcid{0000-0002-3005-5796}\inst{\ref{aff119},\ref{aff105}}
\and A.~Cappi\inst{\ref{aff34},\ref{aff83}}
\and C.~S.~Carvalho\inst{\ref{aff103}}
\and T.~Castro\orcid{0000-0002-6292-3228}\inst{\ref{aff28},\ref{aff37},\ref{aff29}}
\and G.~Ca\~{n}as-Herrera\orcid{0000-0003-2796-2149}\inst{\ref{aff85},\ref{aff120}}
\and K.~C.~Chambers\orcid{0000-0001-6965-7789}\inst{\ref{aff121}}
\and A.~R.~Cooray\orcid{0000-0002-3892-0190}\inst{\ref{aff122}}
\and J.~Coupon\inst{\ref{aff63}}
\and S.~Davini\inst{\ref{aff43}}
\and S.~de~la~Torre\inst{\ref{aff8}}
\and S.~Desai\orcid{0000-0002-0466-3288}\inst{\ref{aff123}}
\and G.~Desprez\inst{\ref{aff124}}
\and A.~D\'iaz-S\'anchez\orcid{0000-0003-0748-4768}\inst{\ref{aff125}}
\and S.~Di~Domizio\orcid{0000-0003-2863-5895}\inst{\ref{aff126}}
\and H.~Dole\orcid{0000-0002-9767-3839}\inst{\ref{aff30}}
\and J.~A.~Escartin~Vigo\inst{\ref{aff26}}
\and S.~Escoffier\orcid{0000-0002-2847-7498}\inst{\ref{aff11}}
\and I.~Ferrero\orcid{0000-0002-1295-1132}\inst{\ref{aff69}}
\and F.~Finelli\inst{\ref{aff34},\ref{aff105}}
\and L.~Gabarra\inst{\ref{aff94},\ref{aff65}}
\and K.~Ganga\orcid{0000-0001-8159-8208}\inst{\ref{aff7}}
\and J.~Garcia-Bellido\orcid{0000-0002-9370-8360}\inst{\ref{aff107}}
\and E.~Gaztanaga\orcid{0000-0001-9632-0815}\inst{\ref{aff2},\ref{aff1},\ref{aff32}}
\and F.~Giacomini\orcid{0000-0002-3129-2814}\inst{\ref{aff39}}
\and G.~Gozaliasl\orcid{0000-0002-0236-919X}\inst{\ref{aff23}}
\and A.~Gregorio\orcid{0000-0003-4028-8785}\inst{\ref{aff115},\ref{aff28},\ref{aff37}}
\and H.~Hildebrandt\orcid{0000-0002-9814-3338}\inst{\ref{aff127}}
\and M.~Huertas-Company\inst{\ref{aff128},\ref{aff56},\ref{aff129},\ref{aff130}}
\and O.~Ilbert\inst{\ref{aff8}}
\and A.~Jimenez~Mu{\~ n}oz\inst{\ref{aff131}}
\and J.~J.~E.~Kajava\orcid{0000-0002-3010-8333}\inst{\ref{aff132},\ref{aff133}}
\and V.~Kansal\inst{\ref{aff14}}
\and C.~C.~Kirkpatrick\inst{\ref{aff16}}
\and L.~Legrand\inst{\ref{aff77}}
\and A.~Loureiro\orcid{0000-0002-4371-0876}\inst{\ref{aff134},\ref{aff111}}
\and J.~Macias-Perez\inst{\ref{aff131}}
\and M.~Magliocchetti\orcid{0000-0001-9158-4838}\inst{\ref{aff64}}
\and R.~Maoli\orcid{0000-0002-6065-3025}\inst{\ref{aff135},\ref{aff51}}
\and C.~J.~A.~P.~Martins\orcid{0000-0002-4886-9261}\inst{\ref{aff136},\ref{aff46}}
\and S.~Matthew\inst{\ref{aff57}}
\and L.~Maurin\orcid{0000-0002-8406-0857}\inst{\ref{aff30}}
\and R.~B.~Metcalf\orcid{0000-0003-3167-2574}\inst{\ref{aff52}}
\and M.~Migliaccio\inst{\ref{aff137},\ref{aff138}}
\and P.~Monaco\inst{\ref{aff115},\ref{aff28},\ref{aff37},\ref{aff29}}
\and G.~Morgante\inst{\ref{aff34}}
\and S.~Nadathur\orcid{0000-0001-9070-3102}\inst{\ref{aff32}}
\and A.~A.~Nucita\inst{\ref{aff139},\ref{aff140},\ref{aff141}}
\and M.~P{\"o}ntinen\orcid{0000-0001-5442-2530}\inst{\ref{aff23}}
\and V.~Popa\inst{\ref{aff92}}
\and C.~Porciani\inst{\ref{aff80}}
\and D.~Potter\orcid{0000-0002-0757-5195}\inst{\ref{aff142}}
\and P.~Reimberg\orcid{0000-0003-3410-0280}\inst{\ref{aff12}}
\and A.~Schneider\orcid{0000-0001-7055-8104}\inst{\ref{aff142}}
\and M.~Sereno\orcid{0000-0003-0302-0325}\inst{\ref{aff34},\ref{aff39}}
\and A.~Shulevski\orcid{0000-0002-1827-0469}\inst{\ref{aff70},\ref{aff143},\ref{aff86}}
\and P.~Simon\inst{\ref{aff80}}
\and A.~Spurio~Mancini\orcid{0000-0001-5698-0990}\inst{\ref{aff35}}
\and J.~Stadel\orcid{0000-0001-7565-8622}\inst{\ref{aff142}}
\and M.~Tewes\orcid{0000-0002-1155-8689}\inst{\ref{aff80}}
\and R.~Teyssier\orcid{0000-0001-7689-0933}\inst{\ref{aff144}}
\and S.~Toft\orcid{0000-0003-3631-7176}\inst{\ref{aff75},\ref{aff145}}
\and M.~Tucci\inst{\ref{aff63}}
\and J.~Valiviita\orcid{0000-0001-6225-3693}\inst{\ref{aff23},\ref{aff24}}
\and M.~Viel\orcid{0000-0002-2642-5707}\inst{\ref{aff29},\ref{aff28},\ref{aff36},\ref{aff37}}
\and I.~A.~Zinchenko\inst{\ref{aff22}}}
                                                                                   
\institute{Institut d'Estudis Espacials de Catalunya (IEEC),  Edifici RDIT, Campus UPC, 08860 Castelldefels, Barcelona, Spain\label{aff1}
\and
Institute of Space Sciences (ICE, CSIC), Campus UAB, Carrer de Can Magrans, s/n, 08193 Barcelona, Spain\label{aff2}
\and
Satlantis, University Science Park, Sede Bld 48940, Leioa-Bilbao, Spain\label{aff3}
\and
Institut d'Astrophysique de Paris, UMR 7095, CNRS, and Sorbonne Universit\'e, 98 bis boulevard Arago, 75014 Paris, France\label{aff4}
\and
Max-Planck-Institut f\"ur Astronomie, K\"onigstuhl 17, 69117 Heidelberg, Germany\label{aff5}
\and
CEA Saclay, DFR/IRFU, Service d'Astrophysique, Bat. 709, 91191 Gif-sur-Yvette, France\label{aff6}
\and
Universit\'e Paris Cit\'e, CNRS, Astroparticule et Cosmologie, 75013 Paris, France\label{aff7}
\and
Aix-Marseille Universit\'e, CNRS, CNES, LAM, Marseille, France\label{aff8}
\and
Port d'Informaci\'{o} Cient\'{i}fica, Campus UAB, C. Albareda s/n, 08193 Bellaterra (Barcelona), Spain\label{aff9}
\and
Centro de Investigaciones Energ\'eticas, Medioambientales y Tecnol\'ogicas (CIEMAT), Avenida Complutense 40, 28040 Madrid, Spain\label{aff10}
\and
Aix-Marseille Universit\'e, CNRS/IN2P3, CPPM, Marseille, France\label{aff11}
\and
Institut d'Astrophysique de Paris, 98bis Boulevard Arago, 75014, Paris, France\label{aff12}
\and
Institut de F\'{i}sica d'Altes Energies (IFAE), The Barcelona Institute of Science and Technology, Campus UAB, 08193 Bellaterra (Barcelona), Spain\label{aff13}
\and
Universit\'e Paris-Saclay, Universit\'e Paris Cit\'e, CEA, CNRS, AIM, 91191, Gif-sur-Yvette, France\label{aff14}
\and
Centre National d'Etudes Spatiales -- Centre spatial de Toulouse, 18 avenue Edouard Belin, 31401 Toulouse Cedex 9, France\label{aff15}
\and
Department of Physics and Helsinki Institute of Physics, Gustaf H\"allstr\"omin katu 2, 00014 University of Helsinki, Finland\label{aff16}
\and
Universit\'e Claude Bernard Lyon 1, CNRS/IN2P3, IP2I Lyon, UMR 5822, Villeurbanne, F-69100, France\label{aff17}
\and
Department of Astronomy and Astrophysics, University of California, Santa Cruz, 1156 High Street, Santa Cruz, CA 95064, USA\label{aff18}
\and
Institute of Physics, Laboratory of Astrophysics, Ecole Polytechnique F\'ed\'erale de Lausanne (EPFL), Observatoire de Sauverny, 1290 Versoix, Switzerland\label{aff19}
\and
Department of Physics, Oxford University, Keble Road, Oxford OX1 3RH, UK\label{aff20}
\and
Jodrell Bank Centre for Astrophysics, Department of Physics and Astronomy, University of Manchester, Oxford Road, Manchester M13 9PL, UK\label{aff21}
\and
Universit\"ats-Sternwarte M\"unchen, Fakult\"at f\"ur Physik, Ludwig-Maximilians-Universit\"at M\"unchen, Scheinerstrasse 1, 81679 M\"unchen, Germany\label{aff22}
\and
Department of Physics, P.O. Box 64, 00014 University of Helsinki, Finland\label{aff23}
\and
Helsinki Institute of Physics, Gustaf H{\"a}llstr{\"o}min katu 2, University of Helsinki, Helsinki, Finland\label{aff24}
\and
Centre de Calcul de l'IN2P3/CNRS, 21 avenue Pierre de Coubertin 69627 Villeurbanne Cedex, France\label{aff25}
\and
Max Planck Institute for Extraterrestrial Physics, Giessenbachstr. 1, 85748 Garching, Germany\label{aff26}
\and
Space physics and astronomy research unit, University of Oulu, Pentti Kaiteran katu 1, FI-90014 Oulu, Finland\label{aff27}
\and
INAF-Osservatorio Astronomico di Trieste, Via G. B. Tiepolo 11, 34143 Trieste, Italy\label{aff28}
\and
IFPU, Institute for Fundamental Physics of the Universe, via Beirut 2, 34151 Trieste, Italy\label{aff29}
\and
Universit\'e Paris-Saclay, CNRS, Institut d'astrophysique spatiale, 91405, Orsay, France\label{aff30}
\and
ESAC/ESA, Camino Bajo del Castillo, s/n., Urb. Villafranca del Castillo, 28692 Villanueva de la Ca\~nada, Madrid, Spain\label{aff31}
\and
Institute of Cosmology and Gravitation, University of Portsmouth, Portsmouth PO1 3FX, UK\label{aff32}
\and
INAF-Osservatorio Astronomico di Brera, Via Brera 28, 20122 Milano, Italy\label{aff33}
\and
INAF-Osservatorio di Astrofisica e Scienza dello Spazio di Bologna, Via Piero Gobetti 93/3, 40129 Bologna, Italy\label{aff34}
\and
Mullard Space Science Laboratory, University College London, Holmbury St Mary, Dorking, Surrey RH5 6NT, UK\label{aff35}
\and
SISSA, International School for Advanced Studies, Via Bonomea 265, 34136 Trieste TS, Italy\label{aff36}
\and
INFN, Sezione di Trieste, Via Valerio 2, 34127 Trieste TS, Italy\label{aff37}
\and
Dipartimento di Fisica e Astronomia, Universit\`a di Bologna, Via Gobetti 93/2, 40129 Bologna, Italy\label{aff38}
\and
INFN-Sezione di Bologna, Viale Berti Pichat 6/2, 40127 Bologna, Italy\label{aff39}
\and
Institut de Physique Th\'eorique, CEA, CNRS, Universit\'e Paris-Saclay 91191 Gif-sur-Yvette Cedex, France\label{aff40}
\and
INAF-Osservatorio Astrofisico di Torino, Via Osservatorio 20, 10025 Pino Torinese (TO), Italy\label{aff41}
\and
Dipartimento di Fisica, Universit\`a di Genova, Via Dodecaneso 33, 16146, Genova, Italy\label{aff42}
\and
INFN-Sezione di Genova, Via Dodecaneso 33, 16146, Genova, Italy\label{aff43}
\and
Department of Physics "E. Pancini", University Federico II, Via Cinthia 6, 80126, Napoli, Italy\label{aff44}
\and
INAF-Osservatorio Astronomico di Capodimonte, Via Moiariello 16, 80131 Napoli, Italy\label{aff45}
\and
Instituto de Astrof\'isica e Ci\^encias do Espa\c{c}o, Universidade do Porto, CAUP, Rua das Estrelas, PT4150-762 Porto, Portugal\label{aff46}
\and
Dipartimento di Fisica, Universit\`a degli Studi di Torino, Via P. Giuria 1, 10125 Torino, Italy\label{aff47}
\and
INFN-Sezione di Torino, Via P. Giuria 1, 10125 Torino, Italy\label{aff48}
\and
INAF-IASF Milano, Via Alfonso Corti 12, 20133 Milano, Italy\label{aff49}
\and
Institute for Theoretical Particle Physics and Cosmology (TTK), RWTH Aachen University, 52056 Aachen, Germany\label{aff50}
\and
INAF-Osservatorio Astronomico di Roma, Via Frascati 33, 00078 Monteporzio Catone, Italy\label{aff51}
\and
Dipartimento di Fisica e Astronomia "Augusto Righi" - Alma Mater Studiorum Universit\`a di Bologna, via Piero Gobetti 93/2, 40129 Bologna, Italy\label{aff52}
\and
INFN section of Naples, Via Cinthia 6, 80126, Napoli, Italy\label{aff53}
\and
Dipartimento di Fisica e Astronomia "Augusto Righi" - Alma Mater Studiorum Universit\`a di Bologna, Viale Berti Pichat 6/2, 40127 Bologna, Italy\label{aff54}
\and
Institut national de physique nucl\'eaire et de physique des particules, 3 rue Michel-Ange, 75794 Paris C\'edex 16, France\label{aff55}
\and
Instituto de Astrof\'isica de Canarias, Calle V\'ia L\'actea s/n, 38204, San Crist\'obal de La Laguna, Tenerife, Spain\label{aff56}
\and
Institute for Astronomy, University of Edinburgh, Royal Observatory, Blackford Hill, Edinburgh EH9 3HJ, UK\label{aff57}
\and
European Space Agency/ESRIN, Largo Galileo Galilei 1, 00044 Frascati, Roma, Italy\label{aff58}
\and
UCB Lyon 1, CNRS/IN2P3, IUF, IP2I Lyon, 4 rue Enrico Fermi, 69622 Villeurbanne, France\label{aff59}
\and
Institut de Ciencies de l'Espai (IEEC-CSIC), Campus UAB, Carrer de Can Magrans, s/n Cerdanyola del Vall\'es, 08193 Barcelona, Spain\label{aff60}
\and
Departamento de F\'isica, Faculdade de Ci\^encias, Universidade de Lisboa, Edif\'icio C8, Campo Grande, PT1749-016 Lisboa, Portugal\label{aff61}
\and
Instituto de Astrof\'isica e Ci\^encias do Espa\c{c}o, Faculdade de Ci\^encias, Universidade de Lisboa, Campo Grande, 1749-016 Lisboa, Portugal\label{aff62}
\and
Department of Astronomy, University of Geneva, ch. d'Ecogia 16, 1290 Versoix, Switzerland\label{aff63}
\and
INAF-Istituto di Astrofisica e Planetologia Spaziali, via del Fosso del Cavaliere, 100, 00100 Roma, Italy\label{aff64}
\and
INFN-Padova, Via Marzolo 8, 35131 Padova, Italy\label{aff65}
\and
INAF-Osservatorio Astronomico di Padova, Via dell'Osservatorio 5, 35122 Padova, Italy\label{aff66}
\and
Dipartimento di Fisica "Aldo Pontremoli", Universit\`a degli Studi di Milano, Via Celoria 16, 20133 Milano, Italy\label{aff67}
\and
INFN-Sezione di Milano, Via Celoria 16, 20133 Milano, Italy\label{aff68}
\and
Institute of Theoretical Astrophysics, University of Oslo, P.O. Box 1029 Blindern, 0315 Oslo, Norway\label{aff69}
\and
Leiden Observatory, Leiden University, Einsteinweg 55, 2333 CC Leiden, The Netherlands\label{aff70}
\and
Jet Propulsion Laboratory, California Institute of Technology, 4800 Oak Grove Drive, Pasadena, CA, 91109, USA\label{aff71}
\and
Department of Physics, Lancaster University, Lancaster, LA1 4YB, UK\label{aff72}
\and
von Hoerner \& Sulger GmbH, Schlossplatz 8, 68723 Schwetzingen, Germany\label{aff73}
\and
Technical University of Denmark, Elektrovej 327, 2800 Kgs. Lyngby, Denmark\label{aff74}
\and
Cosmic Dawn Center (DAWN), Denmark\label{aff75}
\and
Department of Physics and Astronomy, University College London, Gower Street, London WC1E 6BT, UK\label{aff76}
\and
Universit\'e de Gen\`eve, D\'epartement de Physique Th\'eorique and Centre for Astroparticle Physics, 24 quai Ernest-Ansermet, CH-1211 Gen\`eve 4, Switzerland\label{aff77}
\and
NOVA optical infrared instrumentation group at ASTRON, Oude Hoogeveensedijk 4, 7991PD, Dwingeloo, The Netherlands\label{aff78}
\and
University of Applied Sciences and Arts of Northwestern Switzerland, School of Engineering, 5210 Windisch, Switzerland\label{aff79}
\and
Universit\"at Bonn, Argelander-Institut f\"ur Astronomie, Auf dem H\"ugel 71, 53121 Bonn, Germany\label{aff80}
\and
INFN-Sezione di Roma, Piazzale Aldo Moro, 2 - c/o Dipartimento di Fisica, Edificio G. Marconi, 00185 Roma, Italy\label{aff81}
\and
Department of Physics, Institute for Computational Cosmology, Durham University, South Road, DH1 3LE, UK\label{aff82}
\and
Universit\'e C\^{o}te d'Azur, Observatoire de la C\^{o}te d'Azur, CNRS, Laboratoire Lagrange, Bd de l'Observatoire, CS 34229, 06304 Nice cedex 4, France\label{aff83}
\and
California institute of Technology, 1200 E California Blvd, Pasadena, CA 91125, USA\label{aff84}
\and
European Space Agency/ESTEC, Keplerlaan 1, 2201 AZ Noordwijk, The Netherlands\label{aff85}
\and
Kapteyn Astronomical Institute, University of Groningen, PO Box 800, 9700 AV Groningen, The Netherlands\label{aff86}
\and
Department of Physics and Astronomy, University of Aarhus, Ny Munkegade 120, DK-8000 Aarhus C, Denmark\label{aff87}
\and
Waterloo Centre for Astrophysics, University of Waterloo, Waterloo, Ontario N2L 3G1, Canada\label{aff88}
\and
Department of Physics and Astronomy, University of Waterloo, Waterloo, Ontario N2L 3G1, Canada\label{aff89}
\and
Perimeter Institute for Theoretical Physics, Waterloo, Ontario N2L 2Y5, Canada\label{aff90}
\and
Space Science Data Center, Italian Space Agency, via del Politecnico snc, 00133 Roma, Italy\label{aff91}
\and
Institute of Space Science, Str. Atomistilor, nr. 409 M\u{a}gurele, Ilfov, 077125, Romania\label{aff92}
\and
Departamento de Astrof\'isica, Universidad de La Laguna, 38206, La Laguna, Tenerife, Spain\label{aff93}
\and
Dipartimento di Fisica e Astronomia "G. Galilei", Universit\`a di Padova, Via Marzolo 8, 35131 Padova, Italy\label{aff94}
\and
Caltech/IPAC, 1200 E. California Blvd., Pasadena, CA 91125, USA\label{aff95}
\and
Institut f\"ur Theoretische Physik, University of Heidelberg, Philosophenweg 16, 69120 Heidelberg, Germany\label{aff96}
\and
Universit\'e St Joseph; Faculty of Sciences, Beirut, Lebanon\label{aff97}
\and
Institut de Recherche en Astrophysique et Plan\'etologie (IRAP), Universit\'e de Toulouse, CNRS, UPS, CNES, 14 Av. Edouard Belin, 31400 Toulouse, France\label{aff98}
\and
Departamento de F\'isica, FCFM, Universidad de Chile, Blanco Encalada 2008, Santiago, Chile\label{aff99}
\and
Universit\"at Innsbruck, Institut f\"ur Astro- und Teilchenphysik, Technikerstr. 25/8, 6020 Innsbruck, Austria\label{aff100}
\and
Centre for Electronic Imaging, Open University, Walton Hall, Milton Keynes, MK7~6AA, UK\label{aff101}
\and
Infrared Processing and Analysis Center, California Institute of Technology, Pasadena, CA 91125, USA\label{aff102}
\and
Instituto de Astrof\'isica e Ci\^encias do Espa\c{c}o, Faculdade de Ci\^encias, Universidade de Lisboa, Tapada da Ajuda, 1349-018 Lisboa, Portugal\label{aff103}
\and
Universidad Polit\'ecnica de Cartagena, Departamento de Electr\'onica y Tecnolog\'ia de Computadoras,  Plaza del Hospital 1, 30202 Cartagena, Spain\label{aff104}
\and
INFN-Bologna, Via Irnerio 46, 40126 Bologna, Italy\label{aff105}
\and
Junia, EPA department, 41 Bd Vauban, 59800 Lille, France\label{aff106}
\and
Instituto de F\'isica Te\'orica UAM-CSIC, Campus de Cantoblanco, 28049 Madrid, Spain\label{aff107}
\and
CERCA/ISO, Department of Physics, Case Western Reserve University, 10900 Euclid Avenue, Cleveland, OH 44106, USA\label{aff108}
\and
Laboratoire de Physique de l'\'Ecole Normale Sup\'erieure, ENS, Universit\'e PSL, CNRS, Sorbonne Universit\'e, 75005 Paris, France\label{aff109}
\and
Observatoire de Paris, Universit\'e PSL, Sorbonne Universit\'e, LERMA, 750 Paris, France\label{aff110}
\and
Astrophysics Group, Blackett Laboratory, Imperial College London, London SW7 2AZ, UK\label{aff111}
\and
Scuola Normale Superiore, Piazza dei Cavalieri 7, 56126 Pisa, Italy\label{aff112}
\and
Dipartimento di Fisica e Scienze della Terra, Universit\`a degli Studi di Ferrara, Via Giuseppe Saragat 1, 44122 Ferrara, Italy\label{aff113}
\and
Istituto Nazionale di Fisica Nucleare, Sezione di Ferrara, Via Giuseppe Saragat 1, 44122 Ferrara, Italy\label{aff114}
\and
Dipartimento di Fisica - Sezione di Astronomia, Universit\`a di Trieste, Via Tiepolo 11, 34131 Trieste, Italy\label{aff115}
\and
NASA Ames Research Center, Moffett Field, CA 94035, USA\label{aff116}
\and
Kavli Institute for Particle Astrophysics \& Cosmology (KIPAC), Stanford University, Stanford, CA 94305, USA\label{aff117}
\and
Minnesota Institute for Astrophysics, University of Minnesota, 116 Church St SE, Minneapolis, MN 55455, USA\label{aff118}
\and
INAF, Istituto di Radioastronomia, Via Piero Gobetti 101, 40129 Bologna, Italy\label{aff119}
\and
Institute Lorentz, Leiden University, Niels Bohrweg 2, 2333 CA Leiden, The Netherlands\label{aff120}
\and
Institute for Astronomy, University of Hawaii, 2680 Woodlawn Drive, Honolulu, HI 96822, USA\label{aff121}
\and
Department of Physics \& Astronomy, University of California Irvine, Irvine CA 92697, USA\label{aff122}
\and
Dept. of Physics, IIT Hyderabad, Kandi, Telangana 502285, India\label{aff123}
\and
Department of Astronomy \& Physics and Institute for Computational Astrophysics, Saint Mary's University, 923 Robie Street, Halifax, Nova Scotia, B3H 3C3, Canada\label{aff124}
\and
Departamento F\'isica Aplicada, Universidad Polit\'ecnica de Cartagena, Campus Muralla del Mar, 30202 Cartagena, Murcia, Spain\label{aff125}
\and
Dipartimento di Fisica, Universit\`a degli studi di Genova, and INFN-Sezione di Genova, via Dodecaneso 33, 16146, Genova, Italy\label{aff126}
\and
Ruhr University Bochum, Faculty of Physics and Astronomy, Astronomical Institute (AIRUB), German Centre for Cosmological Lensing (GCCL), 44780 Bochum, Germany\label{aff127}
\and
Instituto de Astrof\'isica de Canarias (IAC); Departamento de Astrof\'isica, Universidad de La Laguna (ULL), 38200, La Laguna, Tenerife, Spain\label{aff128}
\and
Universit\'e Paris-Cit\'e, 5 Rue Thomas Mann, 75013, Paris, France\label{aff129}
\and
Universit\'e PSL, Observatoire de Paris, Sorbonne Universit\'e, CNRS, LERMA, 75014, Paris, France\label{aff130}
\and
Univ. Grenoble Alpes, CNRS, Grenoble INP, LPSC-IN2P3, 53, Avenue des Martyrs, 38000, Grenoble, France\label{aff131}
\and
Department of Physics and Astronomy, Vesilinnantie 5, 20014 University of Turku, Finland\label{aff132}
\and
Serco for European Space Agency (ESA), Camino bajo del Castillo, s/n, Urbanizacion Villafranca del Castillo, Villanueva de la Ca\~nada, 28692 Madrid, Spain\label{aff133}
\and
Oskar Klein Centre for Cosmoparticle Physics, Department of Physics, Stockholm University, Stockholm, SE-106 91, Sweden\label{aff134}
\and
Dipartimento di Fisica, Sapienza Universit\`a di Roma, Piazzale Aldo Moro 2, 00185 Roma, Italy\label{aff135}
\and
Centro de Astrof\'{\i}sica da Universidade do Porto, Rua das Estrelas, 4150-762 Porto, Portugal\label{aff136}
\and
Dipartimento di Fisica, Universit\`a di Roma Tor Vergata, Via della Ricerca Scientifica 1, Roma, Italy\label{aff137}
\and
INFN, Sezione di Roma 2, Via della Ricerca Scientifica 1, Roma, Italy\label{aff138}
\and
Department of Mathematics and Physics E. De Giorgi, University of Salento, Via per Arnesano, CP-I93, 73100, Lecce, Italy\label{aff139}
\and
INAF-Sezione di Lecce, c/o Dipartimento Matematica e Fisica, Via per Arnesano, 73100, Lecce, Italy\label{aff140}
\and
INFN, Sezione di Lecce, Via per Arnesano, CP-193, 73100, Lecce, Italy\label{aff141}
\and
Department of Astrophysics, University of Zurich, Winterthurerstrasse 190, 8057 Zurich, Switzerland\label{aff142}
\and
ASTRON, the Netherlands Institute for Radio Astronomy, Postbus 2, 7990 AA, Dwingeloo, The Netherlands\label{aff143}
\and
Department of Astrophysical Sciences, Peyton Hall, Princeton University, Princeton, NJ 08544, USA\label{aff144}
\and
Niels Bohr Institute, University of Copenhagen, Jagtvej 128, 2200 Copenhagen, Denmark\label{aff145}}    

%
%
\abstract
{The European Space Agency's \textit{Euclid} mission is one of a raft of forthcoming large-scale cosmology surveys that will map the large-scale structure in the Universe with unprecedented precision. The mission will collect a vast amount of data that will be processed and analysed by \Euclid's Science Ground Segment (SGS). The development and validation of the SGS pipeline requires state-of-the-art simulations with a high level of complexity and accuracy that include subtle instrumental features not accounted for previously as well as faster algorithms for the large-scale production of the expected \Euclid data products.}
{In this paper, we present the \Euclid SGS simulation framework as it is applied in a large-scale end-to-end simulation exercise named Science Challenge 8. Our simulation pipeline enables the swift production of detailed image simulations for the construction and validation of the \Euclid mission during its qualification phase and will serve as a reference throughout operations.} 
{Our end-to-end simulation framework started with the production of a large cosmological N-body simulation that we used to construct a realistic galaxy mock catalogue. We performed a selection of galaxies down to \IE=26 and 28 mag, respectively, for a Euclid Wide Survey spanning $165\,{\rm deg}^2$ and a $1\,{\rm deg}^2$ Euclid Deep Survey. We built realistic stellar density catalogues containing Milky Way-like stars down to $H<26$ from a combination of a stellar population synthesis model of the Galaxy and real bright stars. Using the latest instrumental models for both the \Euclid instruments and spacecraft as well as \Euclid-like observing sequences, we emulated with high fidelity \Euclid satellite imaging throughout the mission's lifetime.}
{We present the SC8 dataset, consisting of overlapping visible and near-infrared Euclid Wide Survey and Euclid Deep Survey imaging and low-resolution spectroscopy along with ground-based data in five optical bands.  This extensive dataset enables end-to-end testing of the entire ground segment data reduction and science analysis pipeline as well as the \Euclid mission infrastructure, paving the way for future scientific and technical developments and enhancements.}
   {}
    
%
%
%
%
   \titlerunning{Euclid Preparation: XLVIII. Pre-launch \Euclid Simulations}
   \authorrunning{Euclid Collaboration: S. Serrano et al.}
   
   \maketitle
%
%
%
%
   
\section{Introduction}

The accelerated expansion of the Universe is now a well-established fact, corroborated by a large amount of observational evidence. However, the origin of this acceleration is still uncertain. It could either be a cosmological constant in the equation of general relativity, or a mysterious dark energy suggestive of physics beyond the standard model of particle physics \citep{planck2018}. 

To better understand the origin of the Universe's accelerated expansion, the \Euclid space mission was proposed and accepted in 2011 by the European Space Agency (ESA) as a medium-class mission. The spacecraft consists of a 1.2m telescope to observe $15\,000\,{\rm deg}^2$ of extragalactic sky comprising the Euclid Wide Survey \citep{euclid-survey-scaramella2021} and a $50\,{\rm deg}^2 $ Euclid Deep Survey. It is equipped with two instruments to perform single broadband optical imaging with very high resolution in the VISible instrument (VIS), and simultaneous near-infrared (NIR) slitless spectroscopy and imaging with the Near Infrared Spectrometer and Photometer (NISP). These two instruments are part of the innovative design of \Euclid that will acquire measurements of the shapes of 2 billion galaxies up to a redshift of $z\sim2.3$ and 30 million spectroscopic redshifts \citep{euclid-laurejis2011} out to a redshift of $z\sim 2$.
These independent measurements will enable us to reconstruct the expansion history of the Universe using probes in the core science areas of weak gravitational lensing (WL) and galaxy clustering (GC). Moreover, cosmological forecasts for \Euclid predict an increase in the dark energy figure of merit by at least a factor of three when performing a combined analysis using cross-correlations of WL and GC measurements \citep{euclid-forecasts-blanchard2020}.

In order to reach its scientific goals, the following mission level requirements have been proposed \citep{euclid-laurejis2011}: the density of galaxies brighter than $\IE=24.5\,{\rm mag}$ and detected with a $10\sigma$ confidence must be at least 30\,galaxy\,arcmin$^{-2}$.  Their corresponding median redshift should be $z>0.8$ with an error on the mean redshift per bin below $0.002$. The density of galaxies with spectroscopic redshift and an H$\alpha~\rm{\corr{flux}} > 2\times10^{-16}\,{\rm erg}\,{\rm cm}^{-2}\,{\rm s}^{-1}$ is required to be at least $ 3500\,{\rm deg}^{-2}$. The control of systematic errors will be one of the most challenging aspects of the mission. For VIS, the point spread function (PSF) ellipticity must be known to an accuracy better than $2 \times 10^{-3}$, and the stray light should remain less than $20\%$ of the ecliptic zodiacal light.  

Moreover, we aim for the contrast ratio of ghost images — images produced by multiple reflections of the optical system — to be below $10^{-4}$ \citep{cropper2014}. For NIR spectroscopy, we require the purity of the spectroscopic sample ($0.7 < z < 1.8$ and H$\alpha > 2\times10^{-16}\,{\rm erg}\,{\rm cm}^{-2}\,{\rm s}^{-1}$) to be above $80$\%, the completeness higher than $45$\%, and the spectroscopic resolution $R \geq 380$ for the red grisms and $R \geq 260$ for the blue grism. Finally, integration of external (EXT) survey data at the pixel level is essential for \Euclid to reach the required  photometric redshift precision per object: $\sigma(z) / (1+z) \le 0.05$.  These requirements are particularly important, especially in space experiments such as the \Euclid mission, due to their elevated cost and the impossibility of performing repairs once the spacecraft is launched.

Given these tight requirements, a forward modelling approach has been adopted by the Science Ground Segment (SGS) as the main method to aid the development of the data-processing and science analysis pipeline \citep{euclid-laurejis2011}. Within SGS, the Organisation Unit for Simulations (OU-SIM) is tasked with designing a simulation pipeline with added flexibility to deliver expeditiously realistic pixel-level images for the \Euclid mission, following the release of a new instrument model. Our image simulations have two main objectives. The first is to serve as test data for the development of the SGS data-processing pipeline and the computing infrastructure before the arrival of real \Euclid observations. Second, they are to be used to validate the stringent requirements that the SGS has in terms of performance and data quality. In this respect, several end-to-end Science Performance Verification (SPV) tests have been performed throughout the mission preparation, which allowed for the reproduction of certain instrumental issues to assess their impact and guide decision-making.  Pixel-level simulations have been essential in the discovery of alternative solutions to critical problems, such as the non-conformity of one of the three red grisms for the NISP instrument \citep{euclid-survey-scaramella2021}.

The production of simulations is generally planned in advance in orchestrated tests that we call Scientific Challenges \citep[SC;][]{sdcs_2019}. These exercises have been essential for the development of the SGS pipeline and the validation of science and technical requirements. Each challenge has leveraged new algorithms and infrastructure as well as improved instrument models derived during the construction of the spacecraft.  Moreover, the outcomes of each SC have proven to be an effective indicator of the status and readiness of the SGS in addition to the various ESA reviews.

In this paper, we describe our simulation framework developed within OU-SIM with a focus on Scientific Challenge 8 (SC8). SC8 included for the first time the complete ground segment processing chain, enabling a full end-to-end simulation of the so-called level-1 raw images, level-2 detrended and calibrated data products, and level-3 core science-ready measurements (e.g. the cosmic shear and the galaxy two-point correlation function measurements, etc.). A majority of the data-processing was operated entirely by the SGS System Team without manual intervention from the pipeline development teams. The main goals of SC8 were (1) to simulate a large area of the Euclid Wide Survey along with the corresponding ground-based and calibration data for processing through the entire SGS pipeline; (2) to assess the capability of the SGS infrastructure to process a large set of data within a reasonable timescale; (3) to validate SGS operations under nominal conditions; (4) to validate the applicable requirements for the Ground Segment Implementation Review (GSIR); and (5) to benchmark the data quality reference for the third SPV test (SPV03).

All \Euclid photometric imaging and spectroscopic channels were simulated, as well as seven EXT \Euclid-supporting photometric imaging surveys spanning the northern and southern hemispheres. This extensive and challenging simulation effort resulted in 31 TB of raw output images for ${\sim}\,165\,{\rm deg}^2$, which were subsequently processed and analysed with the \Euclid SGS pipeline.

This paper is organised as follows. In Sect.~\ref{s:satellite}, we begin with an overview of the drivers behind the development of our framework; namely, the components of the \Euclid satellite and the organisation of the SGS. Our description of the OU-SIM workflow begins in Sect.~\ref{s:true_universe}, in which we explain our methods of creating synthetic galaxy and stellar catalogues, the inputs into the simulations. Sect.~\ref{s:mdb} follows with an introduction to the mission database, which contains the instrument and survey parameters, and thus provides the specifications for launching a coordinated production of simulations for different surveys (space- and ground-based).  In Sect.~\ref{s:euclid_instrument_simulators}, we describe in detail the instrumental features simulated in SC8.  We outline our workflow for large-scale simulation productions in Sect.~\ref{s:sim_planner}.  The application and performance of our end-to-end framework in SC8 are detailed in Sect.~\ref{s:sc8}. In Sect.~\ref{s:future_perspective}, we summarise ongoing work and the latest improvements to the simulation pipeline. Finally, in Sect.~\ref{s:summ_and_outlook}, we present our conclusions.

\section{The satellite and the Science Ground Segment}
\label{s:satellite}

 The \Euclid satellite is composed of the Service Module (SVM) and the Payload Module (PLM). The SVM\footnote{See figure in \url{https://www.euclid-ec.org/?page_id=2686} for an illustration of the SVM components.} includes the sunshield, the star trackers and gyros, the thrusters, the micro-motions and slews control systems, hydrazine and cold gas tanks, the Attitude and Orbital Control System (AOCS), the solar panel and electric power system, the thermal regulation system, and the downlink communication system. The AOCS jitter will meet the high-quality imaging requirements with a pointing dispersion smaller than $35\,{\rm mas}$ in each exposure, providing a stable attitude through the whole integration time.

The PLM \citep{plm_venancio2014} comprises the telescope, the PLM thermal control system, the Fine Guidance Sensor (FGS), and the VIS and NISP instruments (see Sect.~\ref{s:vis_instrument} and Sect.~\ref{s:nisp_instrument}, respectively, for a more detailed description). The design of the optical system is illustrated in Fig.~8 of \cite{eucliddesign-racca2016}. The telescope follows a Korsch design with a 1.2-m diameter primary mirror providing an optically corrected and unvignetted field of view of $0.79 \times 1.16\,{\rm deg}^2$. The incoming light beam is split with a dichroic plate, reflecting the bluer part into the optical VIS instrument and transmitting the NIR light towards the NISP instrument. A more detailed description of the dichroic optical element is described in \cite{euclid-NISPphotobands-schirmer2022}. This particular design feature allows simultaneous observations to be performed with both instruments. 

Launch occurred in June 2023 on a SpaceX Falcon9 launch vehicle, which departed from Cape Canaveral. The mission will carry out a nominal survey that is expected to be completed in six years. The spacecraft will orbit the Second Sun-Earth Lagrangian point (L2) in a wide halo orbit. In addition to the $15\,000\,{\rm deg}^2$ Euclid Wide Survey, \Euclid will also perform a Euclid Deep Survey with additional observations near both ecliptic poles for a total area of ${\sim}\,50\,{\rm deg}^2$. The Euclid Wide Survey will avoid the Galactic and Ecliptic planes, with high dust extinction and high background, respectively. The VIS instrument will measure extended sources of $m_{\text{AB}} = 24.5$ in the visible band, $\IE$, with a signal-to-noise ratio of at least 10. For the NISP instrument's NIR bands \YE, \JE, and \HE, point-like sources will have a signal-to-noise of 5 or greater at a magnitude $m_{\text{AB}} = 24$. The Euclid Deep Survey will provide 40-53 times more observations (depending on the background level at each field location), reaching 2 magnitudes deeper than the Euclid Wide Survey. Further details are described in \cite{euclid-survey-scaramella2021}.

\subsection{The VISible instrument}
\label{s:vis_instrument}

The VIS instrument is an optical imager featuring a mosaic of 36 ($6\times 6$), $4096\times 4132$ pixel Teledyne e2v CCD detectors, which have been specially optimised for this mission. It operates in the single $\IE$ broadband from 550 to $900\,{\rm nm}$ (equivalent to a combined $r$+$i$+$z$ band) shaped by the reflection of the coating on the optical elements (the dichroic plate and fold mirrors, where FM1, FM2 have a hybrid metal-dielectric coating, and FM3 is silver coated) and the quantum efficiency of the CCD detectors \citep{cropper2014}. The VIS focal plane covers a field of view of $0.57\,{\rm deg}^2$ with a central pixel scale of $\ang{;;0.1}$ per pixel, resulting in an undersampled VIS PSF \corr{which limits the maximum optical resolution of the opto-electrical system}. However, the exquisite image quality and exceptional temporal stability required of the VIS instrument and the telescope will allow measurements of galaxy shapes with enough accuracy to estimate the gravitational-lensing effect caused by the large-scale structures in the Universe on distant background galaxies. 

\subsection{The Near Infrared Spectrometer and Photometer}
\label{s:nisp_instrument}

\begin{figure}
  \centering
  \includegraphics[width=\linewidth]{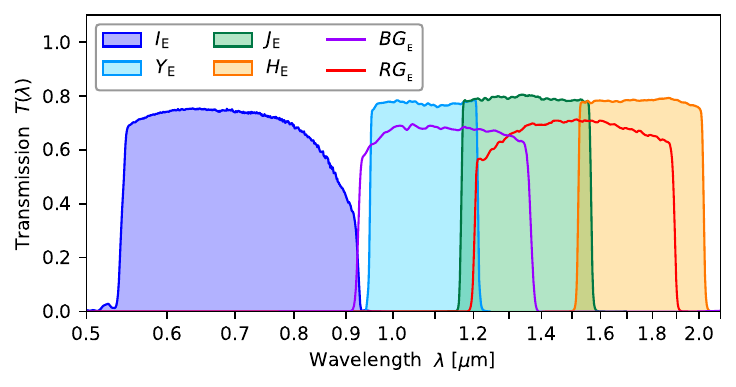}
  \caption{Total transmission of the VIS (\IE) and NISP photometric  (\YE, \JE, \HE), and the NISP Spectroscopic ($BG_E$, $RG_E$) bands.}
  \label{fig:euclid_transmissions}
\end{figure}

The NISP instrument was designed to provide broadband photometry (NISP-P) in three bands as well as low-resolution NIR spectra (NISP-S) divided into two wavelength ranges \citep{maciaszek2016,maciaszek2022}.
The NISP focal plane contains a mosaic of 16 ($4\times 4$), $2048\times2048$ pixel infrared detectors from Teledyne, with a field of view of 0.53 deg$^2$ and a pixel scale of $\ang{;;0.3}$. The NISP-P channel is equipped with three broadband filters (i.e. $\YE$, $\JE$, and $\HE$) that span the range of 950 to $2021\,{\rm nm}$ \citep{euclid-NISPphotobands-schirmer2022}. Figure~\ref{fig:euclid_transmissions} illustrates the broadband transmissions for these filters along with the VIS photometric band.  In combination with the VIS band and ground-based optical photometry, the three NISP broadbands will provide accurate photometric redshifts for billions of galaxies.

The low-resolution spectra acquired by NISP-S are obtained with grisms that disperse all the light without any slit, a technique called slit-less spectroscopy. The advantage of this method is that the design is simpler and more robust than a slit mechanism, requiring no target selection. However, slit-less spectroscopy presents two disadvantages: a higher background that reduces the signal-to-noise ratio and contamination of spectra caused by overlapping sources and intrinsic spatial extent. To optimise NISP-S, the spectroscopic channel contains in total four grisms providing a spectral resolution of $R=480$ for a $\ang{;;0.5}$ diameter source. However, only three of the four grisms will be operational due to the non-conformity of RGS270  \citep{euclid-survey-scaramella2021}..  Two grisms (RGS000 and RGS180) cover the redder portions of the spectrum from 1206 to $1892\,{\rm nm}$ for the Euclid Wide Survey. The third is the ``blue" grism (BGS000) with a wavelength range between 926 and $1266\,{\rm nm}$, which will only be used during the Euclid Deep Survey. The two red grisms cover the same wavelength range with a rotated dispersion orientation (0$^{\circ}$ and 180$^{\circ}$). In order to facilitate the decontamination of overlapping slit-less spectra, they will operate at their nominal positions as well as at orientations of $-4^{\circ}$ and $184^{\circ}$, respectively \citep{maciaszek2022}. With the NISP-S low-resolution spectra, we can obtain spectroscopic redshifts enabling precise measurements of the distribution and clustering of galaxies.

\subsection{The \Euclid Science Ground Segment}

The \Euclid SGS is one half of the so-called \Euclid Ground Segment with the Operations Ground Segment (OGS) comprising the other half \citep{eucliddesign-racca2016}.  The SGS and OGS have different functions. The OGS is in charge of operating the spacecraft, analysing the telemetry, and performing the down-link transmissions of the data. As stated previously, the SGS is responsible for carrying out the entire data-processing and will perform key cosmological measurements, ultimately delivering the main scientific results. 

There are ten organisation units (OUs) comprising the SGS data-processing pipeline.\footnote{A description of each OU is provided in Sect. 7 of \cite{euclid-laurejis2011}.} Each OU is responsible to define, design, and validate a specific analysis of the SGS workflow. Processing the massive data volumes within the \Euclid SGS takes place in a distributed system across the nine Science Data Centres (SDC) in nine countries (Finland, France, Germany, Italy, Netherlands, Spain, Switzerland, United Kingdom, and the United States). While most of the SDCs have a High Throughput Computing (HTC) design, the underlying infrastructure varies across each SDC. The \Euclid SGS has established common tools and guidelines to ensure homogeneity in terms of development, storage, and computing, independent of location \citep{sdcs_2019}. 
%
\begin{figure*}[!t]
  \centering
  \includegraphics[width=\linewidth]{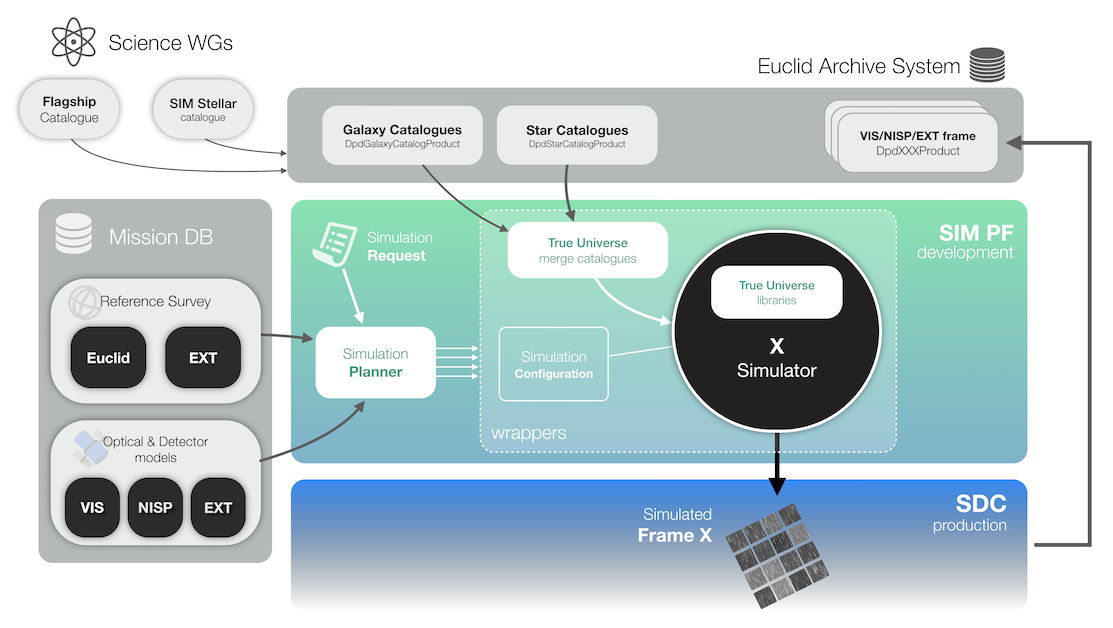}
  \caption{Overview of the SIM pipeline, coordinated by the \texttt{SimPlanner}. The input catalogues provided by the Science Working Groups are ingested in the \Euclid Archive System as well as the output products of SIM. With the models from the Mission Database (described in Sect.~\ref{s:mdb}), the input True Universe catalogues and a simulation request, the \texttt{SimPlanner} produces the corresponding configuration files for each simulator.}
  \label{fig:sim_pipeline}
\end{figure*}
%
We summarise the main elements of the common infrastructure system under which we develop the OU-SIM pipeline to operate within the SGS: (1) The Euclid Archive System (EAS) is a database containing all data products and metadata processed for the \Euclid mission; this includes images, catalogues, processing status, etc. It also manages the transfer of data among the local storage facilities at each SDC\footnote{The EAS is accessible solely to SGS members, who are permitted to transfer, download, index, search, and access information of all \Euclid data products.} \citep[additional information about the EAS is provided in][]{eas-williams2019}. (2) A Common Data Model defines the format of the data and associated metadata to ensure that interfaces between pipelines and with the archive are stable. (3) An Infrastructure Abstraction Layer (IAL) orchestrates the data-processing and adds a common layer allowing jobs to run independently of the underlying IT infrastructure. It defines the Pipeline Processing Order (PPO) which describes the elements, configuration, and inputs of a specific processing task. (4) A common Euclid Development ENvironment (EDEN) establishes the set of libraries and associated versions to be used by any of the \Euclid software. This prevents inconsistencies or changes in the functionality of different libraries between development and production. (5) The COllaborative DEvelopment ENvironment (CODEEN) is a continuous integration and continuous delivery (CI/CD) platform that automates the building, unit testing and distribution of all the scientific software in the SGS. Source code is extracted from a Version Control System (i.e. Gitlab) and run through a CI/CD pipeline to be finally deployed on a distributed file system (CernVM-FS) available on all SDCs.
This system design allows SGS operations to be performed smoothly and efficiently across the nine SDCs, providing the extra advantage of increased computing power and storage capacity.  

\subsection{The simulation framework}
\label{s:sim_framework}

The strict requirements of this mission demand extremely precise simulations with exquisite detail, capturing all known instrumental and environmental models. This is to ensure that all critical data-processing components are tested and validated prior to the launch. Producing image simulations at this high level of quality not only depends on active engagement with the instrument teams, it also requires leveraging our interfaces with all OUs. Each processing function (PF) has its own respective requirements that the simulations must satisfy as well as additional insights and tools to aid the development and validation of our framework. Evidently, liaisons had to be established with the Cosmological Simulations Science Working Group, the main supplier of galaxy catalogues for \Euclid's core science program.

The organisation of the SGS simulation pipeline within the scope of the SGS infrastructure is depicted in Fig.~\ref{fig:sim_pipeline}. The True Universe catalogues contain all input sources and their corresponding parameters, spectra and shape, and are described further in Sect.~\ref{s:true_universe}. The instrument models and the reference survey are introduced into the pipeline from the Mission Database, which we explain in Sect.~\ref{s:mdb}. Within the so-called simulation processing function (SIM PF), there are four image simulators, capable of accurately reproducing imaging and spectral data for the VIS, NISP-P, NISP-S, and EXT survey channels, including the instrumental effects of the two \Euclid instruments and of the ground-based observations; we present each of the four simulators in Sect.~\ref{s:euclid_instrument_simulators}. The operation of the simulation framework is orchestrated by the \texttt{SimPlanner} via a simulation request and is detailed in Sect.~\ref{s:sim_planner}.

\section{The True Universe}
\label{s:true_universe}

The ``True Universe" (TU) catalogues\footnote{The usage of the term ``true" is due to the treatment of sources as noise-free for the simulation even though they may come from noisy observations.  In the simulation, the measurements are altered with physical, environmental and/or instrumental sources of noise.}, consisting of stars and galaxies, are input data products to the SIM workflow.  It is paramount that these catalogues include many realistic features as they serve as the basis for assessing the scientific performance and validating the compliance to the requirements at the different stages of the mission.  In the following subsections, we describe, respectively, the four types of galaxy catalogues used in the simulations, the star catalogue built from real and simulated samples, and the common spectra and thumbnail libraries that allow us to transform galaxy and star parameters into fluxes and images.

\subsection{Galaxy catalogues}
\label{s:euclid_galaxy_cat}

Several types of galaxy samples were used in the simulation (see Table~\ref{table:catalogues_summary}): galaxies in the redshift range ($0<z<2.3$) referred to as standard Flagship galaxies (\textit{Std gals}), \textit{QSOs}, referring to a population of quasi-stellar objects at high redshift ($6 < z < 14$), \textit{High-$z$ gals}, denoting a high-redshift population ($6 < z < 10$) of Lyman break galaxies (LBGs), and \textit{SL}, representing a catalogue of strong lensing systems.

\begin{figure*}
  \centering
  \includegraphics[width=\linewidth]{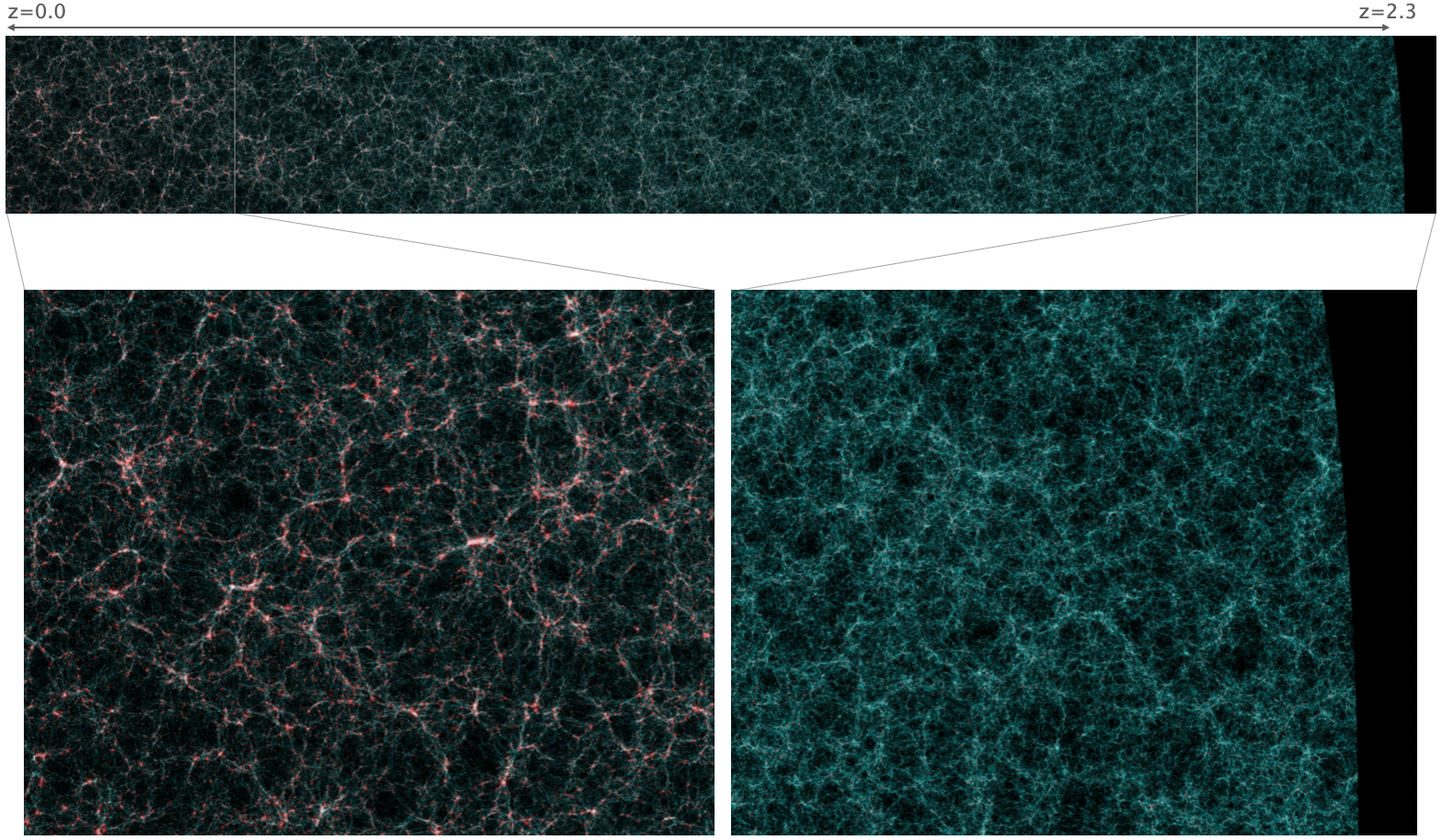}
  \caption{Snapshot of the Flagship galaxy simulation. Top: False colour image showing a slice of 3.800 $h^{-1}$Mpc width and 500 $h^{-1}$Mpc height (roughly $0.3\%$ of the total volume) of the full light-cone Flagship mock. Central galaxies are coloured in green and satellites in red, where filaments from the large-scale structure are seen in great detail. Bottom left: Zoomed-in section with the local Universe ($z=0$). Bottom right: Zoomed-in fraction of the furthest galaxies ($z=2.3$). \corr{The simulation is a 3-dimensional sphere and the dark area at the right of the image corresponds to the end of the simulation where no dark matter particles are simulated.}}
  \label{fig:flagship_image}
\end{figure*}

\subsubsection{\Euclid Flagship mock galaxy catalogues}

The \Euclid Flagship v1.0 Simulation \citep{PKDGRAV3-potter2016} is one of the largest cosmological N-body simulations of the Universe ever produced, with $(12\,600)^3$ particles over a simulation box of $3780\, h^{-1}\,{\rm Mpc}$, leading to a mass resolution of $m_{\rm p} = 2.4 \times 10^9 \,h^{-1}\,M_{\odot}$. Such an unprecedented volume and resolution was required to validate the mission performance at full scale. The simulation was run with cosmological parameters\footnote{$\Omega_{\rm m} = 0.319,\, \Omega_{\rm b} = 0.049,\, \Omega_{\rm CDM} = 0.270,\, \Omega_{\Lambda} = 0.681,\, w = -1.0,\, h = 0.67, \,\sigma_8 = 0.83,\, n_s = 0.96$. These parameters define the density and condition of the energy-matter content of the Universe in matter, $\Omega_{\rm m}$, baryons, $\Omega_{\rm b}$, cold dark matter, $\Omega_{\rm CDM}$ and dark energy, $\Omega_{\Lambda}$, the equation of state parameter of dark energy, $w$, the reduced Hubble constant, $h$, the amplitude of the (linear) power spectrum on the scale of $8\, h^{-1}\,{\rm Mpc}$, and the slope of primordial spectrum of fluctuations, $n_s$.} similar to those of the $Planck$ 2015 cosmology~\citep{planck2015}. The 2 trillion dark matter particle simulation was run with \texttt{PKDGRAV3} \citep{PKDGRAV3-potter2016} on the Piz Daint supercomputer at the Swiss National Supercomputer Center (CSCS). An all-sky particle light-cone up to a redshift of $z=2.3$ was produced in real time. The dark matter halos were identified from the dark matter particle light-cone using the \texttt{ROCKSTAR} halo finder \citep{behroozi2012}, an adaptive hierarchical refinement of a friends-of-friends method able to track substructures and relations with the parent dark matter halos. Following the approach described in \citet{lensing-fosalba2016}, all-sky lensing convergence maps were built, which enable weak lensing effects to be included in the final galaxy catalogue. 

The mock galaxy catalogue was built implementing an improved Halo Occupation Distribution technique (\citealt{micemock-carretero2015}, Castander et al. in preparation) with \texttt{SciPIC}, the Scientific pipeline at Port d'Informació Científica \citep{scipic-carretero2017}, developed to efficiently generate mock galaxy catalogues using as input a dark matter halo population. \texttt{SciPIC} runs on top of the PIC Hadoop cluster using Apache Spark. We present in Fig.~\ref{fig:flagship_image} an example illustration of the mock galaxy catalogue, where a small region of the light-cone is shown, and one can distinguish by eye the emergence of the large-scale structure of the Universe.

For SC8, we selected galaxies up to an AB magnitude of $\IE=26$ for the main Euclid Wide Survey, and $\IE=28$ for the Euclid Deep Survey, which is two magnitudes deeper than the detection requirement. This was essential to maintain the same sources in all instruments (i.e. space and ground) and bands in order to preserve the photometry across the different channels for photometric redshift consistency even if there are sources below the detection threshold in some channels. Lensing measurements are also affected by very low signal-to-noise sources below the detection threshold \citep{WL_sensitivity-hoekstra2017,unresolvedgalaxies-martinet2019}.

Particular care was taken to model the H$\alpha$ emission of the galaxies (following model 3 in \citealt{halpha-pozzetti2016}), a line that is particularly important for NISP to determine spectroscopic redshifts. For the NISP-S simulations, we included only sources with $\HE\le 23$ or sources with an H$\alpha$ emission brighter than $10^{-16}\,{\rm erg}\,{\rm s}^{-1}\,{\rm cm}^{-2}$, as the continuum of sources $\HE>23$ is never detected in the spectroscopic channel. Additionally, the size and shape distributions of the galaxies were updated and improved in \texttt{SciPIC} with respect to previous implementations in the MICE simulation described in \cite{micemock-carretero2015}. The morphological parameter distributions from the Cosmic Assembly Near-infrared Deep Extragalactic Legacy Survey (CANDELS) \citep{candels-grogin2011}, and $Hubble$ Space Telescope (HST) surveys were used to produce highly accurate and realistic galaxy samples (see Sect.~\ref{ss:bulge_disk_model}).

The final mock galaxy catalogue contains the following information that allows the pipeline to reproduce its flux at the pixel level: (1) \textbf{ID}: A unique source ID that also contains information about the parent dark matter halo. (2) \textbf{Positional coordinates}: The source location in equatorial coordinates (RA, Dec) and the lensed positions due to gravitational deflection. (3) \textbf{True Universe kind}: An identifier of the object kind from the various TU inputs such as Flagship central or satellite galaxy, QSO, High-$z$ and strong lensing, as different sources are simulated differently. (4) \textbf{Redshift}: The true and observed redshift of the galaxy, different due to the peculiar velocity. (5) \textbf{Reference magnitude}: An absolute, \corr{instrinsic} and apparent magnitude of the galaxy at a given band. This was used to scale the spectrum at a given brightness. The reference band depends on the object type as some sources are best modelled in the visible bands (stars and low redshift galaxies), while other sources are better characterised in infrared bands (high redshift sources). (6) \textbf{SED template}: A spectral energy distribution (SED) template with dust extinction assigned by sampling the COSMOS galaxy catalogues \citep{cosmosseds-bruzual2003,cosmosseds-polletta2007}, where galaxies are matched to the observed ones using redshift, absolute magnitude, and colour. An additional parameter describes the extinction law to be applied. This was used to reconstruct the spectrum (detailed in Sect.~\ref{s:sim_sedlib}). (7) \textbf{Emission line fluxes}: The fluxes of H$\alpha$, H$\beta$, [\ion{O}{ii}], [\ion{O}{iii}], [\ion{N}{ii}] used in spectroscopic redshift determination. (8) \textbf{Morphological parameters}: Simulating the light profile of a galaxy requires the bulge and disc half-light radius, the ratio of the flux in the bulge component to the total flux (often written B/T), the S\'ersic  index of the bulge profile, the axis ratio defining its intrinsic ellipticity, the inclination angle of the disc, and the position angle with respect to the north (see Sect.~\ref{s:sim_thumbnaillib}, which describes how these parameters are used to reconstruct galaxy shapes). (9) \textbf{Lensing parameters}: The shear and convergence values, $\gamma_1$,  $\gamma_2$, and $\kappa$, were used to distort galaxy shapes and determine their flux magnification.  These three parameters were derived from the \Euclid Flagship dark matter lensing and convergence maps. (10) \textbf{MW extinction}: The optical extinction of the Milky Way in the $V$ band. This was derived from the $Planck$ thermal dust maps at the observed (lensed) coordinates of the galaxy. 

With this list of parameters, we can render all the galaxies in the catalogue in any of the simulation channels, both photometric and spectroscopic, with all their complex instrumental effects. Additionally, we computed the true \corr{reference} fluxes for each galaxy in all the passbands of interest for \Euclid and the ground-based EXT surveys, \corrtwo{including magnification}.

\subsubsection{Other extragalactic sources}
\label{ss:other_extragalactic_sources}

\paragraph{High-redshift sources}

\Euclid's survey area and depth coverage is expected to yield an unprecedented number of new sources at high redshift, which will be catalogued and analysed in its legacy science program: the Primaeval Universe. These are very distant galaxies not contained in the Flagship dark matter run, as its light cone reaches only a redshift of 2.3. The Primaeval Universe Science Working Group provided a sample of 11 million high-redshift galaxies and 2 million quasars at redshifts greater than 6, as is described in \cite{highqso-barnett2019}. They were added to the TU set as an additional set of sources to be simulated on the image, with unlensed  magnitude cuts $\HE<26$ and $\HE<25$ for galaxies and quasars, respectively.

\paragraph{Strong lensing sources}
\label{sec:sl}
\Euclid is expected to also observe tens of thousands of strong lensing sources \citep{collett2015}. For SC8, the Strong Lensing Science Working Group provided a \corrtwo{template} set of 801 sources in the form of high-resolution image thumbnails \corr{that were injected to the simulation} to train algorithms and classify them more efficiently. Strong lensing sources were simulated with the \texttt{GLAMER} ray-tracing code \citep{metcalf2014}, based on the Flagship halo mass properties, and assuming HST-like morphologies for the lensed background sources following the recipe provided in \cite{metcalf2019}. 

\subsection{Stellar catalogue}
\label{s:euclid_stellar_cat}

Observations of stars are an integral part of image processing and analysis. They serve as reference catalogues for all astrometric and photometric calibrations, and are essential for performing the precise PSF modelling required for all lensing measurements. Conversely, they add confusion to the classification of sources and produce numerous complications, such as blending, bleeding, ghost reflections, and scattered light. This is particularly true for the L- and T-type brown dwarf stars, which contaminate high-redshift galaxy and quasar detections. It is critical, then, to have a realistic and accurate distribution of stars at each location of the sky with representative density and stellar types. 

\subsubsection{Besançon stars}

\begin{figure}[!t]
  \centering
  \includegraphics[width=\linewidth]{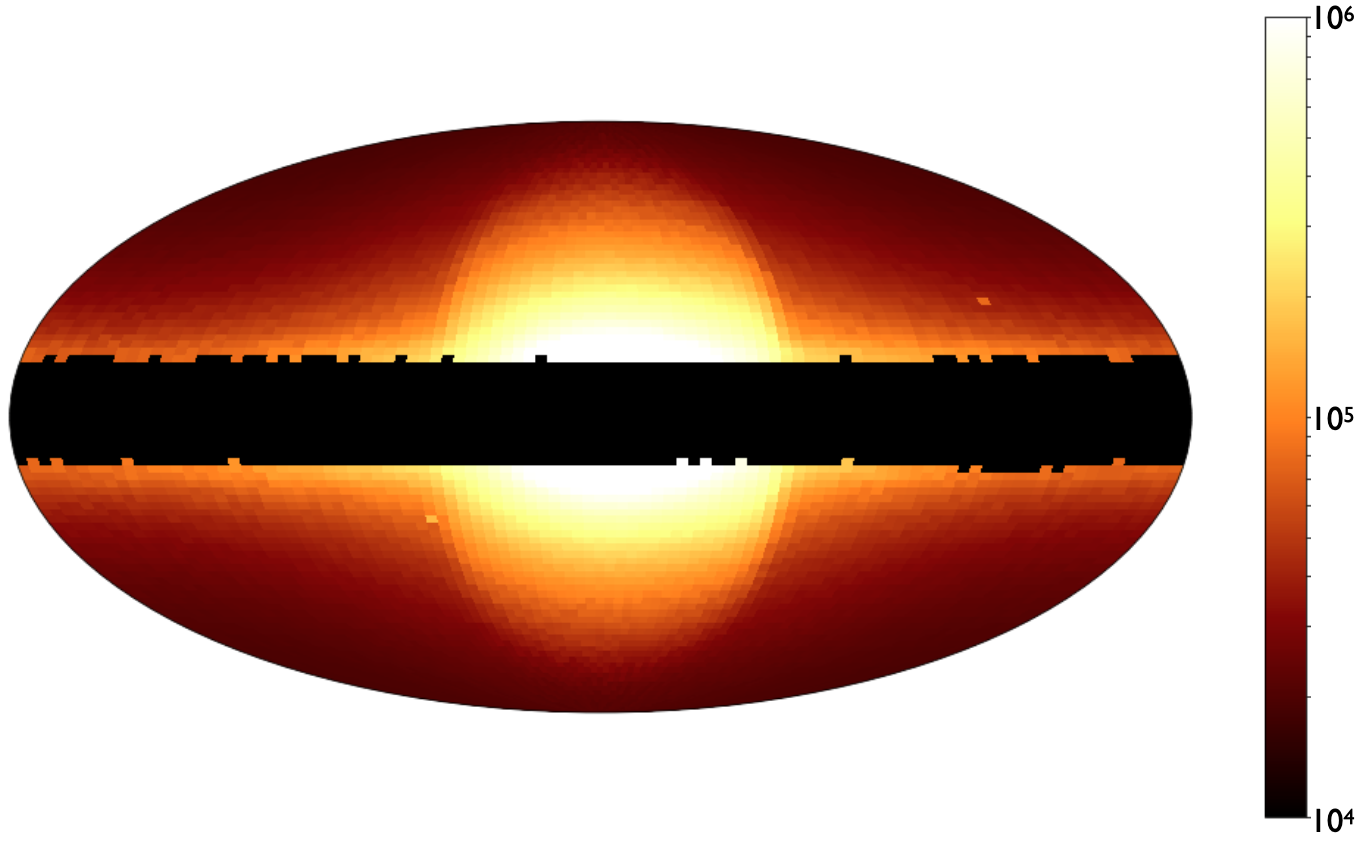}
  \caption{2D histogram plot of the Besançon simulation in galactic coordinates.  It is complete down to $\HE<26$, yielding a total of 3 billion stars. This simulation \corr{excludes} ${\sim}\,10\,{\rm deg}^2$ of the much denser galactic plane, even the reference survey described in \cite{euclid-survey-scaramella2021} describes a more complex and restrictive exclusion of the galactic plane.}
  \label{fig:besancon_sim}
\end{figure}

As \Euclid aims to observe thousands of square degrees in the optical and NIR down to a $\IE$ magnitude of 24.5, we require a synthetic model that accurately represents the distribution of stars in the Milky Way. We adopted the Besançon model \citep{besancon-robin1986, besancon_reference} as our main stellar population synthesis model of the Galaxy. The model provides a realistic representation of populations in the disc and halo down to M-type stars, as well as photometry in various bands. The scaling of the SED for the spectra reconstruction is performed with the 2MASS $H$ band. The model provides the surface temperature, gravity, and metallicity for each star, which we used to derive the stellar template from the Basel 2.2 stellar library \citep{basel-lastennet2002}.  Further details of the library and how we reconstructed spectra are explained in Sect.~\ref{s:sim_sedlib}. We ran an all-sky Besançon model simulation down to $H < 26$, avoiding the Galactic plane as \Euclid will not observe there and it contains too many stars at those magnitudes. Even though the NISP instrument has a defined magnitude limit of $\HE\sim24$, we simulated a deeper catalogue to include sources below the detection threshold into the pixel-level images, as these are also relevant in altering the photometry of other sources of interest, such as the main galaxy sample. This resulted in a massive final stellar catalogue containing 3 billion stars with the following set of parameters: (1) \textbf{ID}: A unique TU ID, which is non-identical to the ID used for galaxy sources. (2) \textbf{Sky coordinates}: The equatorial coordinates of the star in RA and Dec. (3) \textbf{Distance}: The radial distance of the source in kpc. This is used to scale brown dwarfs of L- and T-type for which the reference magnitude is given in absolute magnitude. (4) \textbf{Reference magnitude}: The apparent magnitude in the Vega system for the 2MASS $H$ band, which we use to scale the template spectra to its apparent brightness. (5) \textbf{Extinction}: The Milky Way optical extinction in the $V$ band, derived from 3D dust model maps \citep{3Ddustmaps-schultheis2014}. Like the galaxy catalogue, true fluxes in all \Euclid and EXT bands are also computed for the stellar catalogue. The density map of the Besançon simulation run can be seen in Fig.~\ref{fig:besancon_sim}.

%

\subsubsection{Augmented Tycho2 catalogue}
The design of the Euclid Wide Survey was performed using the reference zodiacal light model and straylight levels from the brightest stars near the galactic plane. For consistency, the brightest end of the stellar catalogue had to be replaced with real stars from an all-sky catalogue. We chose the Tycho2 stellar catalogue \citep{tycho2cat-hog2000}, as we could use the Pickles procedure described in \cite{tycho2match-pickles2010} to match stellar types for each source in the catalogue and then to our reference Basel 2.2 library \citep{basel-lastennet2002}. Our catalogue contains 2.5 million of the brightest stars in the sky down to $V$ magnitude of 11. 

\subsubsection{Brown dwarf stars}
\begin{figure}
    \centering
    \includegraphics[width=\linewidth]{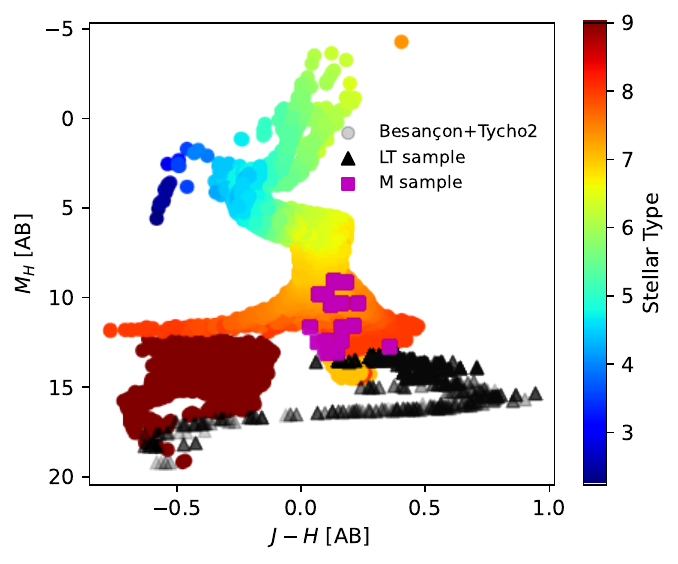}
    \caption{$M_H [\rm{AB}]$ versus $J-H [\rm{AB}]$ diagram of a simulated star catalogue. 
    The colour-coded sample combines real Tycho2 stars with $M_H \lesssim 10$, and a simulated Besan\c{c}on catalogues. LT-dwarfs (black) are simulated as well; however, M-dwarfs (magenta) are not as they are already contained in the Besan\c{c}on catalogues. }
    \label{fig:ccstars}
\end{figure}

One of the challenges in identifying very distant quasars is differentiating them from the point-like brown dwarf stars that have a red spectrum of similar NIR colours. To enable the testing of algorithms designed to perform more complex QSO identification, we supplemented our stellar catalogue with MLT type stars. While M-type stars were already included in the Besançon sample, the redder LT stars were not, so we added new templates to the Basel 2.2 library and included the brown dwarfs following recipes described in \cite{highqso-barnett2019}. Figure~\ref{fig:ccstars} shows the resulting colour-colour diagram of our sample of LT stars in addition to the various stellar types described previously.
%
\subsubsection{Gaia realisation}
\label{s:gaia_realisation}
The first prototypes of the SGS pipeline used the True Stellar catalogue as the astrometric reference for the calibration algorithms. However, this idealised catalogue will not be applied in the astrometric calibration of real observations; instead, \textit{Gaia} will be the reference catalogue used. With the astrometric and photometric performance described in the \textit{Gaia} DR2 \citep{gaia-dr2-2018}, we could replicate a realisation of a \textit{Gaia} catalogue in our TU sample, using realistic astrometric and photometric errors based on the magnitude and stellar type. \corr{Proper motions are not included in this simulation.} For each star in the catalogue, we note the position and magnitudes in $G$, $BP$, and $RP$ bands\footnote{$G$, $BP$, and $RP$ bands cover the wavelength range $400\,\rm{nm} < \lambda < 1\rm{\mu m}$  \citep{montegriffo-gaiadr3-2022}} down to $G \sim 21$, together with their respective errors. This method yields a more realistic Gaia-like astrometric reference catalogue, which is then provided to the respective calibration pipelines for processing VIS, NIR, and EXT data.

\subsection{Spectral energy distribution library}
\label{s:sim_sedlib}
As we shall explain in Sect.~\ref{s:euclid_instrument_simulators}, the VIS, NISP-P, NISP-S, and EXT pixel simulators are implemented as separate independent codes. However, developing different types of software that implement common routines poses some risks that could result in bugs and incompatible results. To minimise possible sources of error in the simulations, we constructed a common spectra reconstruction library tool called \texttt{SimSpectra} that reconstructs observed spectra from the parameters in the catalogue and the corresponding template library. The VIS, NISP-P, and NISP-S pixel simulators reconstruct the incident spectra every time a given source needs to be rendered. This allows the full chromatic information to be used precisely when simulating instrumental effects such as a wavelength-dependent PSF or quantum efficiency. Although it is computationally inefficient to reprocess the spectra for each source in the different dithers and channels, it provides greater practicalities than storing and transferring large catalogues of spectra. Reconstructing the spectra is a quick process (when optimised) that enables low-memory jobs to perform the processing in parallel under distributed architectures. Reconstructed spectra are only stored as a temporary validation product for NISP-S simulations.

In contrast to the \Euclid simulators, simulations for the EXT imaging surveys (described in detail in Sect.~\ref{s:ext_sims}) as well as Gaia(\textit{G}, \textit{BP}, \textit{RP}) are less complex and use pre-computed band fluxes; this is to avoid using spectra in the process of simulating pixels. Indeed, due to \corr{their large fields of view}, ground-based survey simulators usually have to include more galaxies than the \Euclid simulators. Computing spectra for each EXT survey would be computationally expensive and impractical. Instead, we computed the exact values in a matrix, sampling various points in the parameter space, and performed a multi-dimensional interpolation to obtain the integrated fluxes in all bands. This faster interpolation method has an accuracy of $0.005\,{\rm mag}$ root-mean-square (RMS) error in comparison to the full spectra reconstruction algorithm, which is sufficient for all SGS validation applications \citep{micemock-carretero2015}.

\begin{figure}[!h]
  \centering
  \includegraphics[width=\linewidth]{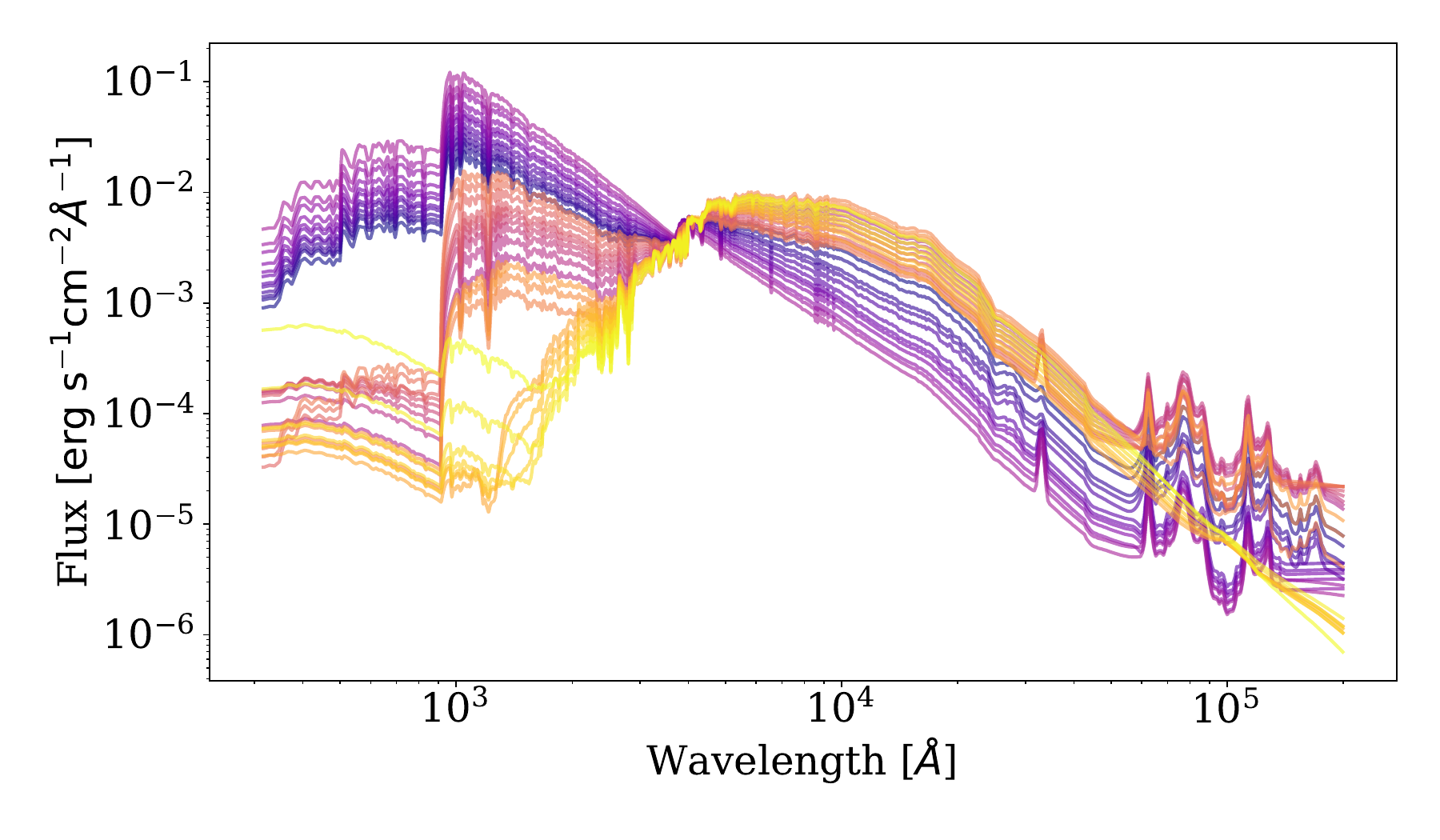}
  \caption{Reference template SEDs, used in combination with extra extinction and emission line prescriptions in order to reconstruct complete spectra.}
  \label{fig:cosmos_seds_dust}
\end{figure}
\begin{figure}
  \centering
  \includegraphics[width=\linewidth]{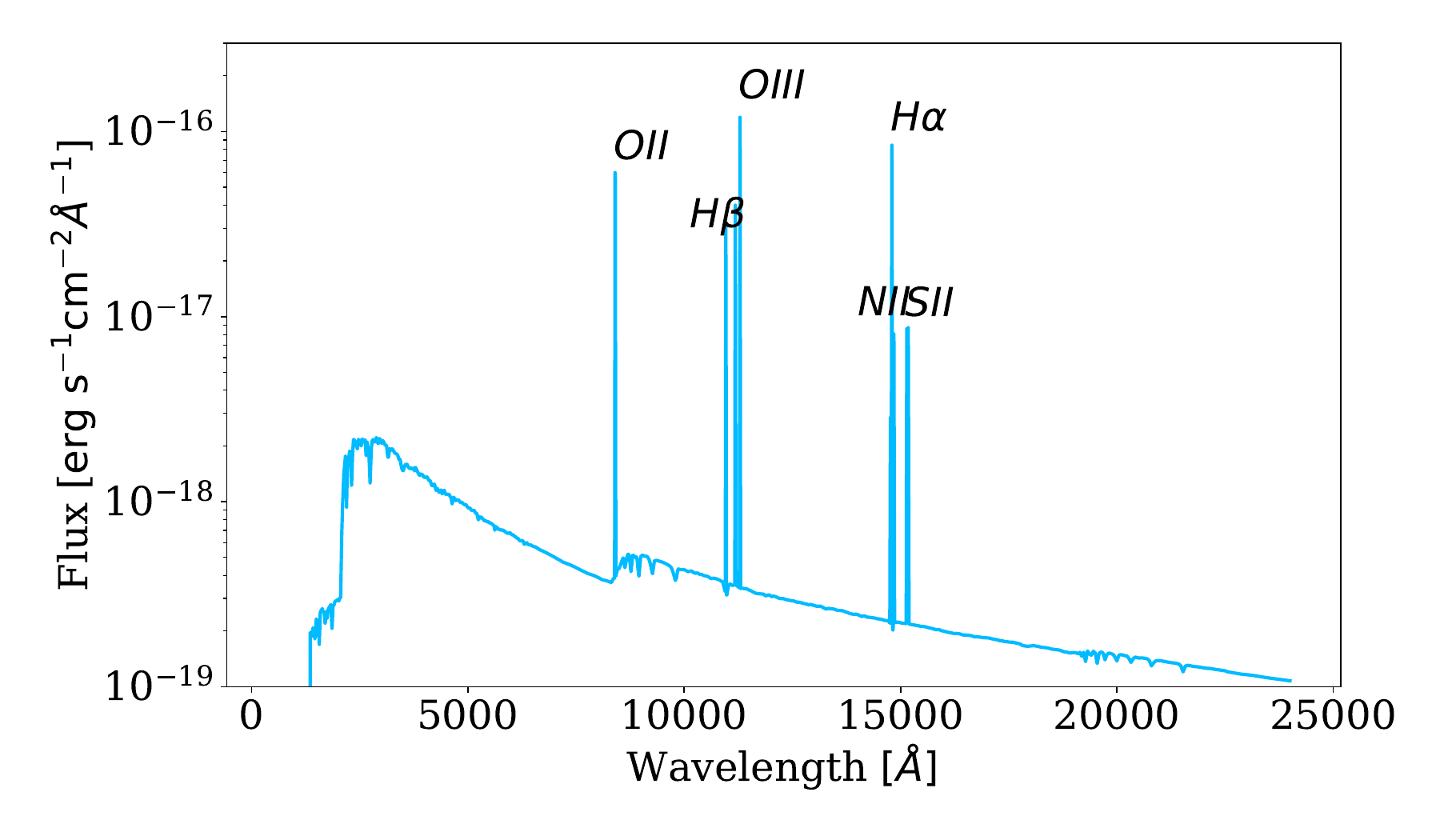}
  \caption{Complete reconstructed spectrum based on the SEDs of Fig. \ref{fig:cosmos_seds_dust} and parameter information from the True Universe catalogue (extinction, emission lines, redshift, lensing magnification)}
  \label{fig:tu_galaxy_spectra}
\end{figure}

\subsubsection{Galaxy spectra reconstruction}

Our base galaxy template library is composed of the COSMOS SED templates used in \cite{cosmos_ilbert2009} that comprise \cite{cosmosseds-bruzual2003} and \cite{cosmosseds-polletta2007}. The SEDs and the intrinsic dust extinction laws were selected from \cite{cosmosdust-prevot1984} and \cite{cosmosdust-calzetti2000}.
We used the luminosity, $g-r$ rest frame colour, and the redshift of each galaxy in the Flagship catalogue to assign the SED and dust extinction of the closest COSMOS galaxy in the \cite{cosmos_ilbert2009} catalogue in the luminosity-colour-redshift space. In order to avoid discrete distributions in the observed photometric properties, we applied scatter to the assigned closest COSMOS template, generating a random realisation of the value of the SED. \corr{The closest SED was identified as an optimised\footnote{\corrtwo{In order to assign the SED, we define a weighted distance between galaxies in the mock catalogue and galaxies in the COSMOS \corrthree{field} using redshift, absolute magnitude, and colour, $dist_{\rm weighted} = w_{\rm col}  | (g-r)_{\rm cosmos} - (g-r)_{\rm cat} |  + w_{Mr}  | Mr_{\rm cosmos} - Mr_{\rm cat} | +  w_{\rm z}  | z_{\rm cosmos} - z_{\rm cat} | $ where $(g-r)$ , $Mr$ and $z$ are the $g-r$ colour, the $r$-band absolute magnitude and redshift, respectively, and the subindex indicates whether it is \corrthree{a COSMOS} galaxy (\corrthree{cosmos}) or \corrthree{a} mock catalogue galaxy (cat). We arbitrarily assign the weights ($w_{\rm col}$, $w_{Mr}$ and $w_{z}$) taking into account the typical values of the quantities and their errors. We choose the SED of the \corrthree{COSMOS} galaxy that has the closest weighted distance ($dist_{\rm weighted}$ ) to the galaxy in the catalogue. To avoid galaxies having the same template, we add scatter to the index and extinction values of the closest template.}}
 function based on colour, luminosity, and redshift}. We assigned the two SEDs closest to the resulting realisation value, weighted by their distance. Each galaxy SED was then constructed as a linear combination of these two SED templates (see Appendix~\ref{a:sed_fluxes}).The galaxy template set used for these simulations is shown in Fig.~\ref{fig:cosmos_seds_dust}.  

We then assigned a star formation rate (SFR) to each galaxy from its unextinguished rest-frame UV luminosity (computed from its SED). \corr{These relations have been derived at low redshift. Most of them depend on physical processes that are not expected to depend on redshift, and therefore have been extrapolated to higher redshifts.} We used the Kennicutt 1998 relation \citep{sfr_kennicutt_1998} to assign an $H\alpha$ luminosity from the SFR with some scatter. We finally corrected the H$\alpha$ luminosity to make the global H$\alpha$ luminosity distribution resemble the \cite{halpha-pozzetti2016} models. We assigned the luminosities of the other most prominent lines from their SFR and H$\alpha$ luminosities using observed correlations. An example of a reconstructed spectrum (including the continnuum SED and the emission lines) from the catalogue parameters is shown in Fig.~\ref{fig:tu_galaxy_spectra}. 

\subsubsection{Stellar spectra reconstruction}
As is stated in Sect.~\ref{s:euclid_stellar_cat}, our stellar library was generated from the Basel 2.2 \citep{basel-lastennet2002} template set. We assigned an SED from both the Besançon model simulation and the Tycho2 catalogue to a particular stellar template. Reconstruction of the stellar spectra is much simpler than for galaxies, since there is no redshift, K correction, or lensing magnification in the reconstruction process. There are only two steps to calibrate the template to the incident calibrated spectrum for each star. First, we needed to scale the flux to our reference $H$ 2MASS band. Second, we corrected for the Galactic extinction. The Tycho2 stars were not corrected for extinction because their flux is observed, and therefore already contains the extinction factor.
In contrast, as Besançon stars are simulated, we computed an accurate reddening with the radial distance information provided by the model and the 3D dust model from \cite{3Ddustmaps-schultheis2014}.
A detailed description of our implementation of stellar spectra reconstruction is given in Appendix~\ref{a:sed_fluxes_stars}. Finally, the stellar library also contains 365 LT star templates that enable the additional brown dwarf stars to be assigned an SED, as the Basel 2.2 library does not contain this class of stars.

\begin{figure*}[t]
  \includegraphics[width=\linewidth]{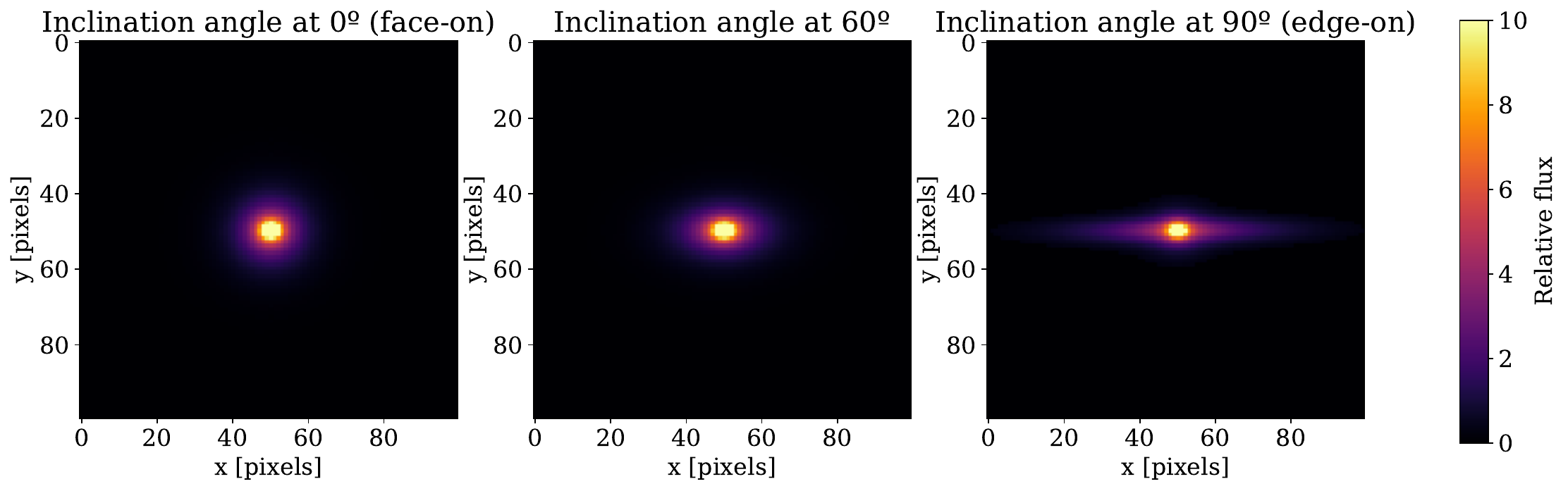}
  \caption{Sample of three spiral galaxies at different inclination angles ($0$\textdegree, $60$\textdegree, and $90$\textdegree from left to right). The inclined exponential model better represents the three-dimensional shape of the disc, with a more realistic thickness than a S\'ersic  profile with index $n=1$.}
  \label{fig:inclined_exp}
\end{figure*}

\subsection{Galaxy profile generators}
\label{s:sim_thumbnaillib}

For our suite of instrument simulators, we built a package called \texttt{SimThumbnails} that transforms galaxy morphological parameters to pixelised galaxy light profiles. In addition, it performs the rendering of the light profiles at the exact pixel scale for each instrument using the `World Coordinate System' (WCS) information of the target image. The core of the library is based on \texttt{GalSim} 2.3.3 \citep{galsim-rowe2015}, which allows plenty of flexibility to simulate different profile configurations. Similar to the \texttt{SimSpectra} library, the \texttt{SimThumbnails} library implies rendering the same galaxies for every pixel simulator, which is inefficient computationally. However, it is a trade-off between pre-computing and storing images for all objects. In addition, every simulator can tune the library parameters to the instrument pixel or sub-pixel scale (see, e.g. Sect.~\ref{s:nispp_sim}).

\subsubsection{Bulge and disc model}
\label{ss:bulge_disk_model}
For the standard Flagship galaxies, we adopted two S\'ersic profiles to trace the bulge and disc components. First, we defined the bulge as a S\'ersic profile:
\begin{equation}
    I_{\rm bulge}(r) \propto \exp ({-b_{n} (r/r_{\rm e})^{1/n}})\, ,
\end{equation}
\noindent where $r_{\rm e}$ is the half-light radius and $n$ is the S\'ersic  index given in the Flagship catalogue. The $b_n$ parameter was calculated to give the correct half-light radius given $n$. The S\'ersic index was determined from an empirical fit to the HST-CANDELS data \citep{candels-grogin2011,candels-koekemoer2011,candels-dimauro2018}, HST-GOODS-South data \citep{welikala2012}, and HST-COSMOS data \citep{cosmos-laigle2016}. The bulge model, $I_{\rm bulge}(r)$, was then sheared to obtain the bulge minor-to-major axis ratio \corr{due to its intrinsic ellipticity} as given in the catalogue.

Next, we initialised the disc component with an exponential thick disc inclined model \citep{vanderkruit1982,bizyaev2007}. The exponential inclined model is a particular S\'ersic profile of index $n=1$ characterised by three parameters: the inclination angle (where $0^{\degree}$ refers to a face-on galaxy and $90^{\degree}$ would indicate it is positioned edge-on), the half-light radius, and the scale height of the disc. We chose a fixed height-to-radius ratio, $h_{s/r} = 0.1$, that reproduces observations well. The relation between the inclination angle and the apparent disc axis ratio, $b/a$, is

\begin{equation}
    \theta_{\rm inclination} = \arcsin{\sqrt{ \frac{1-\left(\frac{b}{a}\right)^2}{1-h_{{\rm s/r}}^2}  }}\,.
\end{equation}
The advantage of this profile is that it accurately replicates the thickness of the disc, which the S\'ersic profile does not properly represent when the galaxy is edge-on. Then, we adjusted the flux of the disc relative to the bulge components with the bulge-to-disc ratio parameter given in the Flagship catalogues. The distribution of this parameter reproduces the ratio measured in the data detailed above \corr{\citep{cardamone2010, candels-grogin2011}}. Finally, the bulge and disc models were summed together and rotated to a position angle on the sky (with a convention of north up, increasing towards the east) given in the catalogues.  Figure~\ref{fig:inclined_exp} shows an example of this bulge and disc model for various inclination angles.

\subsubsection{Gravitational lensing and optical distortions}
We used \texttt{GalSim} to apply the gravitational lensing effect to every galaxy (i.e. the summed bulge and disc components) based on the shear, $\gamma_1$, $\gamma_2$, and convergence, $\kappa$, parameters given in the Flagship catalogues.
\texttt{GalSim} requires the following derived parameters: the reduced shear parameters, $g_1$ and $g_2$, responsible for the distortion effect, as well as the magnification factor, $\mu$, defined, respectively, as
\begin{equation}
    g_i = \frac{\gamma_i}{1-\kappa}
\end{equation}
and
\begin{equation}
\label{eq:magnification_factor}
    \mu = \frac{1}{(1-\kappa)^2 - (\gamma_1^2 + \gamma_2^2)} \, .
\end{equation}
While \texttt{GalSim} can only compute a reduced shear for which $g_1^2 + g_2^2<1$, this is not a limitation as we are mostly interested in the weak lensing regime which falls well below this limit. As was stated above, lensing magnification alters the size of a galaxy. \corr{As True Universe fluxes already include magnification}, we have to apply the inverse of the magnification factor on its incident spectrum (as is explained in Sect.~\ref{s:sim_sedlib}), so that its total integrated flux matches that given in the catalogues. 

The final shape of the galaxy is critical for estimating the weak lensing signal. Instrumental distortions add up to the lensing distortions described above. We accounted for optical distortions of the field as defined by the plate WCS. At every particular position in the focal plane, we associated with each \texttt{GalSim} galaxy object a WCS $2\times 2$ local Jacobian matrix that was subsequently used during the simulation of the galaxy on the image.

\subsubsection{Flux scaling and truncation}
In theory, analytic expressions for the bulge and disc profiles extend to infinity. However, in practice, only limited size stamps can be pasted on the image. This size is automatically computed by \texttt{GalSim}, such that the flux lost outside the stamp is equal to the `folding threshold' parameter, described in \cite{galsim-rowe2015}. The pixel values are then increased to match the desired flux. In our pipeline, we set the folding threshold to 0.5\%, the default \texttt{GalSim} value. In most cases, where the S\'ersic index is high, the stamp contains almost all the flux as it is concentrated in the centre. For low-$n$ S\'ersic profiles where the light decays slowly, the flux loss can be closer to 0.5\%; however, in such extended profiles it will be very difficult to recover the total light with photometric measurements.

\subsubsection{Point spread function convolution}
The convolution of the galaxy profile by the PSF was performed in Fourier space with the FFT method. In \texttt{GalSim}, the profile is centred in the middle of the central pixel by default. Yet, the true position of the object may lie in a slightly shifted fraction of it. The offset to the central pixel was computed and provided for an accurate positioning of the source.

Depending on the instrument, the simulator pipeline allows for a polychromatic PSF model that varies in wavelength, such as in the VIS instrument. Due to limited information in the Flagship galaxy mock catalogue, the galaxy profile is rendered without colour gradients. Therefore, the polychromatic PSF was convolved only once with the galaxy stamp, and collapsed along the wavelength direction. This provides very accurate PSF convolution based on the spectra of the source. 

\subsubsection{Optimisation}
The computing time of our simulations was driven by the processing of the galaxy stamp and PSF convolution. Therefore, optimising these two steps allowed us to perform larger and more complex simulations. We found three ways to speed up this process: first, we enabled a mode whereby only the size of the stamp was provided without requiring any complex calculation; this permitted us to quickly discard sources that did not overlap with the detector pixel array (\corr{from the radial search of the catalogue to the rectangular shape of the pixel array}). As a second step, we rounded the variable S\'ersic index in the galaxy bulges to the first decimal digit, \corr{as the computation of the 2D S\'ersic profile is computationally expensive}. With this approximation, many galaxies share the exact same profile and a look-up table in \texttt{GalSim}, allowing us to reuse the internal 2D profiles to speed up the rendering process, together with the specific shear and PSF model. The impact on the simulated profile is below the required precision on the flux. And finally, faint galaxies entering in the undetected background sample ($\IE^{\mathrm{limit}} \corrthree{<} \IE < \IE^{\mathrm{limit}} + 2$) were replaced by single component models that require less computation. These changes have almost no effect on the quality or the image analysis, reducing the computation time by several orders of magnitude. 

\begin{figure}[!t]
  \centering
  \includegraphics[width=\linewidth]{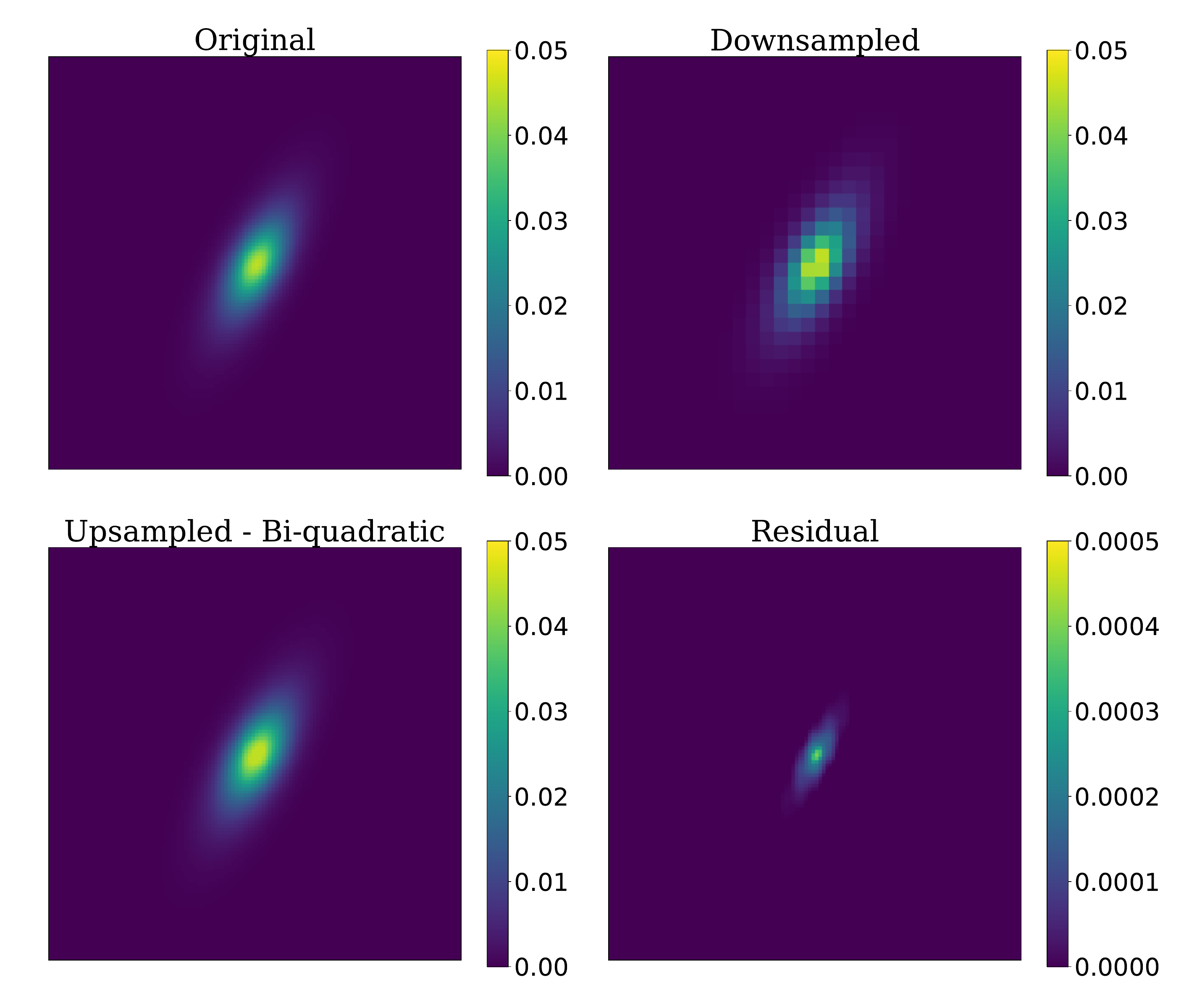}
  \caption{Down-up sampling iterative process used to allow the simulation of very large galaxies. \emph{Top left}: Original high-resolution sample. \emph{Top right}: Downsampled version. \emph{Bottom left}: Upscaled version from bi-quadratic interpolation. \emph{Bottom right}: Residual image between original and upscaled version. In this example, we show a smaller sample of 120 pixels in width for visualisation purposes, but the galaxy stamps that require to be resampled with this process are typically above one thousand pixels wide. Shown is the result of downsampling by a factor of 4; the bi-quadratic 2D interpolation results in a very similar profile in comparison to the original and it minimizes the residual compared to other interpolation methods.}
  \label{fig:downupsample}
\end{figure}

When profiles deliver a very large galaxy stamp (larger than a few thousand of pixels in diameter), the memory necessary to perform the convolution exceeds the limit established by the environment and raises an exception. To handle such cases, we implemented a down-up sampling iterative process. When the memory limit is reached, we down-sample the actual stamp by a factor of 2, providing an equivalent WCS with a scale multiplied by that same factor. Every time the memory limit is reached, we further down-sample the stamp (although most issues have been resolved during the first iteration). Once the stamp is successfully imaged, we up-sample the image with a bi-quadratic interpolation. We minimise the residuals for these types of profiles using this approach. Even though some degradation can occur during the down-up sampling process, this event is rare and happens only for very large Flagship galaxies (above $10"$ effective radius) with a high S\'ersic index. It is not a concern as these sources are not among the list of targets for the main science goals of \Euclid.  We present an example of the performance of the down-up sampling process in Fig.~\ref{fig:downupsample}.

High-$z$ galaxies are drawn with a single S\'ersic profile component of index $n=1.5$. QSO sources are expected to be point-like and we render them using a PSF profile. Finally, strongly lensed galaxies require extra complexity in rendering that is implemented via the generation of stamps at $\ang{;;0.05}$ resolution with the \texttt{GLAMER} code (see Sect.~\ref{sec:sl}). The stamps are then resampled at the instrument resolution by each simulator using the default `Quintic' interpolant in \texttt{GalSim}. 

\section{Mission database}
\label{s:mdb}

In such a complex mission, having a robust data and configuration management system is essential. The instrumental models, reference sky maps, telescope parameters, survey plan, etc., will all evolve through the different phases and maturity of the mission. The \Euclid mission Database (MDB) is a single reference repository designed for the \Euclid mission system; it provides a temporal record of the official representation of the system consistent with the set of reference models. As such, the MDB provides a centralised version control and distribution system of parameters, which is required by all SGS components.

The MDB has the following set of configurations which we select from and assign to the model that we want to simulate: (1) \textbf{as required}: This specifies the worst-case values still within the related requirements. These derive from the top-level requirements that ensure \Euclid can achieve its scientific goals. (2) \textbf{as designed}: This provides the expected performance of a given model with the current design. The model must perform better than the required limit. (3) \textbf{as built}: At this level, the model has been measured in built parts, ideally around the design value. (4) \textbf{current best estimate (CBE)}: This replaces the `as built' model with newer or more accurate measurements that contain our best knowledge of the model. Ideally, the simulation will be done with all CBE models, which best represent the status of the system. It is foreseen that after commissioning and in-flight calibrations, CBE will be updated with more accurate instrument models. (5) \textbf{as simulated}: This is a particular configuration used for testing, where the model is defined with a given value, which may be unrelated to a particular measurement or design. However, we generally avoid implementing unproven, hypothetical models. 

When preparing a production of simulations for the various instruments, we define a Mission Configuration (MC), which gathers a subset of models from the MDB with a unique status on each parameter. The MC is stored in the EAS along with the simulation data products to provide users knowledge of the true models used as input.

\subsection{Models and parameters}
\label{s:mdb_models}

The MDB models and parameters allow us to emulate \Euclid (and EXT surveys) operations during sky observations. As explained previously, the models in the MDB continually evolve with the construction and testing of the instruments and telescope. At the time of pre-launch, we had a sufficiently accurate model of system performance derived from ground tests.  Therefore, we prioritised the models labelled with `current best estimate' (CBE), as these would provide the closest representation of the spacecraft during operations. 

The MDB is organised in sections, each covering a specific component of the mission. The \textit{Environment} section contains physical constants, ensuring all calculations use the exact same reference values for consistency. Other \textit{Environment} models assumed in the definition of the survey include the zodiacal light scattered from interplanetary dust in the Solar System, all sky out-of-field straylight maps induced by stellar light, intergalactic dust extinction maps, and the cosmic ray energy distribution.

The \textit{Space Segment} section contains all known models for the PLM, SVM, VIS, and NISP instruments. The SVM subsection lists models associated with the spacecraft such as the dither steps or slew duration. The PLM subsection stores all telescope related information like wavefront errors, collecting surface areas, or plate scales, as examples. The NISP and VIS instrument sections provide all parameters and models describing the instruments such as detector layout, quantum efficiency, read noise, Charge Transfer Inefficiency (CTI) models, wavelength dispersion and distortions, etc. 

\begin{figure}[!t]
  \centering
  \includegraphics[width=\linewidth]{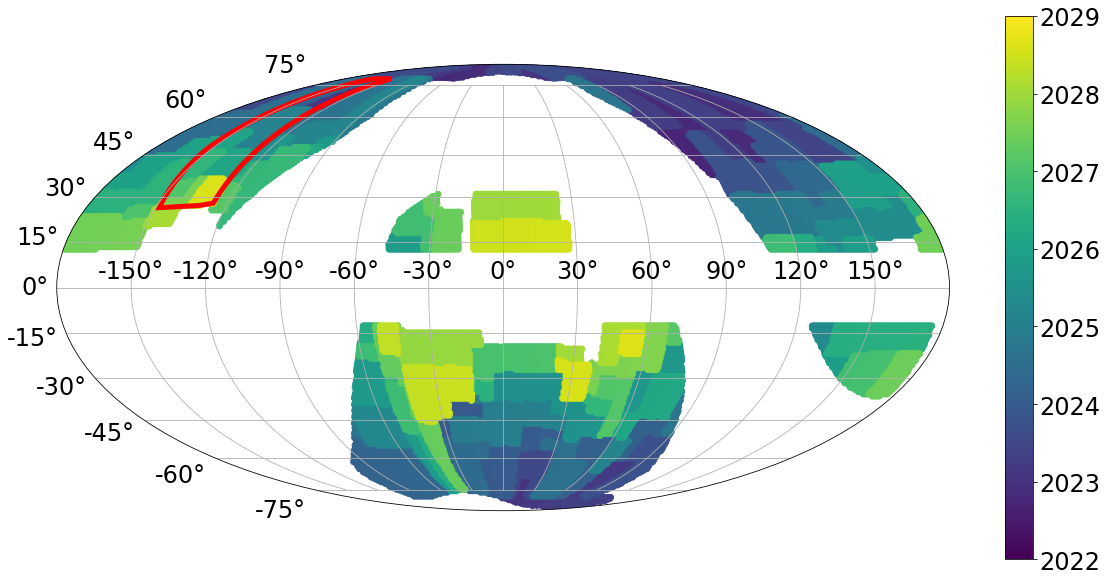}
  \caption{Reference Euclid Wide Survey at SC8 \citep{euclid-survey-scaramella2021} covering $15\,000\,{\rm deg}^2$ of the sky in Ecliptic coordinates, avoiding the Galactic and Ecliptic planes due to the high stellar density and high zodiacal light, respectively. The colour code indicates the observation time where \Euclid is planning to point at each of the fields. The polygon in red shows the original SC8 main area (${\sim}\,600\,{\rm deg}^2$). More recent surveys have been released with further improvements and the updated launch date.}
  \label{fig:wide_survey}
\end{figure}
\begin{figure*}[!t]
  \centering
  \includegraphics[width=\linewidth]{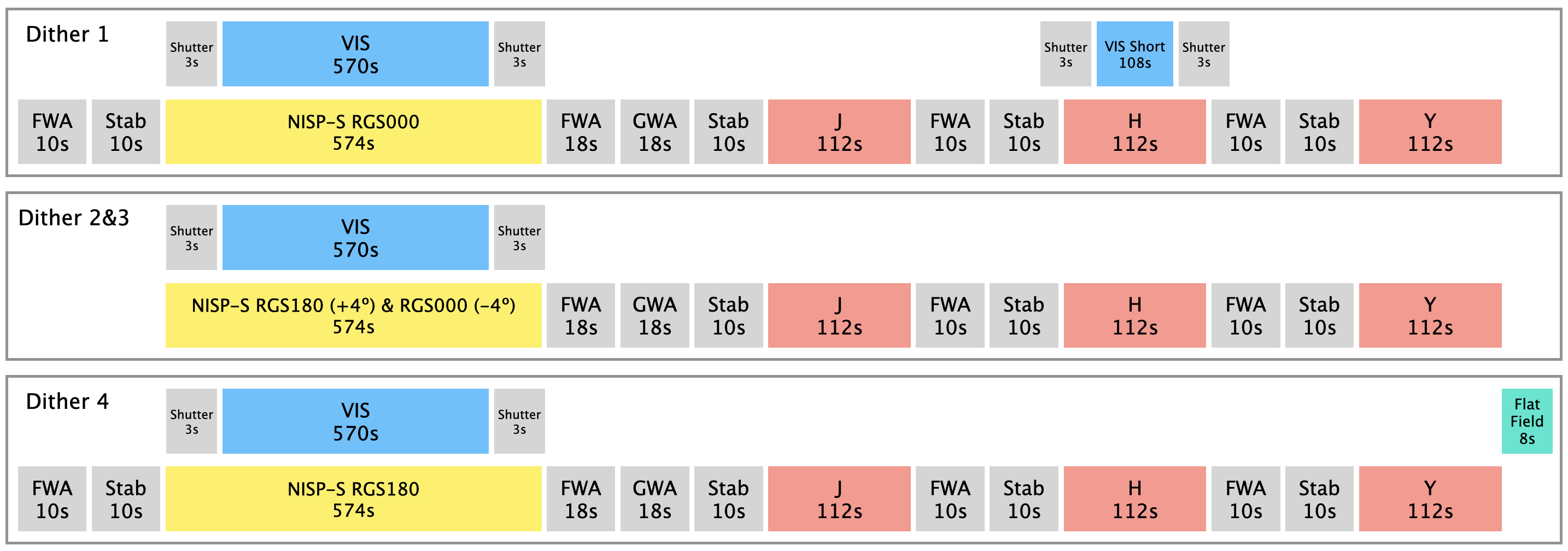}
  \caption{Simplified schema of the reference observing sequence (ROS) in the four dithers for the VIS and NISP channels. At the beginning of each dither VIS and NISP-S observe simultaneously after the Filter Wheel Assembly (FWA) movement and few seconds to allow the telescope to stabilise. After the exposure, the Grism Wheel Assembly (GWA) movement lets the light pass for the filter observations of $\JE$, $\HE$, and $\YE$, \corr{denoted as JHY in red in the figure}. In the meantime, the VIS instrument will perform the readout and processing of the acquired mosaic image. The first dither includes an additional short VIS exposure while the last dither ends with a brief flat field frame.}
  \label{fig:wide_sequence}
\end{figure*}

The \textit{Survey} section defines the Euclid Wide Survey, the Euclid Deep Survey,  and the calibration surveys, as well as the calibration source SEDs for objects such as Vega, a white dwarf, and a planetary nebula.  We describe briefly in the following section the reference survey which we use to emulate realistic \Euclid observing sequences.  

\subsection{The reference survey}
\label{section:ref_survey}
The \Euclid reference survey \citep{euclid-survey-scaramella2021} was defined to detail the set of fields to be observed within the entire \Euclid wide footprint, the dither steps within each field, time of observation of each pointing as well as the various calibration fields and the deep fields during the six-year mission campaign. With this information, OU-SIM can reproduce a realistic sequence of observations covering the area to be simulated. Figure~\ref{fig:wide_survey} shows an image of the fields in the reference survey that cover the Euclid Wide Survey footprint.  We outline in red the region denoting the SC8 footprint.

\paragraph{Fields and calibration frames}
\Euclid has defined four dithers inside each field, where each pointing results in $\IE$, $\YE$, $\JE$, and $\HE$ images, and one spectral exposure. The NISP-S dithers alternate different grisms at different dispersion angles that enable spectral decontamination. We show the reference observing sequence (ROS) we implemented for SC8 in Fig.~\ref{fig:wide_sequence}. We note, however, that the ROS shown is not the definitive one; for example, the true ROS is scheduled to have a second short VIS exposure.  Nevertheless, simulating the ROS provides a realistic representation for spatially varying instrumental effects. 
Moreover, multiple exposures allow us to cover the detector gaps, reach the desired depth, and correct for transitory effects such as cosmic rays. We implemented the dither slews following the pointing accuracy model of the spacecraft with minimal step jumps plus the associated pointing uncertainty of $11"$. The current dithering model \citep{spectra_selfcal_markovic_2017} follows an S shape and implements the following minimal slews:

\begin{itemize}
  \itemsep0em 
  \item  Dither step 1: $\Delta X_{\rm sc}$ = $\ang{;;50}$, $\Delta Y_{\rm sc}$ = \ang{;;100};
  
  \item  Dither step 2: $\Delta X_{\rm sc}$ = $\ang{;;0}$, $\Delta Y_{\rm sc}$ = \ang{;;100};
  
  \item  Dither step 3: $\Delta X_{\rm sc}$ = $\ang{;;50}$, $\Delta Y_{\rm sc}$ = \ang{;;100};
\end{itemize}
plus a random error of $\ang{;;11}$ at 3$\sigma$ after the minimal step. The coordinates $X_{\rm sc}$ and $Y_{\rm sc}$ correspond to the $X$ and $Y$ axis in the spacecraft reference frame. 

\subsection{External surveys}
\label{s:ext_surveys}
The MDB also contains survey, instrumentation, and bandpass information for each EXT survey (i.e. DES, \textit{Rubin}, CFIS, JEDIS, Pan-STARRS, and WISHES, which will supplement the \Euclid mission; see Sect.~\ref{s:ext_sims} for more information). In detail, a survey table containing a list of selected pointings along with additional survey parameters (such as magnitude zeropoints, sky brightness, position angle, seeing FWHM for constant PSFs or PSF model files for variable PSFs) for each EXT survey are stored in the MDB. The instrument file describes the detector layout in the focal plane for each camera, as well as characteristics of the CCDs such as the pixel size, read noise, well capacity, and the gain per CCD amplifier.  Finally, for each respective EXT survey, the MDB keeps a record of the total throughput curves per filter including the atmospheric transmission.  

In the following section, we shall present the various detector, atmosphere, and optical models used for producing realistic pixel images for each of our survey channels: VIS, NISP-P, NISP-S, and EXT.  In Appendix~\ref{a:throughputs}, we mention briefly the methods and resources for acquiring each of the respective transmission curves. More details on the construction of SC8 EXT survey table are shown in Appendix~\ref{a:sc8_main_survey_files}. 

\section{Instrument simulators}
\label{s:euclid_instrument_simulators}

In this section, we describe the implementation of all four simulator codes in the OU-SIM pipeline: \texttt{ELViS} (\Euclid VIS), \texttt{Imagem} (\Euclid NISP-P), \texttt{TIPS} (\Euclid NISP-S), and \texttt{SIM-EXT} (EXT ground-based surveys). These image simulators were originally built for various purposes, in particular, to provide the first evaluations of the mission performance.  However, they have significantly evolved to become the main pixel simulators of the SGS. All simulators integrate the TU star and galaxy catalogues, the common \texttt{SimSpectra}\footnote{\texttt{SIM-EXT} does not integrate \texttt{SimSpectra} as all effects are monochromatic within each band. However, the fluxes imported by \texttt{SIM-EXT} are produced with the \texttt{SimSpectra} library ensuring that all simulated channels have consistent fluxes.}, and \texttt{SimThumbnails} libraries, the reference survey, and the various models in the MDB.

\subsection{VIS simulator}
\label{s:vis_simulator}

\texttt{ELViS} is the \Euclid VIS simulator capable of accurately reproducing the data that will be obtained by the VIS instrument during the \Euclid mission operations, including the required calibration frames; for example, bias, flats, and self-calibration sequences. However, the biggest challenge for this simulator is emulating the optical response of the system and the instrument with the precision required. \Euclid plans to measure with unprecedented precision the weak lensing signal of billions of galaxies.  To achieve such a task, the VIS instrument needs to deliver extraordinarily good and stable image quality. The VIS simulator is required to deliver images with sufficient precision, such that simulation or model limitations do not drive the quality of the image. Moreover, it is the instrument performance as well as the shear measurement algorithms which define how well we can estimate the shape of the galaxies.

\paragraph{Input sources}

\texttt{ELViS} includes all TU stars, comprising the real Tycho2 bright stars, the model Besançon stars, and the MLT type brown dwarfs. Similarly, it renders the TU galaxies from the Flagship simulation, high-$z$ quasars and galaxies as well as strongly lensed galaxies. Recently, we included Solar System Objects (SSO), asteroids that due to their fast and nearby motion appear as traces in the VIS focal plane. Finally, the simulator includes cosmic rays (CR) during integration and readout, which leave substantially different traces. CRs are critical in this simulation due to their high density resulting from the higher levels of radiation in space and to the CTI (explained later in the Detectors subsection). They have a significant effect on image quality, and therefore, impact the ability to measure lensing parameters with high precision. Cosmic rays are simulated in the VIS channel as single pixel thin straight tracks generated from samplings of the posterior probability distributions (PDFs) of the track length and energy deposit. These PDFs are computed using a complex Monte Carlo cosmic rays simulator code \texttt{STARDUST} \citep{stardust_rolland2007} developed at The National Centre for Space Studies (CNES). The STARDUST code input data are the geometrical and physical description of the detector structure, the environment particle spectra, and an aluminum equivalent thickness table describing the shield around the detector. Full scale images are generated out of which the PDFs are evaluated. 

\paragraph{Background models}

There are five main contributions to the background in the VIS channel. (1) Zodiacal light, a diffuse background from sunlight scattered off interplanetary dust. This is based on an MDB model that provides the intensity as a function of ecliptic latitude and its average spectrum. (2) Thermal irradiance, a background signal due to radiation from the heat of the various elements of the telescope and instruments. (3) Diffuse scattered light, unwanted reflections at the telescope and optical path from the stellar sources in the field that illuminate the focal plane resulting in an increase of the background. The intensity of this signal is related to the density of bright stars in the field of view and thus increases at lower Galactic latitude. (4) In-field straylight, while similar to diffuse scattered light, contributes to an increase in the localised background due to very bright stars following a specific path in the optical system. (5) Optical reflections, also known as ghosts, an out-of-focus image of an on-sky source inside or in the immediate vicinity of the instrument field of view. It mainly depends on the light path through the dichroic and telescope baffling and the position and brightness of the sources.  While the first three components are constant across the focal plane and increase in value depending on the telescope's thermal state or its pointing in the sky, the last two create abrupt changes in the background which complicate the calibration and photometry.

\begin{figure}[!t]
  \centering
  \includegraphics[width=0.5\linewidth]{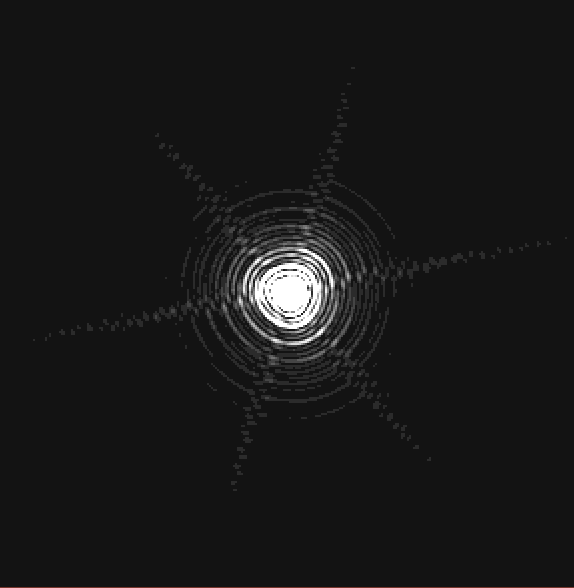}
  \caption{Single monochromatic VIS PSF sample out of the 460\,000 models that cover the entire focal plane at $250 \times 250$ positions at eight wavelengths spanning the VIS passband. Each PSF has a resolution of $300\times 300$ pixels sampled every $4\,\mu{\rm m}$ (a third of a VIS pixel).}
  \label{fig:vis_psf}
\end{figure}
\begin{figure*}[!t]
  \centering
  \includegraphics[width=\linewidth]{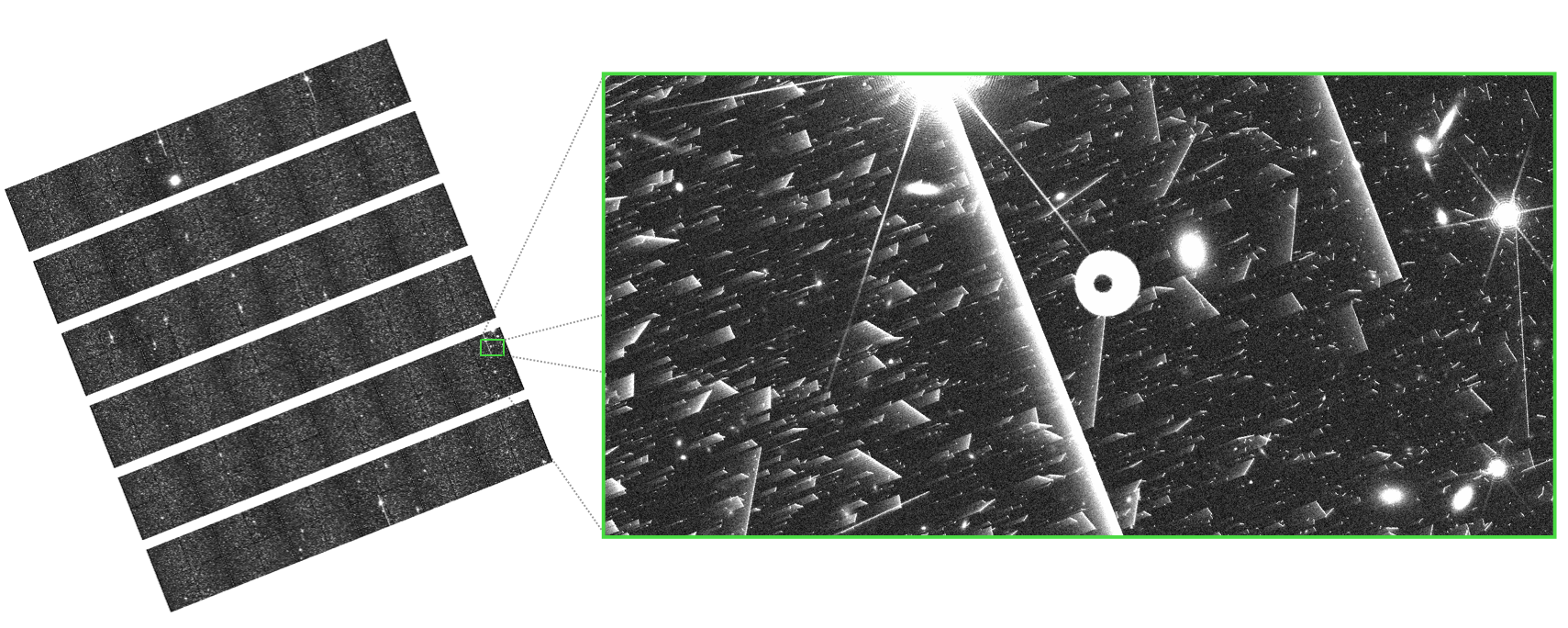}
  \caption{Simulated VIS exposure. The left image shows the entire $6\times 6$ mosaic of detectors and the right image contains a zoomed in fraction of a quadrant, where the multiple effects can be seen such as bright stars, galaxies, cosmic rays, a complex PSF model, a ghost from a nearby star, saturation bleeding, and the trails of the Charge Transfer Inefficiency (CTI) increasingly visible in the readout direction. In this image, the end-of-life model for the CTI was adopted. A linear scale around the median with only 30 ADUs of range was used to visualise the subtle instrumental effects present in the simulation.}
  \label{fig:vis_sim_image}
\end{figure*}
\paragraph{PSF}

As was mentioned, VIS simulations require great care in the reconstruction of the PSF, as it distorts the true shape of the galaxies. This needs to be simulated accurately to estimate the weak lensing signal with the precision required. In the VIS channel, we have four main contributions that affect the final PSF. First, we have the optical PSF, which varies as a function of position in the focal plane and wavelength. The model for the PSF is composed of an array of $240\times 240$ positions in the focal plane, each at eight wavelength positions, comprising a total of 460\,800 PSFs of $300\times 300$ pixels. Figure~\ref{fig:vis_psf} shows an example of a single model at one wavelength. These were constructed using \texttt{SHE\textunderscore PSFToolkit}, a software developed by the Organisation Unit for Shear Data (OU-SHE), which performs forward-modelling and generation of a broadband PSF using exit pupil wavefronts.  The inputs into \texttt{SHE\textunderscore PSFToolkit} were derived from the Zemax\footnote{\url{https://www.zemax.com}} optical model of the telescope (neglecting, however, polishing errors, the effect of the dichroic and the high-frequency components on the surface). Such large and detailed PSF datasets were built to meet the accuracy requirements necessary to measure weak lensing at the target precision. 

The VIS PSF model cannot be interpolated spatially due to its complex shape variation (at small and large scales). Therefore, when we draw a particular star or galaxy, we take the closest PSF to its focal plane position. Each PSF is generated from the Zemax model at eight wavelength values within the VIS passband, and the sampling is incresed by interpolating up to $1\,{\rm nm}$. With the reconstructed spectra from the source, we convolve the light profile with the PSF at each interpolated wavelength position, resulting in a chromatic PSF that varies as a function of the colour for each object.  Secondly, we have the AOCS described earlier in Sect.~\ref{s:satellite}. This complex system performs tiny corrections to keep the spacecraft attitude stable during an exposure, which slightly increases the width of the final PSF. The jitter of the AOCS can be replicated as a time series that mimics how the correction behaves. However, we model its final effect on the VIS PSF by convolving with a Gaussian of $\ang{;;0.025}\,{\rm RMS}$.  The detector degrades the PSF by diffusing the charges in each pixel into its adjacent pixels. This effect can be described in the Pixel Response Function, which gives the distribution of fluxes around each pixel. Due to variations in the CCD thickness, charge diffusion is not constant across the field of view. 

\paragraph{Optics and system}

The combined optical system of the PLM and the VIS instrument is accurately reproduced in \texttt{ELViS}. First, we have the response of the optical system which comprises the reflection of the mirrors and the transmission of the dichroic element. The PLM not only affects the transmission but its geometry distorts the incoming light beam, modifying the plate scale as a function of focal plane position. This results in objects being displaced from their nominal position and a non-uniform illumination of the VIS array. Geometric distortions also affect the shape of the galaxies which then require an adjustment during the drawing process with \texttt{SimThumbnails}. 

Furthermore, we emulate the detector array mount on the focal plane, with accurate gaps, misalignments and rotations as measured in ground tests. This adds further realism and complexity in the astrometric solutions derived from the calibration process. Additionally, we include a calibration unit (VI-CU) with its highly non-flat illumination to replicate flat field calibration images (VIS has up to 6 different LEDs available). Other subtle effects are also included such as the effect of variable illumination and readout trails. The former is caused by the opening and closing of the shutter and the latter effect is caused by the open shutter during the readout process: light coming from astronomical objects create photo-electrons in the opposite direction of the readout.

\paragraph{Detectors}

The 36 Teledyne e2v VIS back-illuminated, red-enhanced CCD detectors will provide a great dynamic range with high precision spatial resolution at $12\,\micron$ square. While these are state-of-the-art detectors, there are numerous effects that degrade the image and need to be accurately simulated. 
The response of the pixels is not uniform due to variations in the gain and in the construction process, intrinsic to the material. This is known as Photon Response Non Uniformity or PRNU. We can obtain realistic PRNU maps from the ground calibration campaigns. Additional cosmetic defects from the process of construction and dust are also included. Furthermore, each pixel does not respond uniformly across its square area, that is the centre of the pixel is more responsive than the corners.  This is an effect we call intra-pixel quantum efficiency. When the number of electrons exceed the full well capacity (the limit of electrons that a pixel can collect), charges overflow in the readout direction into the neighbouring pixels, causing the effect of bleeding. The effect of Dark Current is also included with a dependence on the VIS focal plane temperature.  The final detector effect included is CTI, which creates streaks in the readout direction by electrons that remain in the previous pixel. This effect increases as the radiation from space damages the detectors with time creating charge traps in the silicon chips; it becomes significantly worse towards the end of the mission (and the most relevant degradation that limits the lifetime of the VIS instrument). Our implementation of CTI is based on the model described in \cite{cti-massey2014}.

\paragraph{Electronics}
As is typical with any optical CCD detector, a bias voltage is applied before the Analog-to-Digital Converter (ADC) to increase the signal from the 0 value. This requires models for the electronic bias, the ADC gain conversion, and proper simulation of the prescan and overscan sections. An additional complication is that the amplifier does not respond equally at all flux intensities; an effect that is about 1\% in the lower regime and 5\% when getting closer to the ADC saturation. The ADC has a limited number of bits to sample the charges in each pixel, causing the discretisation of the signal and setting the highest value at which the converter saturates. In the case of the electronics for \Euclid with 16 bits per pixel, it is set to $2^{16}-1$ or $65535\,{\rm ADU}$. In the process of reading the sensor, the electronics add scatter in the measured values, which vary from detector to detector, typically around $5\,{\rm e}^-$ RMS.  As all 36 detectors are being read at the same time, the induction of the current from the wires that transport the signal can produce reflected images on neighbouring amplifiers or detectors (known as electronic crosstalk).

In Fig.~\ref{fig:vis_sim_image}, we provide an example exposure produced by the VIS simulator.  The mosaic of the VIS detectors is seen on the left.  On the right, we show a small region of the exposure which exhibits distinctly the various simulated effects described above such as cosmic rays, CTI, ghosts, and saturation.

\subsection{NISP simulators}
We use two separate simulation softwares, \texttt{Imagem} and \texttt{TIPS}, for the photometric and spectroscopic channels of the NISP instrument, respectively. They share much of the optical path as well as the detector and data-processing units; they also have significant overlap in the models implemented, which we outline in the following before describing the channel specific implementations in Sects.~\ref{s:nispp_sim} and \ref{s:nisps_sim}.

\subsubsection{Common models}
\label{s:nisp_sim_common}

Both the NISP-P and NISP-S simulators implement the same type of astronomical sources and share some common background, PSF, optics, detector, and electronics models, which we explain below. Consistency across the two NISP channels is particularly important as the spectroscopic analysis relies on measurements from the imaging channel.

\paragraph{Sources}
In addition to the sources already included in the VIS simulator discussed in Sect.~\ref{s:vis_simulator}, we simulate additional sources required in the calibration of NISP images.  In particular, we simulate two spectrophotometric standard stars which are used to perform accurate photometric calibrations. For the spectroscopic channel only, we produce simulations of a planetary nebula, which enable wavelength calibration due to its strong and well characterised emission lines. Additionally, we simulate standard stars which provide a photometric calibration reference. The planetary nebulae and the standard stars are simulated in separate specific observations, in addition to the TU sources, at five locations in each detector, providing calibration points through the whole NISP focal plane. Both NISP simulators use a CREME96 model \citep{creme96-tylka1997} \corr{using a representative} in-orbit proton energy distribution to simulate cosmic ray tracks as these particles, arriving from a random direction, traverse a $7\,\micron$ thick HgCdTe layer in the Teledyne H2RG detectors.

\paragraph{Background}
The main common components of the background are zodiacal light, thermal irradiance, and out-of-field diffuse straylight. Even though these have already been explained in the VIS section (see Sect.~\ref{s:vis_simulator}), it is worth noting that due to its redder spectrum, zodiacal light is a significantly more intense source of background for NISP than for VIS.

\paragraph{Point spread function}
The NISP PSF contains the same types of effects as explained for the VIS instrument, although it can be simplified thanks to the less stringent requirements on the instrumental PSF. For SC8, a single model at the centre of each detector was used to represent spatial variations, taking into account the telescope and NISP instrument optical models.

\paragraph{Optics or focal plane array}
Both simulators implement with their corresponding filter/grism transmission and geometric distortion models the PLM transmission (see Fig.~\ref{fig:euclid_transmissions}) and the focal plane array (FPA) metrology that depends on the exact position and tilt of each detector in the focal plane.

\paragraph{Calibration}
Dark and flat simulations are also provided for both NISP channels for the usual data reduction process.  Similar to the calibration unit of VIS, the NISP Calibration Unit (NI-CU) can also illuminate its focal plane with a set of light emitting diodes (LED) at five different SEDs in the infrared. They are additionally set at different intensities to produce flat fields, which are also simulated.

\paragraph{Detectors}
The $4\times 4$ set of Teledyne H2RG infrared detectors are simulated with measured quantum efficiency, dark current, and readout noise maps provided from ground tests \citep{detectorperformance-waczynski2018,detectorcharacterization-secroun2016,detectorgain-secroun2018,detectorcalibration-barbier2018} and the Poisson distributed shot noise from the incident sources and background is accounted for. While a 4-pixel boundary corresponding to the reference pixels is included in the NISP simulation, it is currently set to zero ignoring baseline shifts. For SC8, we did not simulate the effect of NISP persistence; however, we advise the reader to consult Sect.~\ref{s:future_perspective} for a description of our implementation of persistence in our post-SC8 simulations.

\begin{figure}[!t]
  \centering
  \includegraphics[width=1.0\linewidth]{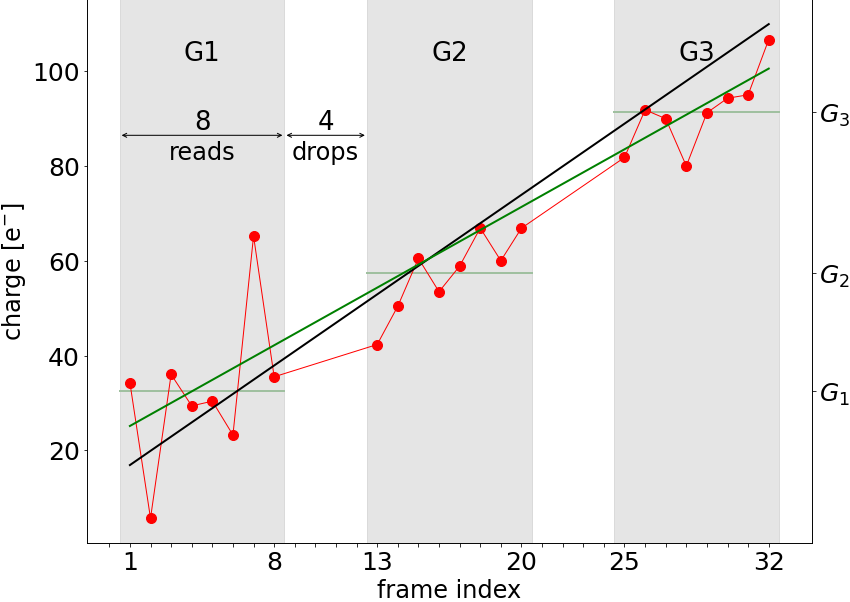}
  \caption{Example of an up-the-ramp readout for MACC(3,8,4) with 3 groups, 8 reads, and 4 drops, with true flux indicated in black, readout values in red and estimated flux in green. The flux error in this illustration largely stems from a comparatively high readout noise of 12 $\mathrm{e}^-$ relative to a flux of 3 $\mathrm{e}^-$/frame. Group average charges $G_i$ are here indicated by horizontal lines for each group as well as on the right y-axis. The actual NISP-P readout mode is MACC(4,16,4) while for NISP-S it is MACC(15,16,11) \corr{and the nominal frame rate is 1.45s}.}
  \label{fig:nip_macc}
\end{figure}

\paragraph{Electronics}
The readout of the H2RG detectors follows the procedure described in \cite{maccread-kubik2016} called multiple accumulated sampling (MACC), where the incoming signal is sampled multiple times during the exposure in an up-the-ramp readout process. In this configuration, we can estimate the ramp of the signal in each pixel, as well as the quality of the fit to the ramp.  This allows us to identify pixels which do not follow a stable rate (like cosmic rays that will produce a sudden jump in the accumulated flux). While NISP photometry and spectroscopy use different MACC readout modes, owing to their different exposure times, each uses an evenly spaced number of groups that are composed of a fixed number of frames read and are separated from the next group by a fixed number of frames that are dropped (see Fig.~\ref{fig:nip_macc}).

For a time $t_{\rm f}$ separating one readout frame from the next and a MACC readout with $n_{\rm g}$ groups, $n_{\rm r}$ reads in each group, and $n_{\rm d}$ drops between groups, the integration time is
\begin{equation}
t_{\rm int}=(n_{\rm g}-1)\,(n_{\rm r}+n_{\rm d})\,t_{\rm f} \, .
\end{equation}

In the following we describe the steps to compute a charge estimate $Q$ that, when divided by the integration time $t_{\rm int}$, directly results in a flux estimate, including the signal but also backgrounds and dark currents. In any given exposure, let $F_{\rm pix}(x,y,i,k)$ be the single frame charge in ADU capturing the accumulated signal and the instrumental effects at frame $k$ in group $i$ for the pixel $(x,y)$, $\sigma_{\rm r}$ the single frame readout noise in ADU and $g$ the gain in $\mathrm{ADU}/\mathrm{e}^-$.

First, we combine the single frame charges in each group $i$ into group averages
\begin{equation}
    G_{i}=\frac{1}{n_{\rm r}}\sum_{k=1}^{n_{\rm r}}F_{\rm pix}(x,y,i,k) \, .
\end{equation}
With these, the overall charge estimate $Q$ in ADU is set as
\begin{multline}
    Q = (n_{\rm g}-1) \\ \times \left[\frac{g(1+\alpha)}{2}\left(\sqrt{1+\frac{4\sum_{i=2}^{n_{\rm g}}(G_{i}-G_{i-1}+\beta)^2}{g^2(n_{\rm g}-1)(1+\alpha)^2}}-1\right)-\beta\right]
\end{multline}
where
\begin{align}
    \alpha &= \frac{1-n_{\rm r}^2}{3n_{\rm r}(n_{\rm r}+n_{\rm d})} \; \rm{and}\\
    \beta  &= \frac{2\sigma_{\rm r}^2}{g n_{\rm r}(\alpha+1)} \, .
\end{align}
%

\begin{figure*}[!h]
  \centering
  \includegraphics[width=\linewidth]{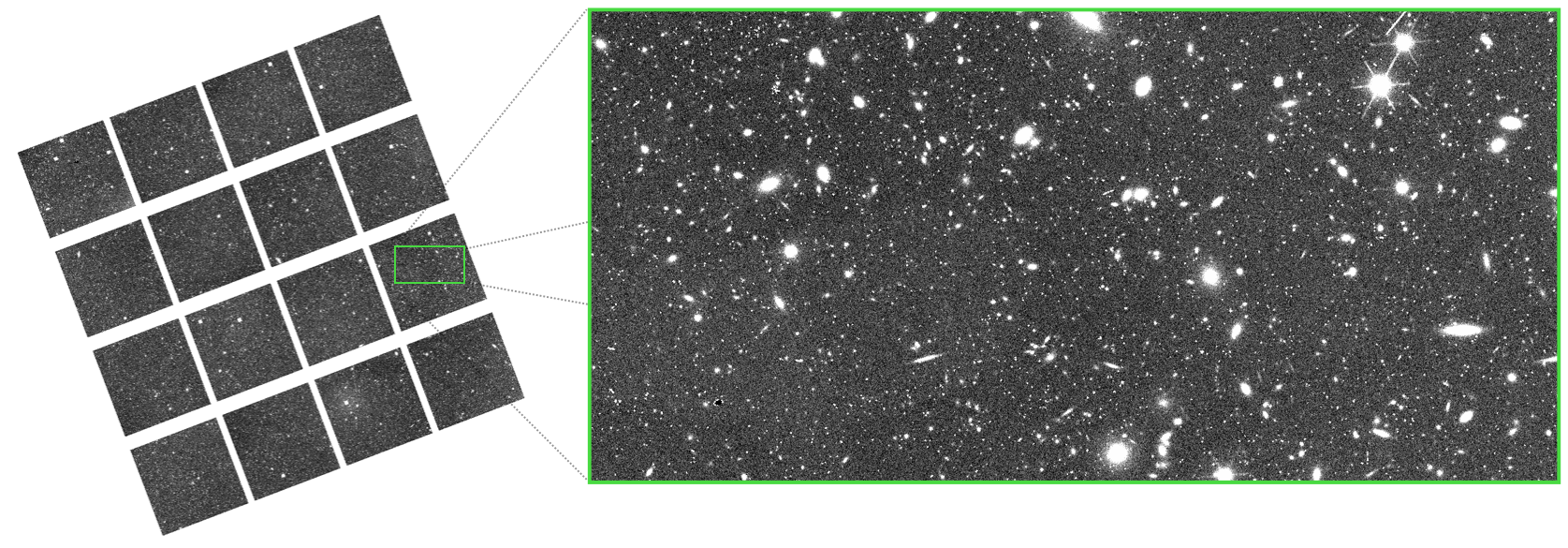}
  \caption{Simulated NISP-P exposure in the $\HE$ band. The left image shows the entire $4\times 4$ mosaic of infrared detectors and the right image contains a zoomed in fraction where the most relevant elements of the simulation can be seen. }
  \label{fig:nip_sim_image}
\end{figure*}
%
The charge estimate's quality factor ${\rm QF}$ in ADU for the same readout ramp is computed as
\begin{equation}\label{eq:qf}
    {\rm QF} = \frac{2}{(1+\alpha)(n_{\rm g}-2)}\left[(n_{\rm g}-1)K-(G_{n_{\rm g}}-G_1)\right]
\end{equation}
with
\begin{equation}
    K = \sqrt{\frac{\sum_{i=2}^{n_{\rm g}}(G_{i}-G_{i-1}+\beta)^2}{n_{\rm g}-1}}-\beta \, .
\end{equation}
This quality factor QF has approximately a $\chi^2$ distribution with $(n_g - 2)$ degrees of freedom; that is, $2$ and $13$ degrees of freedom for NISP-P and NISP-S, respectively. \corr{The QF is later used to mask pixels that do not follow a linear accumulation of counts, such as those affected by cosmic rays.}

\subsubsection{NISP-P simulator}
\label{s:nispp_sim}

The NISP-P channel is simulated with \texttt{Imagem}, which reproduces NIR imaging in the three broadbands of the instrument: $\YE$, $\JE$, and $\HE$. Being the primary source of information for the photometric redshifts in the infrared bands, it is important to accurately simulate the flux of the sources and any possible instrumental effects that could bias the photometric intensity and colour measurements. Below, we highlight the main features included in the NISP-P simulator.

\paragraph{Sources}
In addition to the common sources, the NISP-P simulator adds the specific standard stars with a particularly smooth spectrum, ideal for photometric calibrations where synthetic photometry can be estimated accurately. We used the spectrum of Vega and a white dwarf, scaled to AB magnitude 17.5 in the $Y_{\rm E}$ band, which delivers a good signal-to-noise below the range of the saturation limit or the non-linear regime of the detectors, respectively. 

\paragraph{Background}

In addition to the common backgrounds detailed above, we include localised in-field straylight in the form of faint circular symmetric halos with detailed shapes dependent on the source SEDs.

\paragraph{Point spread function}

We use a chromatic PSF with three wavelength-dependent PSFs for each of the three photometric filters that are integrated with the incident spectra of each TU source and a wavelength-dependent quantum efficiency. NISP-P PSF pixels are sampled at one-sixth of the size of detector pixels in both dimensions. This is to allow us to emulate differences in the projected PSF depending on whether a source is centred; for example, on a pixel centre, edge or corner.  First, we convolve sources with the optical PSF on these sub-pixel scales. After the resulting high resolution stamps are projected onto detector pixels,  we apply an additional convolution with a $3\times3$ matrix on the detector pixels to simulate the effect of inter-pixel capacitance (IPC).

\paragraph{Optics}

Along with the common optical effects like filter and PLM transmission, FPA metrology, and geometric distortions, we also implement the effect of filter wheel tilt variations; in other words, the inaccuracy of the filter wheel to be positioned always at the same exact position, resulting in small angular deviations with a RMS error of \ang{;6;}. These additional tilts impact the final distortions which are partly affected by the optical power of the filter element.

\paragraph{Detector}

For each pixel in a source, electron fluxes are integrated using measured quantum efficiencies for 40 wavelengths from 0.6 to $2.55\,\mu{\rm m}$ for every detector pixel, the source SED and the PSF chromaticity. We simulate each source three times for every exposure to account for the three wavelength-dependent PSFs per filter. Using distinct wavelength dependencies of the quantum efficiency for each pixel can therefore lead to varying pixel sensitivities depending on the source spectrum as well as varying effective PSF shapes according to the source position on the detector. The photometric flat fields, in addition to the calibration unit LEDs spectra, also use the wavelength-dependent quantum efficiency maps.

\paragraph{Electronics}

\texttt{Imagem} simulates the readout process as detailed above in Sect.~\ref{s:nisp_sim_common}, accounting for each intermediate frame in the specific MACC readout mode used of 4 groups, 16 reads and 4 drops for the three photometric bands. This provides an accurate reproduction of the image and the associated on-board flagging based on the charge estimate's quality factor QF, where flagged pixels are transmitted as an additional 1-bit image layer. The resulting image is further modified by pixel dependent non-linearity and a single gain value.

\begin{figure*}[!h]
  \centering
  \includegraphics[width=\linewidth]{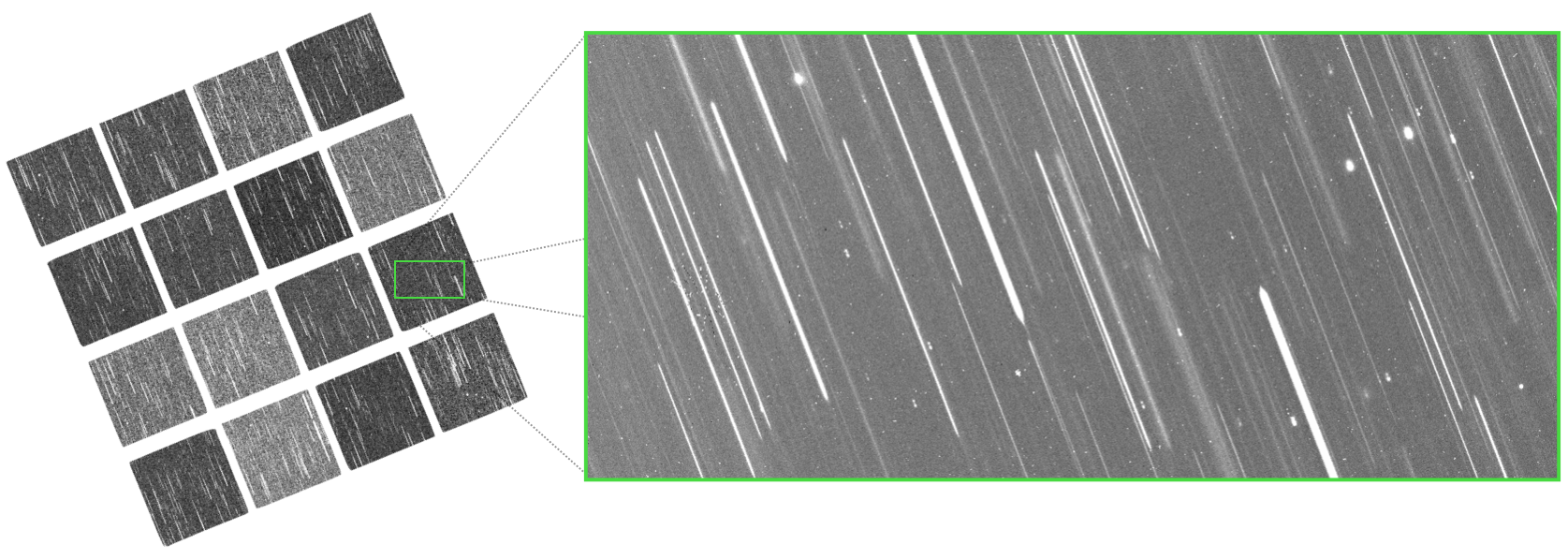}
  \caption{Simulated NISP-S exposure. The left image shows the entire $4\times4$ mosaic of infrared detectors and the right image contains a zoomed in fraction of a single detector, showing a typical slit-less spectroscopy of the NISP-S with 0th, 1st, and 2nd order dispersed spectra. }
  \label{fig:nis_sim_image}
\end{figure*}


Figure~\ref{fig:nip_sim_image} shows an example raw exposure in the $\HE$ band produced by \texttt{Imagem}. The mosaic image for the $4\times4$ detector array for the NISP instrument is displayed on the left.  The right image shows in finer detail a selected area of one of the detectors.

\subsubsection{NISP-S simulator}
\label{s:nisps_sim}

The NISP spectroscopic simulator is implemented in \texttt{TIPS} \citep{tips-zoubian2014}, a program built to simulate the slitless spectroscopy images from the NISP instrument. It is based on the slitless simulation package \texttt{aXeSIM}\footnote{\url{http://axe-info.stsci.edu/axesim/}} \citep{axesim_kuemmel2008} developed by the Space Telescope Science Institute, which performs the core spectra dispersion and convolution with a chromatic PSF of ten wavelength positions, sampled at one-sixth of the native NISP resolution. As stated above, most sources and instrumental effects are shared with the NISP-P channel simulator.  
The performance of the spectroscopic channel is expected to be dominated by the effects that are currently implemented which we shall describe below.

\paragraph{Spectral distortions}
The simulation of the slitless spectra requires a complete characterisation of the trace dispersion, with sensitivity, diffraction coefficients, and PSFs at each dispersion order (expressed as a Taylor series expansion). The dispersion order of interest is the first order, where the main spectrum can be recovered. However, an additional zeroth and second order of dispersion need to be simulated as well, as they provide reference points (for the zeroth order) and contamination in the image. While the average dispersion is $700$, $13.4$, and $6.7\,\AA\,{\rm pixel}^{-1}$ for the 0th, 1st, and 2nd orders of dispersion, respectively, the wavelength steps $\Delta\lambda$ are not constant across the field of view. Thus in the simulation of spectral distortions we included  a spatially varying $\Delta\lambda$. This significantly complicates the extraction of the spectra, which needs to be calibrated precisely to enable the determination of accurate spectroscopic redshifts. 

\paragraph{Point spread function}
The plan initially was to use ZEMAX-issued models for each dispersion order of each grism, in a similar fashion to the NISP-P simulations.  However, this was later discarded because of the challenges it could have presented at that the time for the SGS pipeline, notably at localising with fine precision the zeroth order spectra, an important requirement for the data-processing of spectroscopic data. \corr{This is expected to be calibrated and validated in later processing of the NISP-S pipeline.}
Instead, we used a double Gaussian PSF to fit the ground-based characterisation data acquired from thermal-vacuum balance tests in order to match the experimental radius encircled energies EE50/EE80 measurements \citep{maciaszek2022}.  This provided smoother non-chromatic pixel PSFs. No IPC was not taken into account in NISP-S simulations.

\paragraph{Grism tilt}
Like the filter wheel tilt effect described for the photometric channel, the grism wheel also has a limited positioning accuracy with an error of $0\dotdeg1$ RMS. In this case, the positioning error translates into slightly tilted dispersed spectra with respect to their dispersion direction at nominal angle. To enable an efficient decontamination of overlapping slitless spectra without the RGS270 grism, the instrument implemented a tilt of $+4$\textdegree and $-4$\textdegree in dithers 2 and 3 (as shown in Fig.~\ref{fig:wide_sequence}) in a sequence of $0$\textdegree, $184$\textdegree, $-4$\textdegree, and $180$\textdegree orientation of the dispersion direction. 

\paragraph{Vignetting}
Turning the grating wheel by $\pm$ $4$\textdegree moves the nominal aperture of the grating creating vignetting on the focal plane. This effect is simulated using a basic and conservative model derived from ground calibrations assuming 10\% of vignetting at $+0\dotdeg385$ field angle, which diminishes to $0\%$ at a field angle of $+0{\dotdeg}17$.
We apply vignetting only to the sources to take into account the flux loss; we neglect the effect to the background.

\paragraph{Electronics}
As indicated above in Sect.~\ref{s:nisp_sim_common}, \texttt{TIPS} simulates the readout process for each intermediate frame of the MACC mode in spectroscopy with 15 groups of 16 frames separated by 11 drops. The signal slope and the quality factor of each pixel is then derived following \cite{maccread-kubik2016}, as it is implemented in the onboard payload data-processing. Non-linearity is applied afterwards on the slope by using the inverse function of the non-linearity correction model, but its impact on the quality factor is ignored. The slope, in ADU, is encoded on 16 bits and the quality factor on 8 bit.

We show in Fig.~\ref{fig:nis_sim_image} an example image produced by \texttt{TIPS} for the NISP instrument when emulating readout for the spectroscopic channel. Like the photometric channel, we have the same $4\times4$ detector mosaic on the left. For closer inspection, a small region of a detector is shown on the right containing the characteristic features of slitless spectroscopy.

\subsection{EXT ground-based survey simulator}
\label{s:ext_sims}
Lastly, the ground-based simulator \texttt{SIM-EXT} is a generic simulation software package to produce accurate pixel data for various ground telescope and instrument systems. \Euclid will rely on external optical photometry complementing its own infrared bands in order to obtain accurate photometric redshifts. The necessary photometry in the northern hemisphere will be provided by the Ultraviolet Near Infrared Optical Northern Survey\footnote{\url{https://www.skysurvey.cc}} (UNIONS), a joint effort between the Canada-France Imaging Survey \citep[CFIS;][]{cfis_ibata_2017} for the $u$ and $r$ bands, the Javalambre-\Euclid Deep Imaging Survey (JEDIS) in the $g$ band, the Panoramic Survey Telescope and Rapid Response System \citep[Pan-STARRS;][]{panstarrs_chambers_2016} in the $i$ band, and the Wide Imaging with Subaru Hyper Suprime Camera \citep[HSC;][]{hsc-miyazaki2017} of the \Euclid Sky survey (WISHES) for the $z$ band. For the southern sky, it is anticipated that the Dark Energy Survey \citep[DES;][]{des_2005} and the $Rubin$ Observatory Legacy Survey of Space and Time\footnote{Hereafter, we use \textit{Rubin} or LSST interchangeably when referring to $Rubin$-specific software, hardware and data products.} will provide all $ugriz$ photometry \citep{lsst_science_book2009,lsstcomponents-ivezic2019}. 

\begin{figure}
  \centering
  \includegraphics[width=0.75\linewidth]{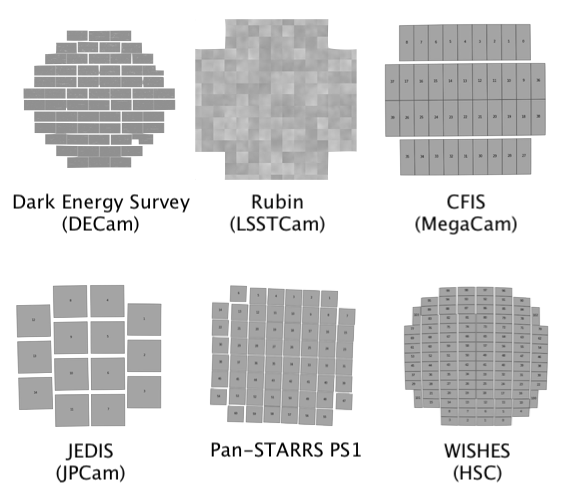}
  \caption{Six ground-based EXT surveys simulated by \texttt{SIM-EXT}. From Top Left to Bottom Right, the detector layout for DECam ($2$\textdegree.$2$ / 570 Mpix), LSSTCam ($3$\textdegree.$5$/ 3.2 Gpix), MegaCam ($1$\textdegree / 380 Mpix), JPCam ($3$\textdegree, 1.2 Gpix), Pan-STARRS PS 1 ($3$\textdegree/ 1.2 Gpix), and HSC ($1$\textdegree.$5$/ 870 Mpix).}
  \label{fig:sim_ext_surveys}
\end{figure}

\begin{figure*}[!t]
  \centering
  \includegraphics[width=\linewidth]{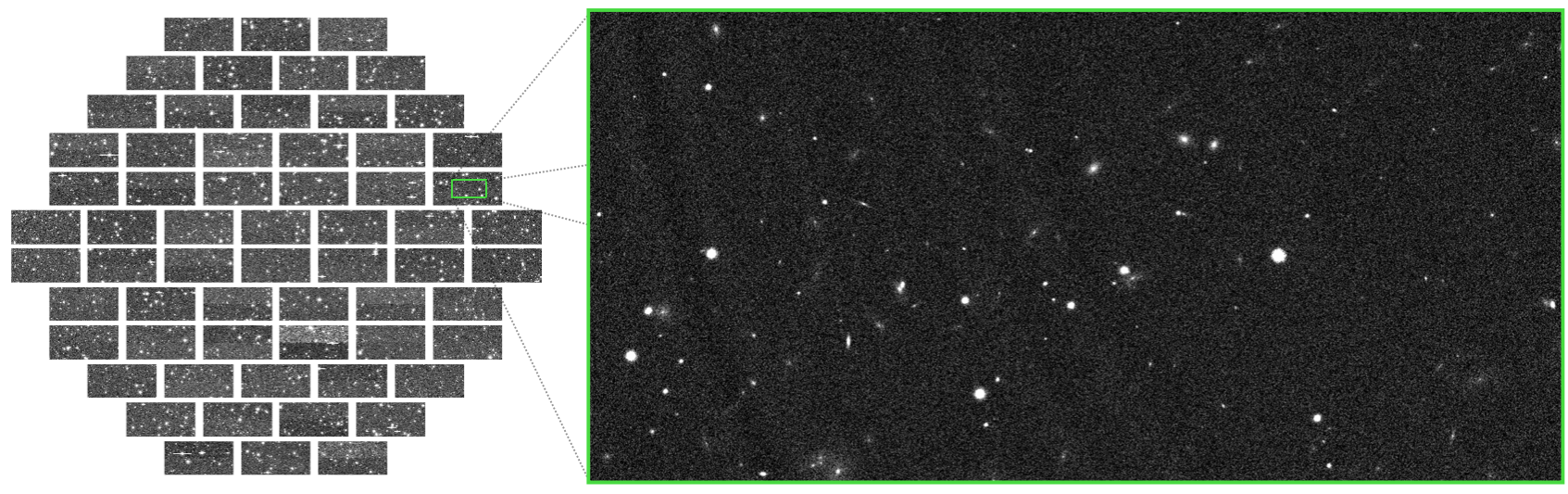}
  \caption{Simulated DES exposure in the $r$ band. The left image shows the entire 62 detector mosaic covering a $2\dotdeg2$ diameter field of view. The right image shows a detailed 10$\arcmin$ wide section of a single detector.}
  \label{fig:ext_sim_image}
\end{figure*}

The function of $\texttt{SIM-EXT}$ differs from that of the other \Euclid instrument simulators. Rather than producing extremely detailed simulations for each EXT survey (a notably costly effort), its purpose is to provide realistic imaging data that can flow inside the ground segment pipeline with the same TU inputs as the other \Euclid channels. The output data products of \texttt{SIM-EXT} vary per EXT survey in accordance with the SGS strategies designated for each one. For example, data-processing of raw DES data will be performed within the \Euclid SGS whereas the detrended and/or calibrated data products produced by the other respective experiments and synergies (e.g. UNIONS and LSST) will be received by the \Euclid SGS when available.  Therefore, for DES we produce outputs consisting of raw images and master calibration frames. For LSST, only raw images are produced with a choice of units: e$^-$ or ADU. Moreover, the FITS format for LSST images produced by \texttt{SIM-EXT} version 2.14.2 is similar to the e-image format defined by the Dark Energy Science Collaboration (DESC) image simulator \texttt{imSim} version 0.1.0\footnote{\url{https://github.com/LSSTDESC/imSim/tree/0.1.0}} described in \cite{descDC1-sanchez2020}. This output format is required for \texttt{SIM-EXT} LSST frames to be processed using the $Rubin$ Observatory LSST Science Pipelines release 16.0\footnote{\url{https://pipelines.lsst.io/releases/v16_0.html}} \citep{lsstdesc-2012,lsstdm-juric2015,lsststack-bosch2018,lsstcomponents-ivezic2019}. This allows the SGS to emulate ingesting official $Rubin$ Observatory calibrated data products for joint pixel processing with the \Euclid data products. For the Northern Surveys, a software package called \texttt{EXT$\_$NS$\_$SIM$\_$AFTERBURNER} developed by the Organisation Unit for External Surveys (OU-EXT) is used in conjunction with \texttt{SIM-EXT} to provide level-2 processed data products consisting of detrended frames, source catalogues, and PSF models.

\paragraph{Sources}
The EXT ground-survey simulator performs the generation of all TU sources such as real and model stars, Flagship, primaeval Universe galaxies, high-$z$, and strongly lensed sources. Using the same \texttt{SimThumbnails} as the \Euclid simulators, it renders the exact same light profiles including weak lensing distortions. As the PSF is assumed to be monochromatic, it does not require the input spectra and uses precomputed band fluxes, thus accelerating the computation of these simulations. 

\paragraph{Background and transmission}
As most of the EXT surveys have already started their observations, we have realistic background magnitudes and zeropoints from their respective reduced image analysis. We use the measurements taken from actual observations contained with the MDB Survey Tables as inputs to the simulation, which provide a very realistic distribution of sky brightness and atmospheric extinction. For the EXT surveys, such as \textit{Rubin}, that have not started their observations, we can model realistic distributions from the available models of the sky brightness \citep{LSST-skybrightness-yoachim2016} and passbands \citep{lsstcam-kahn2010,lsst-syseng-claver2014,opsim-connolly}. 

\paragraph{Point spread function}
The PSF is significantly simpler in most EXT simulations, and it is modelled as either a simple Gaussian or a Moffat profile. From a realistic distribution of seeing FWHM, we can replicate the average optical quality for each survey. In the case of DES, we obtained a large sample of PSF models extracted from real data analysis using \texttt{PSFEx} \citep{psfex-bertin2011}.  
We randomly select a model, enabling variable and complex PSF simulations across the focal plane for the ground-based photometry.

\paragraph{Optics and detectors}
The design of \texttt{SIM-EXT} takes advantage of the common characteristics of the ground-based instruments to perform simulations of various surveys.
These surveys are configurable through the MDB instrument file providing a description of the optics, detector layout and electronics. Figure~\ref{fig:sim_ext_surveys} shows the focal plane layouts for the DES Dark Energy Camera (DECam), \textit{Rubin's} LSST Camera (LSSTCam), CFIS Mega Camera (MegaCam), JEDIS (JPCam), Pan-STARRS PS1, and WISHES HSC, respectively. In detail, the camera detector layout for DECam was derived from \cite{Flaugher_2015}.  The detector focal plane layout for the LSST camera (LSSTCam) was defined with the information provided in \cite{lsstcomponents-ivezic2019}. For simplicity, our camera model is an approximation to LSSTCam, as all 189 detectors in our focal plane are models of either LSSTCam's e2V sensor or LSSTCam's ITL sensor, not a combination of both as will be used during operations.
The detector layout for MegaCam, JPCam, Pan-STARRS PS1, and HSC were each inferred, respectively, from a set of real exposures provided courtesy of the CFIS, JEDIS, Pan-STARRS, and WISHES teams. For a majority of the surveys, we deliver a perfectly calibrated WCS by writing the true WCS in the header of the output images. For DES, we reproduced a certain degree of `un-calibration' by writing a slightly different WCS in the output headers. The input WCS are obtained with the astrometry tool \texttt{SCAMP} \citep{scamp-bertin2006} applied on actual data.  If real data are not available, we apply a simpler estimation of the WCS (without distortions).  Additionally, we use master bias and master flats from the actual observations to include realistic pixel-to-pixel variations and faithful readout patterns. With the same model, we can create individual flat field and bias calibration images that have their own noise realisation.

A sample image of the \texttt{SIM-EXT} simulator configured to produce a DES exposure can be seen in Fig.~\ref{fig:ext_sim_image}.  On the left, we show the 62 detector mosaic and on the right a close-up image of one of the CCDs.  In addition to the features discussed above, we have the additional detector effect of tree rings acquired from sampling from real DES flat data; these are circular patterns due to variations of the silicon dopant concentration in the CCD chip.

With all four instrument simulators presented in this section, we can produce complex and detailed simulations for a large variety of channels and configurations for the same Universe, environment, and common models. In the following section, we present our procedure for launching a large-scale production over segments of the Euclid Wide Survey using our suite of simulation codes.

\section{SC8 simulation processing and results}
\label{s:sc8}

In this section, we present the application of our SIM pipeline framework for SC8 and the corresponding results.  We first begin with an in-depth description of our simulation production workflow as developed for SC8. 

\subsection{Simulation planning and orchestration}
\label{s:sim_planner}

The previously defined elements (i.e. TU catalogues, simulators, etc.) enable the generation of mock observation sets of data to be processed and analysed. However, for the production of large datasets in a distributed and controlled manner these simulation elements need to be connected and coordinated. Here we describe the additional components of our software framework which fulfill this objective: the simulation request, the \texttt{SimPlanner}, and the SIM pipeline.

The simulation request is one of the inputs of the SIM pipeline together with a mission configuration and the catalogues. One can specify some aspects of the simulations such as (i) channel: VIS, NISP-P, NISP-S, EXT-LSST, EXT-DES, (ii) image type: Science, Deep, Flat, Bias. (iii) pointings or area to simulate, and (iv) instrumental effects to be enabled/disabled.

The simulation request is handled by the \texttt{SimPlanner}, a program that constitutes the entry point of the SIM pipeline and allows us to orchestrate large sets of simulations. It parses the information in the simulation request and the survey files and configures the corresponding simulator with the appropriate parameters. The \texttt{SimPlanner} contains additional features such as the generation of particular dithering patterns, in-pointing sequence, enabling complex parameter exploration, configuration of flat field sequences, and other high-level tasks that can be done at this point.

The SIM Pipeline framework allows multilevel splitting and parallelisation of the processing, which is critical to producing large sets of simulations in a reasonable time frame. The processing of the SC8 main production, as well as the other productions in the scope of SC8 (e.g. deep field) was carried out on the \Euclid SGS infrastructure. 

\begin{figure*}[!t]
  \centering
 \includegraphics[width=\linewidth]{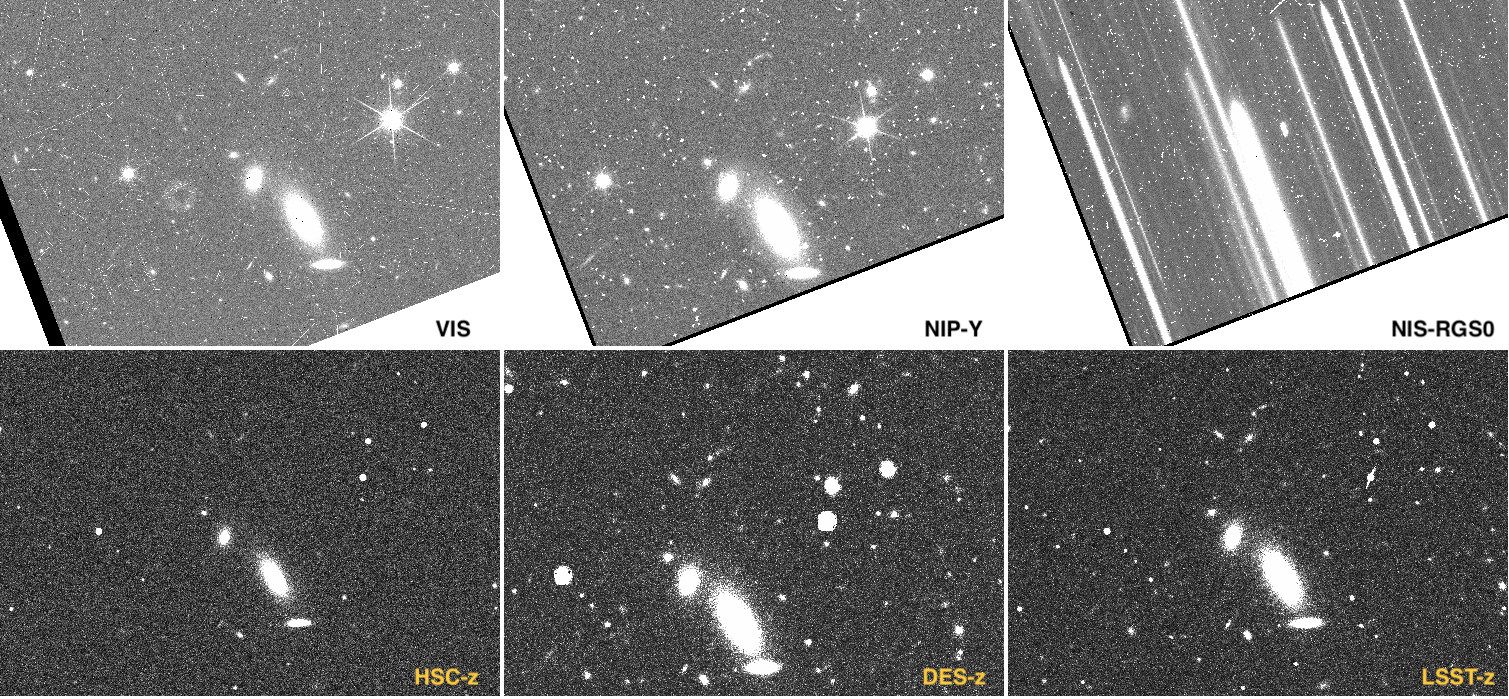}
  \caption{Simulated raw single visit images for each \Euclid imager: VIS, NISP-P (\YE band), NISP-S (NIS-RGS0) with the raw outputs from three of the ground-based survey instruments in the $z$ band: WISHES (HSC-$z$), DES (DES-$z$), and \textit{Rubin} (LSST-$z$) over the same area of sky in \Euclid\/ tile 79171 (Observation ID 18539) of the pilot phase. \corr{The slight misalignment of the DES-z image corresponds to the additional pointing error effect included only in the raw DES image simulation, due to a more realistic reference set of actual observations.}}
  \label{fig:sc8_pilot_pointing}
\end{figure*}

\subsection{Pilot phase} SC8 began with a small test production of simulations in preparation for the main run, called the pilot phase. This production had two primary goals: (1) ensure the validity of the simulations and subsequent data products produced during image processing by each OU in the SGS pipeline; (2) measure the amount of resources, such as number of cores, peak RAM, and maximum wall time, needed for each of the OU-SIM tasks, (3) perform the necessary optimisations to guarantee feasibility of the larger-scale SC8 main run with the available resources.

The pilot phase contained four Science Wide observations covering the so-called \Euclid\/ tile \citep{euclid_tiles_kuemmel_2022} (79171) with an area measuring $\sim$$0.5\,{\rm deg}^2$. \Euclid\/ tiles are non-overlapping ecliptic latitude-longitude rectangles (approximately the size of the \Euclid field of view) and oriented along the meridians.  With no gaps between adjacent tiles, they are essentially the unit of processing for high level OUs to ensure complete coverage of the survey area when constructing the object catalogues for WL and GC analysis.  Even for an area this small, the multiple feedback iterations with the other OUs resulted in a total production of 1866 raw images for all simulation channels. In addition, all the calibration frames required to carry out image processing (such as detrending, astrometric, and photometric calibrations) for the \Euclid instruments were also produced during this phase and later reused for the data reduction during the main run.  The set of VIS calibration products consisted of a sequence of 60 bias images and 60 short-exposure flat-fields using the VIS calibration unit (VIS-CU). Since the shutter will remain open during flat acquisition, there will be objects entering into the field of view of the telescope, and we have simulated the most luminous objects yielding a flux above the photon noise.

For the NISP instrument, we produced a sequence of 600 dark frames and 750 flat-field images with the respective LED intensities and exposure times for the photometric and spectroscopic channels. We also simulated observations of spectrophotometric standard stars and planetary nebulae to enable the performance of NISP photometric and spectroscopic wavelength calibrations.  Finally, we also delivered a self-calibration sequence over a dense stellar field consisting of multiple dithered overlapping images that were to be used to improve the flat-fielding and the astrometric calibration. 

In Table \ref{table:pilot_production}, we list the number of files and the total size for each image data product produced during the pilot phase. We show in Fig.~\ref{fig:sc8_pilot_pointing} raw science images produced by the simulators for the VIS, NISP-P ($\YE$ band), NIS (RGS0), and three EXT ground surveys in the $z$ band: WISHES (HSC), DES, and LSST over the same area of sky in \Euclid\/ tile 79171 for one of the four \Euclid observations (Observation ID 18539).  The simulated \Euclid images show considerably more prominent features (e.g. ghosts, diffraction spikes, and CTI) in comparison to the simplified images produced by the simulator for the EXT ground-based surveys which are known or expected to experience similar effects. 
 
\subsection{Main run}
The original plan of the SC8 main run was to conduct a full end-to-end test of the SGS pipeline using a large and complex simulation of 1286 observations from the Euclid Wide Survey, covering a contiguous area of ${\sim}\,600\,{\rm deg}^2$ with different sky conditions. All known instrumental effects were to be included in the pixel data to enable performance and quality assessments of the level-2 calibration pipelines. Furthermore, the large area would enable the first integration test of the final processing elements of the \Euclid SGS pipeline (i.e. the level-3 cosmological measurements) with enough area to measure the baryon acoustic peak in the galaxy two-point correlation function.

However, as explained in the previous section, the results of the pilot phase indicated that this would present a major challenge as producing and processing large catalogues and realistic images carries a high computational cost. The SC8 main run could not have been achieved in a reasonable time frame with the resources available for the
challenge. Therefore we have reverted to simulate a subset of 331 observations grouped in 11 non-contiguous patches preserving the variety of conditions: mostly stellar density and zodiacal light as shown in Fig.~\ref{fig:star_catalogues_sc8}. The final simulated area measuring ${\sim}\,165\,{\rm deg}^2$ was still sufficient to perform the integration test of the processing elements of both the WL and GC experiments, the topmost priority.

\begin{figure}
  \centering
   \includegraphics[width=\linewidth]{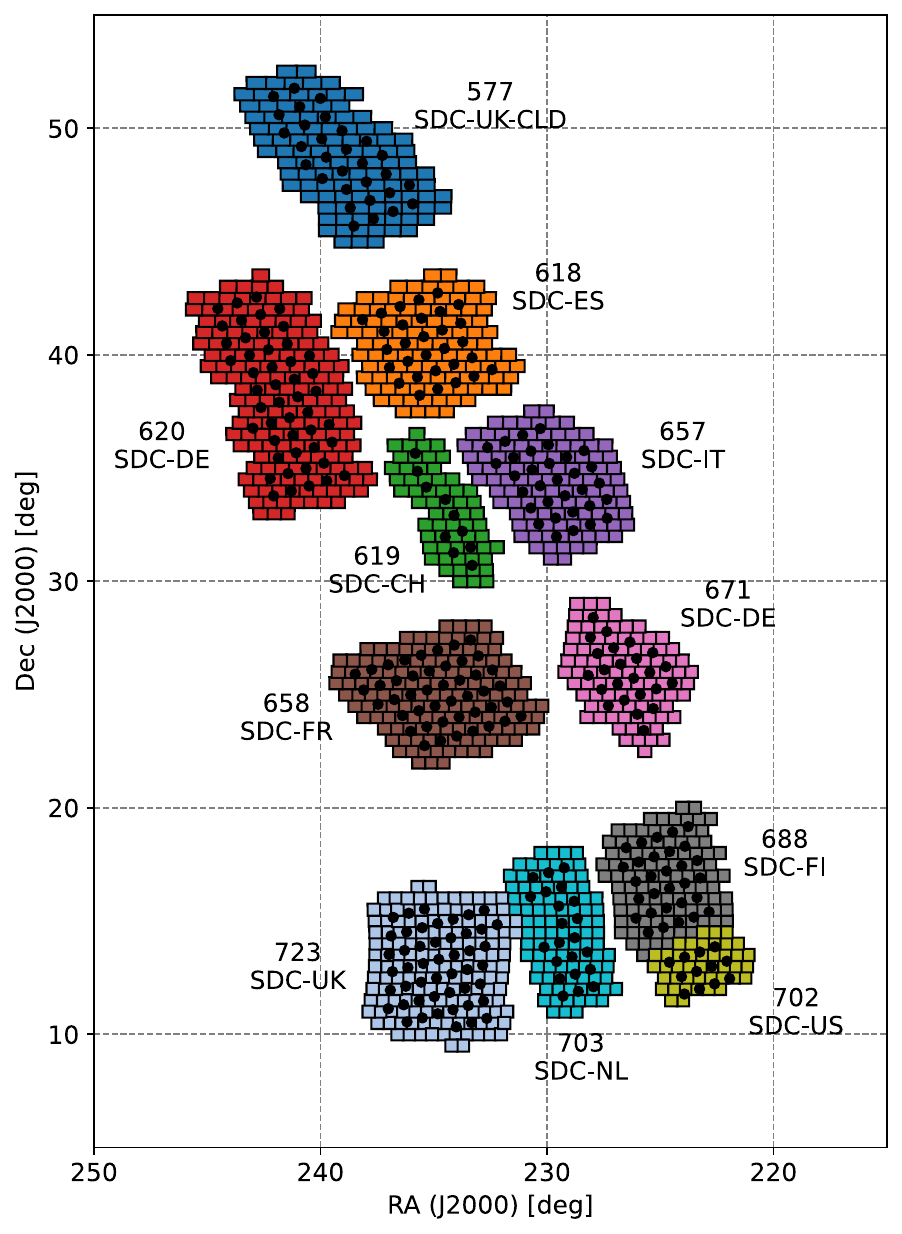}
  \caption{Distribution of the processing for the SC8 main production among the SDCs. The labels in the image indicate the patch identifier and the corresponding SDC responsible of processing the entire patch containing all \Euclid pointings and the overlapping EXT surveys. The black dots denote each of the 331 observations of the Euclid Wide Survey and the squares represent the 1036 Euclid processing tiles.}
  \label{fig:sc8_euclid_obs_sdc}
\end{figure}

Figure~\ref{fig:sc8_euclid_obs_sdc} shows the 11 patches and the position of each $\Euclid$ observation contained within each of them.  The colour of each patch represents the SDC tasked with processing the set of SC8 observations.  The assignment of a region to a SDC was according to their expected contribution and the resources available. Moreover, the strategy of having non-contiguous patches processed at the different SDCs was implemented to avoid data transfers between processing centres.  Each patch is composed of a set of \Euclid\/ tiles.  There are 1039 \Euclid\/ tiles overlapping with the SC8 observations. 

Over the selected \Euclid targets in the sky, we simulated all three \Euclid channels: VIS, NISP-P, and NISP-S, respectively. The main run also included simulations of the EXT ground-based surveys in the northern and southern hemispheres. The target area was partitioned into three non-overlapping patches. All Northern Surveys — CFIS, JEDIS, Pan-STARRS PS1, and WISHES — covered the same region above $30^{\degree}$ in declination. We partitioned the lower rectangular region along the diagonal into two sections: \textit{Rubin} observations spanned the area above the diagonal and DES covered the opposing field below it. We refer the reader to Appendix~\ref{a:sc8_main_survey_files}, which describes with more detail the selections of the pointings for each EXT survey and provides figures illustrating the EXT survey footprints in each band. 

\begin{table}[htbp!]
\centering
\caption{Data products in main SC8 simulation.}

\begin{tabular}{|p{2cm}|p{2cm}|p{1cm}|p{2cm}|p{1cm}|} 
\hline
SC8 Main Channel & Number of products & Storage (TB) & Processing time (days) \\
\hline
\\[-0.9em]
VIS & 1\,324 & 8.7 & 5\,200 \\
NISP-P & 3\,948 & 1.5 & 1\,500 \\
NISP-S & 1\,324 & 5.1 & 4\,700\\
DES &  2\,370 & 2.4 & 600 \\
& & &\\
LSST & 154\,980 & 9.2 & 3\,700\\
CFIS & 55\,440 & 0.7 & 150  \\
JEDIS & 12\,096 & 1.7 & 210  \\
Pan-STARRS & 63\,300 & 1.9 & 570 \\
WISHES & 47\,586 & 0.5 & 100 \\
\hline
\\[-0.9em]
Total & 342\,368 & 31.7 & 16\,730\\
\hline
\end{tabular}
\tablefoot{The datasets for VIS, NISP-P, NISP-S, and DES are stored as exposures, whereas for LSST, CFIS, JEDIS, Pan-STARRS, and WISHES each individual data file corresponds to a detector frame. The processing time corresponds to the sum of all single-thread sub-jobs wall time, as indicated in table \ref{table:observed_resources_1}}

\label{table:sc8_production_summary}
\end{table}

\begin{figure}[!t]
  \centering
 \includegraphics[width=\linewidth]{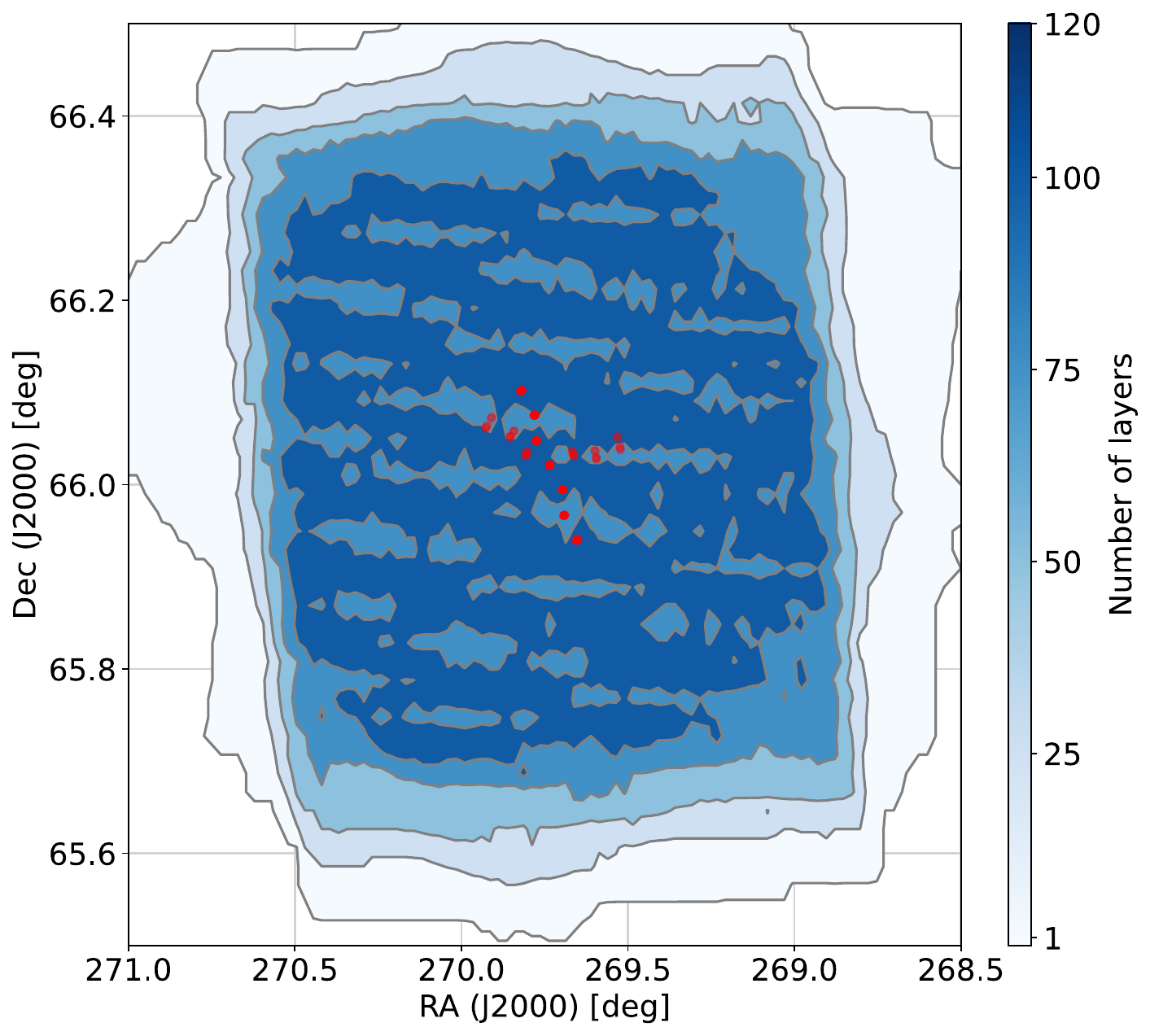}
  \caption{Footprint of the SC8 deep field, including 30 observations (denoted by the red circles), i.e. 120 images, near the ecliptic north pole.}
  \label{fig:sc8_deep_field}
\end{figure}
%
\begin{table}[t]
\caption{Data products in SC8 deep field simulation.}
\begin{tabular}{|l|c|c|}
\hline
SC8 Deep Channel &  Number of products &  Storage (GB)\\
\hline
    VIS   &   120 &   809 \\
    NISP-P &   360 &    140 \\
    NISP-S  &   120 &    469 \\
\hline
\\[-0.9em]
Total &  840 &  1418 \\
\hline
\end{tabular}
\label{table:deep_production_summary}
\end{table}
In Table~\ref{table:sc8_production_summary}, we report the number of data products and the amount of storage required per survey channel participating in the SC8 main run.  We produced a total of $342\,368$ data products amounting to 31.7 TB for the $165\,\rm{deg}^2$ Euclid Wide Survey.  
\begin{figure*}[!h]
  \centering
  \includegraphics[width=\linewidth]{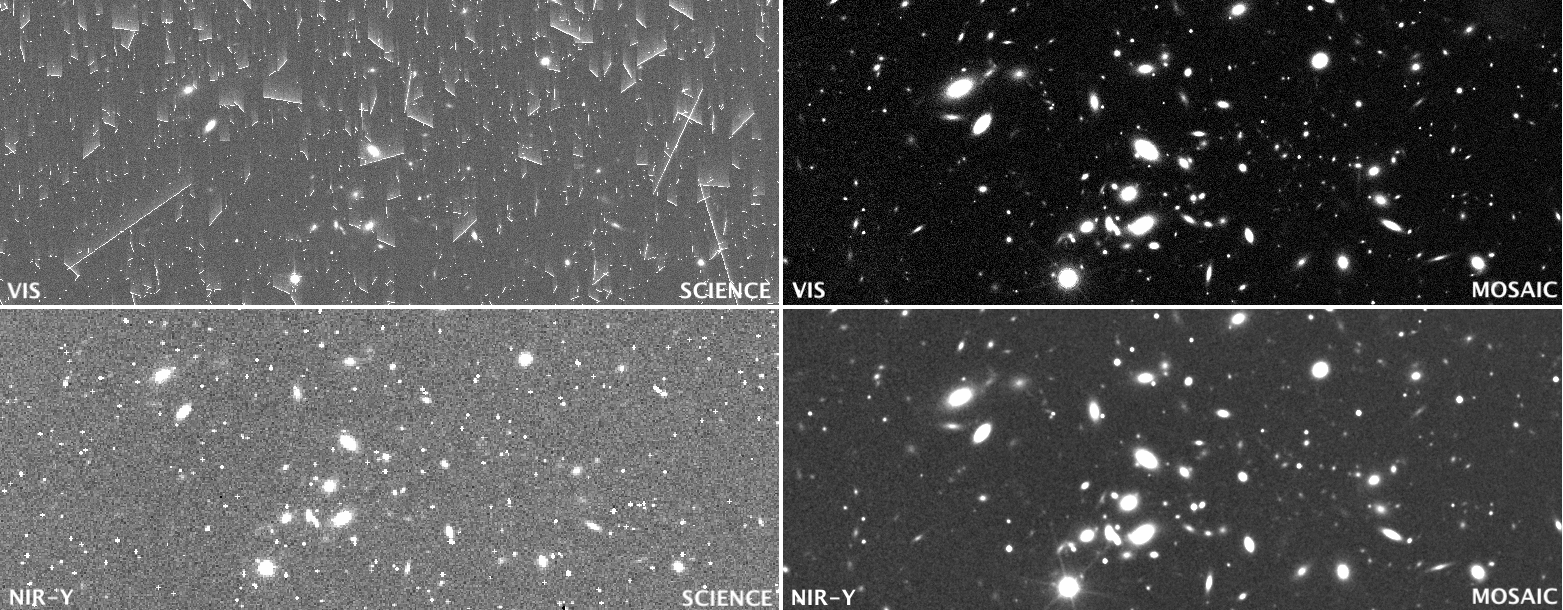}
  \caption{VIS and NIR imaging of a region in the SC8 deep field (\Euclid ile 306182, Observation ID 1914) resulting from the different stages of the SGS processing flow.  The two panels on the left show the simulated raw science frames for the VIS $\IE$ band (top-left), and NISP-P $\YE$ band (bottom-left).  The right panels show the background-subtracted mosaic images (see text for details) of the same field for the VIS $\IE$ band (top-right) and NISP-P $\YE$ band (bottom-right), produced, respectively, during level-2 data-processing by the \corr{Organization Unit for Merging} (OU-MER) pipeline.}
  \label{fig:sc8_sci_mosaic}
\end{figure*}
\begin{figure*}[]
  \centering
  \includegraphics[width=\linewidth]{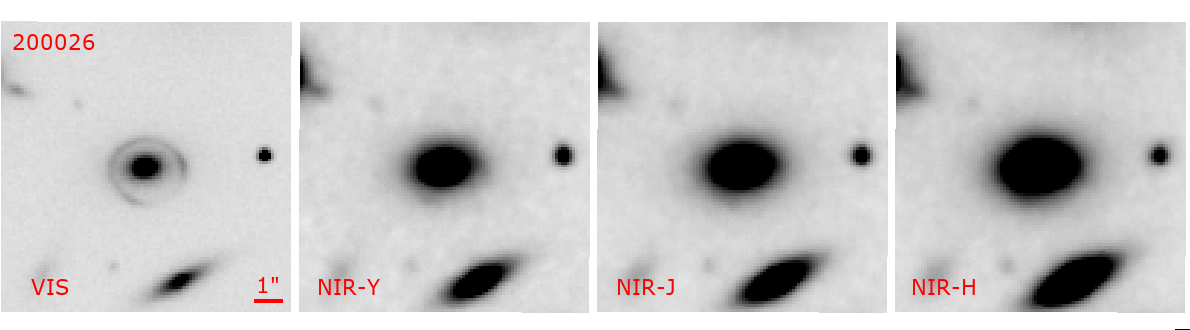}
  \caption{Simulated strong lens in the SC8 deep field (\Euclid\/ tile 306182). Strong Lens 200026 was identified in the VIS $\IE$ band, NISP-P \YE, \JE, and \HE bands by the strong lens detection algorithm developed by the Strong Lensing Science Working Group using the MER background-subtracted mosaic data.}
  \label{fig:sc8_deep_strong_lens}
\end{figure*}
\begin{figure*}[]
  \centering
  \includegraphics[width=\linewidth]{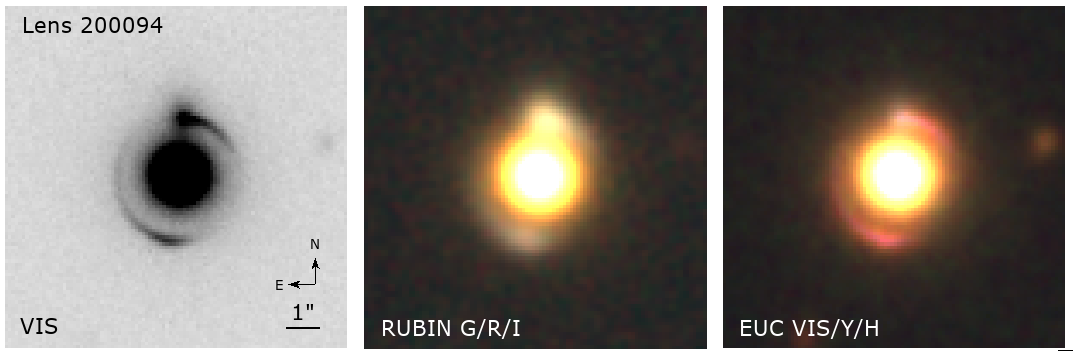}
  \caption{Imaging of a simulated strong lens in the VIS $\IE$ band, $Rubin$ ($g$+$r$+$i$) bands and combined VIS and NISP-P (\IE + \YE + \HE) bands of the SC8 main run.  Strong Lens 200094 was identified in \Euclid\/ tile 74088 by the detection algorithms of the Strong Lensing Science Working Group.}
  \label{fig:sc8_wide_strong_lens}
\end{figure*}
\begin{figure*}[]
  \centering
  \includegraphics[width=\linewidth]{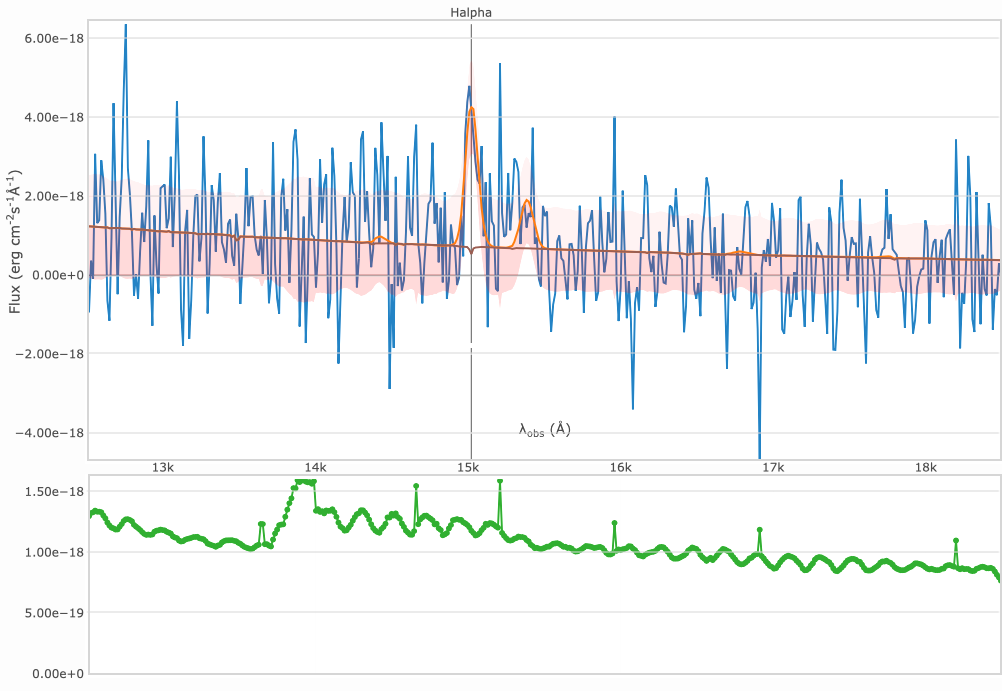}
  \caption{Stacked and processed spectra from simulated SC8 NISP-S images. The colours in the plot are identified as the following: blue line corresponds to the measured spectrum, brown line corresponds to continuum, orange line corresponds to model and green line corresponds to estimated noise. This galaxy was successfully identified at $z=1.29$.}
  \label{fig:sc8_wide_spectra}
\end{figure*}

Notably, even with shallower data than the actual $Rubin$ experiment, the unprecedented enormity of the three billion pixel $Rubin$ camera with its $3\dotdeg5$ ($9.6\,\rm{deg}^2$) field of view, combined with imaging capability in five optical bands,\footnote{We do not simulate the sixth $Rubin$ broadband filter: $Y$, ranging from $948\,{\rm nm}$ to $1060\,{\rm nm}$.} yielded the largest dataset. These results provide insight on the volume of actual EXT ground-based data that has to be processed for the \Euclid mission.  EXT data may require as much storage and computing time (or more) as the \Euclid data themselves, especially if performing pixel-level processing of purely raw data products. Finally, the total wall time to produce the entire main SC8 dataset was $16\,730$ days.  The VIS simulations, which remained full-featured, dominated the processing time with $5\,200$ days followed by NISP-S at $4\,700$ days.

\subsection{Deep run}
\label{s:sc8_deep}

For the SC8 Euclid deep survey, we simulated a small area measuring $\sim 1\,\rm{deg}^{2}$ of the \Euclid deep field around the north ecliptic pole. Figure~\ref{fig:sc8_deep_field} shows the SC8 deep field footprint and the number of layers to reach the desired depth.  We produced sets of full-featured simulations for VIS, NISP-P, and NISP-S with ten times more exposures totalling 840 amounting to $\sim$1.4 TB (as reported in Table~\ref{table:deep_production_summary}) and reaching a flux limit 2 mag fainter than the nominal wide field. For NISP-S, we also ran simulations of the blue grism (BGS00) in addition to the three other spectroscopic channels. We did not produce simulations of any northern EXT ground-based surveys for which no current strategies have been defined at these fainter magnitude depths. Nevertheless, the SC8 deep field run was designed to be used for testing by the level-2 SGS processing chains, which perform image and spectra calibration, photometry, and coaddition for VIS and NISP, as well as PSF modelling and the generation of optical distortion maps.

\subsection{Outcomes of SC8}
\label{s:sc8_results}

Here, we provide a brief summary of the SGS end-to-end performance in SC8 as a more in-depth discussion goes beyond the scope of this work.  Our total production of simulations for the SC8 main and deep surveys yielded 33 TB of data. The entire SC8 dataset was transferred to the EAS and queried by the level-2 OUs to initiate the data reduction pipelines, signaling the start of the end-to-end integration test.

Data-processing by the full SGS pipeline proceeded gradually with data quality inspection by the respective OUs and updates to various software packages. Nearly all level-3 processing elements\footnote{The codes to compute the photometric visibility mask and the covariance matrices for the galaxy two-point correlation and power spectrum (high priority processing functions) were not fully integrated.} were successfully integrated.  The full integration test also served as an aid to identify and resolve performance issues concerning infrastructure, bugs, and bottlenecks in software.  It also provided an impetus to further improve communications between interfacing OUs in order to maximise SGS capabilities as we approach operational readiness.

We conclude this section by highlighting the work enabled by the SC8 simulations such as the testing of  detection algorithms for strong lenses, galaxies high-$z$ quasars and galaxy emission-lines with $\Euclid$. As detailed in Appendix ~\ref{a:sc8_preparation}, along with the standard galaxy and standard stars, the object catalogues also contained a variety of additional sources such as QSOs, high-$z$ galaxies, strong lenses, and LT stars (brown dwarfs).
From the sample of 801 strong lenses, 285 sources were inserted into 216 \Euclid\/ tiles covering the SC8 main field. The strong lensing sample The catalogues containing these sources were provided as inputs to the various simulator codes for the VIS, NISP-P, and EXT instruments (i.e. \texttt{ELViS}, \texttt{Imagem}, \texttt{TiPS}, and \texttt{SIM-EXT}), which introduced instrumental and environmental effects in each band to produce raw images that were then processed by the SGS pipelines.  In detail, VIS and the NISP-P calibrated frames were produced by the respective OU level-2 pipelines for VIS and NIR image reduction.  

The VIS and NIR data products along with their corresponding flag maps (e.g. cosmics) then proceeded through the SGS data-processing chain as inputs to the OU-MER level-2 pipeline, where they were stacked. The MER pipeline then performed the background removal and used the flags to produce fully detrended and background subtracted MER mosaics.

Figure~\ref{fig:sc8_sci_mosaic} presents an example of raw (level-1) and processed (level-2) data products on a zoom-in of \Euclid\/ tile 306182 in the SC8 deep field.  On the left, we show the raw simulated science exposures for VIS (top-left) and NISP-P (bottom-left; for $\YE$ band only) with the characteristic features of these two instruments (described in Sects.~\ref{s:vis_simulator} and \ref{s:nispp_sim}, respectively). The two rightmost panels show the corresponding calibrated, stacked, and background subtracted mosaic images produced by OU-MER after processing the data products generated by OU-VIS and OU-NIR, respectively.  

These MER mosaic images and the source catalogues generated from them were used, respectively, by the Strong Lensing and Primaeval Universe Science Working Groups to perform tests of their corresponding detection algorithms in development. 

Figure~\ref{fig:sc8_deep_strong_lens} shows a zoom-in cut-out of the strong lens 200026 in the VIS and NIR deep simulations.  The Einstein ring can be seen with fine detail by the VIS instrument, whereas it is indiscernible in the $\Euclid$ NIR bands.  Another of the Strong Lensing Science Working Group detections was a strong lens system (200094) in the SC8 main run (\Euclid\/ tile 74088) in both MER processed $\Euclid$ and $Rubin$ mosaic data.  Cutout images of the system are shown in Fig.~\ref{fig:sc8_wide_strong_lens}; from left-to-right, we show a set of images of the Einstein ring as seen in the VIS $\IE$ band, combined Rubin $g$+$r$+$i$ bands, and combined VIS and NISP-P ($\IE$+$\YE$+$\HE$) bands.  These images demonstrate the resolving power of both \Euclid and LSST, two forthcoming contemporary and pivotal science programs.  They also, however, exemplify the exceptional image quality anticipated with the VIS camera alone with its superb capability to deblend compact systems. 

In Fig.~\ref{fig:sc8_wide_spectra}, we show an example of a successfully calibrated, decontaminated, stacked, and modelled galaxy emission-line spectra produced from the raw pixel data of the NISP-S grism simulations in SC8 processed by the Organisation Unit for Near Infrared Slitless Spectroscopy (OU-SIR) and Organisation for Spectral Measurement (OU-SPE), respectively. The results in the figure show a clearly identifying the redshifted H$\alpha$ emission line by the OU-SPE spectroscopic emission line detection algorithm.  The identified emission-lines were then used by OU-SPE to estimate spectroscopic redshifts. 

\section{Future perspective}
\label{s:future_perspective}

Following SC8, new features and improvements have been added to our simulation framework. For example, to produce more realistic profiles, we implement deep generative models, a machine learning framework presented in \citet{ml-stamps-bretoniere2021} that can generate complex galaxy stamps learned from a training set of real $Hubble$ Space Telescope observations. These galaxy thumbnails include spiral arms and other features that cannot be represented with analytical profiles as the ones we currently use. The advantage of this approach is that it can draw complex stamps just from parameterised information such as the one available in the TU galaxy catalogue without complex morphological models in the Flagship galaxy mock. These realistic stamps will enable the development of other complex algorithms to measure shear and to deblend larger objects. The colour gradient and its impact on shear measurments have already been studied \citep{welikala2012} and an implementation in the SIM pipeline is currently in progress. 

For the NISP simulators, we introduced an additional background component caused by persistence: trapped charges from one exposure are released on later image acquisitions. In early ground-based calibrations, a persistence signal was detected more than two hours after the initial exposure for bright sources. Among other dependencies, the charge reached during an exposure, exposure time and history of successive illuminations all affect the release of charges trapped in a pixel. A preliminary model only takes into account the single preceding exposure.  For the persistence signal $I_{\rm persistence}$ that is added to the current signal, we use the following model:
\begin{equation}
    I_{\rm persistence} =  \Big(p_0 + 10^{-5} p_1 I_{-1})(1-{\rm e}^{-I_{-1}/p_2}\Big) \, ,
\end{equation}
where $I_{-1}$ is the previous image and $p_i$ are a set of measured persistence parameters.  Distinct $p_i$ values are used for each pixel, previous exposure time and the elapsed time between exposures in the survey. Additionally, we have released within the Euclid Consortium newer versions of \texttt{SIM-EXT} (2.17+), which feature spatially varying PSFs for CFIS and WISHES.  These are produced using \texttt{PSFEx} models for CFIS $u$ and $r$ bands and WISHES $z$-band data provided to us courtesy of the CFIS and WISHES teams.  Our \textit{Rubin} simulations now include nominal rotation of the camera per pointing acquired from the \texttt{OpSim} v2.0\footnote{\url{https://epyc.astro.washington.edu/~lynnej/opsim_downloads/fbs_2.0/}} survey simulations \citep{lsst-surveysim-jones2020}. Furthermore, these simulations currently implement the FITS format \citep[described in][]{LSE-400} as well as the camera layout and WCS description of DESC \texttt{imSim} version 0.6.2\footnote{\url{https://github.com/LSSTDESC/imSim/tree/v0.6.2}} \citep{descDC2-abolfathi2021}. The file formatting and WCS changes fulfill the requisites of the latest generation LSST Science Pipelines software to perform processing of LSST simulations.  Finally, \texttt{SIM-EXT} 2.17+ now provides simulated data products in the $g$ band for the Waterloo-Hawaii-IfA $G$-band Survey (WHIGS) using the Subaru telescope.

The complete set of calibration exposures for the \Euclid instrument will encompass a large variety of image types. In addition to the standard calibration exposures: bias, dark, flats, we also produce calibration unit exposures, out-of-focus exposures for phase diversity, different readout modes for NISP, injection line frames, trap pumping for VIS CTI calibration, NISP persistence model validation, bleeding and blooming calibration exposures. For the PLM/instrument calibration exposures, we produce straylight checks and dense stellar fields. Finally, for precise astrometric calibration, we generate white dwarf and planetary nebulae observations.

Upon completion of PVPR, the \Euclid Ground Segment will commence its third SPV (SPV3) phase. The focus of this campaign will be on assessing the scientific performance of the \Euclid mission.  In detail, we plan to produce nominal science exposures for the wide and deep areas spanning a few tens of degrees to yield mock WL and GC measurements (e.g. C$_{\ell}$ and power spectrum multipoles) to enable testing of the Euclid Consortium Science Working Groups analysis pipelines.

\section{Summary and outlook}
\label{s:summ_and_outlook}

OU-SIM is responsible for providing detailed simulations to facilitate the development of the SGS data-processing pipeline as well as end-to-end performance tests.  In this work, we have introduced our simulation framework for providing a full pixel simulation pipeline for the entire \Euclid mission and presented its application to SC8, the eighth testing campaign of the \Euclid Ground Segment and its first end-to-end test. The SC8 dataset was generated from galaxy mocks built from the Flagship v1.0 simulation, the largest cosmological $N$-body simulation at the time it was developed. Additional sources were added to complement this massive simulation, such as high-redshift QSOs and strongly lensed galaxies. Our stellar catalogue was constructed from a combination of a very large and deep Besançon model and real all-sky stellar catalogues.  From these catalogues, we produced and delivered simulated images for a simplified Euclid Wide Survey of $165\,{\rm deg}^2$ and a full-featured $1\,{\rm deg}^2$ Euclid Deep Survey. 

The framework we adopted for SC8 follows a highly modular design from the onset, consisting of individual components for the TU catalogue inputs, pixel-level simulators for space- and ground-based instruments, a set of common tools for generating spectra and galaxy shapes, and a software API for orchestrating large sets of simulations at any of the international \Euclid SGS data centres. Furthermore, all instrument models and reference surveys used in SC8 as well as all past or future challenges and production cycles are stored in the MDB repository. We also devised a validation program consisting of suites of tests for our simulation pipeline to ensure the accuracy of the released data products as well as adherence to scientific and technical requirements.

We shall continue to upgrade our simulation framework to enable large-scale testing and improvements to the ground segment pipeline. We plan to provide data products to aid the validation of very complex algorithms (e.g. VIS PSF estimation) and to use in conducting mission performance evaluations throughout this active development phase. Our longer-term goals during \Euclid operations are to produce simulations to calibrate all possible systematics in the data (e.g. shear measurement biases and selection biases) and to interpret discrepancies or surprises found in the in-flight data.

\begin{acknowledgements}
We thank the OU-VIS, -NIR, -SIR, -EXT, and -MER teams for their assistance with validation and providing valuable feedback on the simulations. We are grateful to the System Team for the technical help and continuous support. We thank the Cosmological Simulations Working Group and the Flagship team for the outstanding galaxy mocks and support on their integration into OU-SIM. We thank both VIS and NISP instrument teams for providing instrument models and detailed explanations.  We thank OU-SHE for their support with generating the VIS PSF dataset. We would like to thank Benjamin Clément from the Strong Lensing Science Working Group, who produced the images of the strong lenses detected in the SC8 main and deep fields which were presented in Figures~\ref{fig:sc8_deep_strong_lens} and \ref{fig:sc8_wide_strong_lens}. We thank Vincent Le Brun from the Organisation Unit for Spectral Measurement (OU-SPE) for producing Figure~\ref{fig:sc8_wide_spectra}.  We also express our gratitude to Dominique Boutigny, Jim Chiang, and Phillipe Gris of the Dark Energy Science Collaboration as well as Jim Bosch, Tim Jenness, Lynne Jones, Kian-Tat Lim, and the other members from \textit{Rubin} Observatory Data Management and Survey Strategy teams for providing support, useful insight and sample datasets to aid us with upgrading the SGS Rubin simulations and ensuring compliance with the LSST Science Pipelines. We would also like to thank Jesus Varela and JEDIS, WISHES, Pan-STARRS and WHIGS teams for assisting us with information, data and the models needed to produce the respective image simulations for the Northern Surveys.  We acknowledge all Science Data Centres (SDC-CH: University of Geneva, SDC-DE: Max Planck Computing and Data Facility, SDC-ES: Port d'Informació Científica, SDC-FI: CSC – IT Center for Science, SDC-FR: Centre de calcul de l’IN2P3, SDC-IT: INAF Astronomical Observatory of Trieste/ALTEC S.p.A. in Turin, SDC-NL: University of Groningen, SDC-UK: Institute for Astronomy, University of Edinburgh, and SDC-US: Caltech/IPAC) that contributed to the production of simulations for SC8. A portion of this work also used the resources available at CosmoHub. CosmoHub has been developed by the Port d'Informació Científica, maintained through a collaboration of the Institut de Física d'Altes Energies and the Centro de Investigaciones Energéticas, Medioambientales y Tecnológicas and the Institute of Space Sciences, and was partially funded by the ``Plan Estatal de Investigación Científica y Técnica y de Innovación" program of the Spanish government. The author acknowledges support from the Spanish Research Project PID2021-123012NA-C44 [MICINNFEDER]. This work has been financed by MCINN/AEI and the EU NextGeneration/PRTR project EQC2021-007479-P. This study was supported by MICIIN with funding from European Union NextGenerationEU(PRTR-C17.I1) and by Generalitat de Catalunya.  \AckECol  
\end{acknowledgements}


\begin{thebibliography}{89}
\expandafter\ifx\csname natexlab\endcsname\relax\def\natexlab#1{#1}\fi

\bibitem[{{Abbott} {et~al.}(2018){Abbott}, {Abdalla}, {Allam}, {Amara},
  {Annis}, {Asorey}, {Avila}, {Ballester}, {Banerji}, {Barkhouse}, {Baruah},
  {Baumer}, {Bechtol}, {Becker}, {Benoit-L{\'e}vy}, {Bernstein}, {Bertin},
  {Blazek}, {Bocquet}, {Brooks}, {Brout}, {Buckley-Geer}, {Burke}, {Busti},
  {Campisano}, {Cardiel-Sas}, {Carnero Rosell}, {Carrasco Kind}, {Carretero},
  {Castander}, {Cawthon}, {Chang}, {Chen}, {Conselice}, {Costa}, {Crocce},
  {Cunha}, {D'Andrea}, {da Costa}, {Das}, {Daues}, {Davis}, {Davis}, {De
  Vicente}, {DePoy}, {DeRose}, {Desai}, {Diehl}, {Dietrich}, {Dodelson},
  {Doel}, {Drlica-Wagner}, {Eifler}, {Elliott}, {Evrard}, {Farahi}, {Fausti
  Neto}, {Fernandez}, {Finley}, {Flaugher}, {Foley}, {Fosalba}, {Friedel},
  {Frieman}, {Garc{\'\i}a-Bellido}, {Gaztanaga}, {Gerdes}, {Giannantonio},
  {Gill}, {Glazebrook}, {Goldstein}, {Gower}, {Gruen}, {Gruendl}, {Gschwend},
  {Gupta}, {Gutierrez}, {Hamilton}, {Hartley}, {Hinton}, {Hislop}, {Hollowood},
  {Honscheid}, {Hoyle}, {Huterer}, {Jain}, {James}, {Jeltema}, {Johnson},
  {Johnson}, {Kacprzak}, {Kent}, {Khullar}, {Klein}, {Kovacs}, {Koziol},
  {Krause}, {Kremin}, {Kron}, {Kuehn}, {Kuhlmann}, {Kuropatkin}, {Lahav},
  {Lasker}, {Li}, {Li}, {Liddle}, {Lima}, {Lin}, {L{\'o}pez-Reyes}, {MacCrann},
  {Maia}, {Maloney}, {Manera}, {March}, {Marriner}, {Marshall}, {Martini},
  {McClintock}, {McKay}, {McMahon}, {Melchior}, {Menanteau}, {Miller},
  {Miquel}, {Mohr}, {Morganson}, {Mould}, {Neilsen}, {Nichol}, {Nogueira},
  {Nord}, {Nugent}, {Nunes}, {Ogando}, {Old}, {Pace}, {Palmese},
  {Paz-Chinch{\'o}n}, {Peiris}, {Percival}, {Petravick}, {Plazas}, {Poh},
  {Pond}, {Porredon}, {Pujol}, {Refregier}, {Reil}, {Ricker}, {Rollins},
  {Romer}, {Roodman}, {Rooney}, {Ross}, {Rykoff}, {Sako}, {Sanchez}, {Sanchez},
  {Santiago}, {Saro}, {Scarpine}, {Scolnic}, {Serrano}, {Sevilla-Noarbe},
  {Sheldon}, {Shipp}, {Silveira}, {Smith}, {Smith}, {Smith}, {Soares-Santos},
  {Sobreira}, {Song}, {Stebbins}, {Suchyta}, {Sullivan}, {Swanson}, {Tarle},
  {Thaler}, {Thomas}, {Thomas}, {Troxel}, {Tucker}, {Vikram}, {Vivas},
  {Walker}, {Wechsler}, {Weller}, {Wester}, {Wolf}, {Wu}, {Yanny}, {Zenteno},
  {Zhang}, {Zuntz}, {DES Collaboration}, {Juneau}, {Fitzpatrick}, {Nikutta},
  {Nidever}, {Olsen}, {Scott}, \& {NOAO Data Lab}}]{DR1-DES2018}
{Abbott}, T.~M.~C., {Abdalla}, F.~B., {Allam}, S., {et~al.} 2018, \apjs, 239,
  18

\bibitem[{Abolfathi {et~al.}(2021)Abolfathi, Alonso, Armstrong, Aubourg, Awan,
  Babuji, Bauer, Bean, Beckett, Biswas, Bogart, Boutigny, Chard, Chiang,
  Claver, Cohen-Tanugi, Combet, Connolly, Daniel, Digel, Drlica-Wagner, Dubois,
  Gangler, Gawiser, Glanzman, Gris, Habib, Hearin, Heitmann, Hernandez,
  Hlo{\v{z}}ek, Hollowed, Ishak, Ivezi{\'{c}}, Jarvis, Jha, Kahn, Kalmbach,
  Kelly, Kovacs, Korytov, Krughoff, Lage, Lanusse, Larsen, Guillou, Li,
  Longley, Lupton, Mandelbaum, Mao, Marshall, Meyers, Moniez, Morrison,
  Nomerotski, O'Connor, Park, Park, Peloton, Perrefort, Perry, Plaszczynski,
  Pope, Rasmussen, Reil, Roodman, Rykoff, S{\'{a}}nchez, Schmidt, Scolnic,
  Stubbs, Tyson, Uram, Villarreal, Walter, Wiesner, Wood-Vasey, \&
  Zuntz}]{descDC2-abolfathi2021}
Abolfathi, B., Alonso, D., Armstrong, R., {et~al.} 2021, \apss, 253, 31

\bibitem[{Barbier {et~al.}(2018)Barbier, Buton, Clemens, Conversi, Ealet,
  Ferriol, Fornari, Gillard, Kohley, Kubik, Rosset, Secroun, Serra, Smadja, \&
  Zoubian}]{detectorcalibration-barbier2018}
Barbier, R., Buton, C., Clemens, J.-C., {et~al.} 2018, in High Energy, Optical,
  and Infrared Detectors for Astronomy VIII, ed. A.~D. Holland \& J.~Beletic,
  Vol. 10709, International Society for Optics and Photonics (SPIE), 107090S

\bibitem[{{Behroozi} {et~al.}(2013){Behroozi}, {Wechsler}, \&
  {Wu}}]{behroozi2012}
{Behroozi}, P.~S., {Wechsler}, R.~H., \& {Wu}, H.-Y. 2013, \apj, 762, 109

\bibitem[{{Bertin}(2006)}]{scamp-bertin2006}
{Bertin}, E. 2006, in Astronomical Society of the Pacific Conference Series,
  Vol. 351, Astronomical Data Analysis Software and Systems XV, ed.
  C.~{Gabriel}, C.~{Arviset}, D.~{Ponz}, \& S.~{Enrique}, 112

\bibitem[{{Bertin}(2011)}]{psfex-bertin2011}
{Bertin}, E. 2011, in Astronomical Society of the Pacific Conference Series,
  Vol. 442, Astronomical Data Analysis Software and Systems XX, ed. I.~N.
  {Evans}, A.~{Accomazzi}, D.~J. {Mink}, \& A.~H. {Rots}, 435

\bibitem[{{BGM web-service, OSU
  THETA}(https://model.obs-besancon.fr/)}]{besancon_reference}
{BGM web-service, OSU THETA}. https://model.obs-besancon.fr/, Besan\c{c}on
  model of stellar population synthesis of the Galaxy

\bibitem[{{Bizyaev}(2007)}]{bizyaev2007}
{Bizyaev}, D. 2007, in American Astronomical Society Meeting Abstracts, Vol.
  211, American Astronomical Society Meeting Abstracts, 13.23

\bibitem[{{Bosch} {et~al.}(2018){Bosch}, {Armstrong}, {Bickerton}, {Furusawa},
  {Ikeda}, {Koike}, {Lupton}, {Mineo}, {Price}, {Takata}, {Tanaka}, {Yasuda},
  {AlSayyad}, {Becker}, {Coulton}, {Coupon}, {Garmilla}, {Huang}, {Krughoff},
  {Lang}, {Leauthaud}, {Lim}, {Lust}, {MacArthur}, {Mandelbaum}, {Miyatake},
  {Miyazaki}, {Murata}, {More}, {Okura}, {Owen}, {Swinbank}, {Strauss},
  {Yamada}, \& {Yamanoi}}]{lsststack-bosch2018}
{Bosch}, J., {Armstrong}, R., {Bickerton}, S., {et~al.} 2018, \pasj, 70, S5

\bibitem[{{Bruzual} \& {Charlot}(2003)}]{cosmosseds-bruzual2003}
{Bruzual}, G. \& {Charlot}, S. 2003, \mnras, 344, 1000

\bibitem[{{Calzetti} {et~al.}(2000){Calzetti}, {Armus}, {Bohlin}, {Kinney},
  {Koornneef}, \& {Storchi-Bergmann}}]{cosmosdust-calzetti2000}
{Calzetti}, D., {Armus}, L., {Bohlin}, R.~C., {et~al.} 2000, \apj, 533, 682

\bibitem[Cardamone et al.(2010)]{cardamone2010} Cardamone, 
C.~N., van Dokkum, P.~G., Urry, C.~M., et al.\ 2010, 
\apjs, 189, 270. 

\bibitem[{{Carretero} {et~al.}(2015){Carretero}, {Castander}, {Gazta{\~n}aga},
  {Crocce}, \& {Fosalba}}]{micemock-carretero2015}
{Carretero}, J., {Castander}, F.~J., {Gazta{\~n}aga}, E., {Crocce}, M., \&
  {Fosalba}, P. 2015, \mnras, 447, 646

\bibitem[{{Carretero} {et~al.}(2017){Carretero}, {Tallada}, {Casals}, {Caubet},
  {Castander}, {Blot}, {Alarc{\'o}n}, {Serrano}, {Fosalba}, {Acosta-Silva},
  {Tonello}, {Torradeflot}, {Eriksen}, {Neissner}, \&
  {Delfino}}]{scipic-carretero2017}
{Carretero}, J., {Tallada}, P., {Casals}, J., {et~al.} 2017, in Proceedings of
  the European Physical Society Conference on High Energy Physics. 5-12 July,
  488

\bibitem[{{Chambers} {et~al.}(2016){Chambers}, {Magnier}, {Metcalfe},
  {Flewelling}, {Huber}, {Waters}, {Denneau}, {Draper}, {Farrow}, {Finkbeiner},
  {Holmberg}, {Koppenhoefer}, {Price}, {Rest}, {Saglia}, {Schlafly}, {Smartt},
  {Sweeney}, {Wainscoat}, {Burgett}, {Chastel}, {Grav}, {Heasley}, {Hodapp},
  {Jedicke}, {Kaiser}, {Kudritzki}, {Luppino}, {Lupton}, {Monet}, {Morgan},
  {Onaka}, {Shiao}, {Stubbs}, {Tonry}, {White}, {Ba{\~n}ados}, {Bell},
  {Bender}, {Bernard}, {Boegner}, {Boffi}, {Botticella}, {Calamida},
  {Casertano}, {Chen}, {Chen}, {Cole}, {Deacon}, {Frenk}, {Fitzsimmons},
  {Gezari}, {Gibbs}, {Goessl}, {Goggia}, {Gourgue}, {Goldman}, {Grant},
  {Grebel}, {Hambly}, {Hasinger}, {Heavens}, {Heckman}, {Henderson}, {Henning},
  {Holman}, {Hopp}, {Ip}, {Isani}, {Jackson}, {Keyes}, {Koekemoer}, {Kotak},
  {Le}, {Liska}, {Long}, {Lucey}, {Liu}, {Martin}, {Masci}, {McLean}, {Mindel},
  {Misra}, {Morganson}, {Murphy}, {Obaika}, {Narayan}, {Nieto-Santisteban},
  {Norberg}, {Peacock}, {Pier}, {Postman}, {Primak}, {Rae}, {Rai}, {Riess},
  {Riffeser}, {Rix}, {R{\"o}ser}, {Russel}, {Rutz}, {Schilbach}, {Schultz},
  {Scolnic}, {Strolger}, {Szalay}, {Seitz}, {Small}, {Smith}, {Soderblom},
  {Taylor}, {Thomson}, {Taylor}, {Thakar}, {Thiel}, {Thilker}, {Unger},
  {Urata}, {Valenti}, {Wagner}, {Walder}, {Walter}, {Watters}, {Werner},
  {Wood-Vasey}, \& {Wyse}}]{panstarrs_chambers_2016}
{Chambers}, K.~C., {Magnier}, E.~A., {Metcalfe}, N., {et~al.} 2016, arXiv
  e-prints, 1612.05560

\bibitem[{{Claver} {et~al.}(2014){Claver}, {Selvy}, {Angeli}, {Delgado},
  {Dubois-Felsmann}, {Hascall}, {Lotz}, {Marshall}, {Schumacher}, \&
  {Sebag}}]{lsst-syseng-claver2014}
{Claver}, C.~F., {Selvy}, B.~M., {Angeli}, G., {et~al.} 2014, in Society of
  Photo-Optical Instrumentation Engineers (SPIE) Conference Series, Vol. 9150,
  Modeling, Systems Engineering, and Project Management for Astronomy VI, ed.
  G.~Z. {Angeli} \& P.~{Dierickx}, 91500M

\bibitem[{{Collett}(2015)}]{collett2015}
{Collett}, T.~E. 2015, \apj, 811, 20

\bibitem[{{Connolly} {et~al.}(2014){Connolly}, {Angeli}, {Chandrasekharan},
  {Claver}, {Cook}, {Ivezic}, {Jones}, {Krughoff}, {Peng}, {Peterson}, {Petry},
  {Rasmussen}, {Ridgway}, {Saha}, {Sembroski}, {vanderPlas}, \&
  {Yoachim}}]{opsim-connolly}
{Connolly}, A.~J., {Angeli}, G.~Z., {Chandrasekharan}, S., {et~al.} 2014, in
  Society of Photo-Optical Instrumentation Engineers (SPIE) Conference Series,
  Vol. 9150, Modeling, Systems Engineering, and Project Management for
  Astronomy VI, ed. G.~Z. {Angeli} \& P.~{Dierickx}, 915014

\bibitem[{{Cropper} {et~al.}(2014){Cropper}, {Pottinger}, {Niemi}, {Denniston},
  {Cole}, {Szafraniec}, {Mellier}, {Berth{\'e}}, {Martignac}, {Cara}, {di
  Giorgio}, {Sciortino}, {Paltani}, {Genolet}, {Fourmand}, {Charra},
  {Guttridge}, {Winter}, {Endicott}, {Holland}, {Gow}, {Murray}, {Hall},
  {Amiaux}, {Laureijs}, {Racca}, {Salvignol}, {Short}, {Lorenzo Alvarez},
  {Kitching}, {Hoekstra}, \& {Massey}}]{cropper2014}
{Cropper}, M., {Pottinger}, S., {Niemi}, S.~M., {et~al.} 2014, in Society of
  Photo-Optical Instrumentation Engineers (SPIE) Conference Series, Vol. 9143,
  Space Telescopes and Instrumentation 2014: Optical, Infrared, and Millimeter
  Wave, ed. J.~{Oschmann}, Jacobus~M., M.~{Clampin}, G.~G. {Fazio}, \& H.~A.
  {MacEwen}, 91430J

\bibitem[{Delgado \& Reuter(2016)}]{opsim-delgado}
Delgado, F. \& Reuter, M.~A. 2016, in Observatory Operations: Strategies,
  Processes, and Systems VI, ed. A.~B. Peck, R.~L. Seaman, \& C.~R. Benn, Vol.
  9910, International Society for Optics and Photonics (SPIE), 991013

\bibitem[{{Dimauro} {et~al.}(2018){Dimauro}, {Huertas-Company}, {Daddi},
  {P{\'e}rez-Gonz{\'a}lez}, {Bernardi}, {Barro}, {Buitrago}, {Caro},
  {Cattaneo}, {Dominguez-S{\'a}nchez}, {Faber}, {H{\"a}u{\ss}ler}, {Kocevski},
  {Koekemoer}, {Koo}, {Lee}, {Mei}, {Margalef-Bentabol}, {Primack},
  {Rodriguez-Puebla}, {Salvato}, {Shankar}, \&
  {Tuccillo}}]{candels-dimauro2018}
{Dimauro}, P., {Huertas-Company}, M., {Daddi}, E., {et~al.} 2018, \mnras, 478,
  5410

\bibitem[{{Drimmel} {et~al.}(2003){Drimmel}, {Cabrera-Lavers}, \&
  {L{\'o}pez-Corredoira}}]{drimmel2003}
{Drimmel}, R., {Cabrera-Lavers}, A., \& {L{\'o}pez-Corredoira}, M. 2003, \aap,
  409, 205

\bibitem[{{Euclid Collaboration: Barnett} {et~al.}(2019){Euclid Collaboration:
  Barnett}, {Warren}, {Mortlock}, {Cuby}, {Conselice}, {Hewett}, {Willott},
  {Auricchio}, {Balaguera-Antol{\'\i}nez}, {Baldi}, {Bardelli}, {Bellagamba},
  {Bender}, {Biviano}, {Bonino}, {Bozzo}, {Branchini}, {Brescia}, {Brinchmann},
  {Burigana}, {Camera}, {Capobianco}, {Carbone}, {Carretero}, {Carvalho},
  {Castander}, {Castellano}, {Cavuoti}, {Cimatti}, {Cl{\'e}dassou}, {Congedo},
  {Conversi}, {Copin}, {Corcione}, {Coupon}, {Courtois}, {Cropper}, {Da Silva},
  {Duncan}, {Dusini}, {Ealet}, {Farrens}, {Fosalba}, {Fotopoulou},
  {Fourmanoit}, {Frailis}, {Fumana}, {Galeotta}, {Garilli}, {Gillard},
  {Gillis}, {Graci{\'a}-Carpio}, {Grupp}, {Hoekstra}, {Hormuth}, {Israel},
  {Jahnke}, {Kermiche}, {Kilbinger}, {Kirkpatrick}, {Kitching}, {Kohley},
  {Kubik}, {Kunz}, {Kurki-Suonio}, {Laureijs}, {Ligori}, {Lilje}, {Lloro},
  {Maiorano}, {Mansutti}, {Marggraf}, {Martinet}, {Marulli}, {Massey}, {Mauri},
  {Medinaceli}, {Mei}, {Mellier}, {Metcalf}, {Metge}, {Meylan}, {Moresco},
  {Moscardini}, {Munari}, {Neissner}, {Niemi}, {Nutma}, {Padilla}, {Paltani},
  {Pasian}, {Paykari}, {Percival}, {Pettorino}, {Polenta}, {Poncet},
  {Pozzetti}, {Raison}, {Renzi}, {Rhodes}, {Rix}, {Romelli}, {Roncarelli},
  {Rossetti}, {Saglia}, {Sapone}, {Scaramella}, {Schneider}, {Scottez},
  {Secroun}, {Serrano}, {Sirri}, {Stanco}, {Sureau}, {Tallada-Cresp{\'\i}},
  {Tavagnacco}, {Taylor}, {Tenti}, {Tereno}, {Toledo-Moreo}, {Torradeflot},
  {Valenziano}, {Vassallo}, {Wang}, {Zacchei}, {Zamorani}, {Zoubian}, \&
  {Zucca}}]{highqso-barnett2019}
{Euclid Collaboration: Barnett}, R., {Warren}, S.~J., {Mortlock}, D.~J.,
  {et~al.} 2019, \aap, 631, A85

\bibitem[{{Euclid Collaboration: Blanchard} {et~al.}(2020){Euclid
  Collaboration: Blanchard}, {Camera}, {Carbone}, {Cardone}, {Casas}, {Clesse},
  {Ili{\'c}}, {Kilbinger}, {Kitching}, {Kunz}, {Lacasa}, {Linder}, {Majerotto},
  {Markovi{\v{c}}}, {Martinelli}, {Pettorino}, {Pourtsidou}, {Sakr},
  {S{\'a}nchez}, {Sapone}, {Tutusaus}, {Yahia-Cherif}, {Yankelevich},
  {Andreon}, {Aussel}, {Balaguera-Antol{\'\i}nez}, {Baldi}, {Bardelli},
  {Bender}, {Biviano}, {Bonino}, {Boucaud}, {Bozzo}, {Branchini}, {Brau-Nogue},
  {Brescia}, {Brinchmann}, {Burigana}, {Cabanac}, {Capobianco}, {Cappi},
  {Carretero}, {Carvalho}, {Casas}, {Castander}, {Castellano}, {Cavuoti},
  {Cimatti}, {Cledassou}, {Colodro-Conde}, {Congedo}, {Conselice}, {Conversi},
  {Copin}, {Corcione}, {Coupon}, {Courtois}, {Cropper}, {Da Silva}, {de la
  Torre}, {Di Ferdinando}, {Dubath}, {Ducret}, {Duncan}, {Dupac}, {Dusini},
  {Fabbian}, {Fabricius}, {Farrens}, {Fosalba}, {Fotopoulou}, {Fourmanoit},
  {Frailis}, {Franceschi}, {Franzetti}, {Fumana}, {Galeotta}, {Gillard},
  {Gillis}, {Giocoli}, {G{\'o}mez-Alvarez}, {Graci{\'a}-Carpio}, {Grupp},
  {Guzzo}, {Hoekstra}, {Hormuth}, {Israel}, {Jahnke}, {Keihanen}, {Kermiche},
  {Kirkpatrick}, {Kohley}, {Kubik}, {Kurki-Suonio}, {Ligori}, {Lilje}, {Lloro},
  {Maino}, {Maiorano}, {Marggraf}, {Martinet}, {Marulli}, {Massey},
  {Medinaceli}, {Mei}, {Mellier}, {Metcalf}, {Metge}, {Meylan}, {Moresco},
  {Moscardini}, {Munari}, {Nichol}, {Niemi}, {Nucita}, {Padilla}, {Paltani},
  {Pasian}, {Percival}, {Pires}, {Polenta}, {Poncet}, {Pozzetti}, {Racca},
  {Raison}, {Renzi}, {Rhodes}, {Romelli}, {Roncarelli}, {Rossetti}, {Saglia},
  {Schneider}, {Scottez}, {Secroun}, {Sirri}, {Stanco}, {Starck}, {Sureau},
  {Tallada-Cresp{\'\i}}, {Tavagnacco}, {Taylor}, {Tenti}, {Tereno},
  {Toledo-Moreo}, {Torradeflot}, {Valenziano}, {Vassallo}, {Verdoes Kleijn},
  {Viel}, {Wang}, {Zacchei}, {Zoubian}, \&
  {Zucca}}]{euclid-forecasts-blanchard2020}
{Euclid Collaboration: Blanchard}, A., {Camera}, S., {Carbone}, C., {et~al.}
  2020, \aap, 642, A191

\bibitem[{{Euclid Collaboration: Bretonni{\`e}re} {et~al.}(2022){Euclid
  Collaboration: Bretonni{\`e}re}, {Huertas-Company}, {Boucaud}, {Lanusse},
  {Jullo}, {Merlin}, {Tuccillo}, {Castellano}, {Brinchmann}, {Conselice},
  {Dole}, {Cabanac}, {Courtois}, {Castander}, {Duc}, {Fosalba}, {Guinet},
  {Kruk}, {Kuchner}, {Serrano}, {Soubrie}, {Tramacere}, {Wang}, {Amara},
  {Auricchio}, {Bender}, {Bodendorf}, {Bonino}, {Branchini}, {Brau-Nogue},
  {Brescia}, {Capobianco}, {Carbone}, {Carretero}, {Cavuoti}, {Cimatti},
  {Cledassou}, {Congedo}, {Conversi}, {Copin}, {Corcione}, {Costille},
  {Cropper}, {Da Silva}, {Degaudenzi}, {Douspis}, {Dubath}, {Duncan}, {Dupac},
  {Dusini}, {Farrens}, {Ferriol}, {Frailis}, {Franceschi}, {Fumana}, {Garilli},
  {Gillard}, {Gillis}, {Giocoli}, {Grazian}, {Grupp}, {Haugan}, {Holmes},
  {Hormuth}, {Hudelot}, {Jahnke}, {Kermiche}, {Kiessling}, {Kilbinger},
  {Kitching}, {Kohley}, {K{\"u}mmel}, {Kunz}, {Kurki-Suonio}, {Ligori},
  {Lilje}, {Lloro}, {Maiorano}, {Mansutti}, {Marggraf}, {Markovic}, {Marulli},
  {Massey}, {Maurogordato}, {Melchior}, {Meneghetti}, {Meylan}, {Moresco},
  {Morin}, {Moscardini}, {Munari}, {Nakajima}, {Niemi}, {Padilla}, {Paltani},
  {Pasian}, {Pedersen}, {Pettorino}, {Pires}, {Poncet}, {Popa}, {Pozzetti},
  {Raison}, {Rebolo}, {Rhodes}, {Roncarelli}, {Rossetti}, {Saglia},
  {Schneider}, {Secroun}, {Seidel}, {Sirignano}, {Sirri}, {Stanco}, {Starck},
  {Tallada-Cresp{\'\i}}, {Taylor}, {Tereno}, {Toledo-Moreo}, {Torradeflot},
  {Valentijn}, {Valenziano}, {Wang}, {Welikala}, {Weller}, {Zamorani},
  {Zoubian}, {Baldi}, {Bardelli}, {Camera}, {Farinelli}, {Medinaceli}, {Mei},
  {Polenta}, {Romelli}, {Tenti}, {Vassallo}, {Zacchei}, {Zucca}, {Baccigalupi},
  {Balaguera-Antol{\'\i}nez}, {Biviano}, {Borgani}, {Bozzo}, {Burigana},
  {Cappi}, {Carvalho}, {Casas}, {Castignani}, {Colodro-Conde}, {Coupon}, {de la
  Torre}, {Fabricius}, {Farina}, {Ferreira}, {Flose-Reimberg}, {Fotopoulou},
  {Galeotta}, {Ganga}, {Garcia-Bellido}, {Gaztanaga}, {Gozaliasl}, {Hook},
  {Joachimi}, {Kansal}, {Kashlinsky}, {Keihanen}, {Kirkpatrick}, {Lindholm},
  {Mainetti}, {Maino}, {Maoli}, {Martinelli}, {Martinet}, {McCracken},
  {Metcalf}, {Morgante}, {Morisset}, {Nightingale}, {Nucita}, {Patrizii},
  {Potter}, {Renzi}, {Riccio}, {S{\'a}nchez}, {Sapone}, {Schirmer},
  {Schultheis}, {Scottez}, {Sefusatti}, {Teyssier}, {Tutusaus}, {Valiviita},
  {Viel}, {Whittaker}, \& {Knapen}}]{ml-stamps-bretoniere2021}
{Euclid Collaboration: Bretonni{\`e}re}, H., {Huertas-Company}, M., {Boucaud},
  A., {et~al.} 2022, \aap, 657, A90

\bibitem[{{Euclid Collaboration: Martinet} {et~al.}(2019){Euclid Collaboration:
  Martinet}, {Schrabback}, {Hoekstra}, {Tewes}, {Herbonnet}, {Schneider},
  {Hernandez-Martin}, {Taylor}, {Brinchmann}, {Carvalho}, {Castellano},
  {Congedo}, {Gillis}, {Jullo}, {K{\"u}mmel}, {Ligori}, {Lilje}, {Padilla},
  {Paris}, {Peacock}, {Pilo}, {Pujol}, {Scott}, \&
  {Toledo-Moreo}}]{unresolvedgalaxies-martinet2019}
{Euclid Collaboration: Martinet}, N., {Schrabback}, T., {Hoekstra}, H.,
  {et~al.} 2019, \aap, 627, A59

\bibitem[{{Euclid Collaboration: Scaramella} {et~al.}(2022){Euclid
  Collaboration: Scaramella}, {Amiaux}, {Mellier}, {Burigana}, {Carvalho},
  {Cuillandre}, {Da Silva}, {Derosa}, {Dinis}, {Maiorano}, {Maris}, {Tereno},
  {Laureijs}, {Boenke}, {Buenadicha}, {Dupac}, {Gaspar Venancio},
  {G{\'o}mez-{\'A}lvarez}, {Hoar}, {Lorenzo Alvarez}, {Racca},
  {Saavedra-Criado}, {Schwartz}, {Vavrek}, {Schirmer}, {Aussel}, {Azzollini},
  {Cardone}, {Cropper}, {Ealet}, {Garilli}, {Gillard}, {Granett}, {Guzzo},
  {Hoekstra}, {Jahnke}, {Kitching}, {Maciaszek}, {Meneghetti}, {Miller},
  {Nakajima}, {Niemi}, {Pasian}, {Percival}, {Pottinger}, {Sauvage},
  {Scodeggio}, {Wachter}, {Zacchei}, {Aghanim}, {Amara}, {Auphan}, {Auricchio},
  {Awan}, {Balestra}, {Bender}, {Bodendorf}, {Bonino}, {Branchini},
  {Brau-Nogue}, {Brescia}, {Candini}, {Capobianco}, {Carbone}, {Carlberg},
  {Carretero}, {Casas}, {Castander}, {Castellano}, {Cavuoti}, {Cimatti},
  {Cledassou}, {Congedo}, {Conselice}, {Conversi}, {Copin}, {Corcione},
  {Costille}, {Courbin}, {Degaudenzi}, {Douspis}, {Dubath}, {Duncan}, {Dusini},
  {Farrens}, {Ferriol}, {Fosalba}, {Fourmanoit}, {Frailis}, {Franceschi},
  {Franzetti}, {Fumana}, {Gillis}, {Giocoli}, {Grazian}, {Grupp}, {Haugan},
  {Holmes}, {Hormuth}, {Hudelot}, {Kermiche}, {Kiessling}, {Kilbinger},
  {Kohley}, {Kubik}, {K{\"u}mmel}, {Kunz}, {Kurki-Suonio}, {Lahav}, {Ligori},
  {Lilje}, {Lloro}, {Mansutti}, {Marggraf}, {Markovic}, {Marulli}, {Massey},
  {Maurogordato}, {Melchior}, {Merlin}, {Meylan}, {Mohr}, {Moresco}, {Morin},
  {Moscardini}, {Munari}, {Nichol}, {Padilla}, {Paltani}, {Peacock},
  {Pedersen}, {Pettorino}, {Pires}, {Poncet}, {Popa}, {Pozzetti}, {Raison},
  {Rebolo}, {Rhodes}, {Rix}, {Roncarelli}, {Rossetti}, {Saglia}, {Schneider},
  {Schrabback}, {Secroun}, {Seidel}, {Serrano}, {Sirignano}, {Sirri},
  {Skottfelt}, {Stanco}, {Starck}, {Tallada-Cresp{\'\i}}, {Tavagnacco},
  {Taylor}, {Teplitz}, {Toledo-Moreo}, {Torradeflot}, {Trifoglio}, {Valentijn},
  {Valenziano}, {Verdoes Kleijn}, {Wang}, {Welikala}, {Weller}, {Wetzstein},
  {Zamorani}, {Zoubian}, {Andreon}, {Baldi}, {Bardelli}, {Boucaud}, {Camera},
  {Di Ferdinando}, {Fabbian}, {Farinelli}, {Galeotta}, {Graci{\'a}-Carpio},
  {Maino}, {Medinaceli}, {Mei}, {Neissner}, {Polenta}, {Renzi}, {Romelli},
  {Rosset}, {Sureau}, {Tenti}, {Vassallo}, {Zucca}, {Baccigalupi},
  {Balaguera-Antol{\'\i}nez}, {Battaglia}, {Biviano}, {Borgani}, {Bozzo},
  {Cabanac}, {Cappi}, {Casas}, {Castignani}, {Colodro-Conde}, {Coupon},
  {Courtois}, {Cuby}, {de la Torre}, {Desai}, {Dole}, {Fabricius}, {Farina},
  {Ferreira}, {Finelli}, {Flose-Reimberg}, {Fotopoulou}, {Ganga}, {Gozaliasl},
  {Hook}, {Keihanen}, {Kirkpatrick}, {Liebing}, {Lindholm}, {Mainetti},
  {Martinelli}, {Martinet}, {Maturi}, {McCracken}, {Metcalf}, {Morgante},
  {Nightingale}, {Nucita}, {Patrizii}, {Potter}, {Riccio}, {S{\'a}nchez},
  {Sapone}, {Schewtschenko}, {Schultheis}, {Scottez}, {Teyssier}, {Tutusaus},
  {Valiviita}, {Viel}, {Vriend}, \& {Whittaker}}]{euclid-survey-scaramella2021}
{Euclid Collaboration: Scaramella}, R., {Amiaux}, J., {Mellier}, Y., {et~al.}
  2022, \aap, 662, A112

\bibitem[{{Euclid Collaboration: Schirmer} {et~al.}(2022){Euclid Collaboration:
  Schirmer}, {Jahnke}, {Seidel}, {Aussel}, {Bodendorf}, {Grupp}, {Hormuth},
  {Wachter}, {Appleton}, {Barbier}, {Brinchmann}, {Carrasco}, {Castander},
  {Coupon}, {De Paolis}, {Franco}, {Ganga}, {Hudelot}, {Jullo}, {Lan{\c{c}}on},
  {Nucita}, {Paltani}, {Smadja}, {Strafella}, {Venancio}, {Weiler}, {Amara},
  {Auphan}, {Auricchio}, {Balestra}, {Bender}, {Bonino}, {Branchini},
  {Brescia}, {Capobianco}, {Carbone}, {Carretero}, {Casas}, {Castellano},
  {Cavuoti}, {Cimatti}, {Cledassou}, {Congedo}, {Conselice}, {Conversi},
  {Copin}, {Corcione}, {Costille}, {Courbin}, {Da Silva}, {Degaudenzi},
  {Douspis}, {Dubath}, {Dupac}, {Dusini}, {Ealet}, {Farrens}, {Ferriol},
  {Fosalba}, {Frailis}, {Franceschi}, {Franzetti}, {Fumana}, {Garilli},
  {Gillard}, {Gillis}, {Giocoli}, {Grazian}, {Guzzo}, {Haugan}, {Hoekstra},
  {Holmes}, {Hornstrup}, {K{\"u}mmel}, {Kermiche}, {Kiessling}, {Kilbinger},
  {Kitching}, {Kohley}, {Kunz}, {Kurki-Suonio}, {Laureijs}, {Ligori}, {Lilje},
  {Lloro}, {Maciaszek}, {Maiorano}, {Mansutti}, {Marggraf}, {Markovic},
  {Marulli}, {Massey}, {Maurogordato}, {Mellier}, {Meneghetti}, {Merlin},
  {Meylan}, {Moresco}, {Moscardini}, {Munari}, {Nakajima}, {Nichol}, {Niemi},
  {Padilla}, {Pasian}, {Pedersen}, {Percival}, {Pettorino}, {Pires}, {Poncet},
  {Popa}, {Pozzetti}, {Prieto}, {Raison}, {Rhodes}, {Rix}, {Roncarelli},
  {Rossetti}, {Saglia}, {Sartoris}, {Scaramella}, {Schneider}, {Secroun},
  {Serrano}, {Sirignano}, {Sirri}, {Stanco}, {Tallada-Cresp{\'\i}}, {Taylor},
  {Teplitz}, {Tereno}, {Toledo-Moreo}, {Torradeflot}, {Trifoglio}, {Valentijn},
  {Valenziano}, {Wang}, {Weller}, {Zamorani}, {Zoubian}, {Andreon}, {Bardelli},
  {Boucaud}, {Camera}, {Farinelli}, {Graci{\'a}-Carpio}, {Maino}, {Medinaceli},
  {Mei}, {Morisset}, {Polenta}, {Renzi}, {Romelli}, {Tenti}, {Vassallo},
  {Zacchei}, {Zucca}, {Baccigalupi}, {Balaguera-Antol{\'\i}nez}, {Biviano},
  {Blanchard}, {Borgani}, {Bozzo}, {Burigana}, {Cabanac}, {Cappi}, {Carvalho},
  {Casas}, {Castignani}, {Colodro-Conde}, {Cooray}, {Courtois}, {Crocce},
  {Cuby}, {Davini}, {de la Torre}, {Di Ferdinando}, {Escartin}, {Farina},
  {Ferreira}, {Finelli}, {Fotopoulou}, {Galeotta}, {Garcia-Bellido},
  {Gaztanaga}, {George}, {Gozaliasl}, {Hook}, {Ili{\'c}}, {Kansal},
  {Kashlinsky}, {Keihanen}, {Kirkpatrick}, {Lindholm}, {Mainetti}, {Maoli},
  {Martinelli}, {Martinet}, {Maturi}, {Mauri}, {McCracken}, {Metcalf},
  {Monaco}, {Morgante}, {Nightingale}, {Patrizii}, {Peel}, {Popa}, {Porciani},
  {Potter}, {Reimberg}, {Riccio}, {S{\'a}nchez}, {Sapone}, {Scottez},
  {Sefusatti}, {Teyssier}, {Tutusaus}, {Valieri}, {Valiviita}, {Viel}, \&
  {Hildebrandt}}]{euclid-NISPphotobands-schirmer2022}
{Euclid Collaboration: Schirmer}, M., {Jahnke}, K., {Seidel}, G., {et~al.}
  2022, \aap, 662, A92

\bibitem[{{Faber} \& {Jackson}(1976)}]{velocityrelation-faberjackson1976}
{Faber}, S.~M. \& {Jackson}, R.~E. 1976, \apj, 204, 668

\bibitem[{Flaugher {et~al.}(2015)Flaugher, Diehl, Honscheid, Abbott, Alvarez,
  Angstadt, Annis, Antonik, Ballester, Beaufore, Bernstein, Bernstein, Bigelow,
  Bonati, Boprie, Brooks, Buckley-Geer, Campa, Cardiel-Sas, Castander,
  Castilla, Cease, Cela-Ruiz, Chappa, Chi, Cooper, da~Costa, Dede, Derylo,
  DePoy, de~Vicente, Doel, Drlica-Wagner, Eiting, Elliott, Emes, Estrada, Neto,
  Finley, Flores, Frieman, Gerdes, Gladders, Gregory, Gutierrez, Hao, Holland,
  Holm, Huffman, Jackson, James, Jonas, Karcher, Karliner, Kent, Kessler,
  Kozlovsky, Kron, Kubik, Kuehn, Kuhlmann, Kuk, Lahav, Lathrop, Lee, Levi,
  Lewis, Li, Mandrichenko, Marshall, Martinez, Merritt, Miquel, Mu{\~{n}}oz,
  Neilsen, Nichol, Nord, Ogando, Olsen, Palaio, Patton, Peoples, Plazas, Rauch,
  Reil, Rheault, Roe, Rogers, Roodman, Sanchez, Scarpine, Schindler, Schmidt,
  Schmitt, Schubnell, Schultz, Schurter, Scott, Serrano, Shaw, Smith,
  Soares-Santos, Stefanik, Stuermer, Suchyta, Sypniewski, Tarle, Thaler, Tighe,
  Tran, Tucker, Walker, Wang, Watson, Weaverdyck, Wester, Woods, \&
  and}]{Flaugher_2015}
Flaugher, B., Diehl, H.~T., Honscheid, K., {et~al.} 2015, \aj, 150, 150

\bibitem[{{Fosalba} {et~al.}(2015){Fosalba}, {Gazta{\~n}aga}, {Castander}, \&
  {Crocce}}]{lensing-fosalba2016}
{Fosalba}, P., {Gazta{\~n}aga}, E., {Castander}, F.~J., \& {Crocce}, M. 2015,
  \mnras, 447, 1319

\bibitem[{{Frailis} {et~al.}(2019){Frailis}, {Belikov}, {Benson}, {Bonchi},
  {Dabin}, {Ealet}, {Fumana}, {Grenet}, {Holliman}, {Maggio}, {Maino},
  {McCracken}, {Melchior}, {Piemonte}, {Polenta}, {Poncet}, {Scala}, {Serrano},
  \& {Williams}}]{sdcs_2019}
{Frailis}, M., {Belikov}, A., {Benson}, K., {et~al.} 2019, in Astronomical
  Society of the Pacific Conference Series, Vol. 521, Astronomical Data
  Analysis Software and Systems XXVI, ed. M.~{Molinaro}, K.~{Shortridge}, \&
  F.~{Pasian}, 612

\bibitem[{{Gaia Collaboration} {et~al.}(2018){Gaia Collaboration}, {Brown},
  {Vallenari}, {Prusti}, {de Bruijne}, {Babusiaux}, {Bailer-Jones}, {Biermann},
  {Evans}, {Eyer}, {Jansen}, {Jordi}, {Klioner}, {Lammers}, {Lindegren},
  {Luri}, {Mignard}, {Panem}, {Pourbaix}, {Randich}, {Sartoretti}, {Siddiqui},
  {Soubiran}, {van Leeuwen}, {Walton}, {Arenou}, {Bastian}, {Cropper},
  {Drimmel}, {Katz}, {Lattanzi}, {Bakker}, {Cacciari}, {Casta{\~n}eda},
  {Chaoul}, {Cheek}, {De Angeli}, {Fabricius}, {Guerra}, {Holl}, {Masana},
  {Messineo}, {Mowlavi}, {Nienartowicz}, {Panuzzo}, {Portell}, {Riello},
  {Seabroke}, {Tanga}, {Th{\'e}venin}, {Gracia-Abril}, {Comoretto},
  {Garcia-Reinaldos}, {Teyssier}, {Altmann}, {Andrae}, {Audard},
  {Bellas-Velidis}, {Benson}, {Berthier}, {Blomme}, {Burgess}, {Busso},
  {Carry}, {Cellino}, {Clementini}, {Clotet}, {Creevey}, {Davidson}, {De
  Ridder}, {Delchambre}, {Dell'Oro}, {Ducourant},
  {Fern{\'a}ndez-Hern{\'a}ndez}, {Fouesneau}, {Fr{\'e}mat}, {Galluccio},
  {Garc{\'\i}a-Torres}, {Gonz{\'a}lez-N{\'u}{\~n}ez}, {Gonz{\'a}lez-Vidal},
  {Gosset}, {Guy}, {Halbwachs}, {Hambly}, {Harrison}, {Hern{\'a}ndez},
  {Hestroffer}, {Hodgkin}, {Hutton}, {Jasniewicz}, {Jean-Antoine-Piccolo},
  {Jordan}, {Korn}, {Krone-Martins}, {Lanzafame}, {Lebzelter}, {L{\"o}ffler},
  {Manteiga}, {Marrese}, {Mart{\'\i}n-Fleitas}, {Moitinho}, {Mora}, {Muinonen},
  {Osinde}, {Pancino}, {Pauwels}, {Petit}, {Recio-Blanco}, {Richards},
  {Rimoldini}, {Robin}, {Sarro}, {Siopis}, {Smith}, {Sozzetti}, {S{\"u}veges},
  {Torra}, {van Reeven}, {Abbas}, {Abreu Aramburu}, {Accart}, {Aerts},
  {Altavilla}, {{\'A}lvarez}, {Alvarez}, {Alves}, {Anderson}, {Andrei},
  {Anglada Varela}, {Antiche}, {Antoja}, {Arcay}, {Astraatmadja}, {Bach},
  {Baker}, {Balaguer-N{\'u}{\~n}ez}, {Balm}, {Barache}, {Barata}, {Barbato},
  {Barblan}, {Barklem}, {Barrado}, {Barros}, {Barstow}, {Bartholom{\'e}
  Mu{\~n}oz}, {Bassilana}, {Becciani}, {Bellazzini}, {Berihuete}, {Bertone},
  {Bianchi}, {Bienaym{\'e}}, {Blanco-Cuaresma}, {Boch}, {Boeche}, {Bombrun},
  {Borrachero}, {Bossini}, {Bouquillon}, {Bourda}, {Bragaglia}, {Bramante},
  {Breddels}, {Bressan}, {Brouillet}, {Br{\"u}semeister}, {Brugaletta},
  {Bucciarelli}, {Burlacu}, {Busonero}, {Butkevich}, {Buzzi}, {Caffau},
  {Cancelliere}, {Cannizzaro}, {Cantat-Gaudin}, {Carballo}, {Carlucci},
  {Carrasco}, {Casamiquela}, {Castellani}, {Castro-Ginard}, {Charlot},
  {Chemin}, {Chiavassa}, {Cocozza}, {Costigan}, {Cowell}, {Crifo}, {Crosta},
  {Crowley}, {Cuypers}, {Dafonte}, {Damerdji}, {Dapergolas}, {David}, {David},
  {de Laverny}, {De Luise}, {De March}, {de Martino}, {de Souza}, {de Torres},
  {Debosscher}, {del Pozo}, {Delbo}, {Delgado}, {Delgado}, {Di Matteo},
  {Diakite}, {Diener}, {Distefano}, {Dolding}, {Drazinos}, {Dur{\'a}n},
  {Edvardsson}, {Enke}, {Eriksson}, {Esquej}, {Eynard Bontemps}, {Fabre},
  {Fabrizio}, {Faigler}, {Falc{\~a}o}, {Farr{\`a}s Casas}, {Federici},
  {Fedorets}, {Fernique}, {Figueras}, {Filippi}, {Findeisen}, {Fonti},
  {Fraile}, {Fraser}, {Fr{\'e}zouls}, {Gai}, {Galleti}, {Garabato},
  {Garc{\'\i}a-Sedano}, {Garofalo}, {Garralda}, {Gavel}, {Gavras}, {Gerssen},
  {Geyer}, {Giacobbe}, {Gilmore}, {Girona}, {Giuffrida}, {Glass}, {Gomes},
  {Granvik}, {Gueguen}, {Guerrier}, {Guiraud}, {Guti{\'e}rrez-S{\'a}nchez},
  {Haigron}, {Hatzidimitriou}, {Hauser}, {Haywood}, {Heiter}, {Helmi}, {Heu},
  {Hilger}, {Hobbs}, {Hofmann}, {Holland}, {Huckle}, {Hypki}, {Icardi},
  {Jan{\ss}en}, {Jevardat de Fombelle}, {Jonker}, {Juh{\'a}sz}, {Julbe},
  {Karampelas}, {Kewley}, {Klar}, {Kochoska}, {Kohley}, {Kolenberg},
  {Kontizas}, {Kontizas}, {Koposov}, {Kordopatis}, {Kostrzewa-Rutkowska},
  {Koubsky}, {Lambert}, {Lanza}, {Lasne}, {Lavigne}, {Le Fustec}, {Le
  Poncin-Lafitte}, {Lebreton}, {Leccia}, {Leclerc}, {Lecoeur-Taibi},
  {Lenhardt}, {Leroux}, {Liao}, {Licata}, {Lindstr{\o}m}, {Lister}, {Livanou},
  {Lobel}, {L{\'o}pez}, {Managau}, {Mann}, {Mantelet}, {Marchal}, {Marchant},
  {Marconi}, {Marinoni}, {Marschalk{\'o}}, {Marshall}, {Martino}, {Marton},
  {Mary}, {Massari}, {Matijevi{\v{c}}}, {Mazeh}, {McMillan}, {Messina},
  {Michalik}, {Millar}, {Molina}, {Molinaro}, {Moln{\'a}r}, {Montegriffo},
  {Mor}, {Morbidelli}, {Morel}, {Morris}, {Mulone}, {Muraveva}, {Musella},
  {Nelemans}, {Nicastro}, {Noval}, {O'Mullane}, {Ord{\'e}novic},
  {Ord{\'o}{\~n}ez-Blanco}, {Osborne}, {Pagani}, {Pagano}, {Pailler},
  {Palacin}, {Palaversa}, {Panahi}, {Pawlak}, {Piersimoni}, {Pineau}, {Plachy},
  {Plum}, {Poggio}, {Poujoulet}, {Pr{\v{s}}a}, {Pulone}, {Racero}, {Ragaini},
  {Rambaux}, {Ramos-Lerate}, {Regibo}, {Reyl{\'e}}, {Riclet}, {Ripepi}, {Riva},
  {Rivard}, {Rixon}, {Roegiers}, {Roelens}, {Romero-G{\'o}mez}, {Rowell},
  {Royer}, {Ruiz-Dern}, {Sadowski}, {Sagrist{\`a} Sell{\'e}s}, {Sahlmann},
  {Salgado}, {Salguero}, {Sanna}, {Santana-Ros}, {Sarasso}, {Savietto},
  {Schultheis}, {Sciacca}, {Segol}, {Segovia}, {S{\'e}gransan}, {Shih},
  {Siltala}, {Silva}, {Smart}, {Smith}, {Solano}, {Solitro}, {Sordo}, {Soria
  Nieto}, {Souchay}, {Spagna}, {Spoto}, {Stampa}, {Steele},
  {Steidelm{\"u}ller}, {Stephenson}, {Stoev}, {Suess}, {Surdej}, {Szabados},
  {Szegedi-Elek}, {Tapiador}, {Taris}, {Tauran}, {Taylor}, {Teixeira},
  {Terrett}, {Teyssandier}, {Thuillot}, {Titarenko}, {Torra Clotet}, {Turon},
  {Ulla}, {Utrilla}, {Uzzi}, {Vaillant}, {Valentini}, {Valette}, {van Elteren},
  {Van Hemelryck}, {van Leeuwen}, {Vaschetto}, {Vecchiato}, {Veljanoski},
  {Viala}, {Vicente}, {Vogt}, {von Essen}, {Voss}, {Votruba}, {Voutsinas},
  {Walmsley}, {Weiler}, {Wertz}, {Wevers}, {Wyrzykowski}, {Yoldas},
  {{\v{Z}}erjal}, {Ziaeepour}, {Zorec}, {Zschocke}, {Zucker}, {Zurbach}, \&
  {Zwitter}}]{gaia-dr2-2018}
{Gaia Collaboration}, {Brown}, A.~G.~A., {Vallenari}, A., {et~al.} 2018, \aap,
  616, A1

\bibitem[{{Gaia Collaboration} {et~al.}(2023){Gaia Collaboration},
  {Montegriffo}, {Bellazzini}, {De Angeli}, {Andrae}, {Barstow}, {Bossini},
  {Bragaglia}, {Burgess}, {Cacciari}, {Carrasco}, {Chornay}, {Delchambre},
  {Evans}, {Fouesneau}, {Fr{\'e}mat}, {Garabato}, {Jordi}, {Manteiga},
  {Massari}, {Palaversa}, {Pancino}, {Riello}, {Ruz Mieres}, {Sanna},
  {Santove{\~n}a}, {Sordo}, {Vallenari}, {Walton}, {Brown}, {Prusti}, {de
  Bruijne}, {Arenou}, {Babusiaux}, {Biermann}, {Creevey}, {Ducourant}, {Eyer},
  {Guerra}, {Hutton}, {Klioner}, {Lammers}, {Lindegren}, {Luri}, {Mignard},
  {Panem}, {Pourbaix}, {Randich}, {Sartoretti}, {Soubiran}, {Tanga},
  {Bailer-Jones}, {Bastian}, {Drimmel}, {Jansen}, {Katz}, {Lattanzi}, {van
  Leeuwen}, {Bakker}, {Casta{\~n}eda}, {Fabricius}, {Galluccio}, {Guerrier},
  {Heiter}, {Masana}, {Messineo}, {Mowlavi}, {Nicolas}, {Nienartowicz},
  {Pailler}, {Panuzzo}, {Riclet}, {Roux}, {Seabroke}, {Th{\'e}venin},
  {Gracia-Abril}, {Portell}, {Teyssier}, {Altmann}, {Audard}, {Bellas-Velidis},
  {Benson}, {Berthier}, {Blomme}, {Busonero}, {Busso}, {C{\'a}novas}, {Carry},
  {Cellino}, {Cheek}, {Clementini}, {Damerdji}, {Davidson}, {de Teodoro},
  {Nu{\~n}ez Campos}, {Dell'Oro}, {Esquej}, {Fern{\'a}ndez-Hern{\'a}ndez},
  {Fraile}, {Garc{\'\i}a-Lario}, {Gosset}, {Haigron}, {Halbwachs}, {Hambly},
  {Harrison}, {Hern{\'a}ndez}, {Hestroffer}, {Hodgkin}, {Holl}, {Jan{\ss}en},
  {Jevardat de Fombelle}, {Jordan}, {Krone-Martins}, {Lanzafame},
  {L{\"o}ffler}, {Marchal}, {Marrese}, {Moitinho}, {Muinonen}, {Osborne},
  {Pauwels}, {Recio-Blanco}, {Reyl{\'e}}, {Rimoldini}, {Roegiers}, {Rybizki},
  {Sarro}, {Siopis}, {Smith}, {Sozzetti}, {Utrilla}, {van Leeuwen}, {Abbas},
  {{\'A}brah{\'a}m}, {Abreu Aramburu}, {Aerts}, {Aguado}, {Ajaj},
  {Aldea-Montero}, {Altavilla}, {{\'A}lvarez}, {Alves}, {Anderson}, {Anglada
  Varela}, {Antoja}, {Baines}, {Baker}, {Balaguer-N{\'u}{\~n}ez}, {Balbinot},
  {Balog}, {Barache}, {Barbato}, {Barros}, {Bartolom{\'e}}, {Bassilana},
  {Bauchet}, {Becciani}, {Berihuete}, {Bernet}, {Bertone}, {Bianchi},
  {Binnenfeld}, {Blanco-Cuaresma}, {Boch}, {Bombrun}, {Bouquillon}, {Bramante},
  {Breedt}, {Bressan}, {Brouillet}, {Brugaletta}, {Bucciarelli}, {Burlacu},
  {Butkevich}, {Buzzi}, {Caffau}, {Cancelliere}, {Cantat-Gaudin}, {Carballo},
  {Carlucci}, {Carnerero}, {Casamiquela}, {Castellani}, {Castro-Ginard},
  {Chaoul}, {Charlot}, {Chemin}, {Chiaramida}, {Chiavassa}, {Comoretto},
  {Contursi}, {Cooper}, {Cornez}, {Cowell}, {Crifo}, {Cropper}, {Crosta},
  {Crowley}, {Dafonte}, {Dapergolas}, {David}, {de Laverny}, {De Luise}, {De
  March}, {De Ridder}, {de Souza}, {de Torres}, {del Peloso}, {del Pozo},
  {Delbo}, {Delgado}, {Delisle}, {Demouchy}, {Dharmawardena}, {Diakite},
  {Diener}, {Distefano}, {Dolding}, {Enke}, {Fabre}, {Fabrizio}, {Faigler},
  {Fedorets}, {Fernique}, {Figueras}, {Fournier}, {Fouron}, {Fragkoudi}, {Gai},
  {Garcia-Gutierrez}, {Garcia-Reinaldos}, {Garc{\'\i}a-Torres}, {Garofalo},
  {Gavel}, {Gavras}, {Gerlach}, {Geyer}, {Giacobbe}, {Gilmore}, {Girona},
  {Giuffrida}, {Gomel}, {Gomez}, {Gonz{\'a}lez-N{\'u}{\~n}ez},
  {Gonz{\'a}lez-Santamar{\'\i}a}, {Gonz{\'a}lez-Vidal}, {Granvik}, {Guillout},
  {Guiraud}, {Guti{\'e}rrez-S{\'a}nchez}, {Guy}, {Hatzidimitriou}, {Hauser},
  {Haywood}, {Helmer}, {Helmi}, {Sarmiento}, {Hidalgo}, {H{\l}adczuk}, {Hobbs},
  {Holland}, {Huckle}, {Jardine}, {Jasniewicz}, {Jean-Antoine Piccolo},
  {Jim{\'e}nez-Arranz}, {Juaristi Campillo}, {Julbe}, {Karbevska}, {Kervella},
  {Khanna}, {Kordopatis}, {Korn}, {K{\'o}sp{\'a}l}, {Kostrzewa-Rutkowska},
  {Kruszy{\'n}ska}, {Kun}, {Laizeau}, {Lambert}, {Lanza}, {Lasne}, {Le
  Campion}, {Lebreton}, {Lebzelter}, {Leccia}, {Leclerc}, {Lecoeur-Taibi},
  {Liao}, {Licata}, {Lindstr{\'o}m}, {Lister}, {Livanou}, {Lobel}, {Lorca},
  {Loup}, {Madrero Pardo}, {Magdaleno Romeo}, {Managau}, {Mann}, {Marchant},
  {Marconi}, {Marcos}, {Marcos Santos}, {Mar{\'\i}n Pina}, {Marinoni},
  {Marocco}, {Marshall}, {Martin Polo}, {Mart{\'\i}n-Fleitas}, {Marton},
  {Mary}, {Masip}, {Mastrobuono-Battisti}, {Mazeh}, {McMillan}, {Messina},
  {Michalik}, {Millar}, {Mints}, {Molina}, {Molinaro}, {Moln{\'a}r}, {Monari},
  {Mongui{\'o}}, {Montero}, {Mor}, {Mora}, {Morbidelli}, {Morel}, {Morris},
  {Muraveva}, {Murphy}, {Musella}, {Nagy}, {Noval}, {Oca{\~n}a}, {Ogden},
  {Ordenovic}, {Osinde}, {Pagani}, {Pagano}, {Palicio}, {Pallas-Quintela},
  {Panahi}, {Payne-Wardenaar}, {Pe{\~n}alosa Esteller}, {Penttil{\"a}},
  {Pichon}, {Piersimoni}, {Pineau}, {Plachy}, {Plum}, {Poggio}, {Pr{\v{s}}a},
  {Pulone}, {Racero}, {Ragaini}, {Rainer}, {Raiteri}, {Ramos}, {Ramos-Lerate},
  {Re Fiorentin}, {Regibo}, {Richards}, {Rios Diaz}, {Ripepi}, {Riva}, {Rix},
  {Rixon}, {Robichon}, {Robin}, {Robin}, {Roelens}, {Rogues}, {Rohrbasser},
  {Romero-G{\'o}mez}, {Rowell}, {Royer}, {Rybicki}, {Sadowski}, {S{\'a}ez
  N{\'u}{\~n}ez}, {Sagrist{\`a} Sell{\'e}s}, {Sahlmann}, {Salguero}, {Samaras},
  {Sanchez Gimenez}, {Sarasso}, {Schultheis}, {Sciacca}, {Segol}, {Segovia},
  {S{\'e}gransan}, {Semeux}, {Shahaf}, {Siddiqui}, {Siebert}, {Siltala},
  {Silvelo}, {Slezak}, {Slezak}, {Smart}, {Snaith}, {Solano}, {Solitro},
  {Souami}, {Souchay}, {Spagna}, {Spina}, {Spoto}, {Steele},
  {Steidelm{\"u}ller}, {Stephenson}, {S{\"u}veges}, {Surdej}, {Szabados},
  {Szegedi-Elek}, {Taris}, {Taylor}, {Teixeira}, {Tolomei}, {Tonello}, {Torra},
  {Torra}, {Torralba Elipe}, {Trabucchi}, {Tsounis}, {Turon}, {Ulla}, {Unger},
  {Vaillant}, {van Dillen}, {van Reeven}, {Vanel}, {Vecchiato}, {Viala},
  {Vicente}, {Voutsinas}, {Wevers}, {Wyrzykowski}, {Yoldas}, {Yvard}, {Zhao},
  {Zorec}, {Zucker}, \& {Zwitter}}]{montegriffo-gaiadr3-2022}
{Gaia Collaboration}, {Montegriffo}, P., {Bellazzini}, M., {et~al.} 2023, \aap,
  674, A33

\bibitem[{{Gaspar Venancio} {et~al.}(2014){Gaspar Venancio}, {Laureijs},
  {Lorenzo}, {Salvignol}, {Short}, {Strada}, {Vavrek}, {Vaillon}, {Gennaro},
  {Amiaux}, \& {Prieto}}]{plm_venancio2014}
{Gaspar Venancio}, L.~M., {Laureijs}, R., {Lorenzo}, J., {et~al.} 2014, in
  Society of Photo-Optical Instrumentation Engineers (SPIE) Conference Series,
  Vol. 9143, Space Telescopes and Instrumentation 2014: Optical, Infrared, and
  Millimeter Wave, ed. J.~{Oschmann}, Jacobus~M., M.~{Clampin}, G.~G. {Fazio},
  \& H.~A. {MacEwen}, 91430I

\bibitem[{{Grogin} {et~al.}(2011){Grogin}, {Kocevski}, {Faber}, {Ferguson},
  {Koekemoer}, {Riess}, {Acquaviva}, {Alexander}, {Almaini}, {Ashby}, {Barden},
  {Bell}, {Bournaud}, {Brown}, {Caputi}, {Casertano}, {Cassata}, {Castellano},
  {Challis}, {Chary}, {Cheung}, {Cirasuolo}, {Conselice}, {Roshan Cooray},
  {Croton}, {Daddi}, {Dahlen}, {Dav{\'e}}, {de Mello}, {Dekel}, {Dickinson},
  {Dolch}, {Donley}, {Dunlop}, {Dutton}, {Elbaz}, {Fazio}, {Filippenko},
  {Finkelstein}, {Fontana}, {Gardner}, {Garnavich}, {Gawiser}, {Giavalisco},
  {Grazian}, {Guo}, {Hathi}, {H{\"a}ussler}, {Hopkins}, {Huang}, {Huang},
  {Jha}, {Kartaltepe}, {Kirshner}, {Koo}, {Lai}, {Lee}, {Li}, {Lotz}, {Lucas},
  {Madau}, {McCarthy}, {McGrath}, {McIntosh}, {McLure}, {Mobasher},
  {Moustakas}, {Mozena}, {Nandra}, {Newman}, {Niemi}, {Noeske}, {Papovich},
  {Pentericci}, {Pope}, {Primack}, {Rajan}, {Ravindranath}, {Reddy}, {Renzini},
  {Rix}, {Robaina}, {Rodney}, {Rosario}, {Rosati}, {Salimbeni}, {Scarlata},
  {Siana}, {Simard}, {Smidt}, {Somerville}, {Spinrad}, {Straughn}, {Strolger},
  {Telford}, {Teplitz}, {Trump}, {van der Wel}, {Villforth}, {Wechsler},
  {Weiner}, {Wiklind}, {Wild}, {Wilson}, {Wuyts}, {Yan}, \&
  {Yun}}]{candels-grogin2011}
{Grogin}, N.~A., {Kocevski}, D.~D., {Faber}, S.~M., {et~al.} 2011, \apjs, 197,
  35

\bibitem[{{Hoekstra} {et~al.}(2017){Hoekstra}, {Viola}, \&
  {Herbonnet}}]{WL_sensitivity-hoekstra2017}
{Hoekstra}, H., {Viola}, M., \& {Herbonnet}, R. 2017, \mnras, 468, 3295

\bibitem[{{H{\o}g} {et~al.}(2000){H{\o}g}, {Fabricius}, {Makarov}, {Urban},
  {Corbin}, {Wycoff}, {Bastian}, {Schwekendiek}, \&
  {Wicenec}}]{tycho2cat-hog2000}
{H{\o}g}, E., {Fabricius}, C., {Makarov}, V.~V., {et~al.} 2000, \aap, 355, L27

\bibitem[{{Ibata} {et~al.}(2017){Ibata}, {McConnachie}, {Cuillandre}, {Fantin},
  {Haywood}, {Martin}, {Bergeron}, {Beckmann}, {Bernard}, {Bonifacio},
  {Caffau}, {Carlberg}, {C{\^o}t{\'e}}, {Cabanac}, {Chapman}, {Duc}, {Durret},
  {Famaey}, {Fabbro}, {Gwyn}, {Hammer}, {Hill}, {Hudson}, {Lan{\c{c}}on},
  {Lewis}, {Malhan}, {di Matteo}, {McCracken}, {Mei}, {Mellier}, {Navarro},
  {Pires}, {Pritchet}, {Reyl{\'e}}, {Richer}, {Robin}, {S{\'a}nchez-Janssen},
  {Sawicki}, {Scott}, {Scottez}, {Spekkens}, {Starkenburg}, {Thomas}, \&
  {Venn}}]{cfis_ibata_2017}
{Ibata}, R.~A., {McConnachie}, A., {Cuillandre}, J.-C., {et~al.} 2017, \apj,
  848, 128

\bibitem[{{Ilbert} {et~al.}(2009){Ilbert}, {Capak}, {Salvato}, {Aussel},
  {McCracken}, {Sanders}, {Scoville}, {Kartaltepe}, {Arnouts}, {Le Floc'h},
  {Mobasher}, {Taniguchi}, {Lamareille}, {Leauthaud}, {Sasaki}, {Thompson},
  {Zamojski}, {Zamorani}, {Bardelli}, {Bolzonella}, {Bongiorno}, {Brusa},
  {Caputi}, {Carollo}, {Contini}, {Cook}, {Coppa}, {Cucciati}, {de la Torre},
  {de Ravel}, {Franzetti}, {Garilli}, {Hasinger}, {Iovino}, {Kampczyk},
  {Kneib}, {Knobel}, {Kovac}, {Le Borgne}, {Le Brun}, {Le F{\`e}vre}, {Lilly},
  {Looper}, {Maier}, {Mainieri}, {Mellier}, {Mignoli}, {Murayama}, {Pell{\`o}},
  {Peng}, {P{\'e}rez-Montero}, {Renzini}, {Ricciardelli}, {Schiminovich},
  {Scodeggio}, {Shioya}, {Silverman}, {Surace}, {Tanaka}, {Tasca}, {Tresse},
  {Vergani}, \& {Zucca}}]{cosmos_ilbert2009}
{Ilbert}, O., {Capak}, P., {Salvato}, M., {et~al.} 2009, \apj, 690, 1236

\bibitem[{{Ivezi{\'c}} {et~al.}(2019){Ivezi{\'c}}, {Kahn}, {Tyson}, {Abel},
  {Acosta}, {Allsman}, {Alonso}, {AlSayyad}, {Anderson}, {Andrew}, {Angel},
  {Angeli}, {Ansari}, {Antilogus}, {Araujo}, {Armstrong}, {Arndt}, {Astier},
  {Aubourg}, {Auza}, {Axelrod}, {Bard}, {Barr}, {Barrau}, {Bartlett}, {Bauer},
  {Bauman}, {Baumont}, {Bechtol}, {Bechtol}, {Becker}, {Becla}, {Beldica},
  {Bellavia}, {Bianco}, {Biswas}, {Blanc}, {Blazek}, {Blandford}, {Bloom},
  {Bogart}, {Bond}, {Booth}, {Borgland}, {Borne}, {Bosch}, {Boutigny},
  {Brackett}, {Bradshaw}, {Brandt}, {Brown}, {Bullock}, {Burchat}, {Burke},
  {Cagnoli}, {Calabrese}, {Callahan}, {Callen}, {Carlin}, {Carlson},
  {Chandrasekharan}, {Charles-Emerson}, {Chesley}, {Cheu}, {Chiang}, {Chiang},
  {Chirino}, {Chow}, {Ciardi}, {Claver}, {Cohen-Tanugi}, {Cockrum}, {Coles},
  {Connolly}, {Cook}, {Cooray}, {Covey}, {Cribbs}, {Cui}, {Cutri}, {Daly},
  {Daniel}, {Daruich}, {Daubard}, {Daues}, {Dawson}, {Delgado}, {Dellapenna},
  {de Peyster}, {de Val-Borro}, {Digel}, {Doherty}, {Dubois},
  {Dubois-Felsmann}, {Durech}, {Economou}, {Eifler}, {Eracleous}, {Emmons},
  {Fausti Neto}, {Ferguson}, {Figueroa}, {Fisher-Levine}, {Focke}, {Foss},
  {Frank}, {Freemon}, {Gangler}, {Gawiser}, {Geary}, {Gee}, {Geha}, {Gessner},
  {Gibson}, {Gilmore}, {Glanzman}, {Glick}, {Goldina}, {Goldstein}, {Goodenow},
  {Graham}, {Gressler}, {Gris}, {Guy}, {Guyonnet}, {Haller}, {Harris},
  {Hascall}, {Haupt}, {Hernandez}, {Herrmann}, {Hileman}, {Hoblitt}, {Hodgson},
  {Hogan}, {Howard}, {Huang}, {Huffer}, {Ingraham}, {Innes}, {Jacoby}, {Jain},
  {Jammes}, {Jee}, {Jenness}, {Jernigan}, {Jevremovi{\'c}}, {Johns}, {Johnson},
  {Johnson}, {Jones}, {Juramy-Gilles}, {Juri{\'c}}, {Kalirai}, {Kallivayalil},
  {Kalmbach}, {Kantor}, {Karst}, {Kasliwal}, {Kelly}, {Kessler}, {Kinnison},
  {Kirkby}, {Knox}, {Kotov}, {Krabbendam}, {Krughoff}, {Kub{\'a}nek},
  {Kuczewski}, {Kulkarni}, {Ku}, {Kurita}, {Lage}, {Lambert}, {Lange},
  {Langton}, {Le Guillou}, {Levine}, {Liang}, {Lim}, {Lintott}, {Long},
  {Lopez}, {Lotz}, {Lupton}, {Lust}, {MacArthur}, {Mahabal}, {Mandelbaum},
  {Markiewicz}, {Marsh}, {Marshall}, {Marshall}, {May}, {McKercher}, {McQueen},
  {Meyers}, {Migliore}, {Miller}, {Mills}, {Miraval}, {Moeyens}, {Moolekamp},
  {Monet}, {Moniez}, {Monkewitz}, {Montgomery}, {Morrison}, {Mueller},
  {Muller}, {Mu{\~n}oz Arancibia}, {Neill}, {Newbry}, {Nief}, {Nomerotski},
  {Nordby}, {O'Connor}, {Oliver}, {Olivier}, {Olsen}, {O'Mullane}, {Ortiz},
  {Osier}, {Owen}, {Pain}, {Palecek}, {Parejko}, {Parsons}, {Pease},
  {Peterson}, {Peterson}, {Petravick}, {Libby Petrick}, {Petry},
  {Pierfederici}, {Pietrowicz}, {Pike}, {Pinto}, {Plante}, {Plate}, {Plutchak},
  {Price}, {Prouza}, {Radeka}, {Rajagopal}, {Rasmussen}, {Regnault}, {Reil},
  {Reiss}, {Reuter}, {Ridgway}, {Riot}, {Ritz}, {Robinson}, {Roby}, {Roodman},
  {Rosing}, {Roucelle}, {Rumore}, {Russo}, {Saha}, {Sassolas}, {Schalk},
  {Schellart}, {Schindler}, {Schmidt}, {Schneider}, {Schneider}, {Schoening},
  {Schumacher}, {Schwamb}, {Sebag}, {Selvy}, {Sembroski}, {Seppala}, {Serio},
  {Serrano}, {Shaw}, {Shipsey}, {Sick}, {Silvestri}, {Slater}, {Smith},
  {Smith}, {Sobhani}, {Soldahl}, {Storrie-Lombardi}, {Stover}, {Strauss},
  {Street}, {Stubbs}, {Sullivan}, {Sweeney}, {Swinbank}, {Szalay}, {Takacs},
  {Tether}, {Thaler}, {Thayer}, {Thomas}, {Thornton}, {Thukral}, {Tice},
  {Trilling}, {Turri}, {Van Berg}, {Vanden Berk}, {Vetter}, {Virieux},
  {Vucina}, {Wahl}, {Walkowicz}, {Walsh}, {Walter}, {Wang}, {Wang}, {Warner},
  {Wiecha}, {Willman}, {Winters}, {Wittman}, {Wolff}, {Wood-Vasey}, {Wu},
  {Xin}, {Yoachim}, \& {Zhan}}]{lsstcomponents-ivezic2019}
{Ivezi{\'c}}, {\v{Z}}., {Kahn}, S.~M., {Tyson}, J.~A., {et~al.} 2019, \apj,
  873, 111

\bibitem[{Jones {et~al.}(2020)Jones, Yoachim, Ivezic, Neilsen, \&
  Ribeiro}]{lsst-surveysim-jones2020}
Jones, R.~L., Yoachim, P., Ivezic, Z., Neilsen, E.~H., \& Ribeiro, T. 2020,
  {Survey Strategy and Cadence Choices for the Vera C. Rubin Observatory Legacy
  Survey of Space and Time (LSST)}

\bibitem[{{Juri{\'c}} {et~al.}(2017){Juri{\'c}}, {Kantor}, {Lim}, {Lupton},
  {Dubois-Felsmann}, {Jenness}, {Axelrod}, {Aleksi{\'c}}, {Allsman},
  {AlSayyad}, {Alt}, {Armstrong}, {Basney}, {Becker}, {Becla}, {Biswas},
  {Bosch}, {Boutigny}, {Kind}, {Ciardi}, {Connolly}, {Daniel}, {Daues},
  {Economou}, {Chiang}, {Fausti}, {Fisher-Levine}, {Freemon}, {Gris},
  {Hernandez}, {Hoblitt}, {Ivezi{\'c}}, {Jammes}, {Jevremovi{\'c}}, {Jones},
  {Kalmbach}, {Kasliwal}, {Krughoff}, {Lurie}, {Lust}, {MacArthur}, {Melchior},
  {Moeyens}, {Nidever}, {Owen}, {Parejko}, {Peterson}, {Petravick},
  {Pietrowicz}, {Price}, {Reiss}, {Shaw}, {Sick}, {Slater}, {Strauss},
  {Sullivan}, {Swinbank}, {Van Dyk}, {Vuj{\v{c}}i{\'c}}, {Withers}, \&
  {Yoachim}}]{lsstdm-juric2015}
{Juri{\'c}}, M., {Kantor}, J., {Lim}, K.~T., {et~al.} 2017, in Astronomical
  Society of the Pacific Conference Series, Vol. 512, Astronomical Data
  Analysis Software and Systems XXV, ed. N.~P.~F. {Lorente}, K.~{Shortridge},
  \& R.~{Wayth}, 279

\bibitem[{Kahn {et~al.}(2010)Kahn, Kurita, Gilmore, Nordby, O'Connor,
  Schindler, Oliver, Berg, Olivier, Riot, Antilogus, Schalk, Huffer, Bowden,
  Singal, \& Foss}]{lsstcam-kahn2010}
Kahn, S.~M., Kurita, N., Gilmore, K., {et~al.} 2010, in Ground-based and
  Airborne Instrumentation for Astronomy III, ed. I.~S. McLean, S.~K. Ramsay,
  \& H.~Takami, Vol. 7735, International Society for Optics and Photonics
  (SPIE), 77350J

\bibitem[{{Kennicutt}(1998)}]{sfr_kennicutt_1998}
{Kennicutt}, Robert~C., J. 1998, \araa, 36, 189

\bibitem[{{Koekemoer} {et~al.}(2011){Koekemoer}, {Faber}, {Ferguson}, {Grogin},
  {Kocevski}, {Koo}, {Lai}, {Lotz}, {Lucas}, {McGrath}, {Ogaz}, {Rajan},
  {Riess}, {Rodney}, {Strolger}, {Casertano}, {Castellano}, {Dahlen},
  {Dickinson}, {Dolch}, {Fontana}, {Giavalisco}, {Grazian}, {Guo}, {Hathi},
  {Huang}, {van der Wel}, {Yan}, {Acquaviva}, {Alexander}, {Almaini}, {Ashby},
  {Barden}, {Bell}, {Bournaud}, {Brown}, {Caputi}, {Cassata}, {Challis},
  {Chary}, {Cheung}, {Cirasuolo}, {Conselice}, {Roshan Cooray}, {Croton},
  {Daddi}, {Dav{\'e}}, {de Mello}, {de Ravel}, {Dekel}, {Donley}, {Dunlop},
  {Dutton}, {Elbaz}, {Fazio}, {Filippenko}, {Finkelstein}, {Frazer}, {Gardner},
  {Garnavich}, {Gawiser}, {Gruetzbauch}, {Hartley}, {H{\"a}ussler},
  {Herrington}, {Hopkins}, {Huang}, {Jha}, {Johnson}, {Kartaltepe},
  {Khostovan}, {Kirshner}, {Lani}, {Lee}, {Li}, {Madau}, {McCarthy},
  {McIntosh}, {McLure}, {McPartland}, {Mobasher}, {Moreira}, {Mortlock},
  {Moustakas}, {Mozena}, {Nandra}, {Newman}, {Nielsen}, {Niemi}, {Noeske},
  {Papovich}, {Pentericci}, {Pope}, {Primack}, {Ravindranath}, {Reddy},
  {Renzini}, {Rix}, {Robaina}, {Rosario}, {Rosati}, {Salimbeni}, {Scarlata},
  {Siana}, {Simard}, {Smidt}, {Snyder}, {Somerville}, {Spinrad}, {Straughn},
  {Telford}, {Teplitz}, {Trump}, {Vargas}, {Villforth}, {Wagner}, {Wandro},
  {Wechsler}, {Weiner}, {Wiklind}, {Wild}, {Wilson}, {Wuyts}, \&
  {Yun}}]{candels-koekemoer2011}
{Koekemoer}, A.~M., {Faber}, S.~M., {Ferguson}, H.~C., {et~al.} 2011, \apjs,
  197, 36

\bibitem[{{Kubik} {et~al.}(2016){Kubik}, {Barbier}, {Chabanat}, {Chapon},
  {Clemens}, {Ealet}, {Ferriol}, {Gillard}, {Secroun}, {Serra}, {Smadja}, \&
  {Tilquin}}]{maccread-kubik2016}
{Kubik}, B., {Barbier}, R., {Chabanat}, E., {et~al.} 2016, \pasp, 128, 104504

\bibitem[{{K{\"u}mmel} {et~al.}(2022){K{\"u}mmel}, {Vassallo}, {Dabin}, \&
  {Gracia Carpio}}]{euclid_tiles_kuemmel_2022}
{K{\"u}mmel}, M., {Vassallo}, T., {Dabin}, C., \& {Gracia Carpio}, J. 2022, in
  Astronomical Society of the Pacific Conference Series, Vol. 532, Astronomical
  Society of the Pacific Conference Series, ed. J.~E. {Ruiz},
  F.~{Pierfedereci}, \& P.~{Teuben}, 329

\bibitem[{{K{\"u}mmel} {et~al.}(2009){K{\"u}mmel}, {Walsh}, {Pirzkal},
  {Kuntschner}, \& {Pasquali}}]{axesim_kuemmel2008}
{K{\"u}mmel}, M., {Walsh}, J.~R., {Pirzkal}, N., {Kuntschner}, H., \&
  {Pasquali}, A. 2009, \pasp, 121, 59

\bibitem[{{Laigle} {et~al.}(2016){Laigle}, {McCracken}, {Ilbert}, {Hsieh},
  {Davidzon}, {Capak}, {Hasinger}, {Silverman}, {Pichon}, {Coupon}, {Aussel},
  {Le Borgne}, {Caputi}, {Cassata}, {Chang}, {Civano}, {Dunlop}, {Fynbo},
  {Kartaltepe}, {Koekemoer}, {Le F{\`e}vre}, {Le Floc'h}, {Leauthaud}, {Lilly},
  {Lin}, {Marchesi}, {Milvang-Jensen}, {Salvato}, {Sanders}, {Scoville},
  {Smolcic}, {Stockmann}, {Taniguchi}, {Tasca}, {Toft}, {Vaccari}, \&
  {Zabl}}]{cosmos-laigle2016}
{Laigle}, C., {McCracken}, H.~J., {Ilbert}, O., {et~al.} 2016, \apjs, 224, 24

\bibitem[{{Lastennet} {et~al.}(2002){Lastennet}, {Lejeune}, {Oblak}, {Westera},
  \& {Buser}}]{basel-lastennet2002}
{Lastennet}, E., {Lejeune}, T., {Oblak}, E., {Westera}, P., \& {Buser}, R.
  2002, \apss, 280, 83

\bibitem[{{Laureijs} {et~al.}(2011){Laureijs}, {Amiaux}, {Arduini},
  {Augu{\`e}res}, {Brinchmann}, {Cole}, {Cropper}, {Dabin}, {Duvet}, {Ealet},
  {Garilli}, {Gondoin}, {Guzzo}, {Hoar}, {Hoekstra}, {Holmes}, {Kitching},
  {Maciaszek}, {Mellier}, {Pasian}, {Percival}, {Rhodes}, {Saavedra Criado},
  {Sauvage}, {Scaramella}, {Valenziano}, {Warren}, {Bender}, {Castander},
  {Cimatti}, {Le F{\`e}vre}, {Kurki-Suonio}, {Levi}, {Lilje}, {Meylan},
  {Nichol}, {Pedersen}, {Popa}, {Rebolo Lopez}, {Rix}, {Rottgering},
  {Zeilinger}, {Grupp}, {Hudelot}, {Massey}, {Meneghetti}, {Miller}, {Paltani},
  {Paulin-Henriksson}, {Pires}, {Saxton}, {Schrabback}, {Seidel}, {Walsh},
  {Aghanim}, {Amendola}, {Bartlett}, {Baccigalupi}, {Beaulieu}, {Benabed},
  {Cuby}, {Elbaz}, {Fosalba}, {Gavazzi}, {Helmi}, {Hook}, {Irwin}, {Kneib},
  {Kunz}, {Mannucci}, {Moscardini}, {Tao}, {Teyssier}, {Weller}, {Zamorani},
  {Zapatero Osorio}, {Boulade}, {Foumond}, {Di Giorgio}, {Guttridge}, {James},
  {Kemp}, {Martignac}, {Spencer}, {Walton}, {Bl{\"u}mchen}, {Bonoli},
  {Bortoletto}, {Cerna}, {Corcione}, {Fabron}, {Jahnke}, {Ligori}, {Madrid},
  {Martin}, {Morgante}, {Pamplona}, {Prieto}, {Riva}, {Toledo}, {Trifoglio},
  {Zerbi}, {Abdalla}, {Douspis}, {Grenet}, {Borgani}, {Bouwens}, {Courbin},
  {Delouis}, {Dubath}, {Fontana}, {Frailis}, {Grazian}, {Koppenh{\"o}fer},
  {Mansutti}, {Melchior}, {Mignoli}, {Mohr}, {Neissner}, {Noddle}, {Poncet},
  {Scodeggio}, {Serrano}, {Shane}, {Starck}, {Surace}, {Taylor},
  {Verdoes-Kleijn}, {Vuerli}, {Williams}, {Zacchei}, {Altieri}, {Escudero
  Sanz}, {Kohley}, {Oosterbroek}, {Astier}, {Bacon}, {Bardelli}, {Baugh},
  {Bellagamba}, {Benoist}, {Bianchi}, {Biviano}, {Branchini}, {Carbone},
  {Cardone}, {Clements}, {Colombi}, {Conselice}, {Cresci}, {Deacon}, {Dunlop},
  {Fedeli}, {Fontanot}, {Franzetti}, {Giocoli}, {Garcia-Bellido}, {Gow},
  {Heavens}, {Hewett}, {Heymans}, {Holland}, {Huang}, {Ilbert}, {Joachimi},
  {Jennins}, {Kerins}, {Kiessling}, {Kirk}, {Kotak}, {Krause}, {Lahav}, {van
  Leeuwen}, {Lesgourgues}, {Lombardi}, {Magliocchetti}, {Maguire}, {Majerotto},
  {Maoli}, {Marulli}, {Maurogordato}, {McCracken}, {McLure}, {Melchiorri},
  {Merson}, {Moresco}, {Nonino}, {Norberg}, {Peacock}, {Pello}, {Penny},
  {Pettorino}, {Di Porto}, {Pozzetti}, {Quercellini}, {Radovich}, {Rassat},
  {Roche}, {Ronayette}, {Rossetti}, {Sartoris}, {Schneider}, {Semboloni},
  {Serjeant}, {Simpson}, {Skordis}, {Smadja}, {Smartt}, {Spano}, {Spiro},
  {Sullivan}, {Tilquin}, {Trotta}, {Verde}, {Wang}, {Williger}, {Zhao},
  {Zoubian}, \& {Zucca}}]{euclid-laurejis2011}
{Laureijs}, R., {Amiaux}, J., {Arduini}, S., {et~al.} 2011, arXiv e-prints,
  1110.3193

\bibitem[{Lim(2019)}]{LSE-400}
Lim, K.-T. 2019, {Large Synoptic Survey Telescope (LSST) Header Service
  Interface between the OCS and EFD}

\bibitem[{{LSST Dark Energy Science Collaboration}(2012)}]{lsstdesc-2012}
{LSST Dark Energy Science Collaboration}. 2012, arXiv, 1211.0310

\bibitem[{{LSST Science Collaboration} {et~al.}(2009){LSST Science
  Collaboration}, {Abell}, {Allison}, {Anderson}, {Andrew}, {Angel}, {Armus},
  {Arnett}, {Asztalos}, {Axelrod}, {Bailey}, {Ballantyne}, {Bankert},
  {Barkhouse}, {Barr}, {Barrientos}, {Barth}, {Bartlett}, {Becker}, {Becla},
  {Beers}, {Bernstein}, {Biswas}, {Blanton}, {Bloom}, {Bochanski}, {Boeshaar},
  {Borne}, {Bradac}, {Brandt}, {Bridge}, {Brown}, {Brunner}, {Bullock},
  {Burgasser}, {Burge}, {Burke}, {Cargile}, {Chandrasekharan}, {Chartas},
  {Chesley}, {Chu}, {Cinabro}, {Claire}, {Claver}, {Clowe}, {Connolly}, {Cook},
  {Cooke}, {Cooray}, {Covey}, {Culliton}, {de Jong}, {de Vries}, {Debattista},
  {Delgado}, {Dell'Antonio}, {Dhital}, {Di Stefano}, {Dickinson}, {Dilday},
  {Djorgovski}, {Dobler}, {Donalek}, {Dubois-Felsmann}, {Durech},
  {Eliasdottir}, {Eracleous}, {Eyer}, {Falco}, {Fan}, {Fassnacht}, {Ferguson},
  {Fernandez}, {Fields}, {Finkbeiner}, {Figueroa}, {Fox}, {Francke}, {Frank},
  {Frieman}, {Fromenteau}, {Furqan}, {Galaz}, {Gal-Yam}, {Garnavich},
  {Gawiser}, {Geary}, {Gee}, {Gibson}, {Gilmore}, {Grace}, {Green}, {Gressler},
  {Grillmair}, {Habib}, {Haggerty}, {Hamuy}, {Harris}, {Hawley}, {Heavens},
  {Hebb}, {Henry}, {Hileman}, {Hilton}, {Hoadley}, {Holberg}, {Holman},
  {Howell}, {Infante}, {Ivezic}, {Jacoby}, {Jain}, {R}, {Jedicke}, {Jee},
  {Garrett Jernigan}, {Jha}, {Johnston}, {Jones}, {Juric}, {Kaasalainen},
  {Styliani}, {Kafka}, {Kahn}, {Kaib}, {Kalirai}, {Kantor}, {Kasliwal},
  {Keeton}, {Kessler}, {Knezevic}, {Kowalski}, {Krabbendam}, {Krughoff},
  {Kulkarni}, {Kuhlman}, {Lacy}, {Lepine}, {Liang}, {Lien}, {Lira}, {Long},
  {Lorenz}, {Lotz}, {Lupton}, {Lutz}, {Macri}, {Mahabal}, {Mandelbaum},
  {Marshall}, {May}, {McGehee}, {Meadows}, {Meert}, {Milani}, {Miller},
  {Miller}, {Mills}, {Minniti}, {Monet}, {Mukadam}, {Nakar}, {Neill}, {Newman},
  {Nikolaev}, {Nordby}, {O'Connor}, {Oguri}, {Oliver}, {Olivier}, {Olsen},
  {Olsen}, {Olszewski}, {Oluseyi}, {Padilla}, {Parker}, {Pepper}, {Peterson},
  {Petry}, {Pinto}, {Pizagno}, {Popescu}, {Prsa}, {Radcka}, {Raddick},
  {Rasmussen}, {Rau}, {Rho}, {Rhoads}, {Richards}, {Ridgway}, {Robertson},
  {Roskar}, {Saha}, {Sarajedini}, {Scannapieco}, {Schalk}, {Schindler},
  {Schmidt}, {Schmidt}, {Schneider}, {Schumacher}, {Scranton}, {Sebag},
  {Seppala}, {Shemmer}, {Simon}, {Sivertz}, {Smith}, {Allyn Smith}, {Smith},
  {Spitz}, {Stanford}, {Stassun}, {Strader}, {Strauss}, {Stubbs}, {Sweeney},
  {Szalay}, {Szkody}, {Takada}, {Thorman}, {Trilling}, {Trimble}, {Tyson}, {Van
  Berg}, {Vanden Berk}, {VanderPlas}, {Verde}, {Vrsnak}, {Walkowicz},
  {Wandelt}, {Wang}, {Wang}, {Warner}, {Wechsler}, {West}, {Wiecha},
  {Williams}, {Willman}, {Wittman}, {Wolff}, {Wood-Vasey}, {Wozniak}, {Young},
  {Zentner}, \& {Zhan}}]{lsst_science_book2009}
{LSST Science Collaboration}, {Abell}, P.~A., {Allison}, J., {et~al.} 2009,
  arXiv, 0912.0201

\bibitem[{{Maciaszek} {et~al.}(2022){Maciaszek}, {Ealet}, {Gillard}, {Jahnke},
  {Barbier}, {Prieto}, {Bon}, {Bonnefoi}, {Caillat}, {Carle}, {Costille},
  {Ducret}, {Fabron}, {Foulon}, {Gimenez}, {Grassi}, {Jaquet}, {Le Mignant},
  {Martin}, {Pamplona}, {Sanchez}, {Cl{\'e}mens}, {Caillat}, {Niclas},
  {Secroun}, {Kubik}, {Ferriol}, {Berthe}, {Barri{\`e}re}, {Fontignie},
  {Valenziano}, {Auricchio}, {Battaglia}, {De Rosa}, {Farinelli}, {Franceschi},
  {Medinaceli}, {Morgante}, {Sortino}, {Trifoglio}, {Corcione}, {Capobianco},
  {Ligori}, {Dusini}, {Borsato}, {Dal Corso}, {Laudisio}, {Sirignano},
  {Stanco}, {Ventura}, {Patrizii}, {Chiarusi}, {Fornari}, {Giacomini},
  {Margiotta}, {Mauri}, {Pasqualini}, {Sirri}, {Spurio}, {Tenti}, {Travaglini},
  {Bonoli}, {Bortoletto}, {Balestra}, {Dalessandro}, {Grupp}, {Penka},
  {Steinwagner}, {Hormuth}, {Schirmer}, {Seidel}, {Padilla}, {Casas}, {Lloro},
  {Toledo-Moreo}, {Gomez}, {Colodro-Conde}, {Liz{\'a}n}, {Diaz}, {Lilje},
  {Andersen}, {Andersen}, {S{\o}rensen}, {Hornstrup}, {Jessen}, {Thizy},
  {Holmes}, {Pniel}, {Jhabvala}, {Pravdo}, {Seiffert}, {Waczynski}, {Laureij},
  {Racca}, {Salvignol}, {Boenke}, {Strada}, \& {Mellier}}]{maciaszek2022}
{Maciaszek}, T., {Ealet}, A., {Gillard}, W., {et~al.} 2022, in Society of
  Photo-Optical Instrumentation Engineers (SPIE) Conference Series, Vol. 12180,
  Space Telescopes and Instrumentation 2022: Optical, Infrared, and Millimeter
  Wave, ed. L.~E. {Coyle}, S.~{Matsuura}, \& M.~D. {Perrin}, 121801K

\bibitem[{Maciaszek {et~al.}(2016)Maciaszek, Ealet, Jahnke, Prieto, Barbier,
  Mellier, Beaumont, Bon, Bonnefoi, Carle, Caillat, Costille, Dormoy, Ducret,
  Fabron, Febvre, Foulon, Garcia, Gimenez, Grassi, Laurent, Mignant, Martin,
  Rossin, Pamplona, Sanchez, Vives, Cl{\'e}mens, Gillard, Niclas, Secroun,
  Serra, Kubik, Ferriol, Amiaux, Barri{\`e}re, Berthe, Rosset, Macias-Perez,
  Auricchio, Rosa, Franceschi, Guizzo, Morgante, Sortino, Trifoglio,
  Valenziano, Patrizii, Chiarusi, Fornari, Giacomini, Margiotta, Mauri,
  Pasqualini, Sirri, Spurio, Tenti, Travaglini, Dusini, Corso, Laudisio,
  Sirignano, Stanco, Ventura, Borsato, Bonoli, Bortoletto, Balestra,
  D'Alessandro, Medinaceli, Farinelli, Corcione, Ligori, Grupp, Wimmer,
  Hormuth, Seidel, Wachter, Padilla, Lamensans, Casas, Lloro, Toledo-Moreo,
  Gomez, Colodro-Conde, Liz{\'a}n, Diaz, Lilje, Toulouse-Aastrup, Andersen,
  S{\o}rensen, Jakobsen, Hornstrup, Jessen, Thizy, Holmes, Israelsson,
  Seiffert, Waczynski, Laureijs, Racca, Salvignol, Boenke, \&
  Strada}]{maciaszek2016}
Maciaszek, T., Ealet, A., Jahnke, K., {et~al.} 2016, in Space Telescopes and
  Instrumentation 2016: Optical, Infrared, and Millimeter Wave, ed. H.~A.
  MacEwen, G.~G. Fazio, M.~Lystrup, N.~Batalha, N.~Siegler, \& E.~C. Tong, Vol.
  9904, International Society for Optics and Photonics (SPIE), 99040T

\bibitem[{{Markovi{\v{c}}} {et~al.}(2017){Markovi{\v{c}}}, {Percival},
  {Scodeggio}, {Ealet}, {Wachter}, {Garilli}, {Guzzo}, {Scaramella},
  {Maiorano}, \& {Amiaux}}]{spectra_selfcal_markovic_2017}
{Markovi{\v{c}}}, K., {Percival}, W.~J., {Scodeggio}, M., {et~al.} 2017,
  \mnras, 467, 3677

\bibitem[{{Massey} {et~al.}(2014){Massey}, {Schrabback}, {Cordes}, {Marggraf},
  {Israel}, {Miller}, {Hall}, {Cropper}, {Prod'homme}, \&
  {Niemi}}]{cti-massey2014}
{Massey}, R., {Schrabback}, T., {Cordes}, O., {et~al.} 2014, \mnras, 439, 887

\bibitem[{{Metcalf} {et~al.}(2019){Metcalf}, {Meneghetti}, {Avestruz},
  {Bellagamba}, {Bom}, {Bertin}, {Cabanac}, {Courbin}, {Davies},
  {Decenci{\`e}re}, {Flamary}, {Gavazzi}, {Geiger}, {Hartley},
  {Huertas-Company}, {Jackson}, {Jacobs}, {Jullo}, {Kneib}, {Koopmans},
  {Lanusse}, {Li}, {Ma}, {Makler}, {Li}, {Lightman}, {Petrillo}, {Serjeant},
  {Sch{\"a}fer}, {Sonnenfeld}, {Tagore}, {Tortora}, {Tuccillo},
  {Valent{\'\i}n}, {Velasco-Forero}, {Verdoes Kleijn}, \&
  {Vernardos}}]{metcalf2019}
{Metcalf}, R.~B., {Meneghetti}, M., {Avestruz}, C., {et~al.} 2019, \aap, 625,
  A119

\bibitem[{{Metcalf} \& {Petkova}(2014)}]{metcalf2014}
{Metcalf}, R.~B. \& {Petkova}, M. 2014, \mnras, 445, 1942

\bibitem[{Miyazaki {et~al.}(2017)Miyazaki, Komiyama, Kawanomoto, Doi, Furusawa,
  Hamana, Hayashi, Ikeda, Kamata, Karoji, Koike, Kurakami, Miyama, Morokuma,
  Nakata, Namikawa, Nakaya, Nariai, Obuchi, Oishi, Okada, Okura, Tait, Takata,
  Tanaka, Tanaka, Terai, Tomono, Uraguchi, Usuda, Utsumi, Yamada, Yamanoi,
  Aihara, Fujimori, Mineo, Miyatake, Oguri, Uchida, Tanaka, Yasuda, Takada,
  Murayama, Nishizawa, Sugiyama, Chiba, Futamase, Wang, Chen, Ho, Liaw, Chiu,
  Ho, Lai, Lee, Jeng, Iwamura, Armstrong, Bickerton, Bosch, Gunn, Lupton,
  Loomis, Price, Smith, Strauss, Turner, Suzuki, Miyazaki, Muramatsu, Yamamoto,
  Endo, Ezaki, Ito, Kawaguchi, Sofuku, Taniike, Akutsu, Dojo, Kasumi, Matsuda,
  Imoto, Miwa, Suzuki, Takeshi, \& Yokota}]{hsc-miyazaki2017}
Miyazaki, S., Komiyama, Y., Kawanomoto, S., {et~al.} 2017, \pasj, 70, s1

\bibitem[{{Naghib} {et~al.}(2019){Naghib}, {Yoachim}, {Vanderbei}, {Connolly},
  \& {Jones}}]{lsst-scheduler-naghib}
{Naghib}, E., {Yoachim}, P., {Vanderbei}, R.~J., {Connolly}, A.~J., \& {Jones},
  R.~L. 2019, \aj, 157, 151

\bibitem[{{O'Donnell}(1994)}]{extinction-odonnell1994}
{O'Donnell}, J.~E. 1994, \apj, 422, 158

\bibitem[{{Pickles} \& {Depagne}(2010)}]{tycho2match-pickles2010}
{Pickles}, A. \& {Depagne}, {\'E}. 2010, \pasp, 122, 1437

\bibitem[{{Planck Collaboration: Abergel} {et~al.}(2014){Planck Collaboration:
  Abergel}, {Ade}, {Aghanim}, {Alves}, {Aniano}, {Armitage-Caplan}, {Arnaud},
  {Ashdown}, {Atrio-Barandela}, {Aumont}, {Baccigalupi}, {Banday}, {Barreiro},
  {Bartlett}, {Battaner}, {Benabed}, {Beno{\^\i}t}, {Benoit-L{\'e}vy},
  {Bernard}, {Bersanelli}, {Bielewicz}, {Bobin}, {Bock}, {Bonaldi}, {Bond},
  {Borrill}, {Bouchet}, {Boulanger}, {Bridges}, {Bucher}, {Burigana}, {Butler},
  {Cardoso}, {Catalano}, {Chamballu}, {Chary}, {Chiang}, {Chiang},
  {Christensen}, {Church}, {Clemens}, {Clements}, {Colombi}, {Colombo},
  {Combet}, {Couchot}, {Coulais}, {Crill}, {Curto}, {Cuttaia}, {Danese},
  {Davies}, {Davis}, {de Bernardis}, {de Rosa}, {de Zotti}, {Delabrouille},
  {Delouis}, {D{\'e}sert}, {Dickinson}, {Diego}, {Dole}, {Donzelli},
  {Dor{\'e}}, {Douspis}, {Draine}, {Dupac}, {Efstathiou}, {En{\ss}lin},
  {Eriksen}, {Falgarone}, {Finelli}, {Forni}, {Frailis}, {Fraisse},
  {Franceschi}, {Galeotta}, {Ganga}, {Ghosh}, {Giard}, {Giardino},
  {Giraud-H{\'e}raud}, {Gonz{\'a}lez-Nuevo}, {G{\'o}rski}, {Gratton},
  {Gregorio}, {Grenier}, {Gruppuso}, {Guillet}, {Hansen}, {Hanson}, {Harrison},
  {Helou}, {Henrot-Versill{\'e}}, {Hern{\'a}ndez-Monteagudo}, {Herranz},
  {Hildebrandt}, {Hivon}, {Hobson}, {Holmes}, {Hornstrup}, {Hovest},
  {Huffenberger}, {Jaffe}, {Jaffe}, {Jewell}, {Joncas}, {Jones}, {Juvela},
  {Keih{\"a}nen}, {Keskitalo}, {Kisner}, {Knoche}, {Knox}, {Kunz},
  {Kurki-Suonio}, {Lagache}, {L{\"a}hteenm{\"a}ki}, {Lamarre}, {Lasenby},
  {Laureijs}, {Lawrence}, {Leonardi}, {Le{\'o}n-Tavares}, {Lesgourgues},
  {Levrier}, {Liguori}, {Lilje}, {Linden-V{\o}rnle}, {L{\'o}pez-Caniego},
  {Lubin}, {Mac{\'\i}as-P{\'e}rez}, {Maffei}, {Maino}, {Mandolesi}, {Maris},
  {Marshall}, {Martin}, {Mart{\'\i}nez-Gonz{\'a}lez}, {Masi}, {Massardi},
  {Matarrese}, {Matthai}, {Mazzotta}, {McGehee}, {Melchiorri}, {Mendes},
  {Mennella}, {Migliaccio}, {Mitra}, {Miville-Desch{\^e}nes}, {Moneti},
  {Montier}, {Morgante}, {Mortlock}, {Munshi}, {Murphy}, {Naselsky}, {Nati},
  {Natoli}, {Netterfield}, {N{\o}rgaard-Nielsen}, {Noviello}, {Novikov},
  {Novikov}, {Osborne}, {Oxborrow}, {Paci}, {Pagano}, {Pajot}, {Paladini},
  {Paoletti}, {Pasian}, {Patanchon}, {Perdereau}, {Perotto}, {Perrotta},
  {Piacentini}, {Piat}, {Pierpaoli}, {Pietrobon}, {Plaszczynski},
  {Pointecouteau}, {Polenta}, {Ponthieu}, {Popa}, {Poutanen}, {Pratt},
  {Pr{\'e}zeau}, {Prunet}, {Puget}, {Rachen}, {Reach}, {Rebolo}, {Reinecke},
  {Remazeilles}, {Renault}, {Ricciardi}, {Riller}, {Ristorcelli}, {Rocha},
  {Rosset}, {Roudier}, {Rowan-Robinson}, {Rubi{\~n}o-Mart{\'\i}n}, {Rusholme},
  {Sandri}, {Santos}, {Savini}, {Scott}, {Seiffert}, {Shellard}, {Spencer},
  {Starck}, {Stolyarov}, {Stompor}, {Sudiwala}, {Sunyaev}, {Sureau}, {Sutton},
  {Suur-Uski}, {Sygnet}, {Tauber}, {Tavagnacco}, {Terenzi}, {Toffolatti},
  {Tomasi}, {Tristram}, {Tucci}, {Tuovinen}, {T{\"u}rler}, {Umana},
  {Valenziano}, {Valiviita}, {Van Tent}, {Verstraete}, {Vielva}, {Villa},
  {Vittorio}, {Wade}, {Wandelt}, {Welikala}, {Ysard}, {Yvon}, {Zacchei}, \&
  {Zonca}}]{dust-planck2014}
{Planck Collaboration: Abergel}, A., {Ade}, P.~A.~R., {Aghanim}, N., {et~al.}
  2014, \aap, 571, A11

\bibitem[{{Planck Collaboration: Ade} {et~al.}(2016){Planck Collaboration:
  Ade}, {Aghanim}, {Arnaud}, {Ashdown}, {Aumont}, {Baccigalupi}, {Banday},
  {Barreiro}, {Bartlett}, {Bartolo}, {Battaner}, {Battye}, {Benabed},
  {Beno{\^\i}t}, {Benoit-L{\'e}vy}, {Bernard}, {Bersanelli}, {Bielewicz},
  {Bock}, {Bonaldi}, {Bonavera}, {Bond}, {Borrill}, {Bouchet}, {Boulanger},
  {Bucher}, {Burigana}, {Butler}, {Calabrese}, {Cardoso}, {Catalano},
  {Challinor}, {Chamballu}, {Chary}, {Chiang}, {Chluba}, {Christensen},
  {Church}, {Clements}, {Colombi}, {Colombo}, {Combet}, {Coulais}, {Crill},
  {Curto}, {Cuttaia}, {Danese}, {Davies}, {Davis}, {de Bernardis}, {de Rosa},
  {de Zotti}, {Delabrouille}, {D{\'e}sert}, {Di Valentino}, {Dickinson},
  {Diego}, {Dolag}, {Dole}, {Donzelli}, {Dor{\'e}}, {Douspis}, {Ducout},
  {Dunkley}, {Dupac}, {Efstathiou}, {Elsner}, {En{\ss}lin}, {Eriksen},
  {Farhang}, {Fergusson}, {Finelli}, {Forni}, {Frailis}, {Fraisse},
  {Franceschi}, {Frejsel}, {Galeotta}, {Galli}, {Ganga}, {Gauthier}, {Gerbino},
  {Ghosh}, {Giard}, {Giraud-H{\'e}raud}, {Giusarma}, {Gjerl{\o}w},
  {Gonz{\'a}lez-Nuevo}, {G{\'o}rski}, {Gratton}, {Gregorio}, {Gruppuso},
  {Gudmundsson}, {Hamann}, {Hansen}, {Hanson}, {Harrison}, {Helou},
  {Henrot-Versill{\'e}}, {Hern{\'a}ndez-Monteagudo}, {Herranz}, {Hildebrandt},
  {Hivon}, {Hobson}, {Holmes}, {Hornstrup}, {Hovest}, {Huang}, {Huffenberger},
  {Hurier}, {Jaffe}, {Jaffe}, {Jones}, {Juvela}, {Keih{\"a}nen}, {Keskitalo},
  {Kisner}, {Kneissl}, {Knoche}, {Knox}, {Kunz}, {Kurki-Suonio}, {Lagache},
  {L{\"a}hteenm{\"a}ki}, {Lamarre}, {Lasenby}, {Lattanzi}, {Lawrence}, {Leahy},
  {Leonardi}, {Lesgourgues}, {Levrier}, {Lewis}, {Liguori}, {Lilje},
  {Linden-V{\o}rnle}, {L{\'o}pez-Caniego}, {Lubin}, {Mac{\'\i}as-P{\'e}rez},
  {Maggio}, {Maino}, {Mandolesi}, {Mangilli}, {Marchini}, {Maris}, {Martin},
  {Martinelli}, {Mart{\'\i}nez-Gonz{\'a}lez}, {Masi}, {Matarrese}, {McGehee},
  {Meinhold}, {Melchiorri}, {Melin}, {Mendes}, {Mennella}, {Migliaccio},
  {Millea}, {Mitra}, {Miville-Desch{\^e}nes}, {Moneti}, {Montier}, {Morgante},
  {Mortlock}, {Moss}, {Munshi}, {Murphy}, {Naselsky}, {Nati}, {Natoli},
  {Netterfield}, {N{\o}rgaard-Nielsen}, {Noviello}, {Novikov}, {Novikov},
  {Oxborrow}, {Paci}, {Pagano}, {Pajot}, {Paladini}, {Paoletti}, {Partridge},
  {Pasian}, {Patanchon}, {Pearson}, {Perdereau}, {Perotto}, {Perrotta},
  {Pettorino}, {Piacentini}, {Piat}, {Pierpaoli}, {Pietrobon}, {Plaszczynski},
  {Pointecouteau}, {Polenta}, {Popa}, {Pratt}, {Pr{\'e}zeau}, {Prunet},
  {Puget}, {Rachen}, {Reach}, {Rebolo}, {Reinecke}, {Remazeilles}, {Renault},
  {Renzi}, {Ristorcelli}, {Rocha}, {Rosset}, {Rossetti}, {Roudier},
  {Rouill{\'e} d'Orfeuil}, {Rowan-Robinson}, {Rubi{\~n}o-Mart{\'\i}n},
  {Rusholme}, {Said}, {Salvatelli}, {Salvati}, {Sandri}, {Santos},
  {Savelainen}, {Savini}, {Scott}, {Seiffert}, {Serra}, {Shellard}, {Spencer},
  {Spinelli}, {Stolyarov}, {Stompor}, {Sudiwala}, {Sunyaev}, {Sutton},
  {Suur-Uski}, {Sygnet}, {Tauber}, {Terenzi}, {Toffolatti}, {Tomasi},
  {Tristram}, {Trombetti}, {Tucci}, {Tuovinen}, {T{\"u}rler}, {Umana},
  {Valenziano}, {Valiviita}, {Van Tent}, {Vielva}, {Villa}, {Wade}, {Wandelt},
  {Wehus}, {White}, {White}, {Wilkinson}, {Yvon}, {Zacchei}, \&
  {Zonca}}]{planck2015}
{Planck Collaboration: Ade}, P.~A.~R., {Aghanim}, N., {Arnaud}, M., {et~al.}
  2016, \aap, 594, A13

\bibitem[{{Planck Collaboration: Aghanim} {et~al.}(2020){Planck Collaboration:
  Aghanim}, {Akrami}, {Ashdown}, {Aumont}, {Baccigalupi}, {Ballardini},
  {Banday}, {Barreiro}, {Bartolo}, {Basak}, {Battye}, {Benabed}, {Bernard},
  {Bersanelli}, {Bielewicz}, {Bock}, {Bond}, {Borrill}, {Bouchet}, {Boulanger},
  {Bucher}, {Burigana}, {Butler}, {Calabrese}, {Cardoso}, {Carron},
  {Challinor}, {Chiang}, {Chluba}, {Colombo}, {Combet}, {Contreras}, {Crill},
  {Cuttaia}, {de Bernardis}, {de Zotti}, {Delabrouille}, {Delouis}, {Di
  Valentino}, {Diego}, {Dor{\'e}}, {Douspis}, {Ducout}, {Dupac}, {Dusini},
  {Efstathiou}, {Elsner}, {En{\ss}lin}, {Eriksen}, {Fantaye}, {Farhang},
  {Fergusson}, {Fernandez-Cobos}, {Finelli}, {Forastieri}, {Frailis},
  {Fraisse}, {Franceschi}, {Frolov}, {Galeotta}, {Galli}, {Ganga},
  {G{\'e}nova-Santos}, {Gerbino}, {Ghosh}, {Gonz{\'a}lez-Nuevo}, {G{\'o}rski},
  {Gratton}, {Gruppuso}, {Gudmundsson}, {Hamann}, {Handley}, {Hansen},
  {Herranz}, {Hildebrandt}, {Hivon}, {Huang}, {Jaffe}, {Jones}, {Karakci},
  {Keih{\"a}nen}, {Keskitalo}, {Kiiveri}, {Kim}, {Kisner}, {Knox},
  {Krachmalnicoff}, {Kunz}, {Kurki-Suonio}, {Lagache}, {Lamarre}, {Lasenby},
  {Lattanzi}, {Lawrence}, {Le Jeune}, {Lemos}, {Lesgourgues}, {Levrier},
  {Lewis}, {Liguori}, {Lilje}, {Lilley}, {Lindholm}, {L{\'o}pez-Caniego},
  {Lubin}, {Ma}, {Mac{\'\i}as-P{\'e}rez}, {Maggio}, {Maino}, {Mandolesi},
  {Mangilli}, {Marcos-Caballero}, {Maris}, {Martin}, {Martinelli},
  {Mart{\'\i}nez-Gonz{\'a}lez}, {Matarrese}, {Mauri}, {McEwen}, {Meinhold},
  {Melchiorri}, {Mennella}, {Migliaccio}, {Millea}, {Mitra},
  {Miville-Desch{\^e}nes}, {Molinari}, {Montier}, {Morgante}, {Moss}, {Natoli},
  {N{\o}rgaard-Nielsen}, {Pagano}, {Paoletti}, {Partridge}, {Patanchon},
  {Peiris}, {Perrotta}, {Pettorino}, {Piacentini}, {Polastri}, {Polenta},
  {Puget}, {Rachen}, {Reinecke}, {Remazeilles}, {Renzi}, {Rocha}, {Rosset},
  {Roudier}, {Rubi{\~n}o-Mart{\'\i}n}, {Ruiz-Granados}, {Salvati}, {Sandri},
  {Savelainen}, {Scott}, {Shellard}, {Sirignano}, {Sirri}, {Spencer},
  {Sunyaev}, {Suur-Uski}, {Tauber}, {Tavagnacco}, {Tenti}, {Toffolatti},
  {Tomasi}, {Trombetti}, {Valenziano}, {Valiviita}, {Van Tent}, {Vibert},
  {Vielva}, {Villa}, {Vittorio}, {Wandelt}, {Wehus}, {White}, {White},
  {Zacchei}, \& {Zonca}}]{planck2018}
{Planck Collaboration: Aghanim}, N., {Akrami}, Y., {Ashdown}, M., {et~al.}
  2020, \aap, 641, A6

\bibitem[{{Polletta} {et~al.}(2007){Polletta}, {Tajer}, {Maraschi},
  {Trinchieri}, {Lonsdale}, {Chiappetti}, {Andreon}, {Pierre}, {Le F{\`e}vre},
  {Zamorani}, {Maccagni}, {Garcet}, {Surdej}, {Franceschini}, {Alloin},
  {Shupe}, {Surace}, {Fang}, {Rowan-Robinson}, {Smith}, \&
  {Tresse}}]{cosmosseds-polletta2007}
{Polletta}, M., {Tajer}, M., {Maraschi}, L., {et~al.} 2007, \apj, 663, 81

\bibitem[{{Potter} {et~al.}(2017){Potter}, {Stadel}, \&
  {Teyssier}}]{PKDGRAV3-potter2016}
{Potter}, D., {Stadel}, J., \& {Teyssier}, R. 2017, Computational Astrophysics
  and Cosmology, 4, 2

\bibitem[{{Pozzetti} {et~al.}(2016){Pozzetti}, {Hirata}, {Geach}, {Cimatti},
  {Baugh}, {Cucciati}, {Merson}, {Norberg}, \& {Shi}}]{halpha-pozzetti2016}
{Pozzetti}, L., {Hirata}, C.~M., {Geach}, J.~E., {et~al.} 2016, \aap, 590, A3

\bibitem[{{Prevot} {et~al.}(1984){Prevot}, {Lequeux}, {Maurice}, {Prevot}, \&
  {Rocca-Volmerange}}]{cosmosdust-prevot1984}
{Prevot}, M.~L., {Lequeux}, J., {Maurice}, E., {Prevot}, L., \&
  {Rocca-Volmerange}, B. 1984, \aap, 132, 389

\bibitem[{{Racca} {et~al.}(2016){Racca}, {Laureijs}, {Stagnaro}, {Salvignol},
  {Lorenzo Alvarez}, {Saavedra Criado}, {Gaspar Venancio}, {Short}, {Strada},
  {B{\"o}nke}, {Colombo}, {Calvi}, {Maiorano}, {Piersanti}, {Prezelus},
  {Rosato}, {Pinel}, {Rozemeijer}, {Lesna}, {Musi}, {Sias}, {Anselmi},
  {Cazaubiel}, {Vaillon}, {Mellier}, {Amiaux}, {Berth{\'e}}, {Sauvage},
  {Azzollini}, {Cropper}, {Pottinger}, {Jahnke}, {Ealet}, {Maciaszek},
  {Pasian}, {Zacchei}, {Scaramella}, {Hoar}, {Kohley}, {Vavrek}, {Rudolph}, \&
  {Schmidt}}]{eucliddesign-racca2016}
{Racca}, G.~D., {Laureijs}, R., {Stagnaro}, L., {et~al.} 2016, in Society of
  Photo-Optical Instrumentation Engineers (SPIE) Conference Series, Vol. 9904,
  Space Telescopes and Instrumentation 2016: Optical, Infrared, and Millimeter
  Wave, ed. H.~A. {MacEwen}, G.~G. {Fazio}, M.~{Lystrup}, N.~{Batalha},
  N.~{Siegler}, \& E.~C. {Tong}, 99040O

\bibitem[{{Robin} \& {Creze}(1986)}]{besancon-robin1986}
{Robin}, A. \& {Creze}, M. 1986, \aap, 157, 71

\bibitem[{Rolland {et~al.}(2007)Rolland, Pinheiro~da Silva, Inguimbert, David,
  Ecoffet, \& Auvergne}]{stardust_rolland2007}
Rolland, G., Pinheiro~da Silva, L., Inguimbert, C., {et~al.} 2007, in 2007 9th
  European Conference on Radiation and Its Effects on Components and Systems,
  1--9

\bibitem[{{Rowe} {et~al.}(2015){Rowe}, {Jarvis}, {Mandelbaum}, {Bernstein},
  {Bosch}, {Simet}, {Meyers}, {Kacprzak}, {Nakajima}, {Zuntz}, {Miyatake},
  {Dietrich}, {Armstrong}, {Melchior}, \& {Gill}}]{galsim-rowe2015}
{Rowe}, B.~T.~P., {Jarvis}, M., {Mandelbaum}, R., {et~al.} 2015, Astronomy and
  Computing, 10, 121

\bibitem[{{Schultheis} {et~al.}(2014){Schultheis}, {Chen}, {Jiang}, {Gonzalez},
  {Enokiya}, {Fukui}, {Torii}, {Rejkuba}, \&
  {Minniti}}]{3Ddustmaps-schultheis2014}
{Schultheis}, M., {Chen}, B.~Q., {Jiang}, B.~W., {et~al.} 2014, \aap, 566, A120

\bibitem[{Secroun {et~al.}(2018)Secroun, Barbier, Buton, Cl{\'e}mens, Conversi,
  Ealet, Ferriol, Fornari, Gillard, Kohley, Kubik, Rosset, Serra, Smadja, \&
  Zoubian}]{detectorgain-secroun2018}
Secroun, A., Barbier, R., Buton, C., {et~al.} 2018, in High Energy, Optical,
  and Infrared Detectors for Astronomy VIII, ed. A.~D. Holland \& J.~Beletic,
  Vol. 10709, International Society for Optics and Photonics (SPIE), 1070921

\bibitem[{Secroun {et~al.}(2016)Secroun, Serra, Cl{\'e}mens, Legras, Lagier,
  Niclas, Caillat, Gillard, Tilquin, Ealet, Barbier, Ferriol, Kubik, Smadja,
  Prieto, Maciaszek, \& Sorensen}]{detectorcharacterization-secroun2016}
Secroun, A., Serra, B., Cl{\'e}mens, J.~C., {et~al.} 2016, in High Energy,
  Optical, and Infrared Detectors for Astronomy VII, ed. A.~D. Holland \&
  J.~Beletic, Vol. 9915, International Society for Optics and Photonics (SPIE),
  99151Y

\bibitem[{Sánchez {et~al.}(2020)Sánchez, Walter, Awan, Chiang, Daniel,
  Gawiser, Glanzman, Kirkby, Mandelbaum, Slosar, Wood-Vasey, AlSayyad, Burke,
  Digel, Jarvis, Johnson, Kelly, Krughoff, Lupton, Marshall, Peterson, Price,
  Sembroski, Van Klaveren, Wiesner, Xin, \&
  Collaboration}]{descDC1-sanchez2020}
Sánchez, J., Walter, C.~W., Awan, H., {et~al.} 2020, \mnras, 497, 210

\bibitem[{{Tallada} {et~al.}(2020){Tallada}, {Carretero}, {Casals},
  {Acosta-Silva}, {Serrano}, {Caubet}, {Castander}, {C{\'e}sar}, {Crocce},
  {Delfino}, {Eriksen}, {Fosalba}, {Gazta{\~n}aga}, {Merino}, {Neissner}, \&
  {Tonello}}]{tallada2020}
{Tallada}, P., {Carretero}, J., {Casals}, J., {et~al.} 2020, Astronomy and
  Computing, 32, 100391

\bibitem[{{The Dark Energy Survey Collaboration}(2005)}]{des_2005}
{The Dark Energy Survey Collaboration}. 2005, arXiv, arXiv:astro-ph/0510346

\bibitem[{{Tully} {et~al.}(1975){Tully}, {de Marseille}, \&
  {Fisher}}]{tullyfisher1975}
{Tully}, R.~B., {de Marseille}, O., \& {Fisher}, J.~R. 1975, in \baas, Vol.~7,
  426

\bibitem[{Tylka {et~al.}(1997)Tylka, Adams, Boberg, Brownstein, Dietrich,
  Flueckiger, Petersen, Shea, Smart, \& Smith}]{creme96-tylka1997}
Tylka, A., Adams, J., Boberg, P., {et~al.} 1997, IEEE Transactions on Nuclear
  Science, 44, 2150

\bibitem[{{van der Kruit} \& {Searle}(1982)}]{vanderkruit1982}
{van der Kruit}, P.~C. \& {Searle}, L. 1982, \aap, 110, 61

\bibitem[{Waczynski {et~al.}(2016)Waczynski, Barbier, Cagiano, Chen, Cheung,
  Cho, Cillis, Cl{\'e}mens, Dawson, Delo, Farris, Feizi, Foltz, Hickey, Holmes,
  Hwang, Israelsson, Jhabvala, Kahle, Kan, Kan, Loose, Lotkin, Miko, Nguyen,
  Piquette, Powers, Pravdo, Runkle, Seiffert, Strada, Tucker, Turck, Wang,
  Weber, \& Williams}]{detectorperformance-waczynski2018}
Waczynski, A., Barbier, R., Cagiano, S., {et~al.} 2016, in High Energy,
  Optical, and Infrared Detectors for Astronomy VII, ed. A.~D. Holland \&
  J.~Beletic, Vol. 9915, International Society for Optics and Photonics (SPIE),
  991511

\bibitem[{{Welikala} \& {Kneib}(2012)}]{welikala2012}
{Welikala}, N. \& {Kneib}, J.-P. 2012, arXiv, 1202.0494

\bibitem[{{Williams} {et~al.}(2019){Williams}, {Begeman}, {Boxhoorn}, {Droge},
  {Tsyganov}, {McFarland}, {Valentijn}, {Vriend}, \&
  {Dabin}}]{eas-williams2019}
{Williams}, O.~R., {Begeman}, K., {Boxhoorn}, D., {et~al.} 2019, in
  Astronomical Society of the Pacific Conference Series, Vol. 521, Astronomical
  Data Analysis Software and Systems XXVI, ed. M.~{Molinaro}, K.~{Shortridge},
  \& F.~{Pasian}, 120

\bibitem[{{Yoachim} {et~al.}(2016){Yoachim}, {Coughlin}, {Angeli}, {Claver},
  {Connolly}, {Cook}, {Daniel}, {Ivezi{\'c}}, {Jones}, {Petry}, {Reuter},
  {Stubbs}, \& {Xin}}]{LSST-skybrightness-yoachim2016}
{Yoachim}, P., {Coughlin}, M., {Angeli}, G.~Z., {et~al.} 2016, in \procspie,
  Vol. 9910, Observatory Operations: Strategies, Processes, and Systems VI,
  99101A

\bibitem[{{Zoubian} {et~al.}(2014){Zoubian}, {K{\"u}mmel}, {Kermiche},
  {Apostolakos}, {Chapon}, {Ealet}, {Franzetti}, {Garilli}, {Jullo}, \&
  {Paioro}}]{tips-zoubian2014}
{Zoubian}, J., {K{\"u}mmel}, M., {Kermiche}, S., {et~al.} 2014, in Astronomical
  Society of the Pacific Conference Series, Vol. 485, Astronomical Data
  Analysis Software and Systems XXIII, ed. N.~{Manset} \& P.~{Forshay}, 509

\end{thebibliography}

\begin{appendix}

\section{Throughputs}
\label{a:throughputs}
Spectra and band fluxes were computed using the total transmission curves estimates for VIS (\IE), NISP-P (\YE, \JE, \HE), NISP-S (Red and Blue Grisms), and each EXT survey (e.g. \textit{Gaia}, DES, LSST, and UNIONs) provided by each of the respective instrument teams. We describe below the generation and/or acquisition of the passbands for each instrument and their status for SC8 where applicable: 

\begin{itemize}
    \item \textbf{VIS} The VIS total transmission was computed using the product of the MDB parameters: the VIS quantum efficiency and the PLM VIS transmission curves provided by the VIS Instrument team and Airbus Space and Defense, respectively.
    \item \textbf{NISP}  We used the current best estimates of the NISP photometric passbands (\YE, \JE, \HE) and the transmission curves for the red and blue grisms produced by the NISP Instrument Development Team and validated and crossed-checked by the SGS.  The datasets used for SC8 are stored in the MDB, and the latest current best estimates are described in \cite{euclid-NISPphotobands-schirmer2022}.
    \item \textbf{Gaia} As stated in Sect. \ref{s:gaia_realisation}, \textit{Gaia} band fluxes were derived from the DR2 release \citep{gaia-dr2-2018}\footnote{\url{https://www.cosmos.esa.int/documents/29201/1645651/GaiaDR2_Revised_Passbands_ZeroPoints.zip/54db454f-69cb-ea0c-15be-f1b1f597f191}}.
    \item \textbf{DES} We used the DR1 griz passbands \citep{DR1-DES2018}\footnote{\url{https://noirlab.edu/science/programs/ctio/filters/Dark-Energy-Camera}}.
    \item \textbf{LSST} 
     The baseline total throughput curves (CCDs+Filter+Optics+Atmosphere) at airmass=1.2 \citep{lsstcam-kahn2010,lsst-syseng-claver2014,opsim-connolly} were computed by the LSST system engineering team. For SC8, we used the earlier release-1.4 total throughputs for $ugriz$\footnote{\url{https://github.com/lsst/throughputs/tree/1.4/baseline}}.
    \item \textbf{CFIS}
    The bandpasses for $u$ and $r$ filters, respectively, were generated with the information for the MegaCam filter set\footnote{\url{https://www.cadc-ccda.hia-iha.nrc-cnrc.gc.ca/en/megapipe/docs/filt.html}}.
    \item \textbf{JEDIS} The JEDIS team provided full efficiency curves (CCD+Filter+Optics+Atmosphere) in the $g$ band for 14 CCDs measured in a grid of $12\times 13$ points, which we used to compute an average throughput.
    \item \textbf{Pan-STARRS PS1} The bandpass model for the $i$ band\footnote{\url{http://ipp.ifa.hawaii.edu/ps1.filters/}}.
    \item \textbf{WISHES} The bandpass data for the $z$ band\footnote{\url{https://hsc-release.mtk.nao.ac.jp/doc/index.php/survey-2/}}.

\end{itemize}

\section{Spectra and fluxes}
\label{a:sed_fluxes}
\subsection{Galaxy spectra reconstruction}
\label{a:sed_fluxes_galaxy}


Here we describe the process for reconstructing the observed spectra of the galaxies from the parametrised information in the catalogue.

We begin by homogenising all wavelength-dependent elements into a common uniform sampling to enable direct operation between models that were sampled at different wavelength ranges and resolutions. To ensure the spectra cover the wavelength range of interest and with enough resolution, we set a range between $600\,\AA$ and $24,000\,\AA$, in steps of $1\,\AA$. With a spectral resolution of $R=480$ in NISP-S, we are able to sample properly the dispersed spectra. 

Our reconstruction algorithm commences with the initialisation of a spectrum. The \texttt{SimSpectra} library allows for the linear combination of two galaxy templates, enabling a wider variety of colours and shapes, which gives more realism to the simulation. We compute the combined template as follows:
\begin{equation}
f_{\rm combined}(\lambda) = (1-i_{\rm frac})f_{\rm template}(\lambda)[i] + i_{\rm frac}f_{\rm template}(\lambda)[i-1]  \, ,
\end{equation}
where $i_{\rm frac}$ is the fractional contribution of template with index $i$ in the combined template. Templates are ordered from early to late type galaxy in the SED library. High-$z$ galaxies and QSOs use only one template. Next, we apply one out of 5 dust extinction laws taken from \cite{cosmosdust-prevot1984,cosmosdust-calzetti2000} and presented in \cite{cosmos_ilbert2009}. The dust extinction curve $k(\lambda)$ is scaled by the galaxy colour excess $E(B-V)$ and applied to the combined template, such that 

\begin{equation}
\label{eq:f_ext}
f_{\rm ext}(\lambda) = f_{\rm combined}(\lambda)\, 10^{-0.4 E(B-V) k(\lambda)} \, ,
\end{equation}

\noindent where the extinction curve $k(\lambda)$ is

\begin{equation}
    k(\lambda) = -2.5 \log_{10}\paren{\frac{E_j(\lambda)}{E_0(\lambda)}}  E(B-V)_{0.2}^{-1}\, .
\end{equation}

\noindent where $E_j(\lambda)$ corresponds to the extinction law used \citep{cosmosdust-prevot1984,cosmosdust-calzetti2000,cosmos_ilbert2009}, with $j=0$ corresponding to no extinction, $j=1$ to the Prevot extinction law, $j=2$ to the Calzetti extinction law with no 2175{\AA} bump and $j=3$ and $j=4$ to the Calzetti extinction law with different strengths of the 2175{\AA} bump (see \citealt{cosmos_ilbert2009} for details). For computational ease, we compute all the extinctions in the filter passbands for a value of $E(B-V) = 0.2$ and then re-scale the value to the actual $E(B-V)$ in Eq.~\ref{eq:f_ext}.  

The next step is to shift the spectrum to the observed redshift (including peculiar velocities), $z_{\rm obs}$. This operation is performed by scaling the wavelength array elements by $(1+z_{\rm obs})$ and dividing the fluxes by the same factor $(1+z_{\rm obs})$. With the redshifted spectrum, we can measure the flux in the reference filter, and scale it to the reference apparent AB magnitude. For standard and strongly lensed galaxies, the reference filter is the SDSS $r$ passband redshifted to $z=0.1$. For high-redshift galaxies and QSOs, the template is scaled according to the apparent magnitude at $1500\,(1+z_{\rm obs})\,\AA$ and $1450\,(1+z_{\rm obs})\,\AA$ respectively. 

For standard and strongly lensed galaxies, the reference magnitude to scale the spectrum is determined from the absolute magnitude in the SDSS $r$ band, $M_r$. The $M_r$ magnitude is a product of the semi-analytic recipe in the simulation, and is given in $h$-normalised units. The conversion from absolute to apparent magnitude requires computing the luminosity distance $D_{\rm L}$ and the $K_{\rm Corr}$ correction. 

We compute the luminosity distance with a Flat LCDM cosmology, setting the $Hubble$ constant to $H_{\rm 0}=100\,{\rm km}\, {\rm s}^{-1}\,{\rm Mpc}^{-1}$, and the non-relativistic matter density parameter to the value used in the Flagship simulation $\Omega_{\rm m}=0.319$.

As our reference flux is not bolometric but computed within a particular passband, we need the $K_{\rm Corr}$ correction term to transform from observed to restframe flux to account for the galaxy redshift. This correction term, expressed in magnitudes, is defined as
\begin{equation}
K_{\rm Corr} = 2.5 \log_{\rm 10} \frac{f_{\rm restframe}}{f_{\rm observed}}\, ;
\end{equation}
\noindent where we obtain both the rest-frame, $f_{\rm restframe}$, and the redshifted $f_{\rm observed}$ fluxes integrated in the reference passband using the following formula
\begin{equation}
\label{eq:fnu_integrated_flux}
f_{\rm \nu} = \frac{\int f_{\rm{\lambda}}\, R(\lambda) \frac{\lambda}{c}\, \rm{d} \lambda}{\int R(\lambda)\, \rm{d} \lambda / \lambda} \, ,
\end{equation}
\noindent where $R(\lambda)$ is the system response of the reference magnitude passband in photon counts, resampled at $1\,\AA$ resolution. With the $K_{\rm Corr}$ correction and the luminosity distance $d_{\rm L}$ in parsec, we compute the apparent magnitude,
\begin{equation}
m_r = M_r + 5 \brackets{\log_{10}(d_{\rm L}) - 1} + K_{\rm Corr} \;.
\end{equation}
We note that the absolute magnitude and the luminosity distance are computed for the same value of the $Hubble$ constant and therefore its dependence can be disregarded in the equation making the apparent magnitude independent of its value.

The scaled observed flux in ${\rm erg}\,{\rm cm}^{-2}\,{\rm s}^{-1}\, {\rm Hz}^{-1}$ at the reference passband from the AB definition is then
\begin{equation}
\label{eq:fnu_mag_to_flux}
f_{{\rm \nu},r} = 10^{-0.4 m_r +48.6} \, ,
\end{equation}
and the scaled continuum spectrum is therefore
\begin{equation}
f_{\rm continuum}(\lambda) = f_{\rm ext}(\lambda) \frac{f_{{\rm \nu},r}}{f_{\rm \nu,observed}} \, .
\end{equation}
At this stage, we add the emission lines to the scaled template. As emission lines are already scaled and modelled with extinction, we compute them after applying the dust to the scaled continuum. The central frequency of each line is taken from vacuum or air, depending on whether the channel we simulate is ground or space-based. In Table~\ref{tab:emission_lines}, we list the set of emission lines and their corresponding central wavelength used for both air and vacuum in the simulation.\footnote{\url{https://physics.nist.gov/PhysRefData/ASD/lines_form.html}}

\begin{table}[!ht]
\centering
\caption{Emission lines used in the simulated spectra of galaxies.}
\begin{tabular}{|l|c|c|c|}
\hline
\textbf{Line} & \textbf{Vacuum (\AA)} & \textbf{Air (\AA)} & \textbf{ratio to primary} \\ \hline
H$\alpha$            & 6564.614                & 6562.801             &                                    \\ \hline
H$\beta$            & 4862.721                & 4861.363             &                                    \\ \hline
\ion{O}{ii}           & 3727.090                & 3726.030             &                                    \\ \hline
\ion{O}{ii}*       & 3729.880                & 3728.820             & 1:1                                \\ \hline
\ion{O}{iii}          & 5008.239                & 5006.843             &                                    \\ \hline
\ion{O}{iii}*      & 4960.295                & 4958.911             & 1:3                                \\ \hline
\ion{N}{ii}           & 6585.270                & 6583.450             &                                    \\ \hline
\ion{N}{ii}*       & 6549.860                & 6548.050             & 1:3                                \\ \hline
\ion{S}{ii}          & 6718.290                & 6716.440             &                                    \\ \hline
\ion{S}{ii}*       & 6732.680                & 6730.820             & 1:1                                \\ \hline
\end{tabular}
\tablefoot{Secondary lines are indicated with * and the corresponding flux ratio to the primary is indicated, as no secondary fluxes are explicitly indicated in the True Universe catalogue.}

\label{tab:emission_lines}
\end{table}

We estimate the width of the emission line, $\sigma_{\rm el}$, in ${\rm km}\,{\rm s}^{-1}$ following a luminosity-velocity dispersion relation (e.g. Faber-Jackson \citealt{velocityrelation-faberjackson1976} and Tully-Fisher relation \citealt{tullyfisher1975}) as shown below:
\begin{equation}
\log_{10}(\sigma_{\rm el}) = (-0.10 + 0.01 z_{\rm obs}) (M_r - 3.0) - 0.05 z_{\rm obs} \, .
\end{equation}
\noindent High-redshift galaxies and QSO templates already include emission lines. We limit the approximation to a minimum velocity of $50\,{\rm km}\,{\rm s}^{-1}$ to avoid too narrow lines. We generate all 10 emission lines as Gaussian profiles with the integrated flux as defined in the catalogue at the corresponding central wavelength with a width equal to $\sigma_{\rm el}$. As was done for continuum spectra, the set of reconstructed emission lines are shifted in wavelength by the factor $(1+z_{\rm obs})$. 

The velocity dispersion is converted to wavelength and scaled to the observed frame as
\begin{equation}
\sigma_{\rm el,\lambda}  = \sigma_{\rm el} \frac{\mu_i}{c} (1+z_{\rm obs}) \, .
\end{equation}

Then, we compute the spectrum containing only the emission line contribution, $f_{\rm el}(\lambda)$, as
\begin{equation}
f_{\rm el}(\lambda) = \sum_i^{10} \frac{[f_{\rm el}]_i}{\sigma_{\rm el,\lambda} \sqrt{2 \pi}} \exp{\left(  -\frac{(\lambda-\mu_i)^2}{2\sigma_{\rm el,\lambda}^2} \frac{1}{1+z_{\rm obs}} \right)} \, ,
\end{equation}

where $\mu_i$ and $[f_{\rm el}]_i$ are the central wavelengths and TU line fluxes, respectively, for the ten emission lines in the corresponding medium. 

The resulting line profiles can now be added to the calibrated continuum component of the spectra with the following expression
\begin{equation}
f_{\rm galaxy}(\lambda) = f_{\rm continuum}(\lambda) + f_{\rm el}(\lambda) \, .
\end{equation}
With all the elements from the source galaxy, we introduce the effect of lensing magnification with
\begin{equation}
f_{\rm magnified}(\lambda) = f_{\rm galaxy}(\lambda)\, \mu \, ,
\end{equation}
where $\mu$ is the magnification factor as defined in Eq. \ref{eq:magnification_factor}.

The final remaining step is to apply the Milky Way extinction to the magnified galaxy spectra at the specific location in the sky. From the {\it Planck} dust maps \citep{dust-planck2014}, we obtain a precise colour excess $E(B-V)$, and following an \cite{extinction-odonnell1994} extinction law with the ratio of absolute to differential extinction  $R_V=3.1$, we apply the MW extinction to obtain the final incident spectra:
\begin{equation}
\label{eq:mw_extinction}
f_{\rm incident}(\lambda) = f_{\rm magnified}(\lambda)\, 10^{-0.4\, A(\lambda)} \, ,
\end{equation}
where the extinction law $A(\lambda)$ is the product of the extinction terms $A(\lambda) = E(B-V) R_V$.

\subsection{Stellar spectra reconstruction}
\label{a:sed_fluxes_stars}

With the already available apparent magnitude in the $H$ band we can simply compute the template flux $f_{\rm \nu,template}$ in the observed band as defined in Eq. \ref{eq:fnu_integrated_flux} and the 2MASS $H$-band flux $f_{{\rm \nu},H}$ as Eq. \ref{eq:fnu_mag_to_flux} and scale the star spectrum:
\begin{equation}
f_{\rm star}(\lambda) = f_{\rm template}(\lambda) \frac{f_{{\rm \nu},H}}{f_{\rm \nu,template}} \, .
\end{equation}
The only remaining step is to correct the Galactic extinction. The Tycho2 stars are not extinction corrected and therefore we do not apply further correction. For the additional stars with BaSeL and LT spectra, we apply a procedure similarly to the galaxy spectra as
\begin{equation}
f_{\rm star}(\lambda)_{\rm ext} =  f_{\rm star}(\lambda)\, 10^{-0.4\,A(\lambda)} \ ,
\end{equation}
with the \cite{extinction-odonnell1994} law, $R_{\mathrm V} = 3.1$, and the absorption coefficient $A_{\mathrm V}$  determined assuming the \cite{drimmel2003} 3D Galactic extinction model.

Even though the wavelength range of the stellar templates cover the whole spectrum of interest, the reddest part from $12\,000\,\AA$ onwards has a reduced sampling in the Basel 2.2 library with $50\, \AA$ intervals. The wavelength calibration in NISP-S has to be accurate to better than half a pixel, which translates to approximately $6\, \AA$. Therefore, we would need stellar spectra sampled at $2\,\AA$ to $3\,\AA$ intervals. For this reason we replaced the feature-less low sampled redder part of the spectrum with real high resolution measurements following the continuum as defined in the template. 

With support from OU-SIR, we did a fit for the spectral continuum for each BaSEL spectrum, using functions provided by our redshift and spectra measurements package EZ,\footnote{\url{http://pandora.lambrate.inaf.it/EZ/}} resampled the smooth continuum to a higher resolution, and added the continuum-subtracted spectrum of a star taken from the NASA Infrared Telescope Facility IRTF Spectral Library \footnote{\url{http://irtfweb.ifa.hawaii.edu/~spex/IRTF_Spectral_Library/}} to the high-resolution continuum.

From the full IRTF library we have taken ten stars at random for each spectral class (F, G, K, M - earlier classes O, B, and A are not covered by the library), primarily of luminosity class III (giants stars) and V (main sequence stars), but with at least one star of class either I (supergiants) or IV (subgiants).

The choice of which IRTF continumm-subtracted spectrum to add to the BaSEL continuum-only spectrum was based on broad-band colour: the IRTF template with the $Y-H$ colour closer to the one stored in the original catalog was chosen as the match. There was no attempt to include also the luminosity or other properties in the match. So a very red star was assigned an M type spectrum, without considering if the star luminosity was compatible with an M-type object or not.

In this way we do not alter the colours of the spectra but we allow accurate wavelength calibrations in the simulated NISP-S channel.

\subsection{Synthetic photometry}
\label{a:synthetic_photometry}

We need to generate synthetic photometry from the high resolution spectra to reproduce the passband fluxes that each instrument will measure. This information is provided in the catalogue for the EXT surveys simulations, and for validation purpose only for the \Euclid instruments simulators, because these later make use of the true spectra to take into account all chromatic instrumental models. 

Starting from the high resolution spectra, we compute the integrated flux density of the source $f_\nu$ in ${\rm erg}\,{\rm cm}^{-2}\,{\rm s}^{-1}\,{\rm Hz}^{-1}$ for a specific passband $R(\lambda)$  using Eq.~\eqref{eq:fnu_integrated_flux}. Then, the synthetic magnitudes in the AB system are obtained with the following transformation:
\begin{equation}
m_{\rm syn} = -2.5 \log_{10}(f_{\nu})  -48.6 \,.
\end{equation}

\section{SC8 preparation}
\label{a:sc8_preparation}

\subsection{Input catalogues}

\begin{figure}
  \centering
 \includegraphics[width=\linewidth]{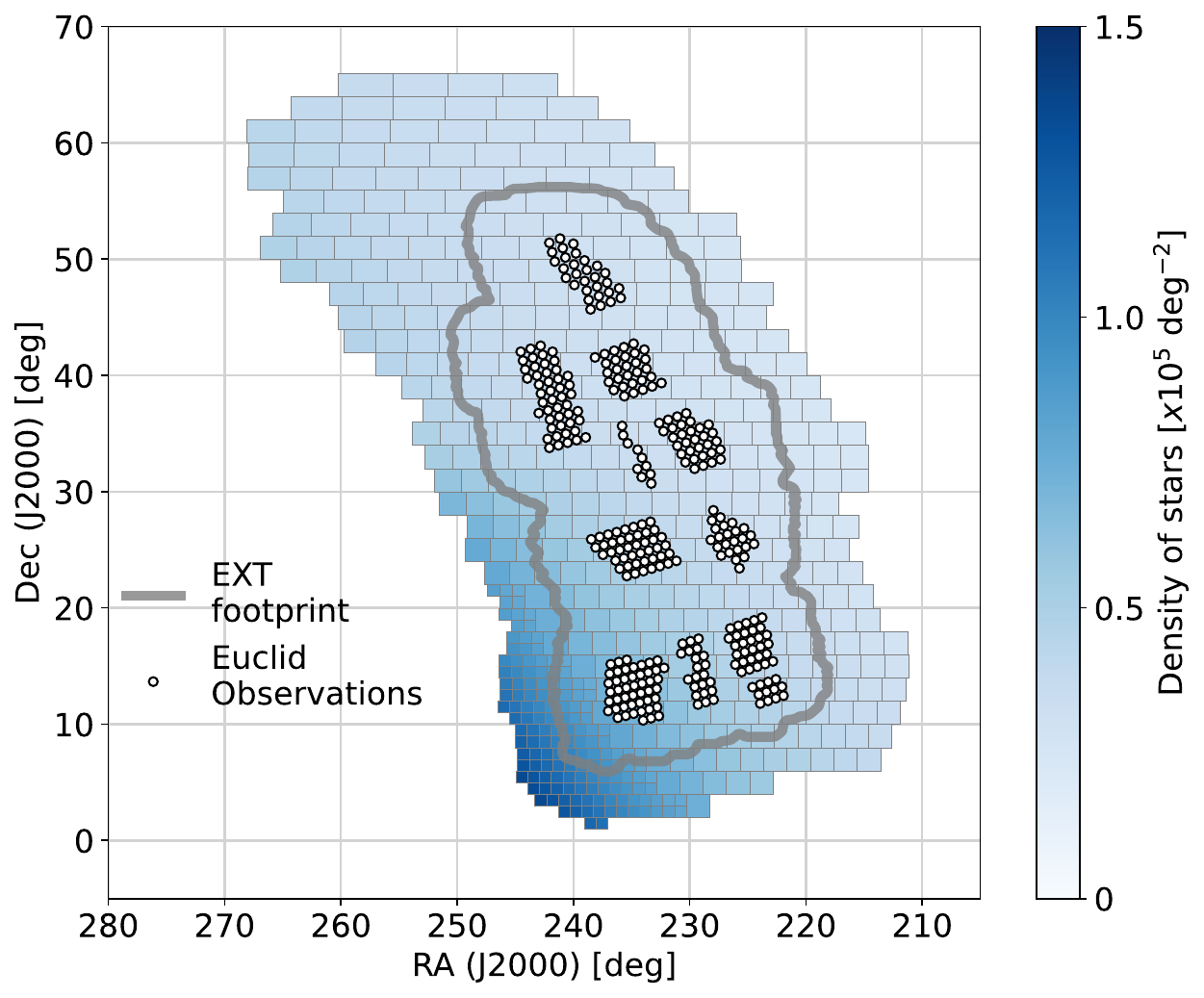}
  \caption{Density of objects (number of stars per ${\rm deg}^2$) in each standard star catalogues file covering SC8 main production area, superimposed with the \Euclid observations and simulated EXT survey area.}
  \label{fig:star_catalogues_sc8}
\end{figure}

\begin{figure}
  \centering
 \includegraphics[width=\linewidth]{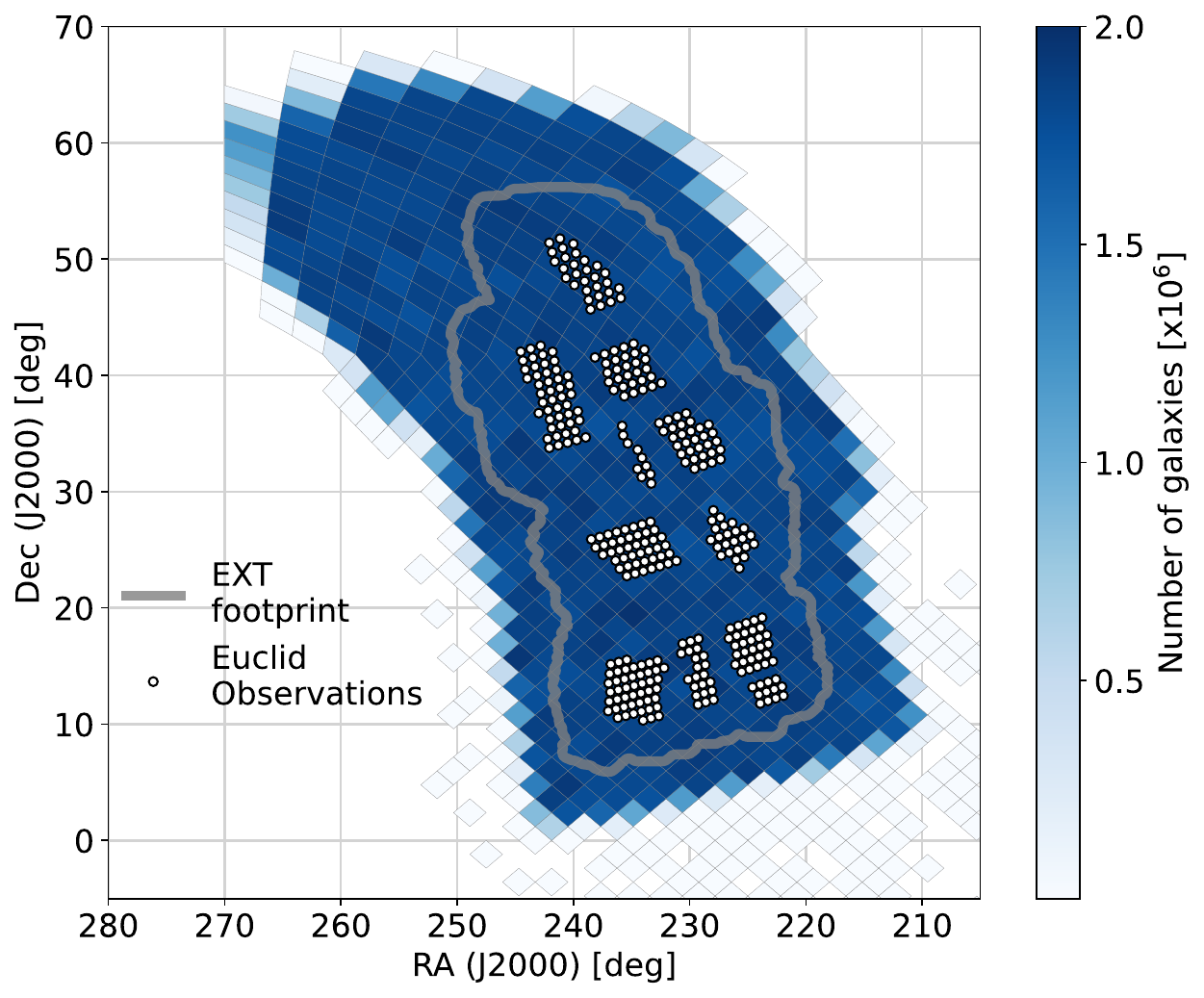}
  \caption{Number of objects in each standard galaxy catalogues file covering SC8 Main production area, superimposed with the \Euclid observations and EXT simulated area.}
  \label{fig:gal_catalogues_sc8}
\end{figure}

 Galaxy and stellar catalogues containing the variety of sources described in Sect.~\ref{s:true_universe} were produced for both the SC8 main (wide) and deep fields and ingested into the EAS.  In particular, for the SC8 main run, the spatial extent of the galaxy and stellar catalogues exceeds ${\sim}\,600\,{\rm deg}^2$, the original wide area survey planned for SC8. We added an additional margin to ensure complete coverage for the EXT ground-based survey simulations. Figure~\ref{fig:gal_catalogues_sc8} shows the footprint of the standard galaxy catalogues and we refer the reader to Figure~\ref{fig:star_catalogues_sc8} in Sect.~\ref{s:euclid_stellar_cat} which shows the footprint of the star catalogues used for the SC8 main run.

Standard galaxies, high-redshift galaxies, and QSOs are directly stored in CosmoHub\footnote{\url{https://cosmohub.pic.es/}} \citep{tallada2020} just after being produced by \texttt{SciPIC}. Taking advantage of the interactive data exploration and fast distribution capabilities of SciPIC, a dedicated team has validated the catalogues. These three catalogues, plus the one containing the strongly lensed galaxies, are then stored as distinct FITS files in a format defined in the \Euclid data model. The catalogues are further divided into different files covering non-overlapping sky areas for each type of object. Each catalogue file covers a \texttt{HEALPix} pixel of $N_{\rm SIDE}=2^{5}$ (corresponding to an area of about $3.35\,{\rm deg}^2$), except for strong lensing galaxies which are all included in a single file. All the FITS files along with their associated metadata are subsequently ingested in the EAS.

Stellar catalogues were stored on a set of non-overlapping FITS files for each kind of object, covering the original SC8 footprint with an area ${\sim}\,600\,{\rm deg}^2$. The catalogues are arranged on a grid with different cell sizes depending on the stellar density of the area. As shown in Fig.~\ref{fig:star_catalogues_sc8}, there is an inverse relation between the area covered by a catalogue file and the density of objects.

In Table~\ref{table:catalogues_summary}, we report the number of objects per galaxy/star type for both wide and deep areas. For the SC8 main run, we produced a total of > 1 billion objects of which 922M were standard Flagship galaxies (described in Sect.~\ref{s:euclid_galaxy_cat}) . The catalogues are divided among 3078 files for a total 254 GB of disc space.

\begin{table}[!htbp]
\caption{Galaxy and stellar catalogues used in SC8.}
\begin{tabular}{|p{0.7cm}|p{1.3cm}|p{1.4cm}|p{1cm}|p{0.7cm}|p{0.7cm}|} 
\hline
 Area & Object Type & Release & Num obj & Size [GB] & Num files \\
 \hline
 \multirow{6}{2em}{SC8 Main} & Std gals & 1.10.18 & 922M & 224 & 747 \\ 
 & QSOs & 1.10.15 & 2M & 0.5 & 562 \\ 
 & High-$z$ & 1.10.19 & 11M & 2.5 & 562 \\  
 & SL gals & SL\_SC8 & 801 & 0.1 & 1 \\ 
 & Std stars & v15 & 77M & 15 & 532 \\  
 & LT stars &  v4\_MLT & 1.8M & 0.4 & 532 \\  \hline
 \multirow{5}{2em}{SC8 Deep} & Std gals & 1.10.22 & 43M & 11 & 50 \\ 
 & QSOs & 1.10.20 & 91K & 0.02 & 36 \\  
 & High-$z$ & 1.10.21 & 477K & 0.1 & 36 \\ 
 & SL gals & SL\_DEEP & 201 & 0 & 1 \\
 & Std stars & v16 & 1.7M & 0.5 & 10 \\  
 & LT stars & v6\_MLT & 150K & 0.03 & 10 \\  \hline
\end{tabular}
\label{table:catalogues_summary}
\end{table}

\subsection{Feasibility analysis and validation}

During the pilot phase, we ran profiling tests of each simulator code and instrument channel for the six EXT surveys. Scaling up the total processing time of the pilot phase to the SC8 main run, we estimated a total simulation time in excess of four million CPU hours to generate full-featured simulations over the SC8 main run target area.  We assessed the cost of each feature identifying those effects for which the computational expense was high, and yet, had minimal impact on image quality and mission performance.  This would enable us to introduce simplifications to carry out the main run production within the appropriated computing budget. Finding the balance between feature complexity and resource usage required several iterations, which thus prolonged the time and effort to validate each resulting product.  
The validation of the pilot phase data products and specific features contained therein for each simulation channel (i.e. VIS, NISP-P, NISP-S, and EXT) was performed by different teams using independent approaches, which we describe further below.

For VIS, short-loop validation of the simulated pilot phase data products were performed by the level-2 OUs and reported any issues identified.  We, in turn, worked to resolve the errors reported, delivering new VIS images to be analysed until no further problems were raised.

For the NISP-P channel, validation of the simulation products was performed by the NISP-P validators in SDC Finland. We performed a separate evaluation of the technical requirements during the pilot phase, which revealed the high computing cost for the NISP-P simulator \texttt{Imagem} would exceed the CPU time constraints for the SC8 main run.  Therefore, we decided to reduce the intra-pixel sampling to speed up the processing by a factor of 4. We verified the astrometric accuracy of all sources to be generated below one milli-arcsecond RMS for both the relative (between images) and absolute astrometry (with respect to the TU catalogue). This ensured that a lower resolution intra-pixel response simulation would not affect the positioning accuracy. The photometry error was less than 0.02 mag with respect to the TU reference magnitude in both aperture and PSF photometry, which also suggested a minimal dependence to lower intra-pixel sampling.

In our validation of the spectroscopic channel for NISP, we verified the $\chi^2$ distribution quality map was consistent with the expected performance of the H2RG detector. Our validation tests confirmed the extracted flux for stellar sources reproduced sufficiently accurately the expected TU flux, with occasional outliers overestimating the flux due to contamination from other sources. This was expected as the validation analysis was not optimised to perform accurate decontamination of nearby spectra.
The polychromatic PSF was simplified by implementing a stacked monochromatic version; improving the computational time by a factor of 7. 

We validated the \texttt{SIM-EXT} science data products for astrometry, photometry, background and PSFs for each EXT survey. We used the \texttt{SourceExtractor} software to identify sources and measure fluxes and compared the derived positions and fluxes with the reference TU catalogue values. The results of our validation tests showed that the generated EXT data products were accurate and consistent with the reference input values.\footnote{It was later discovered during the analysis of the SC8 main data, that the input sky background values received for the Northern Surveys were incorrect, leading to shallower depths than expected. The following simulation production cycle, which came after SC8, included the appropriate sky brightness values.}
Finally, to help reduce computational costs, we decided to simplify the galaxy simulation across all simulation channels; rendering very small galaxies below $\ang{;;0.01}$ with a simpler exponential profile, instead of the more complex inclined S\'ersic profile.

\subsection{Pilot phase processing}
\label{a:pilot_phase_processing}

The main objective of the pilot phase was to profile and validate the OU-SIM pipeline in preparation for the SC8 main run. Extrapolations of the pilot phase computing costs revealed the resources available for SC8 would not be sufficient to generate full-featured simulations for a Euclid Wide Survey of $600\,\rm{deg}^2$ within the appropriated time frame. While strategies were being implemented to lower computing costs such as the reduction to a smaller area of 165$\,\rm{deg}^2$ and the simplification of certain features for the NISP instruments (e.g. intrapixel sampling and monochromatic stacked PSFs), as many as 19 new features and data products (e.g. a cosmic ray map, a ghost map, stars-only image, etc.) were requested to be simulated for VIS to test and validate the OU-VIS processing and analysis software. Thus, while the pilot phase area only included four \Euclid observations overlapping one \Euclid\/ tile and a small sample of EXT pointings per band, multiple productions were launched with each new release due to new and/or modified features. 
In Table~\ref{table:pilot_production}, we list all products produced for a given instrument channel during the pilot phase, noting the number of images per production, number of production sequences and the final number of products and storage requirements.  For the \Euclid instruments, as each observation has 4 dithers a total of 16 images were to be simulated for VIS, 48 images for NISP-P, and 16 images for NISP-S for the pilot phase production.  However, we performed 12, 14, and 6 productions, respectively for VIS, NISP-P, and NISP-S channels yielding 192, 672, and 96 data products.  As seen in Table~\ref{table:pilot_production}, multiple productions were also carried out for each EXT survey.  
A total of 3,614 exposures were produced during the pilot phase. Furthermore, for $Rubin$ and the Northern Surveys, each simulated detector frame comprising an image was saved as an individual file. Thus, the total number of data products ingested into the EAS was 64,679 amounting to 7.1 TB.

\begin{table*}[t]
 \caption{Data products produced for SC8 pilot phase field production.}
\begin{tabular}{|l|l|c|c|c|c|c|c|}
\hline
Channel &     Image type &  \makecell{Original number \\ of images} & \makecell{Number of \\ productions} & \makecell{Final number \\ of images} & \makecell{Number of \\ products \\ per image} & \makecell{Number of \\ products} &  \makecell{Storage \\(GB)}\\
\hline
    VIS & Science Wide &    16 & 12 & 192 & 1 &     192 &   1294 \\
    VIS &         Bias &     60 & 1 & 60 & 1 &     60 &    404 \\
    VIS &         Flat &     60 & 1 & 60 &  1 &    60 &    404 \\
    NISP-P & Science Wide &  48 & 14 & 672 & 1 &    672 &    263 \\
    NISP-P &         Dark &    600 & 1 & 600 & 1 &    600 &    234 \\
    NISP-P &         Flat &    750 & 1 & 750 &  1 &   750 &    293 \\
    NISP-P &     Self Cal &    180 & 1 & 180 & 1 &    180 &     70 \\
    NISP-P &    Std Stars &    24 & 1 & 24 & 1 &     24 &     9 \\
    NISP-S & Science Wide &    16 & 6 & 96 & 1 &     96 &    375 \\
    NISP-S &    Std Stars &     8 & 1 & 8 & 1 &      8 &     32 \\
    NISP-S &           PN &     16 & 1 & 16 & 1 &     16 &     62 \\
    DES & Science Wide &    76 & 5 & 380 & 1 &    380 &    408 \\
    DES &         Flat &     40 & 1 & 40 & 1 &     40 &     43 \\
    DES &         Bias &     10 & 1 & 10 & 1 &     10 &     11 \\
   LSST & Science Wide &  65 & 4 &  260 & 189 &  49\,140 &   3004 \\
   CFIS & Science Wide &    14 & 4 & 56 & 40 &   2240 &     30 \\
  JEDIS & Science Wide &    23 & 3 & 69 & 14 &    966 &    138 \\
    Pan-STARRS & Science Wide &  40 & 3 &  120 & 60 &   7200 &    221 \\
 WISHES & Science Wide &     7 & 3 & 21 &  103 &  2163 &     23 \\
 \hline
 \end{tabular}
 \tablefoot{The calibration frames were reused from previous challenges.}
\label{table:pilot_production}
\end{table*}

For each production sequence, we characterised the resources needed for each processing element (i.e. memory, CPU, processing time, and storage).  This was to assess whether the resources originally requested to the HTC systems at each SDC for each payload job would suffice for the SC8 Main run.  In Table \ref{table:observed_resources_1}, we report the final profiling results of the main processing elements of the SIM pipeline described in Sect.~\ref{s:euclid_instrument_simulators} for each instrument.
\begin{table}[!ht]
\centering
\caption{Detector frame simulation resources.}
\begin{tabular}{|l|c c|c c|c c|} 
 \hline
\multirow{2}{*}{job} & \multicolumn{2}{|c|}{\makecell{wall time \\ (h:mm)}} & \multicolumn{2}{|c|}{\makecell{RSS \\ (GB)}} & \multicolumn{2}{|c|}{\makecell{VMEM\\(GB)}} \\
 & avg & max & avg & max & avg & max \\ 
 \hline
 \\[-0.9em]
VIS detector & 2:37 & 4:49 & 2.8 & 5.9 & 11.2 & 14.6 \\
NISP-P detector & 0:34 & 1:14 & 4.3 & 5.9 & 4.8 & 6.6 \\
NISP-S detector & 5:21 & 8:40 & 1.1 & 7.2 & 1.3 & 13.5 \\
SIM\_EXT detector & & & & & &  \\
\hspace{3mm}DES & 0:06 & 0:16 & 0.5 & 1.0 & 0.8 & 1.4 \\
\hspace{3mm} LSST & 0:34 & 1:50 & 0.3 & 3.8 & 0.5 & 6.5 \\
\hspace{3mm}CFIS & 0:03 & 0:15 & 0.6 & 1.6 & 1.1 & 2.2 \\
\hspace{3mm}JEDIS & 0:16 & 0:49 & 1.2 & 1.8 & 1.5 & 2.2 \\
\hspace{3mm}Pan-STARRS & 0:08 & 1:18 & 0.6 & 1.4 & 0.9 & 2.0 \\
\hspace{3mm}WISHES & 0:03 & 0:12 & 0.7 & 1.9 & 1.2 & 2.6 \\
NS Afterburner & & & & & & \\
\hspace{3mm}CFIS & 0:01 & 0:01 & 0.6 & 0.9 & 1.9 & 3.8 \\
\hspace{3mm}JEDIS & 0:09 & 0:23 & 4.9 & 6.9 & 5.8 & 7.4 \\
\hspace{3mm}Pan-STARRS & 0:05 & 0:58 & 1.4 & 2.0 & 2.1 & 5.3\\
\hspace{3mm}WISHES & 0:00 & 0:02 & 0.6 & 0.8 & 1.2 & 4.8 \\
\hline
\end{tabular}
 \tablefoot{Resources required to simulate a given detector frame using the corresponding simulator, observed during SC8 pilot phase: Wall time, Resident Set Size (RSS), and Virtual Memory (VMEM).}

\label{table:observed_resources_1}
\end{table}

We used the upper-bounds reported to issue a request for resources to the HTC system, which had to be sufficiently higher than the observed maximum but somewhat conservative to optimise SDC performance and capacity.  Table~\ref{table:observed_resources_2} lists the resource characterisations for each software release and the configuration of each processing element used to produce image simulations for the SC8 main run. 
 
\begin{table*}[!ht]
\centering
\caption{Program and version used for each of the main jobs in the SIM pipeline and resources requested to run on an HTC system}
\begin{tabular}{|l|l|l|c c c|} 
 \hline
 job & project & version & n cpu & RSS (GB) & wall time (h) \\ 
 \hline
VIS detector & ELViS & 1.20.3 & 1 & 8 & 6 \\
NISP-P detector & SIM\_\texttt{Imagem} & 1.9.11 & 1 & 8 & 10 \\
NISP-S detector & SIM\_TIPS\_Simulator & 5.4.4 & 1 & 4 & 50 \\
SIM\_EXT detector & SIM\_EXT & 2.14.2 & 1 & 4 & 6 \\
NS Afterburner & EXT\_NS\_AFTERBURNER & 2.0.0 & 1 & 8 & 2 \\
\hline
\end{tabular}
\label{table:observed_resources_2}
\end{table*}

\subsection{SC8 main survey files}
\label{a:sc8_main_survey_files}

\begin{figure}
  \centering
 \includegraphics[width=\linewidth]{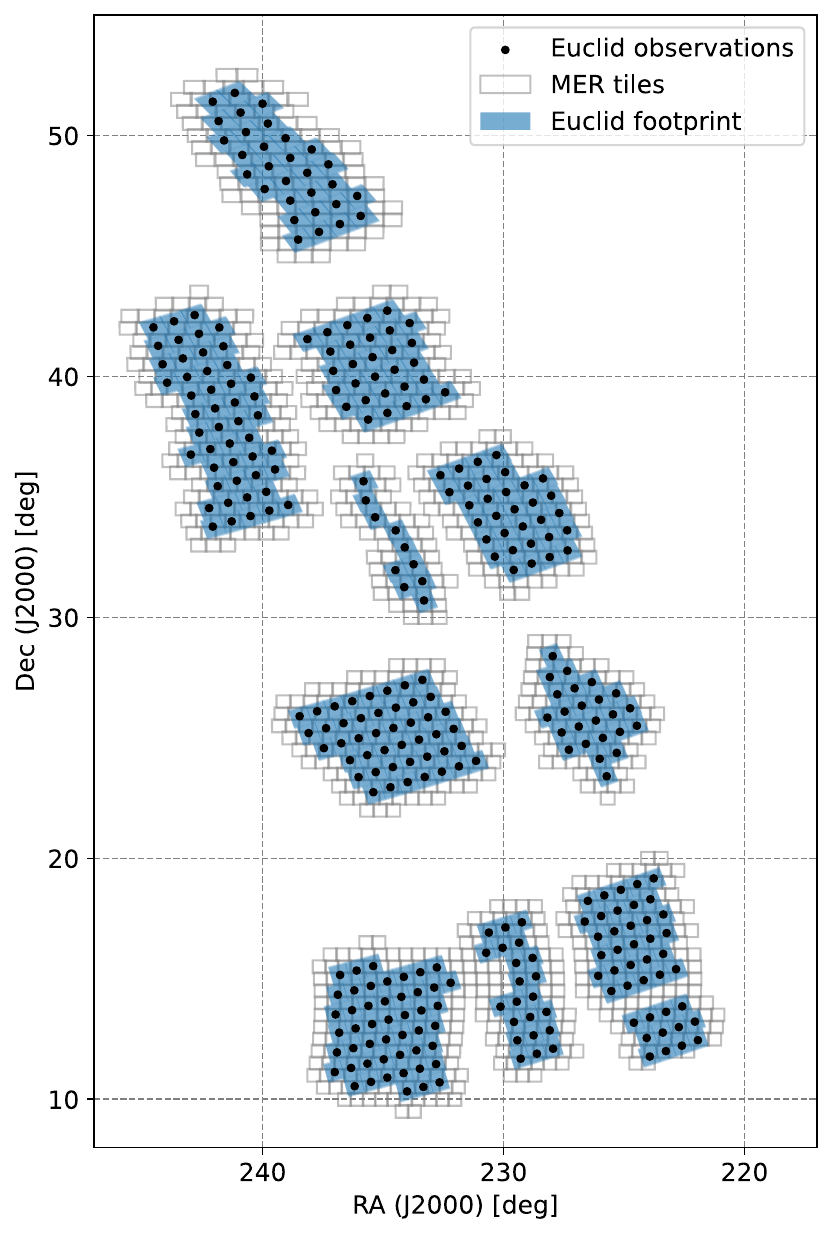}
  \caption{Footprint of the \Euclid observations selected for main SC8 production.}
  \label{fig:euclid_footprint}
\end{figure}

\begin{figure}
  \centering
 \includegraphics[width=\linewidth]{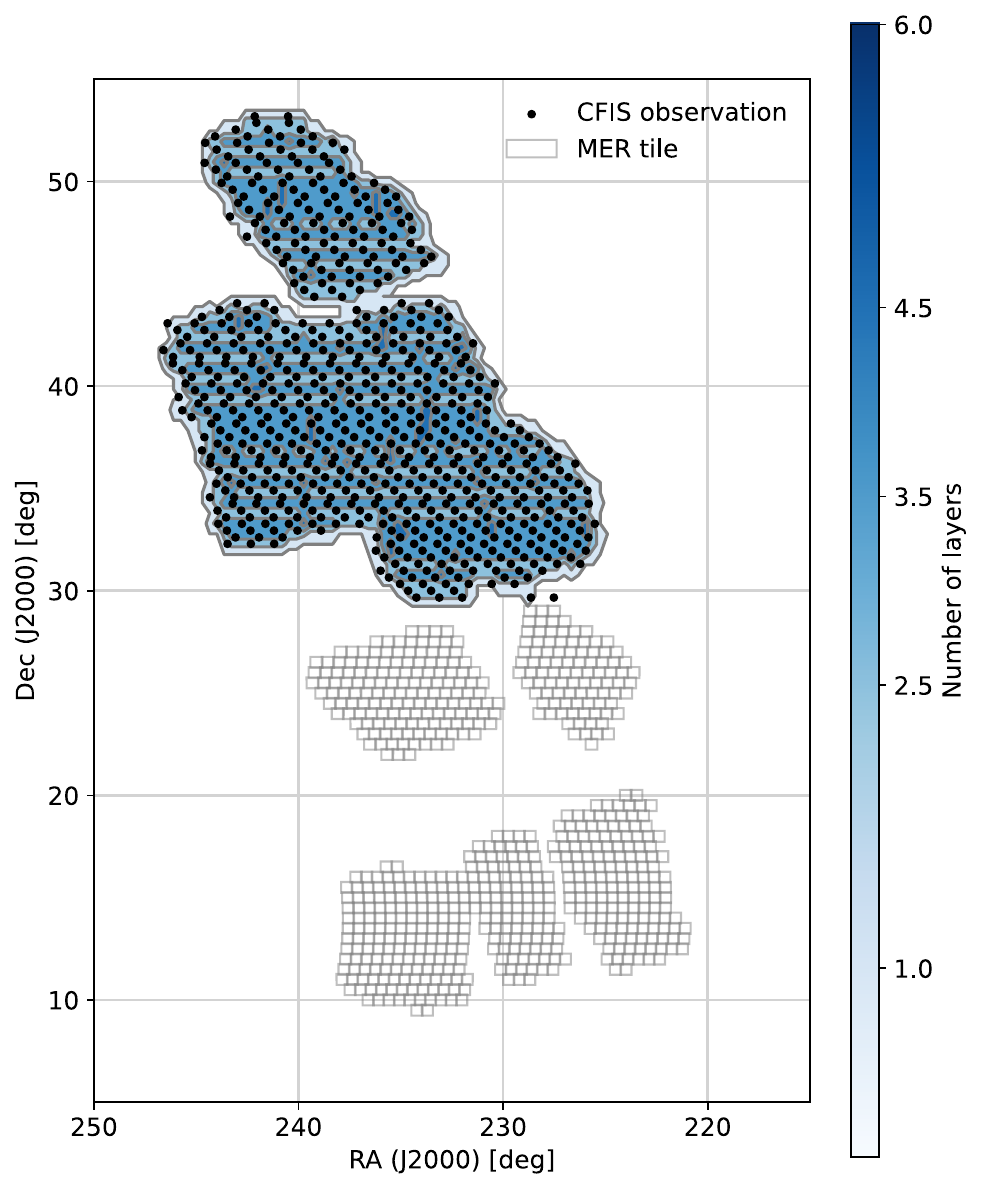}
  \caption{Positions of the exposures and the number of layers for the CFIS $u$ and $r$ bands.}
  \label{fig:cfis_footprint}
\end{figure}

\begin{figure}
  \centering
 \includegraphics[width=\linewidth]{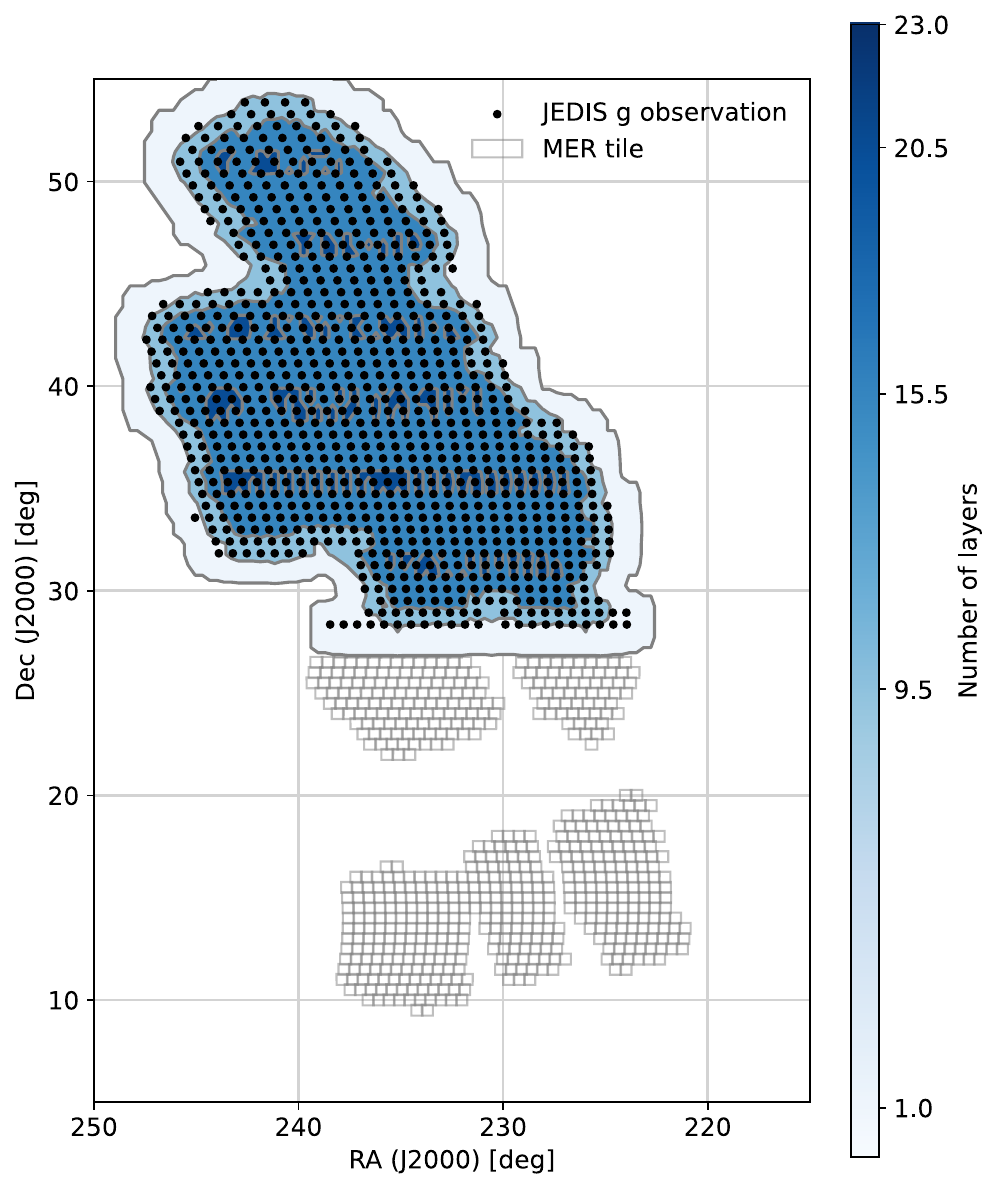}
  \caption{Positions of the exposures and the number of layers for the JEDIS $g$ band.}
  \label{fig:jedis_footprint}
\end{figure}

\begin{figure}
  \centering
 \includegraphics[width=\linewidth]{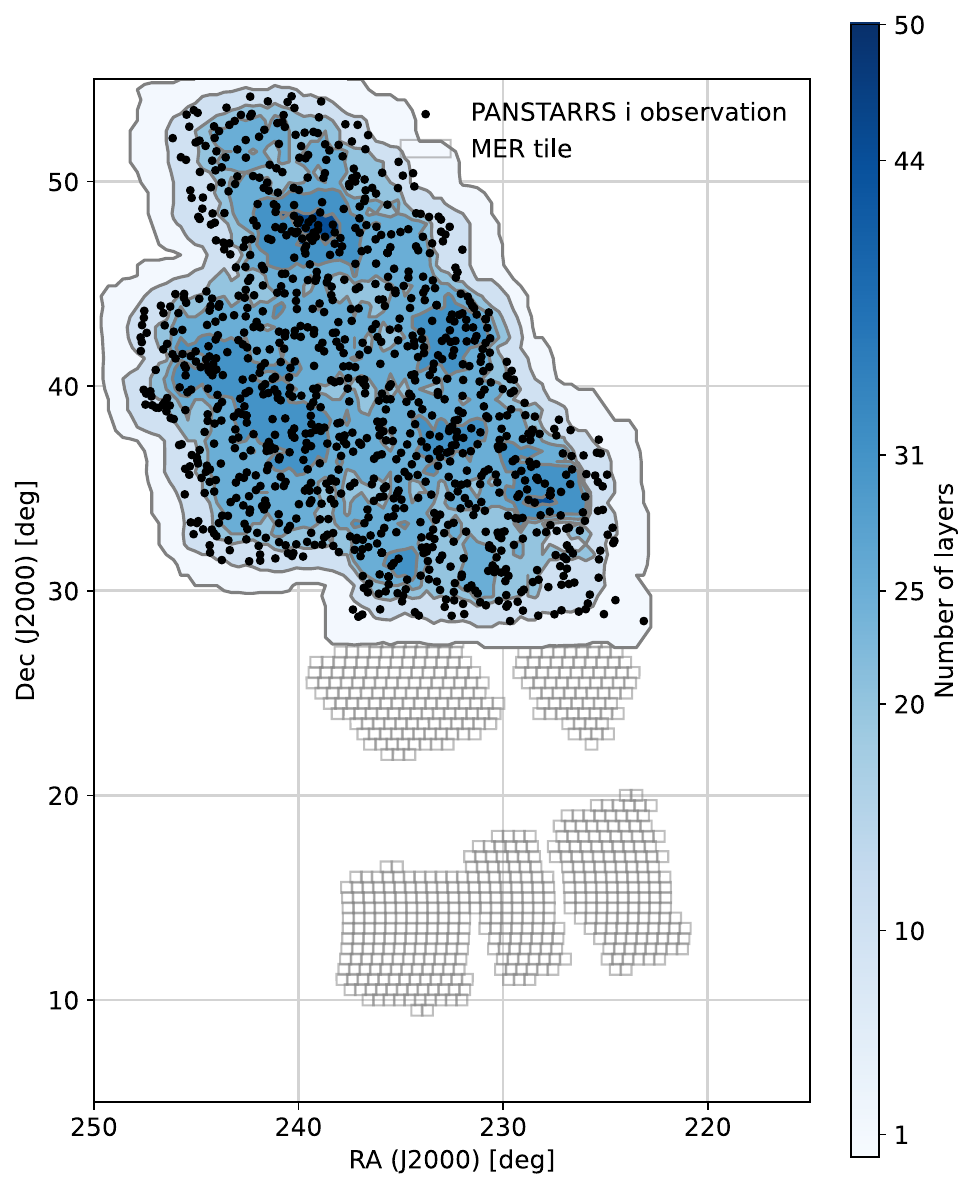}
  \caption{Positions of the exposures and the number of layers for the Pan-STARRS $i$ band.}
  \label{fig:ps_footprint}
\end{figure}

\begin{figure}
  \centering
 \includegraphics[width=\linewidth]{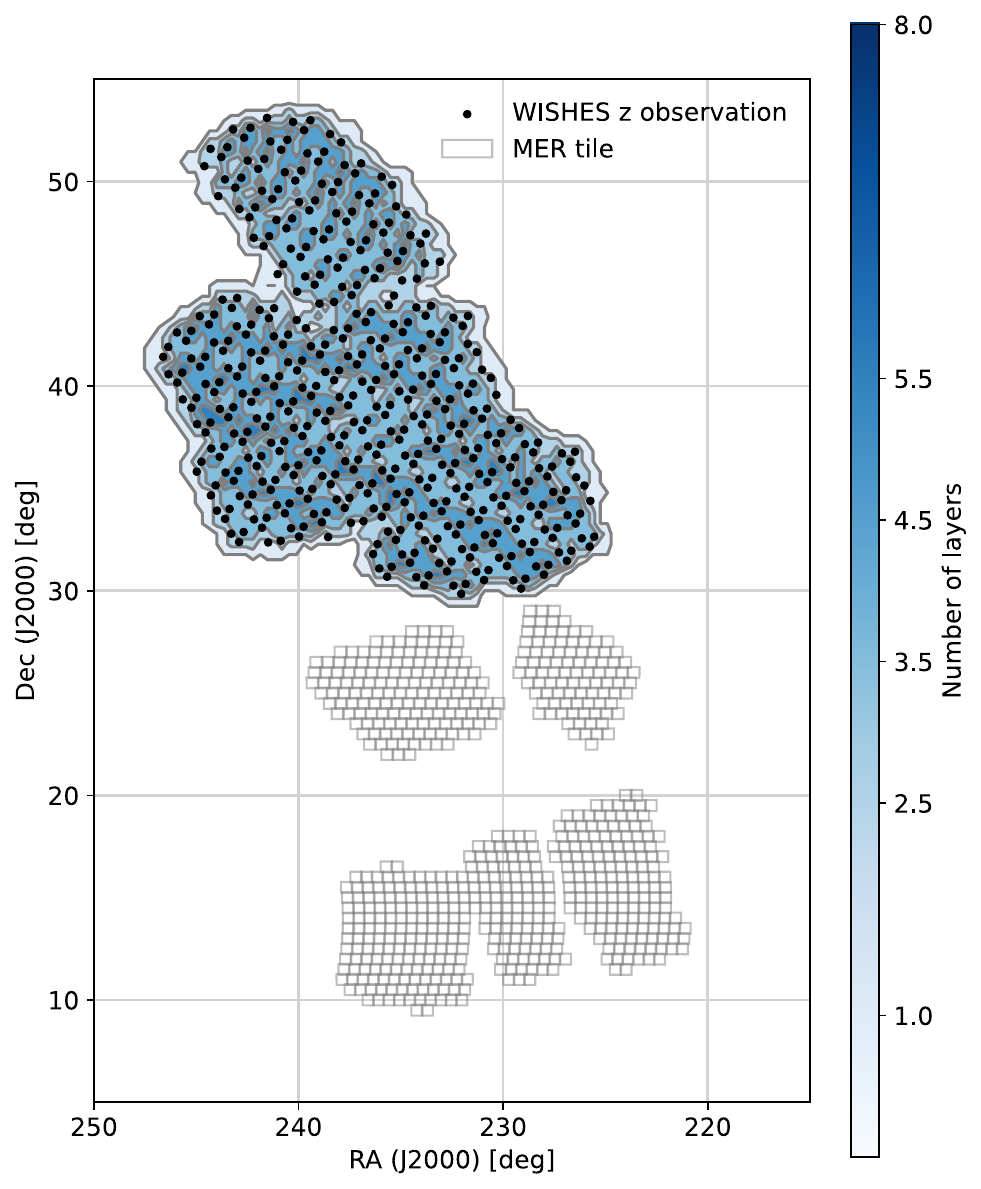}
  \caption{Positions of the exposures and the number of layers for the WISHES $z$ band.}
  \label{fig:wishes_footprint}
\end{figure}

\begin{figure}
  \centering
 \includegraphics[width=\linewidth]{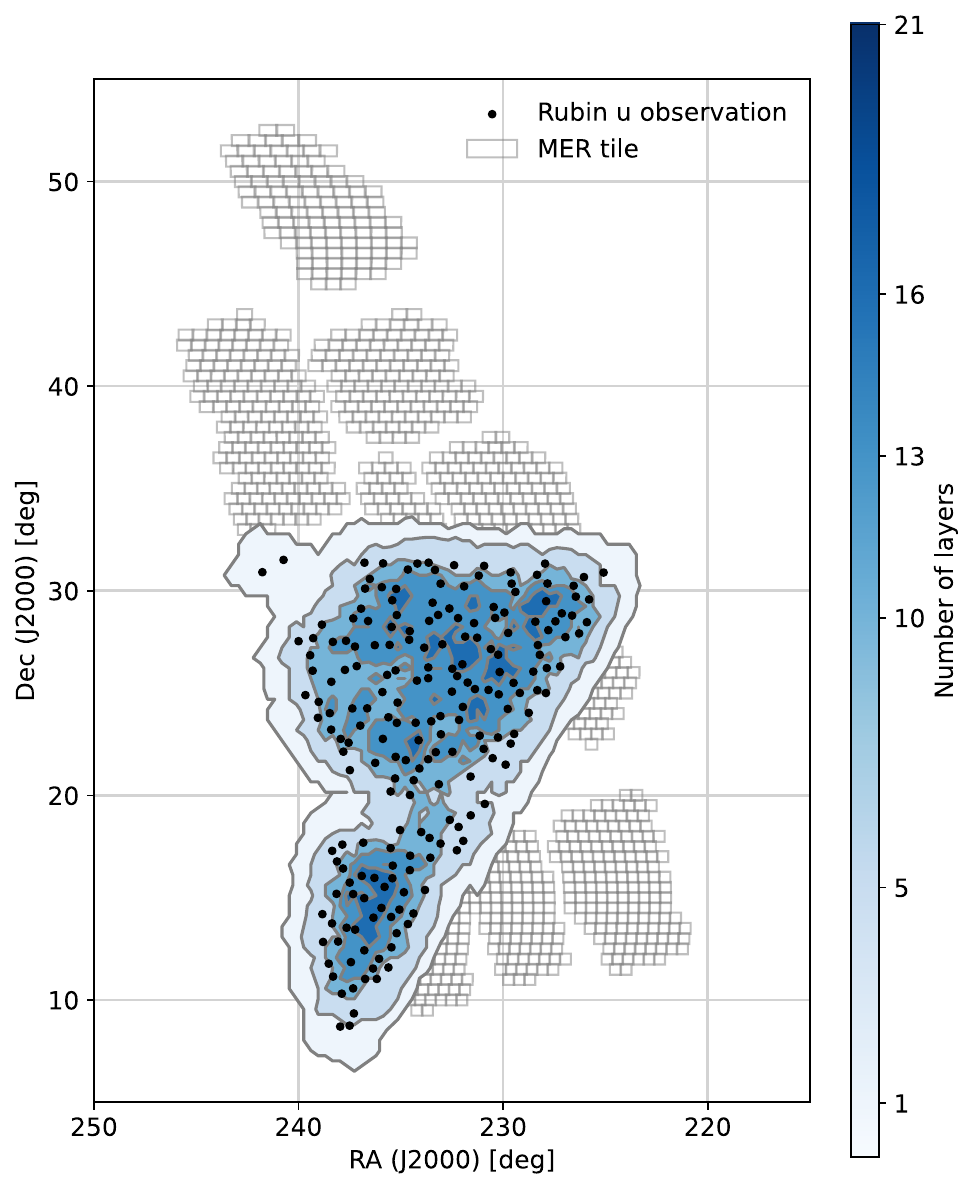}
  \caption{Positions of the exposures and the number of layers for the LSST $u$ band. The number of exposures and layers for the $u$ and $z$ bands are roughly equal.}
  \label{fig:rubin_u_footprint}
\end{figure}

\begin{figure}
  \centering
 \includegraphics[width=\linewidth]{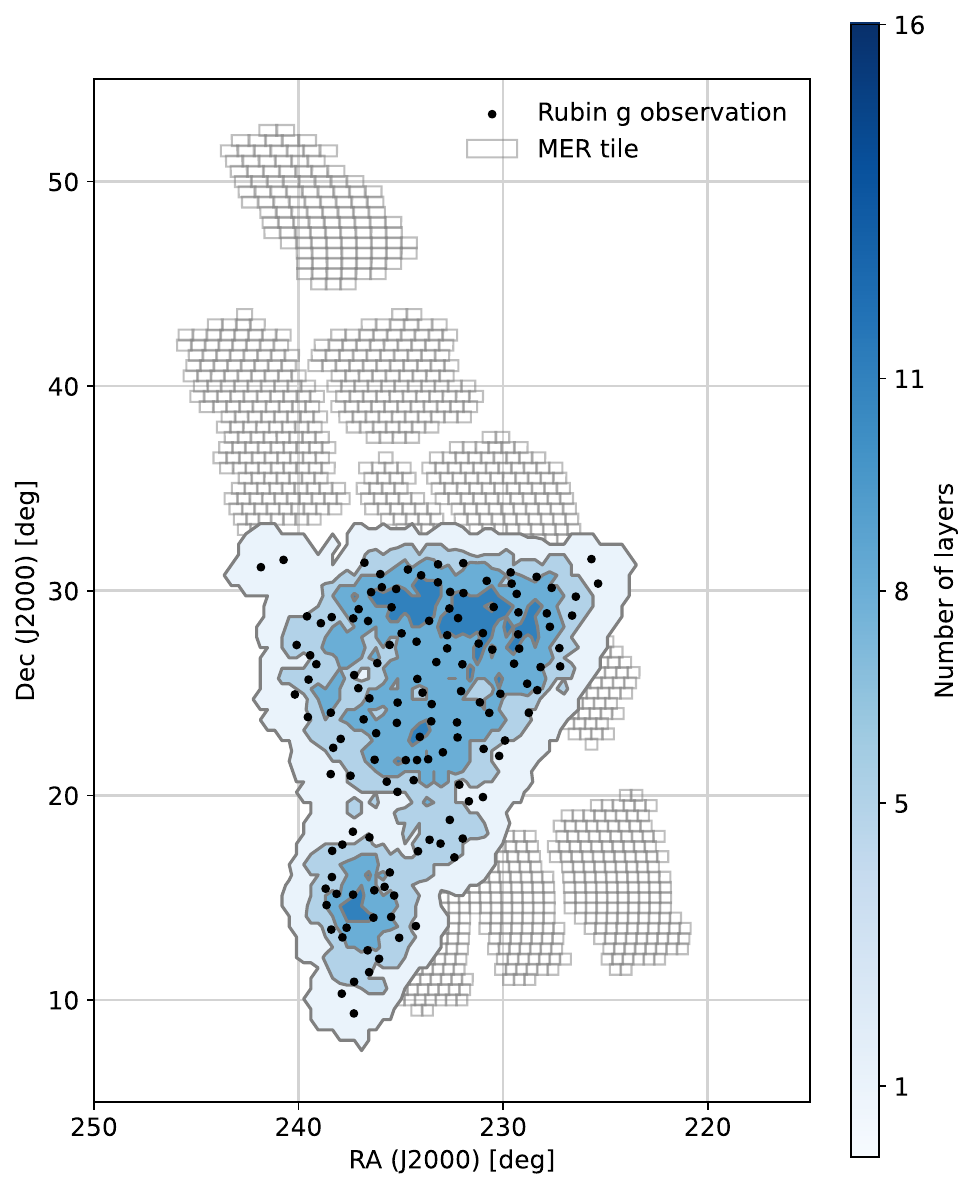}
  \caption{Positions of the exposures and the number of layers for the LSST $g$ band. The number of exposures and layers in $g$, $r$, and $i$ bands are also approximately the same.}
  \label{fig:rubin_g_footprint}
\end{figure}

\begin{figure}
  \centering
 \includegraphics[width=\linewidth]{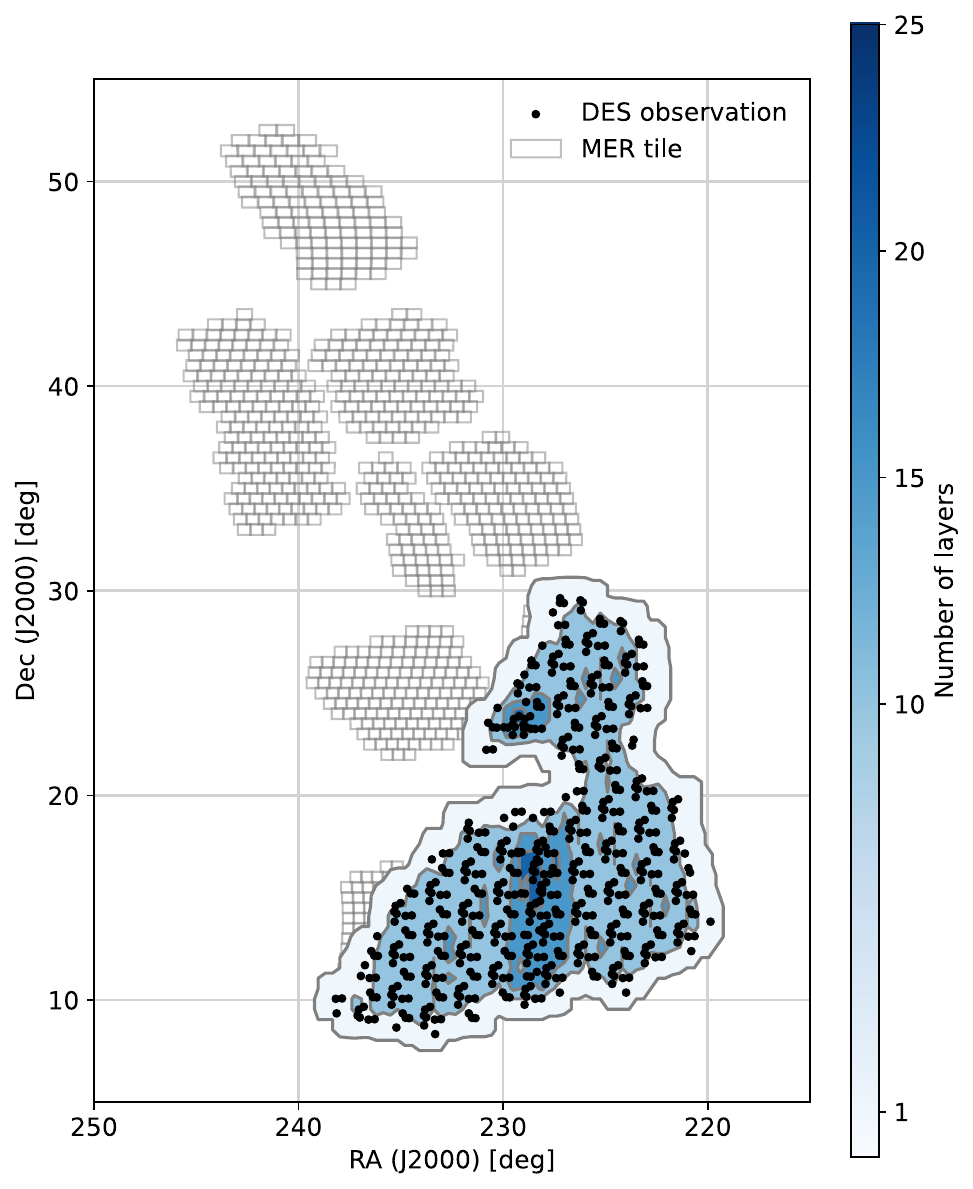}
  \caption{Observations and the number of layers in DES SC8 main survey. Exposures of different bands ($g$, $r$, $i$, and $z$) are at the same positions.}
  \label{fig:des_footprint}
\end{figure}

The SC8 main area was reduced from a continuous region measuring $600\,\rm{deg}^2$ to 11 patches with a total area of approximately $165\,\rm{deg}^2$. Figure~\ref{fig:euclid_footprint} shows the distribution of \Euclid observations among 11 discontiguous regions. For the EXT surveys, we subdivided the original SC8 main area into three non-overlapping patches. The area above $30^{\degree}$ in Declination was covered by the Northern Surveys in certain broadbands: CFIS ($u$ and $r$ bands), JEDIS ($g$ band), Pan-STARRS ($i$ band), and WISHES ($z$ band). The rectangular region below $30^{\degree}$ in Declination was further sub-divided into two sections along a diagonal to allow a range of sky conditions in both segments. The area above the diagonal contained the set of \textit{Rubin} pointings for each broadband filter (e.g. $u$, $g$, $r$, $i$, and $z$ bands), whereas the region below the diagonal contained only DES pointings for the bands ($g$, $r$, $i$, and $z$ bands). OU-EXT provided a set of files containing the respective observing cadence strategies for each survey from which we selected exposures that intersect the SC8 main area for processing.
The preparation of each survey file is described below and their corresponding file contents are summarised in Table \ref{table:ext_surveys}:

\begin{itemize}

    \item \textbf{CFIS} The pointings from the available CFIS data for SC8 preparation did not cover the whole SC8 area of the sky. Therefore, the pointing pattern from another area of the sky with similar declination was translated to the SC8 area as shown in Figure~\ref{fig:cfis_footprint}.
    
    \item \textbf{JEDIS} By the start of SC8, JEDIS observations had yet to be performed. The pointing pattern was generated following procedures outlined by the JEDIS team. For any given sky location, the pattern consists of 14 overlapping pointings with a dithering step of the size of the JEDIS CCD, so that each point in the sky is imaged with all the 14 CCDs of the camera. The JEDIS SC8 footprint and depth in layer count can be seen in Figure~\ref{fig:jedis_footprint}.
    
    \item \textbf{Pan-STARRS} A pre-selection of all pointings overlapping the SC8 area was performed by OU-EXT.  They reduced the original number of pointings to 6.5 times of the total in order to minimise data storage. To reach the required depths, the exposure times were increased from $240.0\,{\rm s}$ to $292.5\,{\rm s}$.  The saturation limit of the CCDs was increased accordingly in order to retain the saturation limit of the stars. Figure~\ref{fig:ps_footprint} shows the Pan-STARRS PS1 SC8 footprint and the depth indicated by the contours denoting the layer count.
    
    \item \textbf{WISHES} The survey file in the $z$ band was built from real WISHES pointings overlaying the SC8 area (as seen in Figure~\ref{fig:wishes_footprint} which includes the corresponding number of layers).

  \item \textbf{Rubin} The LSST scheduler \citep{lsst-scheduler-naghib} and associated simulated pointing histories (\texttt{OpSim} outputs) provide many examples of survey strategy options \citep{opsim-delgado, opsim-connolly}. For this work, a survey table was derived courtesy of the Dark Energy Science Collaboration (DESC) using the wide-fast-deep output from the v1.5 simulation, \texttt{footprint\_sky\_dustv1.5\_10yrs} \citep{lsst-surveysim-jones2020}, which we downsampled in order to reach the desired limiting magnitudes in $ugriz$ for the Euclid Wide Survey. We show in Figures~\ref{fig:rubin_u_footprint} and ~\ref{fig:rubin_g_footprint}, respectively, the number of layers in the $u$ and $g$ bands for the $Rubin$ SC8 main survey. Magnitude zero points estimates for each bandpass were provided by DESC.  Furthermore, the same position angle was used for all visits; therefore, we did not not account for rotation of the camera denoted by the \texttt{OpSim} parameter `rottelpos'.
  
        \item \textbf{DES} Similar to CFIS, the SC8 region is also outside of the DES footprint, observations were extracted from the same range of Declination in the southern celestial hemisphere at a shifted Right Ascension and transformed to cover the SC8 area. An additional margin of 1.05 deg, half of DECam FOV, \citep{Flaugher_2015}, was applied to the selection to ensure that all exposures overlapping the target area were included.  Fig.~\ref{fig:des_footprint} shows the resulting DES SC8 footprint and the blue contours indicate the number of layers to reach the desired depth for bands $griz$.  
  
\end{itemize}

\begin{table}[!ht]
\centering
\caption{Number of exposures selected for processing and average number of layers in SC8 main area for each of the survey-band pairs}
\begin{tabular}{|l|c|c|c|} 
\hline
Survey & Bands & \makecell{Num \\exposures} & \makecell{Num \\layers} \\
\hline
\Euclid & \makecell{\IE, \JE, \HE, \\ $\YE$, and Spectro} & 331 & 4 \\ 
\hline
DES & $g$, $r$, $i$, $z$ & 593 & 10--15 \\ \hline
\multirow{5}{2em}{LSST} & $u$ & 201 & 10--15 \\
& $g$ & 133 & 5--10 \\
& $r$ & 130 & 5--10 \\
& $i$ & 139 & 5--10 \\
& $z$ & 197 & 10--15 \\ \hline
CFIS & $u$, $r$ & 673 & ${\sim}\,4$ \\
JEDIS & $g$ & 864 & 15--20 \\
Pan-STARRS & $i$ & 1055 & 20--30 \\
WISHES & $z$ & 462 & ${\sim}\,4$ \\ \hline
\end{tabular}

\label{table:ext_surveys}
\end{table}

\end{appendix}

\end{document}